\documentclass[12pt]{article}
\usepackage{graphicx,amsmath,setspace,amssymb}
\usepackage{units}
\usepackage{color}
\usepackage{cite}
\usepackage{hyperref}
\usepackage{multirow}
\usepackage{pbox}
\usepackage{array}
\usepackage{float}
\usepackage{lineno}
\usepackage{hyperref}
\usepackage[caption=true]{subfig}
\usepackage[font=small,labelfont=bf]{caption}

\newlength{\dinwidth}
\newlength{\dinmargin}
\def\lapproxeq{\lower .7ex\hbox{$\;\stackrel{\textstyle                                                    
<}{\sim}\;$}}                                                    
\def\gapproxeq{\lower .7ex\hbox{$\;\stackrel{\textstyle                                                    
>}{\sim}\;$}}                                                    
\def\be{\begin{equation}}                                                    
\def\ee{\end{equation}}                                                    
\def\bea{\begin{eqnarray}}                      
\def\eea{\end{eqnarray}}

\def\GeV{\rm GeV}
\def\TeV{\rm TeV}

\def\sh{\hat s}
\def\sh2{{\hat s}^2}

\begin{document}



\begin{center}
{\Large \bf Parton distributions from LHC, HERA, \\ \vspace*{0.2cm} Tevatron and fixed target data:}\\ 
\vspace*{0.5cm}{\Large \bf MSHT20 PDFs}\\

\vspace*{1cm}
S. Bailey$^a$, T. Cridge$^b$, L. A. Harland-Lang$^{a}$, A. D. Martin$^c$, 
and R.S. Thorne$^b$\\                                               
\vspace*{0.5cm}                                                    
   
$^a$ Rudolf Peierls Centre, Beecroft Building, Parks Road, Oxford, OX1 3PU   \\  
$^b$ Department of Physics and Astronomy, University College London, London, WC1E 6BT, UK \\           
$^c$ Institute for Particle Physics Phenomenology, Durham University, Durham, DH1 3LE, UK                   \\                                 
                                                    

\begin{abstract} 
\noindent We present the new MSHT20 set of parton distribution functions (PDFs) of the 
proton, determined from global analyses of the available hard scattering data.
The PDFs are made available at NNLO, NLO, and LO, and supersede the MMHT14 sets. They are obtained 
using the same basic framework, but the parameterisation is now adapted and extended, 
and there are 32 pairs of eigenvector PDFs. We also include a large number of new data sets:
from the final HERA combined data on total and 
heavy flavour structure functions, to final Tevatron data, and in particular a significant number of new LHC 7 and 
8~TeV data sets on vector boson production, inclusive jets and top quark distributions. 
We include up to NNLO QCD corrections for all data sets that play a major role in the fit, and NLO EW corrections where relevant.
We find that these updates have an important impact on the PDFs,  
 and for the first time the NNLO fit is strongly favoured over the NLO, reflecting the wider range and in particular increased precision of data included in the fit.
There are some changes to central values and a significant reduction in the uncertainties 
of the PDFs in many, though not all, cases.
Nonetheless, the PDFs and the resulting predictions are generally within one standard deviation of the MMHT14 results. The major changes are the $u-d$ valence quark 
difference at small $x$, due to the improved parameterisation and new precise data, 
the $\bar d, \bar u$ difference at small $x$, due to a much improved parameterisation,  
and the strange quark PDF due to the effect of 
LHC $W,Z$ data and inclusion of new NNLO corrections for dimuon production in neutrino DIS.  We discuss the phenomenological impact of our results, and in general find reduced uncertainties in predictions for processes such as Higgs, top quark pair and $W,Z$ production at post LHC Run--II energies.

\end{abstract}                                                        
\vspace*{0.5cm}

{\it \footnotesize{We dedicate this paper to the memory of James Stirling and Dick Roberts, both of whom sadly died in the past two years.  They were founding members of the collaboration of our PDF global analyses, which started with the presentation of the first ever NLO PDF sets in 1987.}}                                       
                                                    
\end{center}

\begin{spacing}{0.8}
\clearpage

\tableofcontents
\clearpage
\end{spacing}

\section{Introduction  \label{sec:1}} 

The parton distribution functions (PDFs) of the proton are determined from 
fits to the world data on deep inelastic scattering (DIS), and more recently from the rapidly increasing 
variety of related hard scattering 
processes at hadron colliders. For examples of the most up-to-date PDFs using a variety of
approaches and more or less comprehensive choices of input data see 
\cite{MMHT14,NNPDF3.1,CT18,ABMP16,HERAcomb,CJ15}. 
More than five years have 
elapsed since MMHT published the results of the global PDF analysis entitled 
`Parton distributions in the LHC era: MMHT14 PDFs' \cite{MMHT14}.  Since then there have 
been extensive improvements in the data, in particular from the LHC, but also 
important final analyses from HERA and the Tevatron. It is therefore important to 
present a major update of the MMHT14 PDFs, to take account of both the improvements and 
extensions in data and the accompanying improvements in our theoretical framework. We denote
these new PDFs by MSHT20.  

We have assessed the nomenclature that we apply to our latest update of 
the PDFs. Historically the name of our PDFs has reflected the authorship of 
the articles. This has led to only a slow evolution in the naming, partially
due to a relatively slow change in personnel, but also partially due 
to the happy accident that some new surname initials have been the same as
previous or existing ones. However, we have now reached the stage where 
this no longer seems feasible and where the PDF  name should instead reflect
something with more permanence. However, we also want the name to remain 
familiar to previous ones and to reflect to some extent the history of the 
group. Hence we have decided to call this and future incarnations the MSHT parton 
distributions.  This can be taken to stand for ``Mass Scheme Hessian 
Tolerance''. We were the first group to obtain the PDFs via the use of a 
general mass variable flavour number scheme in \cite{MRST} and have continued 
to do this ever since. The uncertainties on the PDFs have always been derived 
and presented using the Hessian framework, though it is now known how one can 
convert to an equivalent Monte Carlo framework \cite{WattThorne}. 
In addition, we determine the size 
of these uncertainties by using a dynamic tolerance procedure to inflate the 
value of $\Delta \chi^2$ in a manner determined by the fit, rather than input 
by hand \cite{MSTW}. This reflects the strong evidence that the standard $\Delta \chi^2=1$ approach does not fully 
account for uncertainties due to tensions between different data sets, 
limitations in fixed order perturbation theory, limits in flexibility of 
input PDFs and potentially other sources. The name MSHT clearly also 
incorporates the initial of a number of the current and previous group 
members. 

There have been a number of intermediate updates between the MMHT14 PDFs and the MSHT20
PDFs that we present here. Soon after the MMHT14 PDFs appeared the HERA collaboration released its 
final combination of total cross section measurements \cite{HERAcomb}. An analysis of the 
effect of these was quickly produced \cite{MMHT2015}, and it was concluded that while the 
improvement in PDFs was significant it was hardly dramatic, and a full update was not required. 
A small amount of new LHC data was also considered in \cite{MMHTDIS2016}, and then 
increasing amounts of new data and accompanying procedural improvements appeared in 
\cite{MMHTDiff2016, MMHTKrakow,MMHTDIS2017,MMHTjets,MMHTDIS2018,Thorne:2019mpt}. 
To begin with changes and improvements in the PDFs were incremental \cite{MMHT2015,MMHTDIS2016}, 
with predictions for new data sets all good, and only relatively minor changes being required. 
However, with increasing amounts of new data some problems appeared and changes in procedure 
and more significant changes in PDFs were required. In roughly chronological 
order the most striking of these were: difficulties in fitting some jet data \cite{ATLAS7jets}, studied in \cite{MMHTKrakow,MMHTjets}, associated with correlated uncertainties (similar
issues later also appearing in differential top data \cite{Bailey:2019yze}); a degree of tension
between the strange quark required to best fit new precise ATLAS $W$, $Z$ data \cite{ATLASWZ7f}
and older dimuon structure function data studied in \cite{MMHTDIS2017,Thorne:2019mpt}; and a general
need for the PDFs to have a more flexible parameterisation, particularly for the valence quarks and 
$\bar u,\bar d$ difference in order to best fit a variety of new data \cite{Thorne:2019mpt}. 
Overall, this has resulted in a cumulative change compared to the MMHT14 PDFs mainly in the 
flavour sector, i.e. the strange quarks, down valence quark and $\bar u,\bar d$ difference, made
possible by a considerable extension of the parameterisation flexibility. 
These changes have been driven largely by new LHC precision data on processes with $W$, $Z$ bosons in the
final state, but also by the final D{\O} $W$ asymmetry data \cite{D0Wasym}. The previously 
best constrained PDFs, i.e. the up quark and the gluon distribution (particularly at low $x$), 
remain largely determined by data on structure functions and their evolution, and so the 
central values are similar to those in the MMHT14 PDFs. However, new data is playing a role and we see generally reduced uncertainties in the PDFs at both intermediate $x$ values and higher $x$. Similarly, benchmark processes also have reduced errors relative to those obtained from the MMHT14 PDFs. Nonetheless, we note that as in previous major updates, the improvement in parameterisation can mean that, despite extra data in the fit, the PDF uncertainty can increase in a few places, though this is mainly for $x$ values 
where the data constraints still remain relatively weak. 

We note that in the past we have accompanied the main PDF article with more specific studies
on the relation to the strong coupling constant \cite{MMHTas} and heavy quark masses \cite{MMHTmass}.
These dedicated studies will soon appear in relation to the MSHT20 PDFs, though we will discuss the most 
important findings in this article. We also note that we have recently released PDFs with 
electromagnetic corrections to the evolution and cross sections, and the inclusion of a photon
PDF \cite{MMHTQED}. Again, a set to accompany the MSHT20 PDFs will soon appear. 
However, in this article we focus very much on the central PDF study with each of these further 
complications and extensions left in the background for the moment.  

The outline of the paper is as follows.
In Section \ref{sec:theory} we describe the improvements that have been made 
to our theoretical procedures since the MMHT14 analysis was performed.
We discuss the considerable extension and modification of the parameterisation of the 
input PDFs, in particular the replacement of a parameterisation of $\bar d - \bar u$ with one for
$\bar d/\bar u$. We briefly discuss the treatment of 
 deuteron and nuclear corrections, of the heavy flavour PDFs and of 
 the experimental errors of the data, even though there are no major changes in these cases. We discuss in more depth 
the inclusion, for the first time, of full NNLO theory for neutrino-produced dimuon production
\cite{NNLOdimuon}, as well as the extension to full NNLO for various other data sets. Finally we briefly discuss
theoretical uncertainties, though we do not attempt to provide them in this article.   
In Section \ref{sec:preLHC}  we discuss the non-LHC data which have been added since the MMHT14 
analysis. This is in itself quite significant since it includes the final HERA analyses of both 
total \cite{HERAcomb} and heavy flavour cross sections \cite{HERAhf}, as well as
the final D{\O} asymmetry data \cite{D0Wasym}. Section \ref{sec:4} describes in detail the 
 substantial new collection of LHC data that are now 
included in the fit. This includes not only updates of the type of data we have previously considered
such as single and double differential Drell-Yan (DY) data, inclusive jet data (not previously included at 
NNLO) and inclusive $t\overline{t}$ cross sections, but also new types of data, i.e. $W+$ jets, $W+$ charm,
$Z$ boson $p_T$ distributions, and both single and double differential top quark pair production cross sections. We only include
data taken at 7 and 8 TeV, and not at 13 TeV. This is partially due to the relative lack of real 
precision data constraining PDFs so far at this energy, but also to allow for predictions made at 13~TeV 
using the PDFs to be completely uncontaminated by data at this energy. 

\begin{figure} 
\begin{center}
\includegraphics[scale=0.9]{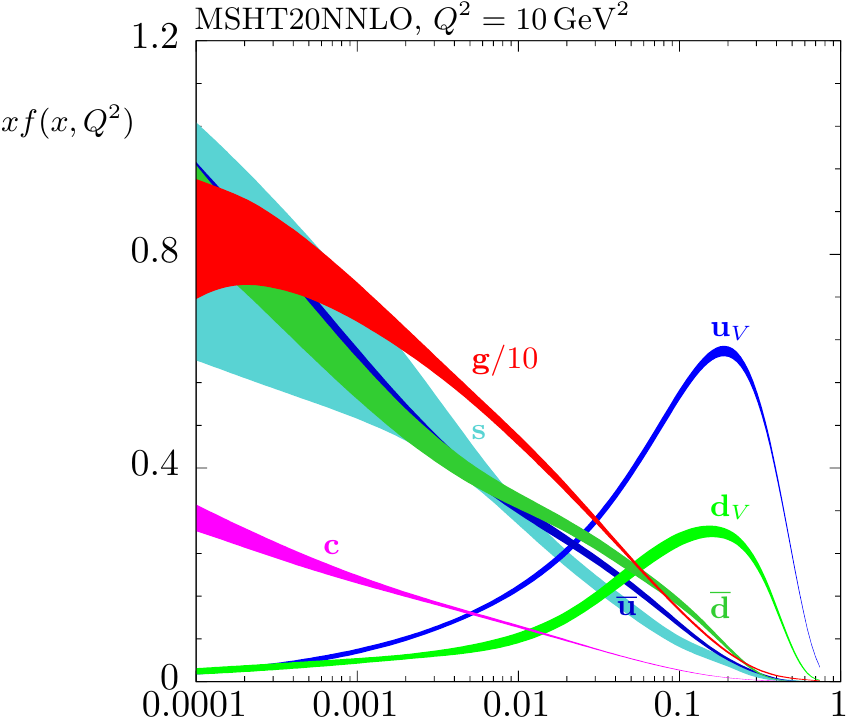}
\includegraphics[scale=0.9]{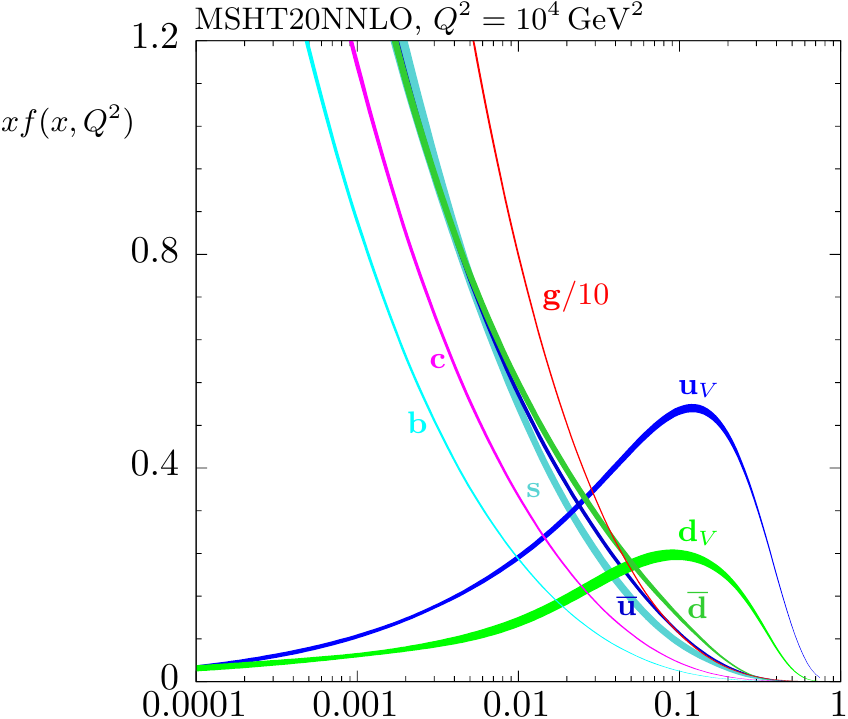}
\caption{\sf MSHT20 NNLO PDFs at $Q^2=10~\GeV^2$ and $Q^2=10^4~\GeV^2$, with associated 68$\%$ confidence-level uncertainty bands.}
\label{fig:NNLOpdfs}
\end{center}
\end{figure}

\begin{figure} 
\begin{center}
\includegraphics[scale=0.9]{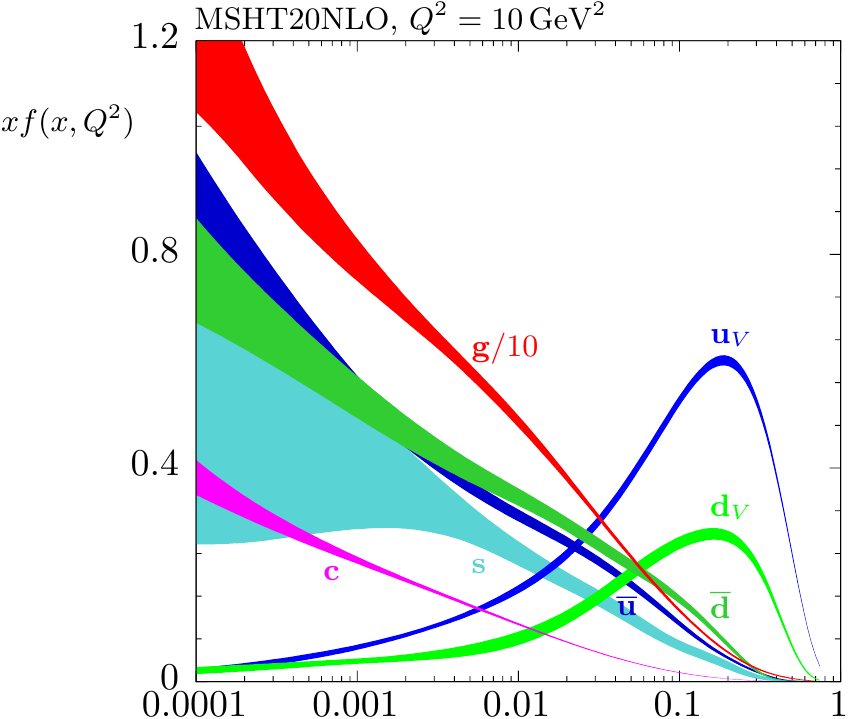}
\includegraphics[scale=0.9]{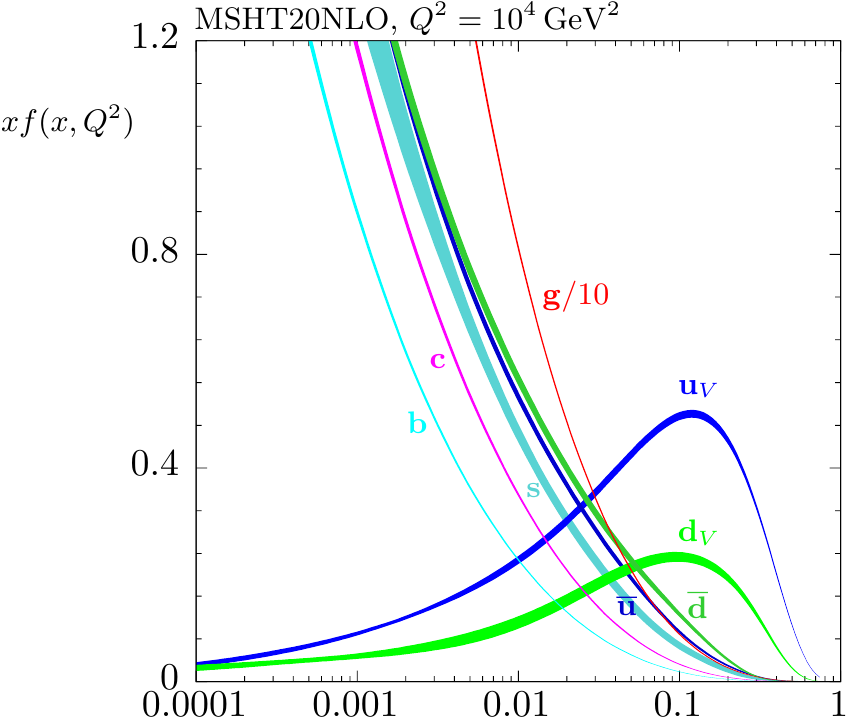}
\caption{\sf As in Fig.~\ref{fig:NNLOpdfs}, but at NLO.}
\label{fig:NLOpdfs}
\end{center}
\end{figure}

The results of the global analysis can be found in Section \ref{sec:5}. This section starts with a 
discussion of the treatment of the QCD coupling, which is in principle treated as a free parameter in our fit. At NNLO the preferred value of $\alpha_S(M_Z^2)$ is very close to 
the default value of $\alpha_S(M_Z^2)=0.118$ and to the world average value. At NLO a value of  $\alpha_S(M_Z^2)=0.120$ is preferred, but we also make available a set at $\alpha_S(M_Z^2)=0.118$.
The fit quality is presented, and we find that for the first time the NNLO fit is of far better quality than that at NLO. This was not evident in previous fits, which were dominated by structure function data, and is driven by the abundance of high precision LHC data in the current fit. 
The quality of the fit to the data at LO is enormously worse than that at NLO and NNLO. 
We release this set for completeness, and potential use in LO Monte Carlo generators, if required. We then present the NNLO, NLO and LO 
PDFs and their uncertainties, together with the 
values of the input parameters (except for LO).  These sets of PDFs are the ultimate products of the analysis -- 
the grids and interpolation code for the PDFs can be found at 
\cite{UCLsite} and will be available at \cite{LHAPDF}. 
A summary of the PDFs appears in Fig. \ref{fig:NNLOpdfs} and Fig. \ref{fig:NLOpdfs} which 
respectively show the NNLO and NLO PDFs at scales 
of $Q^2=10~\GeV^2$ and $Q^2=10^4~\GeV^2$, including the associated one-sigma (68$\%$) confidence-level 
uncertainty bands.

In Section \ref{sec:6} we compare the MSHT20 PDFs with those of 
MMHT14 \cite{MMHT14} at NNLO and NLO. In Section \ref{sec:7} we compare the MSHT20 PDFs at NLO and NNLO,
with the results highlighting that these are intrinsically different quantities. We also concentrate on 
those PDFs that have changed most, and the reasons for the changes. We end this section by briefly presenting the LO PDFs. In Section \ref{sec:8} this is examined both in terms of 
the change in procedures, i.e. parameterisation and/or improved accuracy and precision in calculations, 
and in terms of the impact of certain data. Both these issues lead us also to consider the degree of 
tension between data sets in the fit, highlighting which data sets provide the  greatest tension. 

In Section \ref{sec:9} we make predictions for various benchmark processes at the LHC (and Tevatron),
focussing on the standard candles of $W$, $Z$, Higgs boson and $t\overline{t}$ production. In general a good reduction in the PDF errors for these processes is observed in comparison to the previous MMHT14 release, with the central values remaining relatively stable and within uncertainties.

In Section \ref{sec:10} we discuss a selection of other data sets that are available at the LHC which constrain 
the PDFs, but that are not included in the present global fit. In particular we consider:
CMS 13~TeV data on $W+c$ production \cite{CMSWc13}, which tests predictions particularly dependent on 
the strange quark; the ratios of $Z$ and $t\bar{t}$ cross sections at 8~TeV and 13~TeV at ATLAS~\cite{Aad:2019hzw}; the CMS measurements of single-top production~\cite{Sirunyan:2019hqb,Sirunyan:2018rlu}; the potential impact of LHCb exclusive $J/\psi$ production data~\cite{Aaij:2014iea,Aaij:2018arx}, as accounted for in the analysis of~\cite{Flett:2020duk}, and LHCb data on $D$ meson production~\cite{Aaij:2016jht,Aaij:2014iea,Aaij:2015bpa}, as accounted for in the analysis of~\cite{Bertone:2018dse}.
In Section \ref{sec:11} we compare our MSHT PDFs with those of the other most recent global analyses 
of PDFs -- NNPDF3.1~\cite{NNPDF3.1} and CT18~\cite{CT18}, and also with older sets of PDFs 
of other collaborations. In Section \ref{sec:access} we summarise the availability of the MSHT20 PDF sets and their delivery. In Section \ref{sec:12} we present our conclusions.

\section{Changes in the theoretical procedures \label{sec:theory}}

As in the case of MMHT14, we present PDF sets at LO, NLO and NNLO in $\alpha_S$. In the latter case we use the splitting functions calculated in~\cite{Moch:2004pa,Vogt:2004mw} and for structure function data, the massless coefficient functions calculated in~\cite{vanNeerven:1991nn,Zijlstra:1991qc,Zijlstra:1992kj,Zijlstra:1992qd,Moch:2004xu,Vermaseren:2005qc}. There are however, a significant number of changes in our theoretical 
description of the 
data, compared to that used in the MMHT14 analysis.  We present these in this section, and when appropriate we also 
mention some of the main effects on the PDFs resulting from these improvements. 

\subsection{Input distributions}\label{sec:inputPDF}

In MMHT14 we began to use parameterisations for the input distributions based on Chebyshev polynomials. Following the detailed study in \cite{MMSTWW}, we take for most PDFs a parameterisation of the form
\be
xf(x,Q_0^2)~=~A(1-x)^\eta x^\delta \left( 1+\sum^n_{i=1} a_i T^{\rm Ch}_i(y(x)) \right),
\label{eq:1}
\ee
where $Q_0^2=1~\GeV^2$ is the input scale, and $T^{\rm Ch}_i(y)$ are Chebyshev polynomials in $y$, with $y=1-2x^k$, where we take  $k=0.5$. 

In the MMHT14 study we took $n=4$ in general, though used a slightly 
different parameterisation for the gluon and used more limited 
parameterisations for $\bar d - \bar u$ and $s- \bar s$ (`$s_{-}$'), since these were less 
well constrained  by data, whilst for similar reasons two of the $s+ \bar s$ (`$s_{+}$') Chebyshevs and its low $x$ power were tied to those of the light sea, $S(x)= 2(\bar u(x)+\bar d(x)) +s(x)+\bar s(x)$. However, with the substantial increase in the amount of LHC and other data included in MSHT20, we can now extend the parameterisation of the PDFs significantly. We therefore take $n=6$ by default in MSHT20, allowing a fit of better than 1\% precision over the vast majority of the $x$ range \cite{MMSTWW}. The MSHT20 set of input distributions are now\footnote{As is usual in PDF definitions, there is an implicit $x$ preceding the input distributions in their definitions in equations~(\ref{eq:uv})-(\ref{eq:dbaroverubar}), so that they are in reality like the left-hand side of (\ref{eq:1}), this also applies to figures and other uses throughout the rest of the paper.}:
\begin{equation} \label{eq:uv}
u_V (x,Q_0^2) = A_u (1-x)^{\eta_u}x^{\delta_u}\left(1+\sum_{i=1}^{6}a_{u,i}T_i(y(x))\right)
\end{equation}
\begin{equation} \label{eq:dv}
d_V (x,Q_0^2) = A_d (1-x)^{\eta_d}x^{\delta_d}\left(1+\sum_{i=1}^{6}a_{d,i}T_i(y(x))\right)
\end{equation} 
\begin{equation} \label{eq:Sea}
S (x,Q_0^2) = A_S (1-x)^{\eta_S}x^{\delta_S}\left(1+\sum_{i=1}^{6}a_{S,i}T_i(y(x))\right)
\end{equation}
\begin{equation} \label{eq:s+}
s_+ (x,Q_0^2) = A_{s_+} (1-x)^{\eta_{s+}}x^{\delta_{S}}\left(1+\sum_{i=1}^{6}a_{{s_+},i}T_i(y(x))\right)
\end{equation}
\begin{equation} \label{eq:g}
g (x,Q_0^2) = A_{g} (1-x)^{\eta_{g}}x^{\delta_{g}}\left(1+\sum_{i=1}^{4}a_{{g},i}T_i(y(x))\right) + A_{g-}(1-x)^{\eta_{g-}}x^{\delta_{g-}}
\end{equation}
\begin{equation} \label{eq:s-}
s_- (x,Q_0^2) = A_{s_-} (1-x)^{\eta_{s-}} (1-x/x_0)x^{\delta_{s-}}
\end{equation}
\begin{equation} \label{eq:dbaroverubar}
(\bar{d}/\bar{u})(x,Q_0^2) = A_{\rho}(1-x)^{\eta_{\rho}}\left(1+\sum_{i=1}^{6}a_{{\rho},i}T_i(y(x))\right)
\end{equation}
The departures from the general form in~\eqref{eq:1} with $n=6$ come, as before, in the gluon, where $n=4$ but the additional term proportional to $A_{g-}$ includes 3 additional parameters and allows for a better fit to the small-$x$ and $Q^2$ HERA data, as first shown in \cite{MRSTfirstNNLO}. For $s_+$ there are now 6 Chebyshev polynomials used and, whilst the high $x$ power is separate from the sea, the low $x$ power remains set to the same value as the sea, $\delta_S$. Meanwhile, there is still insufficient data to allow an extended parameterisation of the strangeness asymmetry, $s_-$, so its form remains that used in MMHT14, with $x_0$ giving a switch between positive and negative values. 

Finally, the major change in the PDF parameterisation comes in the first generation antiquark asymmetry. With MSHT20 we make the decision to now parameterise the ratio $\rho = \bar{d}/\bar{u}$ rather than the difference ($\bar{d}-\bar{u}$) and we allow 6 Chebyshev polynomials for this ratio. There is also no low $x$ power for this ratio as we assume it must tend to a constant as $x \to 0$. This allows for an improved central fit, whilst also giving a better description of the error bands on the asymmetry in the very low $x$ region, as illustrated later in Fig.~\ref{dbarminusubar} (left). 

An analysis of the effects of these changes on the global fit was performed. The main improvements come from the extension of the $\bar{d}/\bar{u}$ to 6 Chebyshev polynomials, which enabled an improvement in the global chi-squared of $-\Delta\chi^2_{tot} \approx 20$. Additionally extending the down valence enabled the cumulative global chi-squared improvement to be $-\Delta\chi^2_{tot} \approx 35$, the gluon extension moves this to $-\Delta\chi^2_{tot} \approx 50$, while finally the changes to the sea ($S$) and $s_+$ result in the total improvement of $-\Delta\chi^2_{tot} \approx 75$. More detail on each of the PDF distributions, and on the improvements due to the changes in parameterisation, will be given later in Sections~\ref{centralpdfs} and \ref{newparameterisationeffects}. 

Overall, these changes in the input distribution represent an increase of 2 parameters for each of the $u_V$, $d_V$, $S$, $g$, with an additional 4 parameters in the $\bar{d}/\bar{u}$ relative to the previous asymmetry ($\eta_{\rho}$ is free whilst $\eta_{\Delta} = \eta_S + 2$ in MMHT14), 4 further parameters in $s_+$ and no change in the $s_-$. With the usual constraints on the integral of the valence quark distributions, the conservation of total momentum, and the integral of the strangeness asymmetry ($s_-$) set to 0, we now have a total 52 parton parameters to fit, with the strong coupling $\alpha_S(M_Z^2)$ also allowed to be free when the best fit is obtained. A subset of these parameters are then formed into a set of 32 eigenvectors (64 eigenvector directions) in the determination of the PDF uncertainty bands, as described later in Section~\ref{centralpdfs}.

\subsection{Deuteron and heavy nuclei corrections   \label{sec:2.2}}

The increase and improvement in data from the LHC is not yet such that we are 
able to remove the constraints obtained from deep inelastic data using deuteron \cite{BCDMS,NMC,NMCn/p,E665,SLAC,SLAC1990} and heavier nuclei \cite{NuTeV,CHORUS,Dimuon} as targets. 
The former are still required to fully separate the $u$ and $d$ distributions at moderate and high 
values of $x$. The latter, being obtained via charged-current scattering,
provide complementary constraints on flavour decomposition, and in particular
on the strange quark distribution in the case of dimuon final states.   

Hence, we still consider the correction factor $c(x)$  
applied to the deuteron data
\be
F^d(x,Q^2)~=~c(x) \left[F^p(x,Q^2)+F^n(x,Q^2) \right]/2,
\ee
where we assume $c$ is independent of $Q^2$, and where $F^n$ is obtained 
from $F^p$ by swapping up and down quarks, and antiquarks; that is, isospin 
asymmetry is assumed\footnote{This is no longer assumed in QED partons
\cite{MMHTQED}.}.

In \cite{MMSTWW} we studied the deuteron correction factor in detail.  
We introduced the following flexible parameterisation of $c(x)$, which followed 
 the theoretical expectations of shadowing while allowing the 
precise deuteron correction factor to be determined by the data:
\bea
c(x)~&=&~(1+0.01N)~[1+0.01c_1 {\rm ln}^2(x_p/x)], ~~~~~~~~~~~~~~~~~~~~~~~~~~~~   x<x_p,\\
c(x)~&=&~(1+0.01N)~[1+0.01c_2 {\rm ln}^2(x/x_p)+0.01c_3{\rm ln}^{20}(x/x_p)], ~~~   x>x_p,
\eea
where $x_p$ is a `pivot point' at which the normalisation is $(1+0.01N)$. 

In practice, $x_p$ is chosen to be equal 
to $0.05$ at NLO, but a slightly smaller value of $x_p=0.03$ is marginally 
preferred at NNLO. We use the same parameterisation again in this study, and 
the values of the parameters are shown in Table \ref{tab:1}.  

\begin{table} 
\begin{center}
\begin{tabular}{|l|c|c|c|c|}\hline
 PDF fit &   $N$~~~ & $c_1$ &  $c_2$ & $c_3\times 10^8 $ \\ \hline
  MSHT20 NLO  & $-0.080 \pm 0.276$  & $-0.467 \pm 0.212$ & $-0.473 \pm 0.089$ & $3.79 \pm 0.50$\\
  MSHT20 NNLO  & $0.656 \pm 0.305$ & $0.102 \pm 0.400$ & $-0.343 \pm 0.0597$ & $0.0900\pm 0.0157$\\
\hline
\end{tabular}
\end{center}
\caption{\sf The values of the parameters for the deuteron correction factor  found in the present global fits.}
\label{tab:1}
\end{table}

The correlation matrices for the 
4 deuteron parameters $N, c_1, c_2,c_3$ for the NLO and NNLO analyses are, respectively, 
\be
c_{ij}^{\rm NLO} = 
\begin{pmatrix}  
1.000  &  0.103 &  -0.594 &  0.103 \\
0.103 &  1.000 &  0.236 & -0.084 \\
-0.594 &  0.236 &  1.000 & -0.353 \\
 0.123 & -0.084 & -0.353 &  1.000 \\
\end{pmatrix},
\ee

\be
c_{ij}^{\rm NNLO} = 
\begin{pmatrix}  
1.000  &  0.075 & -0.687 &  0.133 \\
0.075 &  1.000 &  0.175 & -0.055 \\
-0.687 &  0.175 &  1.000 & -0.351 \\
 0.133 & -0.055 & -0.351 &  1.000 \\
\end{pmatrix}.
\ee 
In the MMHT analysis \cite{MSTW} we applied the nuclear corrections $R_f$, 
defined as
\be
f^A(x,Q^2)~=~R_f(x,Q^2,A)~f(x,Q^2).
\ee
The $f^A$ are defined to be the PDFs of a proton bound in a nucleus of mass number $A$. 
In the present analysis we use the results of de Florian et al., 
which are shown in Fig.~14 of \cite{deF}. 
We multiply the nuclear corrections by a $3$-parameter 
modification function, Eq.~(73) in \cite{MSTW}, which allows a penalty-free
change in the details of the normalisation and shape. As in \cite{MSTW}, the 
free parameters choose values such that they prefer modifications of 
only a couple of percent at most away from the default values.

\subsection{General Mass - Variable Flavour Number Scheme (GM-VFNS)}

As always we employ a general mass variable flavour number scheme, and 
continue to  use the `optimal' scheme \cite{Thorne}, based on 
\cite{TR,TR1}, which improves 
the smoothness of the transition region 
where the number of active flavours is increased by one. Note that at NNLO
this still requires some degree of approximation in the vicinity of 
$Q^2 \sim m_h^2$ as the ${\cal O}(\alpha_S^3)$ heavy flavour coefficient
functions in the fixed flavour number scheme (FFNS) are still not known exactly, though the leading small-$x$ term~\cite{Catani:1990eg} and threshold logarithms~\cite{Laenen:1998kp,Kawamura:2012cr} have been calculated.

We use as default the quark masses $m_c=1.4~\GeV$ and $m_b=4.75~\GeV$, where both are defined as the pole mass. 
These are the same values as for MMHT14. 
As with the MMHT14 PDFs \cite{MMHTmass} we will make PDF sets available with varying masses, and will also present a study of the variation 
of the fit quality and PDFs with varying mass in a future publication. However, as a summary we note that the best global fits are achieved with 
values very slightly lower than these values, and the default values are chosen as a compromise between the most likely pole mass values of 
$m_c\sim1.5~\GeV$ and $m_b\sim 4.9~\GeV$ obtained by conversion from the better known $\overline{\rm MS}$ values and the best fit values. However, there is now distinctly less tension between the two 
than for MMHT14, with the $\chi^2$ values for masses lower than the default being only very marginally better. 

\subsection{Treatment of uncertainties}

All data sets which are common to the MMHT14 and the present analysis
are treated in the same manner in both, except that the shift corresponding 
to the luminosity 
uncertainty in each data set is now determined analytically rather than 
via numerical $\chi^2$ minimisation. This results in only minuscule changes.   

If only the final covariance matrix for a given set of data is provided, then 
we use the expression 
\be 
\chi^2 = \sum_{i=1}^{N_{\rm pts}} \sum_{j=1}^{N_{\rm pts}}
(D_i-T_i) (C^{-1})_{ij} (D_j-T_j),
\ee
where $D_i$ are the data 
values, $T_i$ are the parameterised 
predictions, and $C_{ij}$ is the covariance matrix.

In the case where the $N_{\rm corr}$ individual sources of correlated errors are provided
the goodness-of-fit, $\chi^2$, including the full correlated error 
information, is defined as 
\be
\chi^2=\sum_{i=1}^{N_{\rm pts}}\left(\frac{D_i+\sum_{k=1}^{N_{\rm corr}}
r_k\sigma_{k,i}^{\rm corr}-T_i}{\sigma_i^{\rm uncorr}}\right)^2+\sum_{k=1}^{N_{\rm corr}}r_k^2,
\ee 
where $D_i+\sum_{k=1}^{N_{\rm corr}}r_k\sigma_{k,i}^{\rm corr}$ are the data 
values allowed to shift by some multiple $r_k$ of the systematic error
$\sigma_{k,i}^{\rm corr}$ in order to give the best 
fit, and where $T_i$ are the parameterised predictions. 

The last term on the right is the penalty for the shifts of 
data relative to theory for each source of correlated uncertainty. 
The errors are combined multiplicatively, that is 
$\sigma_{k,i}^{\rm corr}= \beta_{k,i}^{\rm corr}T_i$,  where
$\beta_{k,i}^{\rm corr}$ are the percentage errors. 

In some cases we have found that the fit is very poor unless one relaxes some
of the correlation between uncertainties. We only do this where some 
over-estimation of the correlation is deemed likely, or at least possible, 
and where relaxation 
seems justified. We will discuss individual cases later in the article. We note, however, that this approach is only possible if the full breakdown of correlated errors is provided, rather than simply the covariance matrix. As such we would strongly recommend the former information is provided in experimental analyses relevant for PDF fits.

\subsection{Fit to dimuon data} \label{sec:dimuonfit}

Information on the $s$ and $\bar{s}$ quark distributions comes from dimuon 
production in $\nu_\mu N$ and ${\bar \nu}_\mu N$ scattering \cite{Dimuon},
where (up to Cabibbo mixing) an incoming muon (anti)neutrino scatters off a
(anti)strange quark to produce a charm quark, which is detected via the decay
of a charmed meson into a muon. 
Until recently the massive cross section for this process was only available at NLO. The calculation has now been extended to NNLO in
the FFNS \cite{NNLOdimuon}. 

\begin{figure}
\begin{center}
\includegraphics[scale=0.7]{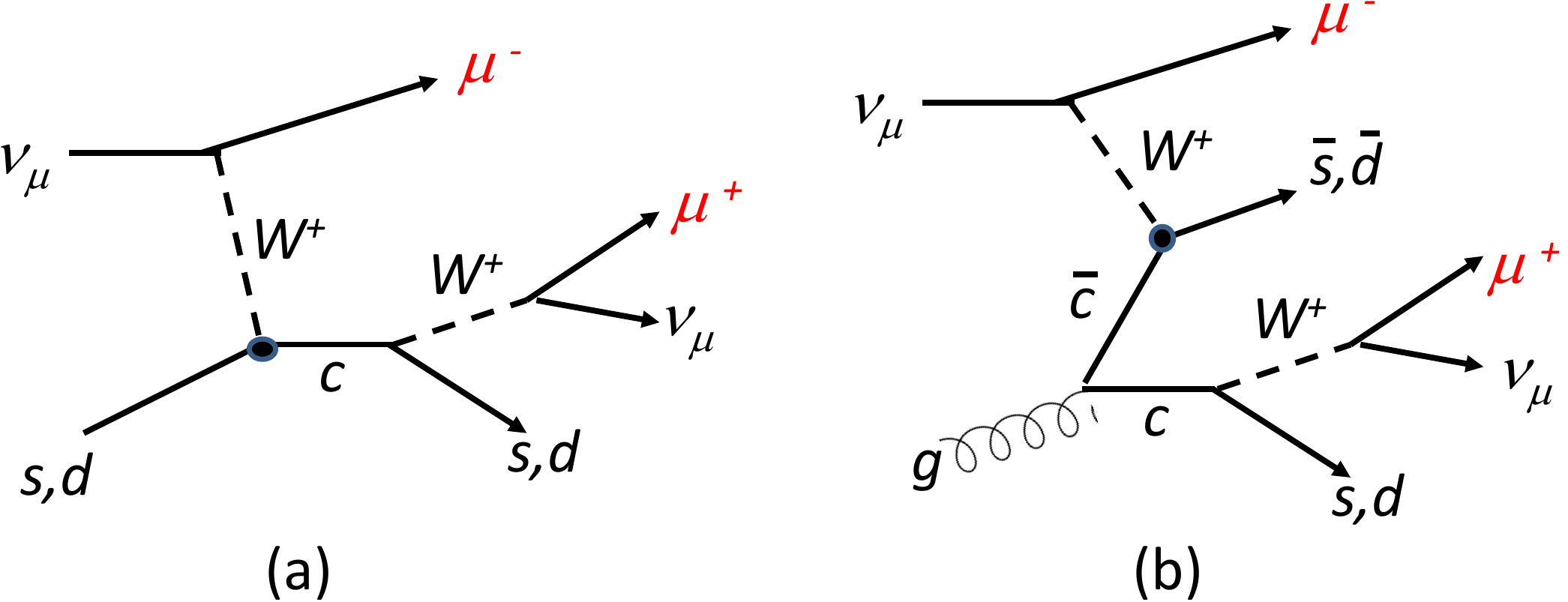}
\caption{\sf Diagrams for (a) light quark and (b) charm--initiated dimuon production in $\nu_\mu N$ scattering.}
\label{fig:dimuon}
\end{center}
\end{figure}

These results were applied within the framework of our GM-VFNS to make
the first fully NNLO analysis of the dimuon data in a global PDF fit in~\cite{Thorne:2019mpt} (see also~\cite{Faura:2020oom} for a subsequent study by the NNPDF group).
For the NNLO FFNS contribution, we simply make use of the publicly available implementation provided in~\cite{NNLOdimuon}. For the variable flavour number scheme (VFNS) corrections, at NNLO we have charm/bottom--initiated contributions to topology (a) via the $c,b \to g \to s,d$ splitting, though the effect of this is very small. Indeed, the principle source of correction here is in fact not as in this standard topology, but rather the charm quark--initiated topology shown in Fig.~\ref{fig:dimuon} (b), where in both cases `quark' is taken to indicate either quark or antiquark, depending on whether the beam is $\nu$ or $\bar{\nu}$. As discussed in~\cite{MMHT14}, this latter class of diagrams, where the charm quark is produced away from the interaction point, is not subtracted in the corresponding acceptance corrections for the dimuon data, and hence should be included. From the point of view of VFNS contributions, along with the FFNS gluon--initiated diagram shown in Fig.~\ref{fig:dimuon} (b), which enters at NLO, we also include the corresponding anticharm--initiated diagram, which enters at LO, along with the appropriate subtraction terms. Although the corresponding Feynman diagrams in this case do not contain an explicit charm quark in the final state, this is always there implicitly, as the initiating anticharm is produced via a $g\to c\bar{c}$ splitting; the effect of this, and other splittings, being resummed in the DGLAP evolution of the anticharm PDF. Moreover, for the fixed--target process under consideration, we can expect there to still be reasonable acceptance for the muon from the charmed hadron decay even when this gluon splitting occurs relatively collinearly with respect to the scale of the hard scatter. Thus, we can reasonably expect the inclusion of these contributions to give a more accurate prediction for this class of diagrams. We will however investigate the impact of not including these diagrams in Section~\ref{NNLOdimuoneffects}, where we find that their impact is small.

When accounting for the above VFNS corrections, a minor error in the previous implementations was noticed, which has now been corrected. In particular, the charm/anticharm initiated contributions were in fact included according to the ACOT, rather than TR' scheme. This results in a different form of the structure functions, most notably the charm-initiated contribution to  $F_L$, which is no longer zero at LO. As well as being inconsistent with the VFNS applied for other DIS processes in the fit, the TR' scheme was applied to calculate the subtraction terms, leading to an incorrect cancellation between the corresponding terms and even a very mild discontinuity in the structure functions. This has been corrected, with the charm-initiated diagrams calculated in the appropriate TR' scheme. 
 
Finally, we note that though we now include the corresponding theory at NNLO, strictly speaking the data are extracted using acceptance corrections derived from a NLO Monte Carlo generator~\cite{Kretzer:2001tc}. In principle, one might argue that a re--analysis of the data is in order, with the acceptance corrections evaluated using NNLO theory, and indeed in~\cite{Hou:2019efy} the NLO precision of the acceptance corrections is taken as an argument for not including the NNLO theory predictions. However, we would argue that the size of such effects should in general be less significant than the impact of using NNLO rather than NLO theory for the PDF fit itself, and hence taking NNLO theory is more appropriate. Moreover, this situation is of course commonplace in e.g. LHC analyses, where the Monte Carlo (MC) generators used for the data unfolding do not generally match the NNLO precision of the theory one might use in a PDF fit. 

In general, the NNLO correction is small at high $x$, but at the lowest $x$ values for
the dimuon data, i.e. $x \sim 0.01$, the corrections are about $10\%$ and negative. This implies that a larger 
strange quark (and antiquark) cross section will be required to fit dimuon data at small $x$, and that the 
NNLO corrections may help relieve tension between the strange quark preferred by dimuon data and the fit to $W$, $Z$ data 
at the LHC, the latter preferring a higher strange quark. We will comment on this in detail in Section~\ref{NNLOdimuoneffects}.

\subsection{Collider data: theory updates}\label{sec:theoryupdate}

Previously we used either threshold improvements to the full NNLO
jet cross sections in the case of Tevatron data, where the kinematics are not 
too far from threshold, or in the case of LHC jet data, we only included them 
in the global fit at NLO. Now the full NNLO calculations to inclusive jet data 
are in principle available \cite{NNLOjets}. In practice these are still time consuming to 
produce and have been provided on a case-by-case basis. We apply the NNLO corrections to all
LHC jet data included in the fit via $K$-factors. 
In contrast to most other studies we do not use point-by-point $K$-factors, with the MC uncertainty accounted for by an uncorrelated error for each data point, judging that this is likely to 
overestimate the real theoretical freedom. Given that the true $K$-factors are smooth functions of the jet 
transverse momentum, $p_T$,
for each rapidity bin for a given set of data we fit the point-by-point $K$-factors to a smooth functional form,
e.g. with a 4-parameter smooth cubic fit performed to the calculated $K$-factors and 4 corresponding systematic uncertainties on the parameters of this fit then included (with correlations taken into account). Examples can be 
found in \cite{MMHTjets}. For the Tevatron jet data we still use the threshold approximations for the NNLO corrections,
as these data now carry relatively little weight in the fit, and the lower centre-of-mass energy at the Tevatron means that
the high-$p_T$ jet data is overall much nearer to threshold than the LHC data. 

We also use full NNLO corrections for collider data with final state electroweak bosons. As with jet data 
these are applied using smooth $K$-factors. In the NNLO fit we also apply electroweak corrections if these 
are at all significant. We do not 
include the photon explicitly as a parton, i.e. we only include QCD evolution effects. The MSHT20
PDFs will soon be followed by an accompanying set with full QED corrections, based on the procedure 
outlined in \cite{MMHTQED}, i.e. determining the input photon distribution using the approach developed in 
\cite{LUXQED1,LUXQED2}.  On the other hand, for the small number of processes where photon--initiated (PI) production may be important,  and it is not already subtracted from the data, we do include this contribution. The predictions are all provided using the structure function approach described in~\cite{Harland-Lang:2019eai}, which provides a direct high precision calculation of PI production in hadronic collisions.

Currently, the cases where PI production may be relevant correspond to certain Drell--Yan data sets which extend below or above the $Z$ peak region.  For the ATLAS 7~TeV  precision $W$, $Z$ boson production data~\cite{ATLASWZ7f} the photon--initiated contribution is already subtracted from the measurement, and hence we do not include any correction in our fit. We do include corrections for the CMS 7~TeV double differential Drell--Yan~\cite{CMS-ddDY} and ATLAS 8~TeV high--mass DY~\cite{ATLASHMDY8}  and 8~TeV DY~\cite{ATLAS8Z3D} cross sections. These then in effect correspond to correcting the data back to a `QCD--only' cross section, which we would argue is a more reasonable baseline to compare against in a fit excluding QED effects. The impact of these is relatively mild but not completely negligible: for example in the NNLO fit -- with $\alpha_S$ free -- we find that the inclusion of PI corrections leads to somewhat larger than a $\sim 0.1$ per point improvement in the fit quality to the 8~TeV DY data, and a little less than a $\sim 0.1$ per point deterioration in the fit quality to the CMS 7~TeV data, with the high mass DY remaining rather stable. Finally we note that in principle, as discussed in~\cite{Harland-Lang:2019eai}, PI corrections may also play a non--negligible role in measurements of lepton pair $p_\perp$ distribution, as in~\cite{ATLASZpT}. However a full calculation of all of the corresponding diagrams in this case is still in progress, and hence we do not currently include these corrections.

The data on inclusive top quark pair production is calculated at NNLO using the 
code from \cite{NNLOtop}. All $t\overline{t}$ differential cross sections are also 
calculated at NNLO using either the results of \cite{NNLOtopdiff} and the 
grids created by the procedure described in \cite{TopfastNLO}, or the grids and results presented in \cite{CzakonCMS8ttDDtables}. For single 
differential distributions we also include electroweak corrections \cite{NNLOtopEW}.

\subsection{Theoretical uncertainties}\label{sec:theoryunc}

We do not consider these in detail in this article, though this will be addressed in a future publication. One example of a study using scale variations can be found in 
\cite{NNPDFscales}, though we have highlighted the delicacy of the 
relationship between scale variations in fits and predictions in
\cite{HLTscales}. 
When there is considerable scale dependence of the theoretical 
prediction, even at NNLO, then we investigate a range of scales, and generally choose 
as default one that corresponds to better fit quality. We will mention the 
cases where we do this, and also refer to potential sensitivity, when 
discussing fits to particular data sets. However, we note that in the 
majority of cases  the sensitivity of the fit quality and the resulting PDFs 
to scale choice is less than the effect of the correlated uncertainties on 
the data, and in some cases, less than the ambiguity in exactly how one might 
treat the details of the correlated systematics. 

\section{Non-LHC data included since  MMHT14  \label{sec:preLHC}}

In this section we list the changes and additions to the non-LHC data sets  in the 
present analysis. 
All the data sets included in the MMHT14 analysis are still included, 
unless the update is explicitly mentioned below. We continue to use the 
same cuts on structure function data, i.e. $Q^2>2~\GeV^2$ and $W^2>15~\GeV^2$ 
and we imposed a stronger $W^2>25~\GeV^2$ and $Q^2> 5~\GeV^2$ cut on 
$F_3(x,Q^2)$ structure 
function data due to the expected larger contribution from 
higher-twist corrections
in $F_3(x,Q^2)$ than in $F_2(x,Q^2)$, see e.g. \cite{renormalon}. 
We do not impose any $x$-dependent cut. 

\subsection{Final inclusive HERA cross section data}

We replace the previously used HERA run I neutral and charged current 
combined data \cite{H1+ZEUS} by the final HERA run I+II data obtained 
using a variety of beam energies \cite{HERAcomb}. These  were released soon after 
the MMHT14 PDFs, and the effect of their inclusion was presented in detail 
in \cite{MMHT2015}. The fit is good in general but there are some clear exceptions.
The most notable of these is the low $x$ and $Q^2$ regime where the elasticity $y= Q^2/(xs)$ 
is high. In this region the contribution to the total cross section from the longitudinal structure
function can be significant. In \cite{MMHT2015} and also in \cite{Abt} it was shown that a 
larger $F_L(x,Q^2)$ results in a better fit and specific higher twist contributions were 
investigated. It has also subsequently been shown that small-$x$ resummation results in a larger $F_L(x,Q^2)$ in this 
region \cite{NNPDFsx,xfittersx} (as already indicated earlier, in e.g. \cite{White:2006yh}), 
and similarly improves the fit quality. There is also a distinct preference 
from the $e^-p$ HERA charged current data for a 
higher up quark in the $x\sim 0.3$ region than that obtained by a global fit, as highlighted in \cite{MMHT2015}. 
These features persist in the MSHT20 fit, and the overall fit quality for the HERA data deteriorates by $\sim 30$
units, due to some tensions with new LHC data in the global fit. Further analysis of the effect of the HERA data (inclusive and heavy flavour) on the MSHT20 global fit is given later in Section~\ref{noHERA}.

\subsection{Final heavy flavour HERA cross section data}

\begin{figure} 
\begin{center}
\includegraphics[scale=0.31]{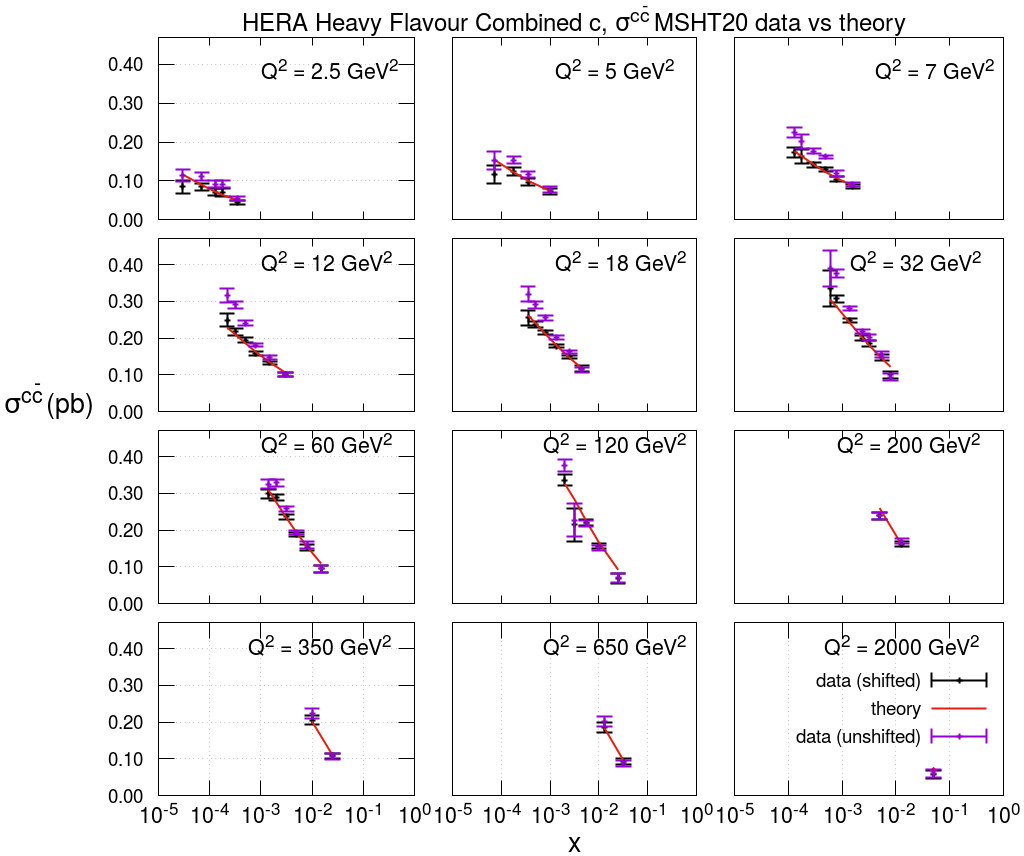}
\caption{\sf Fit quality for the combined HERA heavy flavour data for the charm quark. Purple represents the unshifted data, black are the shifted data and the red line is the MSHT20 theory prediction with these data included in the fit. The errors plotted are the total uncorrelated errors for each point.}
\label{HERA_HF_charm_datavstheory_withshifts}
\end{center}
\end{figure}

\begin{figure} 
\begin{center}
\includegraphics[scale=0.31]{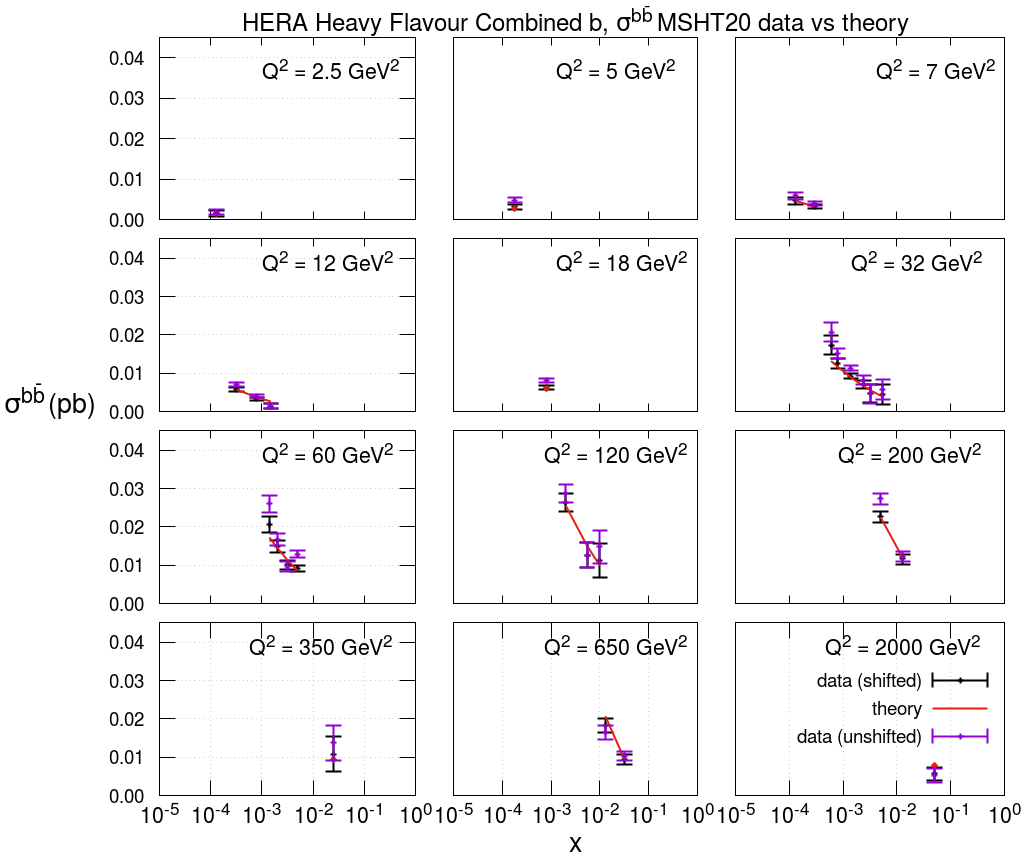}
\caption{\sf As in Fig.~\ref{HERA_HF_charm_datavstheory_withshifts}, but for the bottom quark.}
\label{HERA_HF_bottom_datavstheory_withshifts}
\end{center}
\end{figure}

We remove the combined HERA data on  $F_c(x,Q^2)$ 
\cite{H1+ZEUScharm} and use the final combined data on both 
$F_c(x,Q^2)$ and $F_b(x,Q^2)$ including full information on the statistical and systematic correlations 
between them \cite{HERAhf}. The fit quality, with $\chi^2/N_{\rm pts}= 1.68$ for 79 points at NNLO, is rather higher than 
one might expect. However, this appears to be similar to predictions from other groups and 
from fits within the HERAPDF framework in Table 4 of \cite{HERAhf}. We find that nearly half the 
$\chi^2$ comes from the penalty required to shift the data relative to the theory, in order to obtain 
the correct shape with $Q^2$ in a number of $x$ bins. This can be seen in the relatively large shifts in some $Q^2$ bins in the data and theory comparison in Figs.~\ref{HERA_HF_charm_datavstheory_withshifts} and \ref{HERA_HF_bottom_datavstheory_withshifts}. We find no significant improvement in the fit to the HERA 
heavy flavour structure function data by varying the quark masses, or by changing the unknown parameters in 
our approximation for the ${\cal O}(\alpha_S^3)$ contribution in the FFNS (the latter 
only having any significant impact for the lowest $Q^2$).

\subsection{Tevatron asymmetry data}\label{Tevatronasymdata}

In the present analysis we include the CDF and D{\O} data previously included in 
the MMHT14 study. We have modified the application of correlated 
uncertainties for the CDF $W$ charge asymmetry data 
\cite{CDF-Wasym}, following the recommendation in \cite{Camarda}.
We now also include the final
D{\O} electron charge asymmetry data, with $p_T>25~\GeV$,
based on 19.7 ${\rm fb}^{-1}$ \cite{D0easym}. 
We choose to fit these D{\O} data in the form of $W$ 
charge asymmetry data\cite{D0Wasym}. We summarise the argument for this change here briefly. At the Tevatron the $W^{+/-}$ are preferentially produced in the direction of the proton/antiproton due to the larger momentum fraction of the up/antiup than the antidown/down quark in $u+\bar{d} \rightarrow W^+$ and $\bar{u}+d \rightarrow W^-$. However, when the asymmetry itself is measured in terms of the charged leptons produced, the original $W$ asymmetry is  convoluted with the $V-A$ structure of the $W \to l \nu$ vertex. As a result the lepton/antilepton is preferentially emitted in the direction opposite to the $W^+/W^-$, thereby washing out the asymmetry partially. This effect is particularly prevalent at high absolute rapidities, see Fig.~1 of \cite{D0easym}. As a result, leptons at a specific rapidity originate from $W$ bosons across a wide range of rapidities, in turn corresponding to a range of parton $x$ values at which we wish to constrain the PDFs. Therefore the constraining power of the data on the PDFs is reduced by the statistical errors, inherent in the lepton asymmetry. Instead, the lepton asymmetry can be mapped back into a $W$ asymmetry assuming a given PDF set, with the cost of the addition of relatively small PDF errors, but in turn reducing the relatively large statistical errors, as the asymmetry is no longer washed out. Consequently more precise constraints may then be obtained, as indeed we observe, upon interpreting the data as a $W$ asymmetry. Further details are given in \cite{D0easym,D0Wasym}, whilst we discuss the influence of these data on our PDFs in more detail in Section~\ref{D0Wasymeffects}. 

\section{LHC data included in the present fit \label{sec:4}}

We now discuss the inclusion of the new  LHC data in the PDF fit. 
This includes a variety of data on $W$ and $Z/\gamma^*$ production\footnote{We will in some cases refer to $Z/\gamma^*$ as simply $Z$ production for brevity, though the photon contribution is always implied. In addition, we will use the term Drell-Yan interchangeably with this.}, over a range of invariant masses and usually differential in rapidity, but now also including
$W$+jets data, $Z$ boson $p_T$ distributions and $W+c$ data. 
We also include not only total $t\overline{t}$ cross section data, but single and double differential top quark 
distributions. As in previous fits we include inclusive jet production, 
but now both at NNLO as well as NLO, and with a much larger and more precise collection of jet data.  

\begin{table}[t]
\begin{center}
\begin{tabular}{|c|c|c|c|}
\hline
Data set &Points& NLO $\chi^2/N_{pts}$&NNLO $\chi^2/N_{pts}$\\ \hline
D{\O} $W$ asymmetry & 14 & 0.94 (2.53) & 0.86 (14.7)\\ 
$\sigma_{t\overline{t}}$ \cite{Tevatron-top}-\cite{CMS-top8} & 17 & 1.34 (1.39) & 0.85 (0.87)\\ 
LHCb 7+8 TeV $W+Z$  \cite{LHCbZ7,LHCbWZ8}  & 67 & 1.71 (2.35) & 1.48 (1.55)\\
LHCb 8 TeV $Z\to ee$ \cite{LHCbZ8} & 17 & 2.29 (2.89) & 1.54 (1.78)\\
CMS 8 TeV $W$  \cite{CMSW8} & 22 & 1.05 (1.79) & 0.58 (1.30)\\
CMS 7 TeV $W+c$ \cite{CMS7Wpc} & 10 & 0.82 (0.85) & 0.86 (0.84)\\
ATLAS 7 TeV jets $R=0.6$ \cite{ATLAS7jets} & 140 & 1.62 (1.59) & 1.59 (1.68)\\
ATLAS 7 TeV $W+Z$  \cite{ATLASWZ7f}  & 61 & 5.00 (7.62) & 1.91 (5.58)\\
CMS 7 TeV jets $R=0.7$ \cite{CMS7jetsfinal} & 158 & 1.27 (1.32) & 1.11 (1.17)\\
ATLAS 8 TeV $Z$ $p_T$ \cite{ATLASZpT} & 104 & 2.26 (2.31) & 1.81 (1.59)\\
CMS 8 TeV jets $R=0.7$ \cite{CMS8jets} & 174 & 1.64 (1.73) & 1.50 (1.59)\\
ATLAS 8 TeV $ t \bar t \to l +j$ sd \cite{ATLASsdtop} & 25 & 1.56 (1.50) & 1.02 (1.15) \\
ATLAS 8 TeV $ t \bar t \to l^+l^-$ sd \cite{ATLASttbarDilep08_ytt} & 5 & 0.94 (0.82) & 0.68 (1.11) \\
ATLAS 8 TeV high-mass DY  \cite{ATLASHMDY8}s& 48 & 1.79 (1.99) & 1.18 (1.26)\\
ATLAS 8 TeV $W^+W^- +$ jets  \cite{ATLASWjet}  & 30 & 1.13 (1.13) & 0.60 (0.57) \\
CMS 8 TeV $(d\sigma_{\bar t t}/dp_{T,t}dy_t)/\sigma_{\bar t t}$  \cite{CMS8ttDD} & 15 & 2.19 (2.20)& 1.50 (1.48)\\
ATLAS 8 TeV $W^+W^-$ \cite{ATLASW8} & 22 & 3.85 (13.9) & 2.61 (5.25)\\
CMS 2.76 TeV jets \cite{CMS276jets} & 81 & 1.53 (1.59)& 1.27 (1.39)\\
CMS 8 TeV $\sigma_{\bar t t}/dy_t$ \cite{CMSttbar08_ytt}  & 9 & 1.43 (1.02) & 1.47 (2.14)\\
ATLAS 8 TeV double differential $Z$  \cite{ATLAS8Z3D} & 59 & 2.67 (3.26) & 1.45 (5.16)\\ \hline
Total, LHC data in MSHT20 & 1328 & 1.79 (2.18) & 1.33 (1.77) \\ \hline
Total, non-LHC data in MSHT20 & 3035 & 1.13 (1.18) & 1.10 (1.18) \\ \hline
Total, all data & 4363 & 1.33 (1.48) & 1.17 (1.36) \\ \hline
\hline
\end{tabular}
\vspace{-0.1cm}
\caption{\sf $\chi^2/N_{pts}$ at NLO and NNLO for the fit to the new LHC and Tevatron data included in the MSHT20 fit. The corresponding fit qualities are also given for the total LHC and non-LHC data included in MSHT20, as well as the overall fit across all data. In brackets are the predictions obtained using the MMHT14 PDFs (also at $\alpha_S(M_Z^2) = 0.118$).}
\label{tab:TabnewLHC}
\end{center}
\end{table}

We will present full details of the fit quality and the PDFs in the next 
section, but first we present the results of the fit to each of the 
different types of LHC data.  A summary is provided in Table \ref{tab:TabnewLHC}, alongside the totals for the LHC and non-LHC data included in the fit.
We see that in general the MMHT14 prediction is quite good, though in most cases an 
improvement is achieved after refitting. This improvement is most marked for the data sets with very high 
precision, i.e. the inclusive $W$ and Drell-Yan cross sections differential in rapidity, and in the case of 
Drell-Yan data, in different mass bins. For these sets the improvement after refitting is often 
considerable, and the prediction from the MMHT14 PDFs can be very poor. The 
improvement with refitting for these data sets is mainly achieved by changes in the details of the flavour 
content of the quarks and antiquarks. However, overall improvement also results from changes in the 
gluon distribution and in the common shape of the quark distributions as a function of $x$. In most cases the fit quality is clearly better at NNLO than at NLO, with the data sensitive to 
the fine detail of the shape corrections in both the PDFs and the hard cross sections at NNLO. It is clear from the totals for the LHC and non-LHC data, that the description of the former is clearly improved from NLO to NNLO, whereas the latter improves only marginally.

We now discuss individual data sets in turn.   

\subsection{Drell--Yan data}

In this section we discuss the range of  $W$ and $Z$ data from the ATLAS, CMS and LHCb experiments that are included in the fit. In all cases, the NLO theory predictions are provided by \texttt{MCFM}~\cite{Campbell:2002tg,Campbell:2004ch} interfaced to \texttt{APPLGrid}~\cite{Carli:2010rw}, supplemented with NNLO $K$-factors produced with \texttt{NNLOjet}~\cite{NNLOZpT,Bizon:2018foh} (ATLAS 8 TeV $Z$ and $Z$ boson $p_\perp$ distribution), \texttt{MCFM 8.3}~\cite{Boughezal:2016wmq} (CMS 8 TeV $W^{\pm}$ production), $N_{\rm jetti}$~\cite{Njetti1,Njetti2} (ATLAS $W+{\rm jets}$), and \texttt{FEWZ}~\cite{Li:2012wna} and/or \texttt{DYNNLO}~\cite{Catani:2010en} (all other data sets).

\subsubsection{ATLAS 7~TeV $W$ and $Z$ data}\label{AT7WZdata}

These precision data\cite{ATLASWZ}, in particular due to the correlation between the $W$ and $Z$ cross sections, 
provide a strong constraint on the strange quark, one of the few data sets in 
the global fit to do so. As one can see from Table~\ref{tab:TabnewLHC} the MMHT14 PDFs give a poor description at both NLO and NNLO. However, given the very precise nature of 
these data, with uncorrelated uncertainties of in some cases less than $0.1\%$, large improvements in 
$\chi^2$ can be obtained from changes in the fine details of the PDFs. Most notably the simultaneous fit of 
the $W$ and $Z$ data can be improved by an increase in the strange quark. The largest impact is at
 central rapidities, where $x \approx m_{Z,W}/\sqrt{s} \approx 0.01$, and hence the increase in the strange quark is focussed on
this region. The prediction for the relative rates of $W^+$ and $W^-$ data at central rapidity can 
also be improved, in this case by an increase of $u_V$ relative to $d_V$ for $x\sim 0.1$. These changes 
allow for a significant improvement in the fit quality at NNLO, though this is still rather poor, with $\chi^2/N_{\rm pt}\sim 1.9$. Part of the poor fit is due to the low mass bin for $Z$ data, where there are some large 
fluctuations. At NLO we do not obtain a similarly significant improvement in the fit 
quality, and the implications for the changes in the PDFs are different, see Section~\ref{sec:7}. We note that we find the best fit is obtained for a common choice of 
$\mu_{R,F}=m_{ll}/2$, and therefore at NNLO we use this choice for all $W$ and Drell-Yan 
rapidity distributions in the fit. If we instead take $\mu_{ll}$ ($2 \mu_{ll}$) the fit quality quality for the 7 TeV data deteriorates by $\sim 20$ ($26$) points, while the impact on the 8 TeV datasets is relatively mild. In addition, there is only a very limited improvement with respect to the baseline PDF predictions upon refitting, indicating that the corresponding PDFs will be relatively unaffected.

\subsubsection{Other ATLAS $W$ and $Z,\gamma^\star$ data.}

\begin{figure} 
\begin{center}
\includegraphics[scale=0.21]{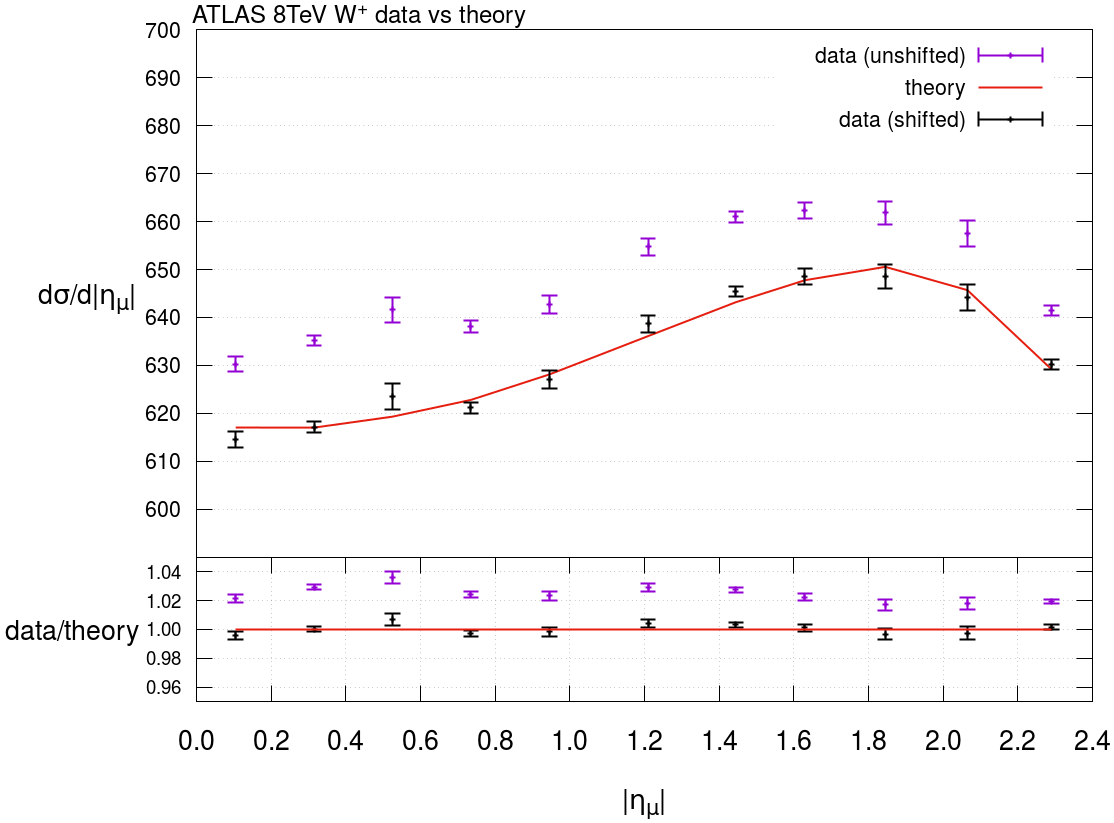}
\includegraphics[scale=0.21]{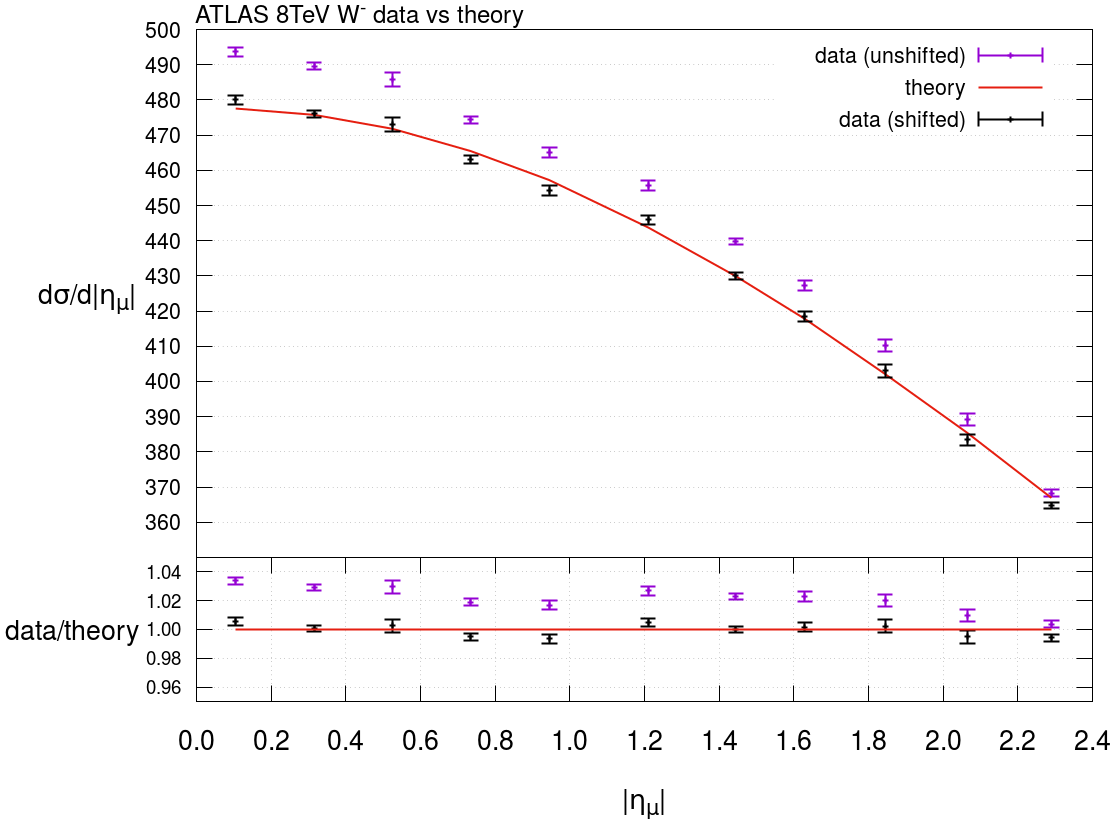}
\caption{\sf Data vs. MSHT20 NNLO theory for the ATLAS 8~TeV $W^{\pm}$ data with $W^+$ ($W^-$) in the left (right) plots. The purple represents the unshifted data, the black the data after shifting via correlated and uncorrelated systematics and other error sources, and the red line is the MSHT20 theory prediction, with these data included in the fit. The errors plotted here are the total uncorrelated errors for each point.}
\label{ATLAS8Wppmu_datavstheory}
\end{center}
\end{figure}

\begin{figure}
\begin{center}
\includegraphics[scale=0.3]{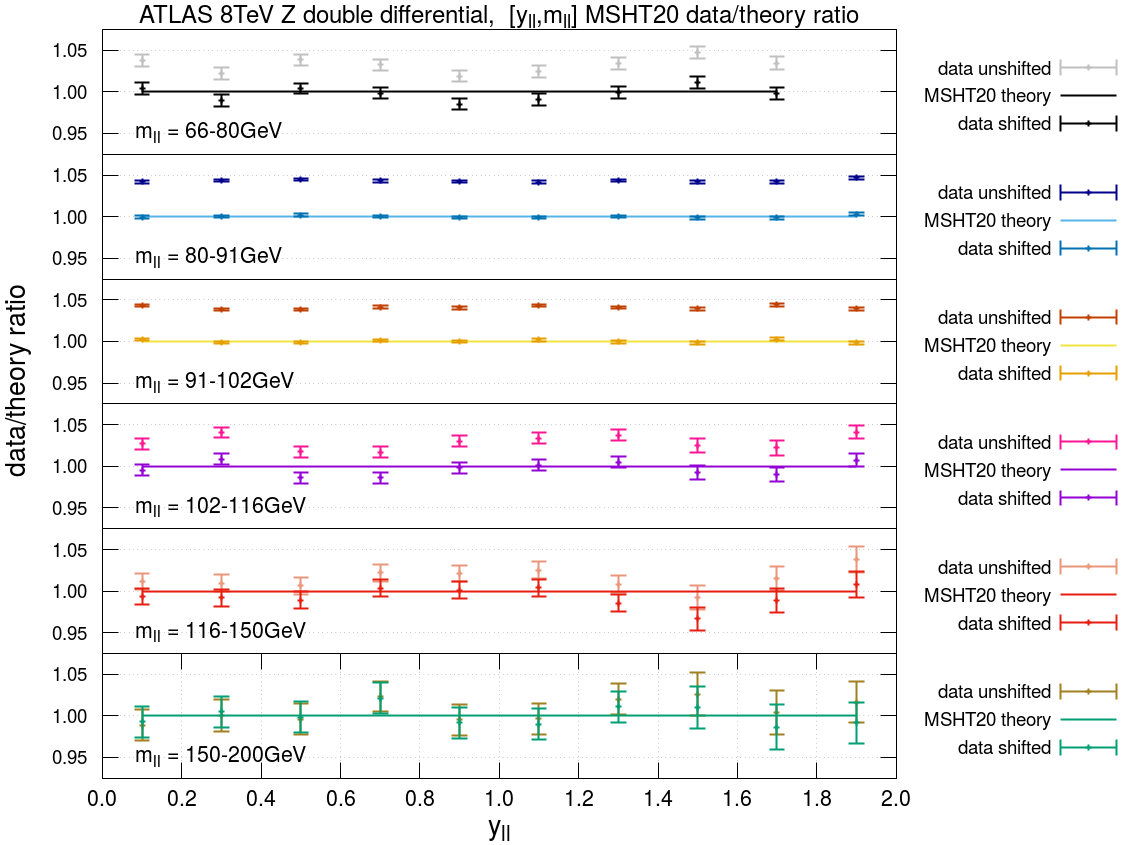}
\caption{\sf Ratio of Data to MSHT20 NNLO theory  for the ATLAS double differential 8~TeV $Z$ data, differential in $[y_{ll},m_{ll}]$. The errors plotted here are the total uncorrelated errors for each point which is the quadrature sum of the statistical and uncorrelated systematic errors. Both the unshifted and shifted data are shown.}
\label{ATLAS8Z3D_datavstheory}
\end{center}
\end{figure}

As well as the very precise 7~TeV ATLAS $W$ and Drell-Yan data we also 
include more recent but similar data at 8~TeV on $W^{+,-}$ production\cite{ATLASW8} and
8~TeV Drell-Yan data \cite{ATLAS8Z3D}, which is presented triple differentially in $m_{ll}$, $y_{ll}$ and $\cos\theta^\star$. In the latter case we integrate the data and theory over 
$\cos \theta^\star$ in order to avoid sensitivity to $\sin^2 \theta_W$. For the data we exclude those $\cos\theta^*$ bins in this combination (and in the corresponding theory prediction) for which the NNLO theory prediction has less than $ 95\%$ acceptance with respect to the experimental event selection, in order to avoid regions of phase space where the cross section is largely or entirely non--zero only at NLO. 
This reduces the data set from 89 to 59 $(m_{ll},y_{ll}$) bins. The missing 30 points have much higher 
statistical uncertainty than the 59 we include, and the
fit is not sensitive to their omission. We also find that fitting instead to the triple differential data directly results in a rather similar fit. We will discuss both of these points in more detail in Section~\ref{ATLASDYeffects}.

The ATLAS 8~TeV $W$ data are given as a function of the decay muon pseudorapidity, and as usual we choose to fit both the $W^+$ and $W^-$ distributions rather than the asymmetry. Our treatment of this data set mirrors that of the ATLAS analysis. 
In the standard global fit, as seen in Table 
\ref{tab:TabnewLHC}, this data set is relatively poorly fitted even at NNLO, with $\chi^2/N_{\rm pts} \sim 2.6$ for 22 points, with $\sim 0.8$ coming from the penalty term due to the systematic errors. The data theory comparison for this data set is shown in Fig.~\ref{ATLAS8Wppmu_datavstheory}. We can  see that before shifting the data by the systematic errors the MSHT20 theory predictions undershoot the data over the entire range by up to 4\%. As a result, in our fit we find a shift of the data by the luminosity error of around 2$\sigma$, where the luminosity error itself is 1.9\%. This shift is found to be consistent with that seen for the ATLAS 8~TeV $Z$ data in Fig.~\ref{ATLAS8Z3D_datavstheory}, as one might expect. A similar effect is also present in the 7~TeV $W$, $Z$ data. Nonetheless, even after shifting the data we find that whilst the normalisation is now reasonable, there is still some disagreement remaining from bin to bin, which results in part in the comparatively poor fit quality for this data set. 

This relatively poor fit quality reflects both the high precision of the measurements and also tensions with other data in the global fits.
Notably we see clear tensions between the ATLAS 7 and 8~TeV $W$, $Z$ data sets with both the BCDMS data and also the new D{\O} $W$ asymmetry data and several of the new LHC data sets including the ATLAS 8~TeV $Z$ $p_T$ data. In particular, if we remove these data sets from the fit we obtain a lower value of $\chi^2/N_{\rm pts} \sim 2.3$. 
It is interesting to note that the ATLAS 8~TeV $W^{\pm}$ and $Z$ data and ATLAS 7~TeV $W$ and $Z$ data pull in the same direction, with the 7~TeV data showing $\Delta \chi^2 \approx 5$ improvement when the 8~TeV $W^{\pm}$ data are added, and a further $\Delta \chi^2 \approx 5$ improvement when the ATLAS 8~TeV $Z$ data are added. 
Moreover, the tensions observed for the 8~TeV $W^{\pm}$ data also apply to the 7~TeV data, with the latter improving by $\Delta \chi^2 = 11.8$ once the BCDMS, D{\O} $W$ asymmetry and new LHC data sets described above are removed. Details of these tensions are elaborated upon later in Section~\ref{tensions} and Table~\ref{tab:BCDMSD0WasymATLASZpt_delchisqtable}.

\subsubsection{CMS $W$ data}

We include the $W^+$ and $W^-$ rapidity distributions measured at 8~TeV by CMS 
\cite{CMSW8}. Unlike earlier CMS $W^+,W^-$ data the results are presented as absolute distributions (with 
full information about correlations) as opposed to asymmetry data alone. 
The data are for central and low rapidities, and hence are sensitive to PDFs 
in the region $0.001 < x < 0.1$. The fit quality is good, particularly at NNLO.

\subsubsection{LHCb $W$ and $Z$ data}

\begin{figure} 
\begin{center}
\includegraphics[scale=0.23]{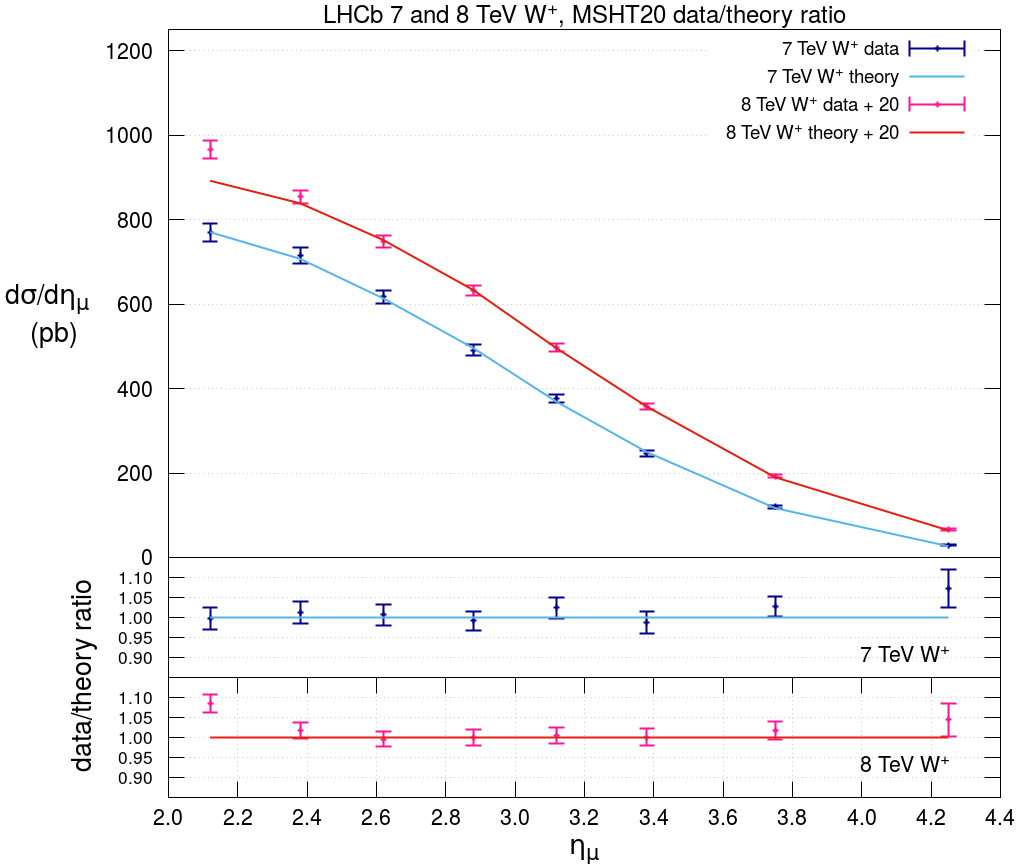}
\includegraphics[scale=0.23]{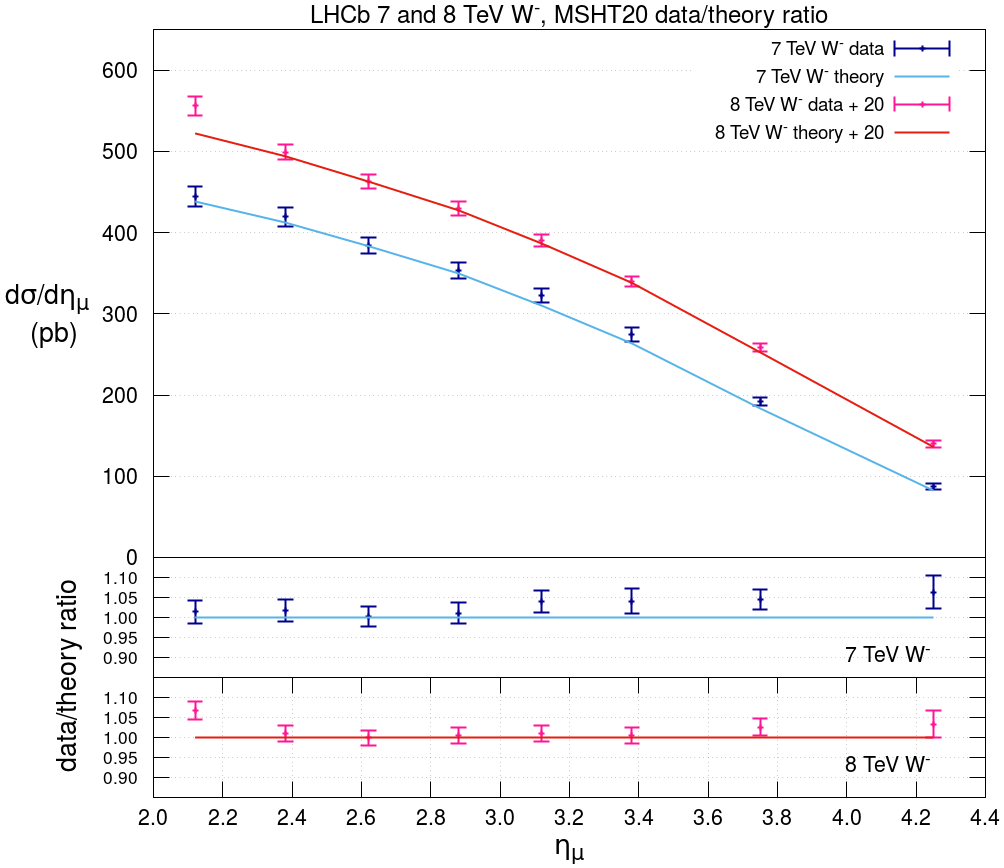}
\caption{\sf Data vs. MSHT20 NNLO theory  for the LHCb 7 and 8~TeV (left) $W^+$ and (right) $W^-$ data differential in the muon pseudorapidity, $\eta_{\mu}$. The errors plotted here are the total uncorrelated errors for each point.}
\label{LHCb15W_datavstheory}
\end{center}
\end{figure}

\begin{figure} 
\begin{center}
\includegraphics[scale=0.23]{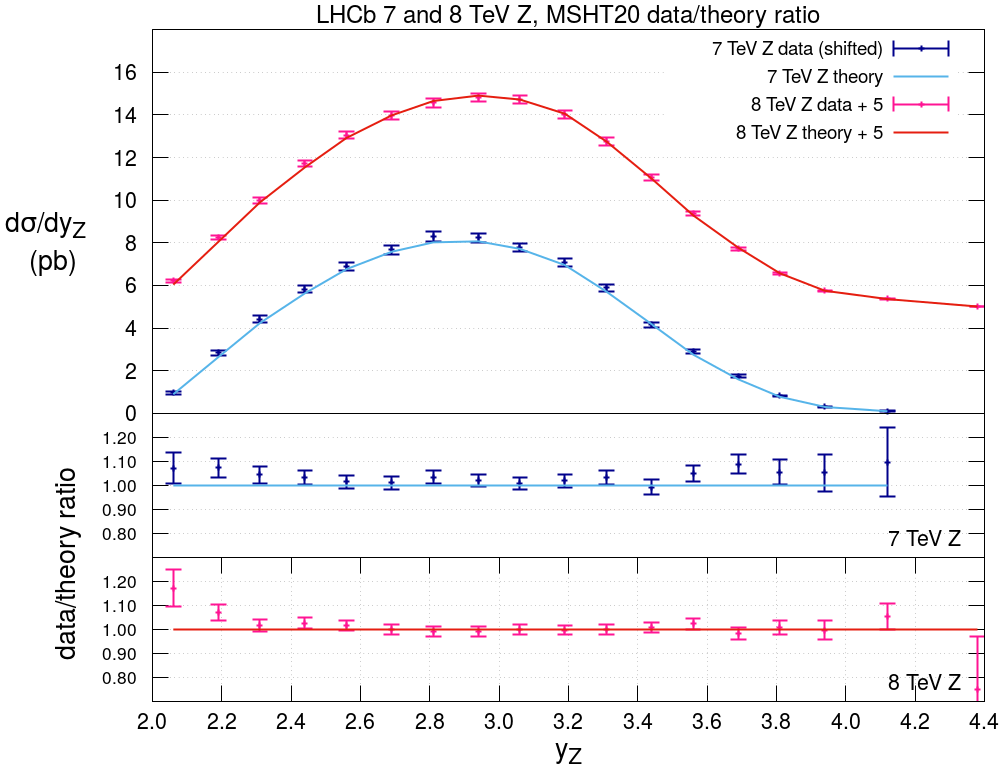}
\includegraphics[scale=0.23]{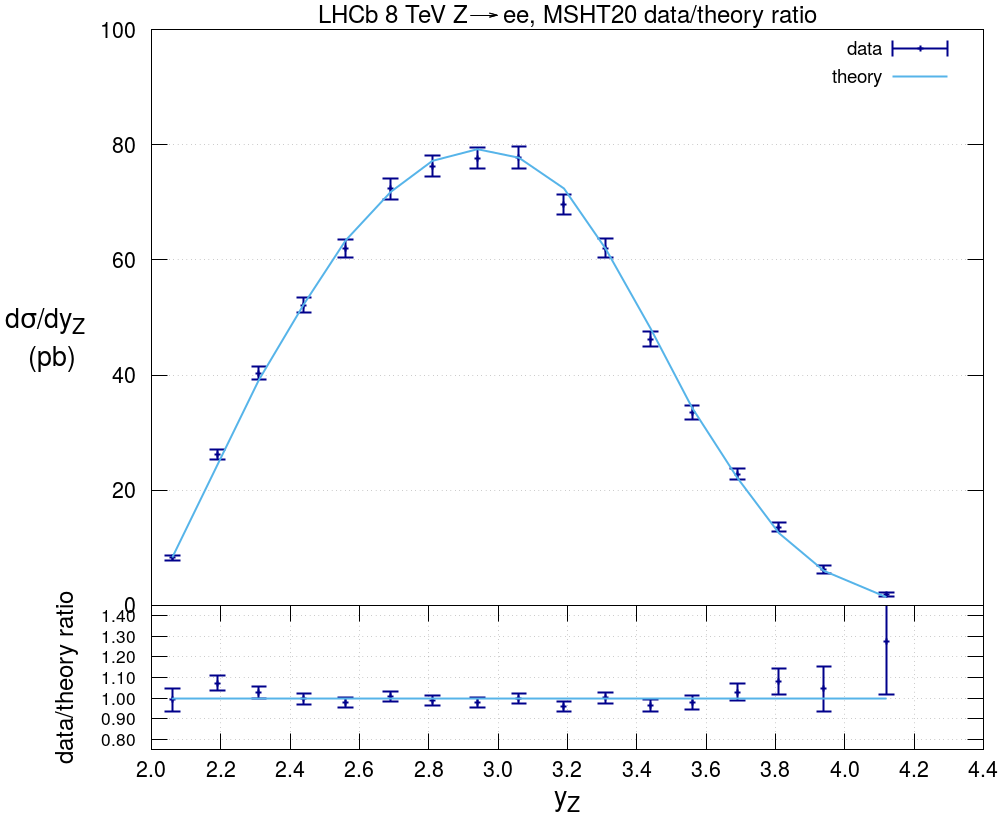}
\caption{\sf Data vs. MSHT20 NNLO theory for the (left) LHCb 7 and 8~TeV $Z$ data in the muon channel and (right) the LHCb 8~TeV $Z$ in the electron channel, both differential in $y_{Z}$. The errors plotted here are the total uncorrelated errors for each point.}
\label{LHCb15Z_8Zee_datavstheory}
\end{center}
\end{figure}

The LHCb vector boson production data exist for pseudorapidities between 2 and 4.5, and hence  probe valence quarks at higher $x$ and sea quarks at lower $x$ than the corresponding ATLAS and CMS measurements. We fit the $W$ and $Z$ distributions in the muon channel
\cite{LHCbZ7,LHCbWZ8}, maintaining all correlations between the data sets, and also the $Z$ distributions
in the electron channel \cite{LHCbZ8}. In the former case the fit is generally good, though
there  are some issues with undershooting the data in the lowest rapidity bins in the 8~TeV case, as seen in Figs.~\ref{LHCb15W_datavstheory} and \ref{LHCb15Z_8Zee_datavstheory} (left). 
For the 8~TeV $Z \to e^+e^-$ data there are no issues of undershooting at the lower rapidity end and instead the larger $\chi^2$ and poorer than average fit quality in the global fit appears to just be due to fluctuations, as shown in Fig.~\ref{LHCb15Z_8Zee_datavstheory} (right).

\subsubsection{CMS double differential and ATLAS high-mass Drell-Yan data}

We include the ATLAS high mass Drell-Yan data at 8 TeV \cite{ATLASHMDY8}, 
which were already studied in detail in \cite{MMHTQED}. These are well fit, as already discussed in 
\cite{MMHTQED} and show no real tensions with other data. We continue to fit CMS 7~TeV double differential 
Drell-Yan data \cite{CMS-ddDY}. 
We do not include 
the CMS 8~TeV Drell-Yan data \cite{CMSDDDY8}. We in particular find a very poor fit quality (also 
observed by NNPDF~\cite{Guffantipriv}) but  
also have identified internal inconsistencies in the data uncertainties, which appear to indicate underlying issues in the published results.

\subsection{ATLAS $W$ + jets.}

We  fit the ATLAS measurement of the $W+{\rm jets}$ production at 8~TeV, presented differentially in the transverse momentum of the $W$ boson \cite{ATLASWjet}. This process is sensitive to up and down quarks and antiquarks as well as the gluon, particularly at high $x$, however in practice it provides little constraint on quark decomposition relative to other data in the fit. As for our precise implementation of this data set in MSHT20, uncorrelated systematics dominate over the statistical errors for this data set and we treat them exactly as indicated by ATLAS. We however do not include statistical correlations between bins as their effect is negligible. Non-perturbative corrections from SHERPA~2.2.1 \cite{SHERPA} were applied, as indicated by the ATLAS study \cite{ATLASWjet}. The $p_{T}^{W}$ spectra for $W^+$ and $W^-$ bosons were fit separately. We exclude the first of the $p_{T}^W$ bins ($0 < p_{T}^W < 25$~GeV) as we expect resummation and other effects to be present here; indeed the NNLO $K$-factor and non-perturbative corrections provided for this bin are very much larger than for the remaining bins. If we include the first bins the fit quality deteriorates significantly, by $\sim 0.6$ per point, indicating the issues with including these.

\begin{figure} 
\begin{center}
\includegraphics[scale=0.21]{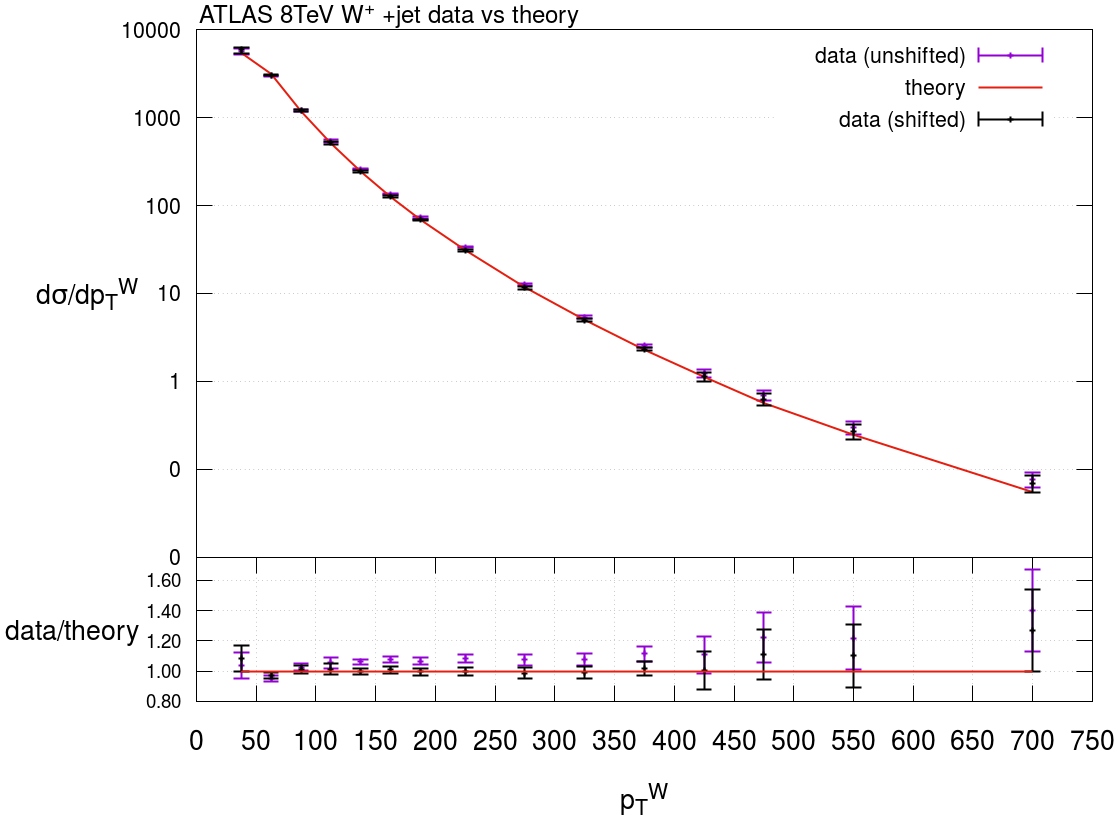}
\includegraphics[scale=0.21]{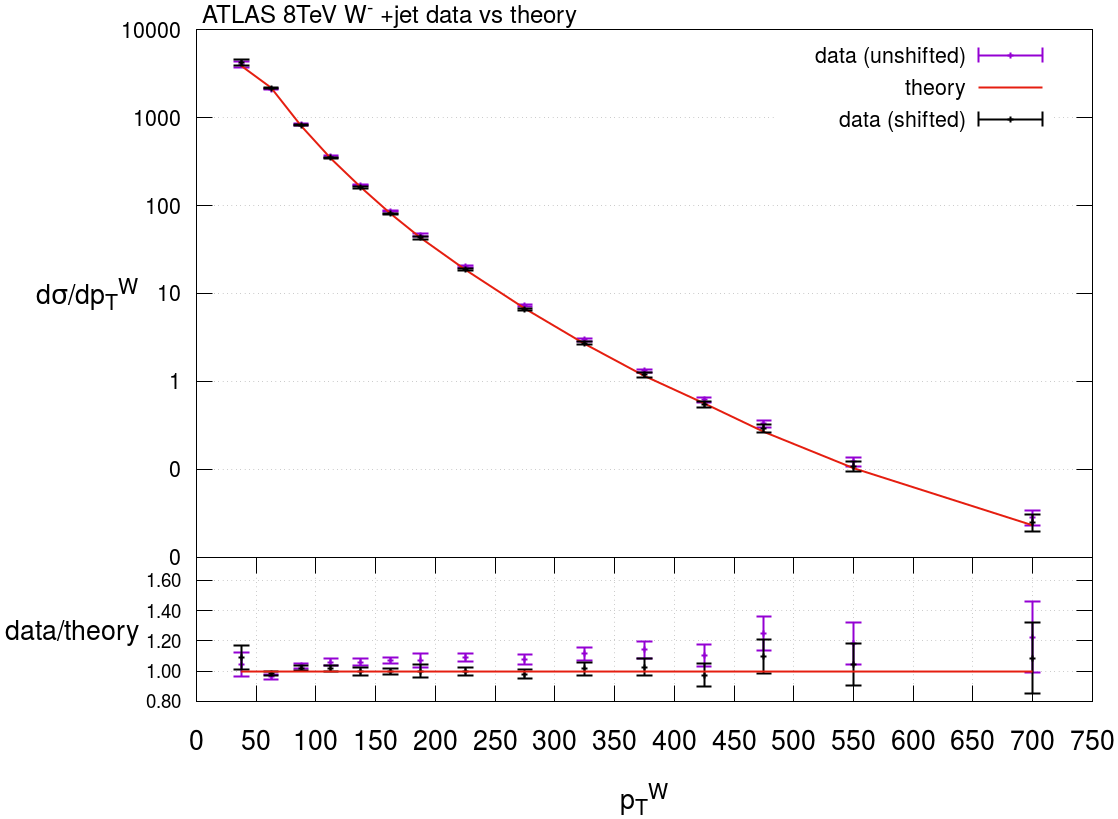}
\caption{\sf Data vs. MSHT20 NNLO theory  for the ATLAS 8~TeV $W^{\pm} + $ jets data with $W^+$  ($W^-$) in the left (right) plots. The purple represents the unshifted data, the black the data after shifting via correlated and uncorrelated systematics and other error sources, and the red line is the MSHT20 theory prediction, with these data included in the fit. The errors plotted here are the total uncorrelated errors for each point.}
\label{ATLAS8Wjet_datavstheory}
\end{center}
\end{figure}

Upon refitting, there is only very small change in the $\chi^2$, showing that these data will have little impact on our PDFs. Marginal improvements in other LHC data sets were seen upon refitting with the $W+{\rm jets}$ data, although the description of the CMS 8~TeV $W$ data deteriorated slightly. The data and theory comparison for this data set in the MSHT20 NNLO fit is shown in Fig.~\ref{ATLAS8Wjet_datavstheory}.

\subsection{CMS $W+c$ data.}

We include CMS data on the production of $W+$ charm jets. In principle one might think that this is not 
appropriate as there is no NNLO calculation of this process included at present, with the full NNLO corrections to the dominant CKM-diagonal contribution only recently calculated \cite{Czakon:2020coa}. However, we choose to include it 
essentially as a cross check on the impact of other data, given the degree of tension observed between the dimuon and ATLAS $W$, $Z$ data. In particular, it provides 
a very direct constraint on the strange quark and strange antiquark (when one considers $W^+$ and $W^-$ data
separately).
In practice, the data are fit very well with a strange (and antistrange) quark distribution which is a compromise 
between the dimuon and ATLAS $W$, $Z$ data, but given the small number of data points and relatively low 
statistical precision, the pull of the data is not very strong. The NNLO corrections from \cite{Czakon:2020coa} 
appear to imply a positive correction in the cross section of perhaps up to 10\% relative to the 
NLO included in MSHT20, hence this suggests they may provide a marginal additional pull downwards on the 
strange PDF. It should be noted however that the NNLO calculations are performed using a flavour-$k_\perp$ algorithm, which is not used in the measurement, and hence the precise value of this $K$--factor is not yet firmly established.

\subsection{LHC data on jets} \label{LHCjets}

We now include a far more extensive collection of collider jet data. In \cite{MMHT14} we only included Tevatron jet data \cite{CDFjet,D0jet} at NNLO, 
using the threshold approximation to NNLO corrections \cite{Kidonakis}. As mentioned, we still apply
this approximate NNLO correction to these data sets, as the data are all quite near threshold and now carry 
very little weight in the fit. However, we now also include far more precise LHC jet data, which moreover, 
have a much wider kinematic coverage. For all of these data we include full NNLO corrections (in the form 
of $p_T$ dependent $K$-factors), using the results of \cite{NNLOjets}, with the NLO theory provided by \texttt{NLOjet++}~\cite{Nagy:2003tz} interfaced to \texttt{APPLGrid}~\cite{Carli:2010rw} or \texttt{FastNLO}~\cite{Kluge:2006xs,Britzger:2012bs}. In all cases we use
the scale choice $\mu=p_T$ for both renormalisation scale and factorisation scale, though as demonstrated 
in \cite{MMHTjets} the significant correlated uncertainties on the data allow very similar fit quality and 
PDFs independent of scale choice. For the  ATLAS 7~TeV data, we in addition include EW corrections, as provided by and described in~\cite{ATLAS7jets}. Such corrections were not included in~\cite{MMHTjets}, and hence we will examine their impact below.

\begin{table}
\begin{center}
\begin{tabular}{|c|c|c|c|c|}
\hline
 & No decor. &  Ref.~\cite{MMHTjets} decor. & Smooth decor.& Full decor.  \\ \hline
 ${\rm LO}_{\rm EW}$ &2.00 & 1.09& 1.48 & 0.81\\ \hline
${\rm NLO}_{\rm EW}$ &1.80  & 1.15 & 1.57&0.92\\
\hline
\end{tabular}
\caption{\sf $\chi^2/N_{\rm pts}$ for the ATLAS 7~TeV inclusive jet data, as in the global MSHT20 NNLO fit ($\alpha_S$ free). Results including various decorrelation scenarios, explained in the text, as well including/excluding EW corrections, are shown.}
\label{table:decorjets}
\end{center}
\end{table}

For ATLAS we fit the 7~TeV inclusive jet distributions \cite{ATLAS7jets} (these supersede the 
much lower statistics 7~TeV and 2.76~TeV jet data \cite{ATLAS7jetsold,ATLAS276jets} used in MMHT14), and use the higher jet radius $R=0.6$.
A detailed discussion of the inclusion of these data in a global   fit has already appeared in 
\cite{MMHTjets}, where the difficulty in fitting all rapidity bins simultaneously was highlighted, and 
the possibility of solving this by allowing a very small number of systematic uncertainties to be 
decorrelated across rapidity bins explored. The focus here was in particular on the decorrelation of so--called `two--point' systematic uncertainties, which are based on the difference between two alternative MC treatments, and therefore may not be expected to provide a reliable guide to the true error correlation, see also~\cite{Bailey:2019yze} for further discussion. In principle the approach of~\cite{MMHTjets} may be too drastic, as it allows the corresponding shifts in each rapidity bin to vary in a manner that may not be particularly smooth (though in practice it is far from guaranteed that the preferred variation will not be smooth), whereas we expect the correction due to these effects to vary smoothly. This was addressed in~\cite{ATLAS8jets} where a set of smoother potential decorrelation scenarios were presented. 

We now perform our fit using an approach based on~\cite{Bailey:2019yze,ATLAS8jets}, which allows a suitably smooth variation and focuses on allowing data points that are distant in $(y_j, p_\perp^j)$ space to in principle have different variations. We in particular define
\be
x_{p_\perp} = \frac{\log(p_\perp^j) -  \log(p_{\perp,{\rm min}}^j)}{ \log(p_{\perp,{\rm max}}^j)- \log(p_{\perp,{\rm min}}^j)}\;,\qquad x_y = \frac{y_j - y_{j,{\rm min}}}{y_{j,{\rm max}}-y_{j,{\rm min}}}\;,
\ee
and then
\be
r= \frac{1}{\sqrt{2}} \left( x_{p_\perp}^2 + x_y^2\right) \;,\qquad \phi = {\rm arctan}\left(\frac{x_y}{x_{p_\perp}}\right)\;.
\ee
Defining
\be\label{eq:ltrig}
L_{\rm trig} (z,z_{\rm min},z_{\rm max}) = \cos \left[ \pi\left( \frac{z-z_{\rm min}}{z_{\rm max} - z_{\rm min}}\right)\right]\;,
\ee
as in~\cite{Bailey:2019yze}, we use:
\begin{align}
\beta_i^{(1)}&= L_{\rm trig} (r,0,1)\cdot  L_{\rm trig} \left(\phi,0,\frac{\pi}{2}\right)\beta_i^{\rm tot}\;,\\
\beta_i^{(2)}&= \sqrt{1- L_{\rm trig} (r,0,1)^2}\cdot  L_{\rm trig} \left(\phi,0,\frac{\pi}{2}\right)\beta_i^{\rm tot}\;,\\
\beta_i^{(3)}&= L_{\rm trig} (r,0,1)\cdot  \sqrt{1-L_{\rm trig} \left(\phi,0,\frac{\pi}{2}\right)^2}\beta_i^{\rm tot}\;,\\
\beta_i^{(4)}& = \sqrt{1- L_{\rm trig} (r,0,1)^2}\cdot   \sqrt{1-L_{\rm trig} \left(\phi,0,\frac{\pi}{2}\right)^2}\beta_i^{\rm tot}\;.
\end{align}
Although it may seem preferable to instead use a factor of $\pi/2$ in \eqref{eq:ltrig} to preserve symmetry, the above choice provides a better description of the desired shifts and provided a higher quality fit in the case of~\cite{Bailey:2019yze} and so is taken here as well. We in addition choose a different set of systematics uncertainties to decorrelate, taking three in total, due to the multi--jet balance asymmetry (as in~\cite{MMHTjets}), the jet flavour response and the multi--jet fragmentation. All of these correspond to genuine MC two--point uncertainties which we are therefore justified in decorrelating in this way.

The impact on the fit quality is given in Table~\ref{table:decorjets}. `No decor' corresponds to applying the published ATLAS systemic error correlation, `smooth' is the approach described above and `full' to allowing all systematic error shifts (i.e. not just the three mentioned above) to vary independently in each rapidity bin. The latter case in particular corresponds in effect to fitting a single rapidity bin, given that in that case one is dropping the requirement that the preferred shifts from that individual rapidity bin should match those from other bins. A fit to a single rapidity bin is done in~\cite{NNPDF3.1}, where it is found that the fit to any individual rapidity bin is independent of the bin that is chosen. We in addition show results with and without NLO EW corrections.

We can see that broadly the smooth decorrelation leads to some improvement in the $\chi^2$, with the effect being less dramatic than the approach of~\cite{MMHTjets}. Interestingly, the inclusion of EW corrections, while leading to a relatively mild improvement in the $\chi^2$ when using default systematic errors, is such that the relative improvement from our smooth decorrelation is quite a bit smaller than without. Certainly the difference between the default ATLAS systematics errors of $1.8$ per point, and the result with our decorrelation, of $\sim 1.6$ per point is not particularly significant. However, we note that for other choices of jet scale, and jet radius, where the baseline fit quality tends to be worse, the relative impact may well be larger, as seen in~\cite{MMHTjets}. We have investigated further decorrelation of e.g. the non--perturbative corrections, but find that these improve the fit quality by only $\sim 0.1$ per point.

\begin{figure}
\begin{center}
\includegraphics[scale=0.24]{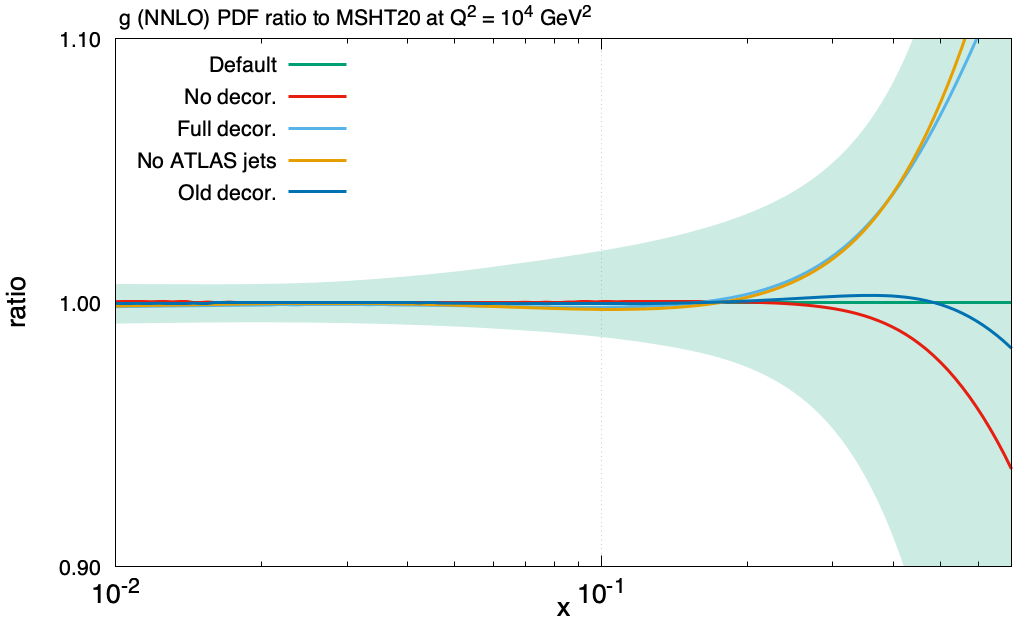}
\caption{\sf Ratio of gluon PDFs to MSHT20 baseline, with $\alpha_S$ free, at NNLO at $Q^2=10^4~\GeV^2$. The result of a fit to the ATLAS 7~TeV jet data, with the standard experimental correlated systematics errors (No decor.), a full decorrelation of all systematic errors across each rapidity bin (Full decor.), and with the jet data removed from the fit, are shown.}
\label{fig:at7jets}
\end{center}
\end{figure}

However, as discussed in~\cite{MMHTjets}, the more important question arguably relates to the impact of the above variations on the extracted gluon. We show this in Fig.~\ref{fig:at7jets}, and we can see that the difference between the fit with no decorrelation and our baseline smooth decorrelation is rather small, only larger than $\sim 1\%$ above $x \gtrsim 0.5$, where the PDF uncertainty is significantly larger and the constraints from the jet data are small. The result for the decorrelation of~\cite{MMHTjets} is also shown, and we can see that it lies very close to our baseline decorrelation across the entire $x$ region. On the other hand, when we take a full decorrelation of errors, which we can see from Table~\ref{table:decorjets} leads to a $\chi^2/N_{\rm pts}\sim 0.9$, the difference is more significant, though clearly within error bands of the baseline. More specifically, the pull of the ATLAS data on the gluon in this case is smaller, as by throwing away all information about the systematic correlations we loosen the constraining power of the data. To demonstrate this, we show the result of performing a fit but with the ATLAS jet data excluded, and we can see that this lies very close to the fully decorrelated case.
Thus the impact of throwing all of the experimental information contained in the systematic errors, most of which are known very precisely, is non--negligible. 

We also include 2.76~TeV, 7~TeV and 8~TeV inclusive jet data from CMS in our fit 
\cite{CMS276jets,CMS7jetsfinal,CMS8jets}. We use the same scale choice as for the ATLAS data, and again
chose the larger available jet radius, in this case $R=0.7$ (for 2.76~TeV data this is the only choice), 
in order to minimise nonperturbative corrections. Unlike the ATLAS data there is no problem in obtaining 
a good quality fit for each of the CMS data sets. The fit is consistently better at NNLO than at NLO for the CMS and ATLAS jets data sets.

\begin{figure} 
\begin{center}
\includegraphics[scale=0.26]{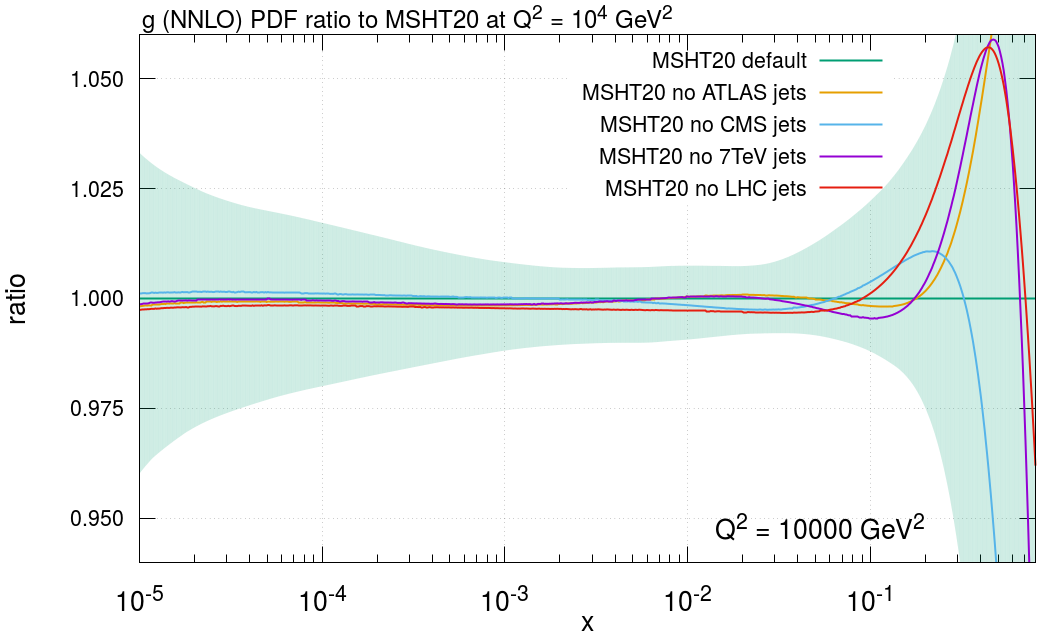}
\caption{\sf Gluon PDF at NNLO at $Q^2=10^4~\GeV^2$  comparing the MSHT20 global fit with the same fit upon removal of various subsections of the LHC jet data.}
\label{gluon_q210000_NNLO_novariousjets}
\end{center}
\end{figure}

We find some tensions between the different sets of jet data included in the global fit. These can be illustrated by removing subsets of the jet data sets from the fit and determining the impact on the gluon PDF, as shown in Fig.~\ref{gluon_q210000_NNLO_novariousjets}. This illustrates their different pulls on the high $x$ gluon, and their relative overall importance at given $x$ values can be seen by comparing to the fit with no LHC jet data included. There is clearly a slight tension between the CMS jet data and the ATLAS jet data in the region $0.3 \lesssim x \lesssim 0.5$, with the former pulling the gluon up (and so upon its removal the gluon is reduced) and the latter pulling the gluon down in this region (again therefore upon its removal in Fig.~\ref{gluon_q210000_NNLO_novariousjets} the high $x$ gluon raises). The default MSHT20 gluon in this region is then a balance of these competing effects, although it is important to note that all the resulting gluon PDFs shown in Fig.~\ref{gluon_q210000_NNLO_novariousjets} are within the MSHT20 error bands across the whole $x$ range. Given the behaviour of the gluon in this region upon removal of the 7~TeV data (both ATLAS and CMS) is similar to the behaviour when the ATLAS jets data are removed, this implies both that the ATLAS 7~TeV data dominates over the CMS 7~TeV data, and that it is most likely the CMS 8~TeV data which are in tension with the ATLAS jets data. It is also interesting to note that the impact of removing the LHC jet data as a whole is similar to the impacts of removing the ATLAS jet data or of removing the 7~TeV data, again implying that the ATLAS 7~TeV data have the largest pull, this supports observations seen previously in \cite{MMHTjets} for the 7~TeV data alone. Nonetheless, the CMS 8~TeV data in particular still remain important, with the behaviour in the region of $ 0.08 \lesssim x \lesssim 0.2$ a balance between its effect pulling the gluon down (and so upon its removal the gluon raises here) and the effect of the 7~TeV and ATLAS data in pulling the gluon up here (lowering the gluon when it is removed). Indeed in this $x$ range the no LHC jets fit is arguably closest to the no CMS jets data, suggesting it (and specifically its 8~TeV data set) has the largest pull there.

\subsection{$Z$ boson $p_T$ distributions}

For the first time we include data on the $Z$ boson $p_T$ distribution. We fit the ATLAS 8~TeV distributions \cite{ATLASZpT}, choosing the absolute, rather than normalized
cross sections. The data in the mass bin containing the $Z$ peak, i.e. $[66,116]~\GeV$, are 
double-differential in the lepton pair $p_T$ and rapidity.  
All other mass bins are presented as distributions single differential in $p_T$. We choose to fit the 
maximal amount of data, i.e.  the double-differential data for the $Z$-peak mass bin, and for all
other mass bins the single differential data. We cut all data with $p_T<30~\GeV$ due to the likely large 
influence of resummation and nonperturbative corrections. We do not impose a cut to exclude data at high $p_\perp$, but do include the electroweak corrections as given in~\cite{Boughezal:2017nla}.

The data act as a constraint on the gluon distribution, and are also sensitive to the value of the 
strong coupling $\alpha_S$. We find a quite high value of $\chi^2/N_{\rm pt}\sim 1.8$ for the data, due in part to significant 
tensions with a number of other data sets, including the HERA inclusive structure function data and 
the ATLAS 7~TeV $W$, $Z$ data (seen in more detail later in Section~\ref{tensions}, and in this case the ``no $Z$ $p_T$'' column of Table~\ref{tab:BCDMSD0WasymATLASZpt_delchisqtable}). In addition, the NNLO corrections are quite important in this case, and the data 
very precise, so it is possible that even beyond NNLO further theory corrections might result in a markedly improved
data/theory comparison.

We note that  CT18 and NNPDF3.1 have fit a variant of this data set, both finding rather lower values of $\chi^2/N_{\rm pt}\sim 1$. In the NNPDF case, a cut is imposed of $p_\perp < 150~\GeV$ in the $Z$ peak region, it is argued in order to remove sensitivity to the region where EW corrections are important. However, we (and NNPDF) do include such corrections to the theory, so we do not believe there is a strong rationale for removing this region. However, cutting out these 12 data points has relatively little impact on the fit quality, and hence is not a major source of difference, though arguably another motivation for not cutting these points out. On the other hand, as originally described in~\cite{Boughezal:2017nla}, NNPDF do include an additional $1\%$ fully uncorrelated source of uncertainty in order to account for a combination of MC errors on the $K$-factors, other theory uncertainties and potentially underestimated experimental errors. We do not see any clear evidence that the MC uncertainty is as large as $1\%$ and indeed as described in Section~\ref{sec:theoryupdate}, a more accurate way to account for this source of uncertainty would be to fit the $K$-factors according to a smooth distribution. In terms of the other two possible sources, we would argue that any contributions to these  should be clearly identified before including such a source of error and dealt with in a systematic way across all data sets. Other theory uncertainties could for example fall under the more general treatment that is required as described in Section~\ref{sec:theoryunc}, while clearly firm evidence is needed before concluding that the experimental uncertainties might be underestimated. Indeed, in the latter case we have discussed numerous examples in this paper where this may be the case, in particular in terms of the degree of correlation in various systematic uncertainties, and have taken care to identify these.

Taking our $K$-factors and excluding the correlated error sources associated with our smooth fit leads to a deterioration in the fit quality to $\sim 2.1$ per point, and thus the improvement from these is $\sim 0.3$ per point. However, if we then include a $1\%$ source of uncorrelated uncertainty (still excluding the correlated error source from the smooth fit), the impact is dramatic, giving $\chi^2/N_{\rm pt}\sim 1.1$. (1.2 if we take the $K$-factor values directly, without smoothing). While we do not use the same NNLO $K$-factors as in the NNPDF analysis, rather taking the \texttt{NNLOjet} calculation~\cite{NNLOZpT,Bizon:2018foh}, the quoted MC uncertainties (and central values within these errors) are comparable to the theory used in~\cite{NNPDF3.1,Boughezal:2017nla}. 

Turning to the CT18 fit, a $0.5\%$ uncorrelated uncertainty is included to account for the MC uncertainties in the same \texttt{NNLOjet} $K$-factors that we use. The size of this is more consistent with the corresponding MC errors, and indeed we find that taking this rather than our smooth fit gives a rather similar fit quality. In addition, CT fit to a much more limited subset of the ATLAS data, corresponding only to the $m_{ll} > 46~\GeV$ bins, and with the rapidity integrated, $|y_{ll}|<2.4$, data in the $Z$ peak region. A more restrictive region of $45 < p_\perp < 150$ GeV is also taken for all mass bins. As we fit to the double differential distribution in the $Z$ peak region, it is difficult to make a direct comparison, though requiring  $m_{ll} > 46~\GeV$ and  $45 < p_\perp < 150~\GeV$  leads to a relatively mild improvement to $\chi^2/N_{\rm pt}\sim 1.7$. It is natural to expect that the fit quality to the rapidity integrated $Z$ peak distribution may be better, given this places less constraint on the PDFs, but without a completely like--for--like comparison it is unclear whether this is the source of the difference.

We have also tried to fit to the 8~TeV CMS $Z$  $p_T$ data \cite{CMSZpT}, which are double differential in 
$p_T$ and rapidity. However, as in \cite{NNPDF3.1, Boughezal:2017nla} we find that although the fit in most rapidity 
bins is acceptable, in the highest bin, $1.6 \leq y \leq 2$, the fit quality becomes very poor. Although 
it is not possible to make a direct comparison with ATLAS data to determine compatibility, it is clear 
conventional PDFs cannot describe the data in this rapidity bin. Lacking an understanding of this we choose
to omit the whole CMS data set rather than fit some subset of it, given that if there are apparent issues in some rapidity bins there is no reason to believe these might not affect all bins.

\subsection{Total $t\bar{t}$ cross section data}

A number of measurements of the total $t\overline{t}$ cross section from ATLAS, CMS and the Tevatron were included in the MMHT14 fit. In the current fit our focus is on the more constraining single and double differential top quark data, which we discuss in detail below. Thus, while we do include four additional total cross section measurements at 8~TeV from ATLAS~\cite{Aad:2015pga} and CMS~\cite{Chatrchyan:2013faa,Khachatryan:2014loa,Khachatryan:2015fwh}, this is by no means exhaustive of all the available data at this energy. We defer a more detailed study of the impact of all such data points to future studies. As in~\cite{MMHT14}  we fit with a default value of $m_t = 172.5~\GeV$ and allow the cross section to vary, applying a $\chi^2$ penalty with one sigma corresponding to $3\%$, which corresponds to a $1~\GeV$ change in the mass. At NNLO the best fit corresponds to $m_t=172.9~\GeV$, in very good agreement with the measured value of the top pole mass, $173.2 \pm 0.9~\GeV$~\cite{PDG2019}. As in~\cite{MMHT14} at NLO a lower value is preferred; we find $m_t = 169.9~\GeV$.

\subsection{Data on $t\bar{t}$ single differential pair production}

We include single differential top data at 8~TeV from ATLAS in the lepton + jet~\cite{ATLASsdtop} and dilepton~\cite{ATLASttbarDilep08_ytt} channels, and from CMS~\cite{CMSttbar08_ytt} in the lepton + jet channel. The effect of these data sets with respect to a baseline close to MMHT14 were studied in detail in~\cite{Bailey:2019yze}, and we only highlight the key elements here. 

The ATLAS lepton+jet data are given differentially in the top quark pair invariant mass, $m_{t\overline{t}}$, and rapidity, $y_{t\overline{t}}$, and the individual top quark/antiquark transverse momentum, $p_\perp^t$, and rapidity, $y_t$, with full statistical correlations provided~\cite{ATL-PHYS-PUB-2018-017}, allowing all four distributions to be fit simultaneously. However, as discussed in~\cite{Bailey:2019yze}, and seen also by~\cite{ATL-PHYS-PUB-2018-017,Hou:2019efy,Amoroso:2020lgh}, the fit quality is in this case extremely poor. This is found to be driven primarily by the most significant sources of systematic error, and their correlations. These uncertainties, which correspond to the parton shower (p.s.), ISR/FSR and hard scattering uncertainties, are in particular 2--point errors evaluated using two choices of Monte Carlo (MC) generator or generator inputs, and for which the corresponding degree of correlation in the error source between and within the four distributions is certainly not precisely known. In particular, it is assumed that any correction factor evaluated by this two--point procedure should be applied in a fully correlated way across all bins.

Such an assumption is certainly too strong, may well bias the fit, and indeed as we see in the current case, it leads to a very poor fit quality. We therefore, as in~\cite{Bailey:2019yze}, choose to decorrelate the parton--shower error across the four distributions, while to be conservative we in addition take the decorrelation of \eqref{eq:ltrig} to split this uncertainty source for each distribution into two separate sources; this corresponds to our baseline fit. The impact of this is dramatic, as can be see in Table~\ref{table:ttbar1}: whereas the baseline experimental correlations gives $\chi^2/N_{\rm pts} = 6.84$, simply decorrelating the parton--shower uncertainty across the distributions gives a sizeable reduction to $\chi^2/N_{\rm pts} = 1.69$, while in addition allowing some freedom within the distributions gives $\chi^2/N_{\rm pts} = 1.04$. A further `maximal' decorrelation, where the above procedure is also followed for the ISR/FSR and hard scattering uncertainties (in principle a reasonable possibility), gives some further mild reduction, but the dominant effect is accounted for by the parton--shower error. Further investigation shows that there is a large degree of degeneracy in these three error sources, and hence decorrelation of any of the three error sources alone gives sufficient freedom to describe the data well.

\begin{table}
\begin{center}
\begin{tabular}{|c|c|c|c|}
\hline
 Baseline & No decor. &  parton shower across & Max decor.  \\ \hline
  1.04 & 6.84 & 1.69 & 0.81\\
\hline
\end{tabular}
\caption{\sf $\chi^2/N_{\rm pts}$ for the ATLAS 8~TeV differential top data, with $\alpha_S$ free, in the lepton+jet channel. Results for different correlation scenarios, explained in the text, are given.}
\label{table:ttbar1}
\end{center}
\end{table}

\begin{figure} 
\begin{center}
\includegraphics[scale=0.23]{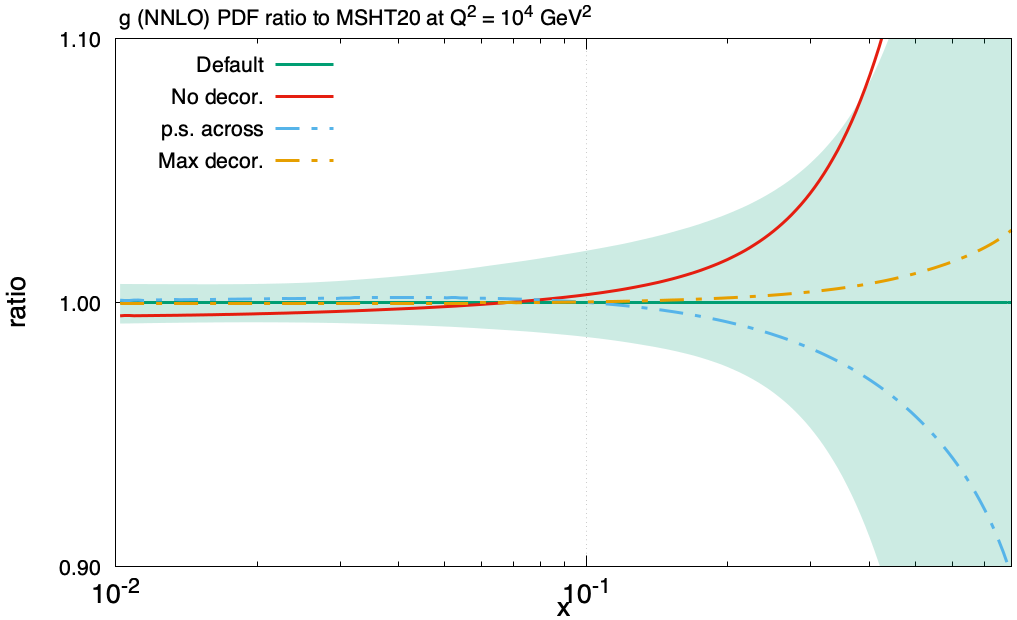}
\includegraphics[scale=0.23]{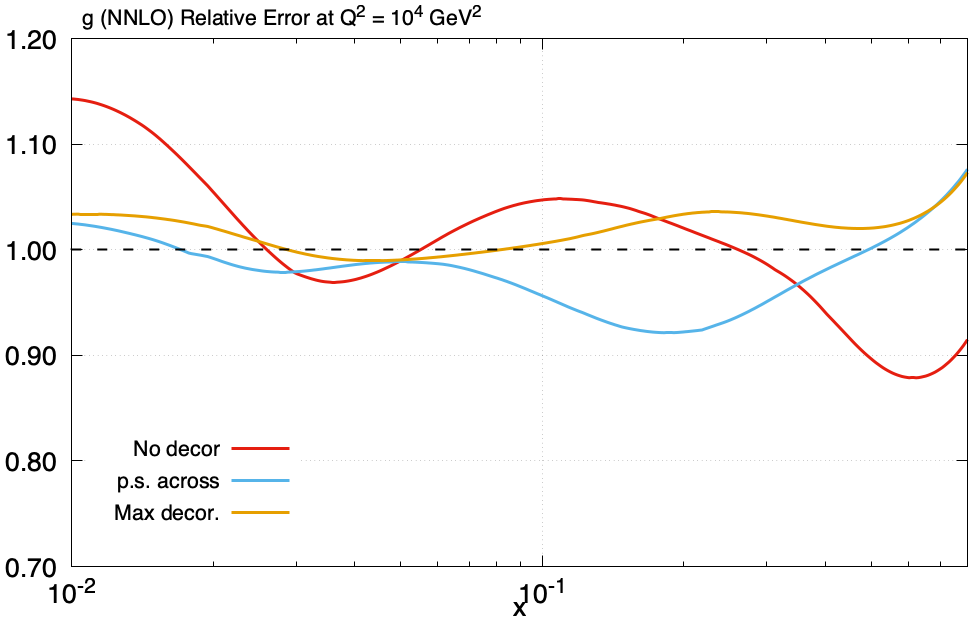}
\caption{\sf (Left) Ratio of gluon PDFs to MSHT20 baseline, with $\alpha_S$ free, at NNLO at $Q^2=10^4~\GeV^2$. The result of simultaneous fits to four absolute distributions within the ATLAS 8~TeV single differential top quark data set are shown: with no systematic error decorrelation (No decor), the parton shower error decorrelated across, but not within, the four distributions, and with the parton shower, ISR/FSR and hard-scattering decorrelated across and within all distributions (Max decor). (Right) Fraction symmetrised error with respect to the baseline fit.}
\label{fig:ttbar1}
\end{center}
\end{figure}

\begin{figure}
\begin{center}
\includegraphics[scale=0.23]{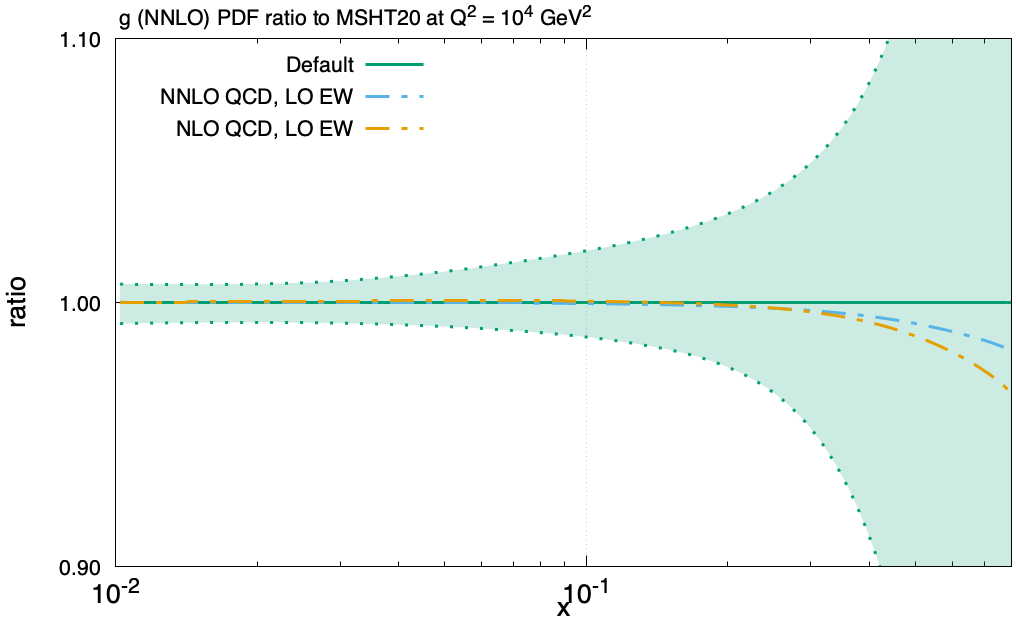}
\caption{\sf As in Fig.~\ref{fig:ttbar1} (left), but for fits with baseline systematic error treatment, but LO EW and NLO QCD + LO EW theory used in the theory matrix elements.}
\label{fig:ttbar2}
\end{center}
\end{figure} 

\begin{table}
\begin{center}
\begin{tabular}{|c|c|c|}
\hline
 Baseline & NNLO QCD, LO EW &  NLO QCD, LO EW \\ \hline
  1.04 & 0.92 & 1.66 \\
\hline
\end{tabular}
\caption{\sf $\chi^2/N_{\rm pts}$ for the ATLAS 8~TeV differential top data, with $\alpha_S$ free, in the lepton+jet channel. Results using different choices of theory for the matrix element calculation are shown.}
\label{table:ttbar2}
\end{center}
\end{table}

The impact on the gluon PDF is shown in Fig.~\ref{fig:ttbar1}. We can see that, although generally within PDF errors, there is a quite large difference between our baseline fit and the fit with the default systematic errors.  If we only decorrelate the parton shower uncertainty across distributions (`p.s. across'), the resulting gluon shows some reasonable deviation from the baseline at higher $x$, while within the maximal decorrelation scenario it is broadly the same.  The fractional symmetrised uncertainty with respect to the baseline is also shown: broadly, the default error gives a larger uncertainty apart from at the highest $x$ values, while the p.s. across decorrelation gives a smaller uncertainty in some regions, and the maximal decorrelation gives a slightly larger, though comparable, error band. Thus, we can see that, in contradiction to the case of the ATLAS jet data discussed in Section~\ref{LHCjets}, the treatment of the correlations of the experimental systematic errors does have a non--negligible impact on the extracted gluon. This is in  agreement with the findings of~\cite{Bailey:2019yze}. Indeed, as shown in Table~\ref{table:ttbar2}, while the impact of e.g. using NLO theory for the matrix element calculation leads to a quite large deterioration in the fit quality, as can be seen in Fig.~\ref{fig:ttbar2}, the impact on the extracted gluon is relatively mild, quite similar to changing the order of the EW corrections, and significantly smaller than that due to the treatment of the experimental error correlations.

Nonetheless, we consider our treatment to be the most reliable procedure one can follow in this case.  In particular, we have only taken those uncertainty sources for which there is a clear physical motivation for loosening the default correlations. After doing this a good data/theory description is achieved, and the data set can play a useful and reliable role in determining the gluon at high $x$. In~\cite{Amoroso:2020lgh} a similarly very poor description of the combined absolute distributions is found, but for normalized distributions a lower value of $\chi^2/N_{\rm pts} \sim 2.3$ is presented, with a reasonably large difference in the extracted gluon between the two cases found. It is argued that on this basis one should instead fit to the normalized distributions. However, we consider this to be a  rather problematic recommendation, as by fitting the normalized distributions alone one is simply discarding a 
potentially significant degree of correlation of the systematic uncertainties in the data, namely anything which is tied to the data normalisation, with no control over what is being removed\footnote{In~\cite{Amoroso:2020lgh} the total cross section is included in the fit, but not the cross correlations between this and the normalized distributions, which if included correctly would by construction simply correspond to a fit to the absolute cross section.}. In effect, one is allowing the various systematic shifts associated with the different sources of systematic error (which are the same in the absolute and normalized cases) to take values, which if translated to a prediction for the absolute distributions, would certainly give a poor fit quality. Thus one is effectively decorrelating the experimental systematic errors, but such that control over how this is done is lost. Clearly further analysis is needed to determine the extent to which this procedure and the one we outline above agree, or do not, in terms of the extracted gluon PDF, but this is certainly not guaranteed. We leave a detailed analysis of this to future studies.

Finally, for the ATLAS dilepton data, we find the fit quality to the $y_{tt}$ distribution is very good, although as no statistical correlations are provided we cannot investigate how this might change if a combined fit to the $y_{tt}$ and $m_{tt}$ distribution were performed. We fit the CMS $y_{tt}$ distribution (only given as normalized) in the lepton+jet channel, taking the systematic errors as completely uncorrelated; as discussed in~\cite{Bailey:2019yze}, it is unclear how one should treat the systematic errors in this case, but the quoted values, being all positive, certainly cannot be consistently interpreted as correlated errors for such a normalized distribution. We in addition remove the final bin from the fit so that the covariance matrix corresponding to this normalised distribution is non–singular.

\subsection{Data on $t\bar{t}$ double differential pair production}

We include CMS data for top quark pair production in the dilepton channel, presented double differentially in a variety of variables \cite{CMS8ttDD}: including the transverse momentum and rapidity of the top, $p_T(t)$, $y(t)$; the invariant mass, rapidity and transverse momentum of the top pair, $M(t\bar{t})$, $y(t\bar{t})$ and $p_T(t\bar{t})$; as well as the rapidity and angular separations of the top and antitop, $\Delta\eta(t,\bar{t})$ and $\Delta\phi(t,\bar{t})$. Six pairs of these variables are formed and normalized distributions are presented double differentially: $[p_T(t), y(t)]$, $[y(t),M(t\bar{t})]$, $[y(t\bar{t}),M(t\bar{t})]$, $[\Delta\eta(t,\bar{t}),M(t\bar{t})]$, $[p_T(t\bar{t}), M(t\bar{t})]$ and $[\Delta\phi(t,\bar{t}),M(t\bar{t})]$. Following the analysis of CT \cite{CTCMS8ttDD} which showed the first 3 of these had the best fit (achieving $\chi^2 \approx 1.25$) we investigated the same  3 distributions and found a similar behaviour. Ultimately we fit the $[p_T(t), y(t)]$ data and find a fit quality of $\chi^2 = 22.5$ for 15 points. The theory results are generated at up to NNLO using the \texttt{Fastnlo} grids provided in~\cite{CzakonCMS8ttDDtables}. The statistical correlations between the bins are included and so we must drop a bin of the normalised data to avoid singularities in the matrix inversion; we therefore drop the final bin which has both the highest $p_T(t)$ and highest $y(t)$ and so contains less information due to the relatively large statistical uncertainty of this bin. We treat the experimental PDF systematic uncertainty as fully uncorrelated, whilst we also partially decorrelate the $b$-tagging, hadronisation and hard scattering errors each into 4 components following a trigonometric decorrelation similar to that proposed for the ATLAS jet data in \cite{ATLAS8jets} and Section~\ref{LHCjets}. This is performed as each of these systematic errors do not preserve the normalisation of the data, not summing to 0 across all bins. Moreover, in the case of the PDF uncertainties it is clear that this cannot be correctly interpreted as a two--point fully correlated uncertainty alone.

\begin{figure}
\begin{center}
\includegraphics[scale=0.23]{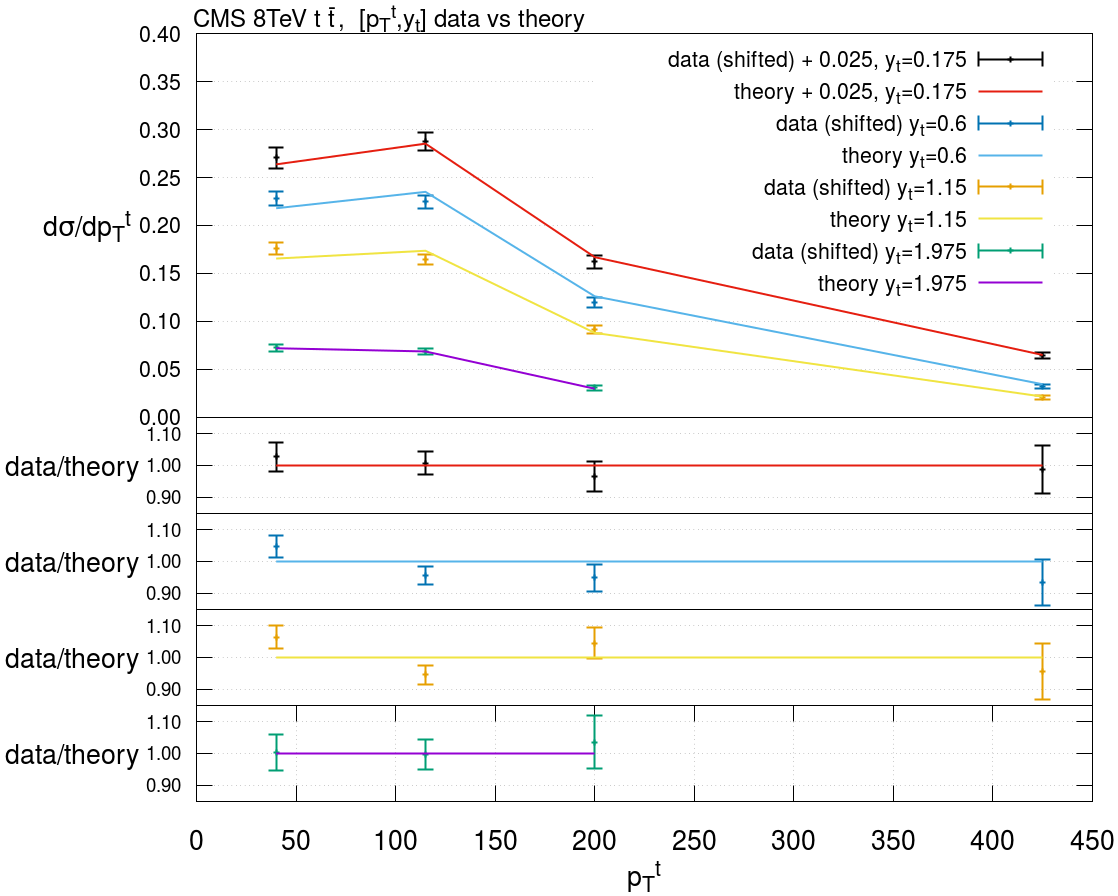}
\caption{\sf Data vs. MSHT20 NNLO theory for the CMS 8~TeV $t\bar{t}$ data differential in $[p_T^t,y_t]$. The errors plotted here are the total uncorrelated errors for each point which is the quadrature sum of the statistical and PDF errors for that point.}
\label{CMS8ttDD_datavstheory}
\end{center}
\end{figure}

The data versus theory comparison is shown in Fig.~\ref{CMS8ttDD_datavstheory}. The agreement is reasonable, with in general no clear systematic offset. One possible exception to this is in the second rapidity bin, where the theory undershoots in the first $p_\perp$ bin and then overshoots in the higher bins, although this effect is small and of order the errors. Given the data are already shifted by correlated systematics in these figures, the interpretation of this effect is not straightforward.

\section{Results for the MSHT20 global analysis \label{sec:5}}

In this section we discuss the overall fit quality and resulting PDFs, comparing with MMHT14 where relevant. 

\subsection{The values of the QCD coupling, $\alpha_S(M^2_Z)$}

We define the default NLO and NNLO PDFs to correspond to  a value $\alpha_S(M_Z^2)=0.118$, 
as for MMHT14.  This is the common value that is taken in the PDF4LHC15 combination\cite{PDF4LHC15}, and is consistent with the world average value of $\alpha_S(M_Z^2)=0.1179\pm 0.0010$~\cite{PDG2019}.
However, we also allow the value of $\alpha_S(M_Z^2)$ to vary as a free parameter in the 
fit at both NLO and at NNLO. We find that at NLO $\alpha_S(M_Z^2)=0.1203$ while at 
NNLO, $\alpha_S(M_Z^2)=0.1175$. Hence, we find as in \cite{MMHT14}, and similar to numerous
other $\alpha_S(M_Z^2)$ determinations from PDF fits, see e.g. \cite{ABMP16,ABMP16NLO,NNPDFas}, 
that the value found at NLO is about $0.003$
higher than at NNLO. At NNLO the best fit at $\alpha_S(M_Z^2)=0.1175$ is only 2 units in $\chi^2$ better than 
the default fit, while at NLO the best fit at $\alpha_S(M_Z^2)=0.1203$ is 50 units better. Whilst in our detailed comparisons here we use the default NLO fit at $\alpha_S(M_Z^2)=0.118$, a fit at $\alpha_S(M_Z^2)=0.120$ is also available, see Section~\ref{sec:access} for details.

We will discuss the details of the variation with $\alpha_S(M_Z^2)$ and the uncertainty in  
a PDF fit determination in a future publication, as we did for the MMHT study \cite{MMHTas}.

\subsection{The fit quality}

\begin{table}
\begin{center}
\begin{tabular}{l|c|c}\hline \hline
  Data set & NLO & NNLO \\ \hline
  BCDMS $\mu p$ $F_2$ \cite{BCDMS} &  169.4/163 & 180.2/163  \\ 
  BCDMS $\mu d$ $F_2$ \cite{BCDMS} &  135.0/151 & 146.0/151 \\ 
  NMC $\mu p$ $F_2$ \cite{NMC} &  142.9/123 & 124.1/123\\ 
  NMC $\mu d$ $F_2$ \cite{NMC} &  128.2/123 & 112.4/123\\ 
  NMC $\mu n/\mu p$ \cite{NMCn/p} & 127.8/148 & 130.8/148  \\ 
  E665 $\mu p$ $F_2$ \cite{E665} & 59.5/53 & 64.7/53 \\ 
  E665 $\mu d$ $F_2$ \cite{E665} &  50.3/53 & 59.7/53\\   
  SLAC $e p$ $F_2$ \cite{SLAC,SLAC1990} &  29.4/37 & 32.0/37 \\   
  SLAC $e d$ $F_2$ \cite{SLAC,SLAC1990} &  37.4/38 & 23.0/38\\     
  NMC/BCDMS/SLAC/HERA $F_L$ \cite{NMC,BCDMS,SLAC1990,H1FL,H1-FL,ZEUS-FL}  & 79.4/57 & 68.4/57\\ \hline
  E866/NuSea $pp$ DY \cite{E866DY} &  216.2/184 & 225.1/184\\
  E866/NuSea $pd/pp$ DY \cite{E866DYrat} &  10.6/15 & 10.4/15\\ \hline
  NuTeV $\nu N$ $F_2$ \cite{NuTeV}  & 43.7/53 & 38.3/53\\
  CHORUS $\nu N$ $F_2$ \cite{CHORUS}  & 27.8/42 & 30.2/42\\
  NuTeV $\nu N$ $x F_3$ \cite{NuTeV}  & 37.8/42 & 30.7/42\\  
  CHORUS $\nu N$ $x F_3$ \cite{CHORUS}  & 22.0/28 & 18.4/28\\
  CCFR $\nu N \rightarrow \mu \mu X$ \cite{Dimuon} &  73.2/86 & 67.7/86\\
  NuTeV $\nu N \rightarrow \mu \mu X$ \cite{Dimuon} &  41.0/84 & 58.4/84\\ \hline
  HERA $e^+ p$ CC \cite{H1+ZEUS} &  54.3/39 & 52.0/39\\
  HERA $e^- p$ CC \cite{H1+ZEUS} &  80.4/42 & 70.2/42\\
  HERA $e^+ p$ ${\rm NC}~820$~GeV \cite{H1+ZEUS} & 91.6/75 & 89.8/75\\ 
  HERA $e^+ p$ ${\rm NC}~920$~GeV \cite{H1+ZEUS} & 553.9/402 & 512.7/402\\
  HERA $e^- p$ ${\rm NC}~460$~GeV \cite{H1+ZEUS} & 253.3/209 & 248.3/209\\
  HERA $e^- p$ ${\rm NC}~575$~GeV \cite{H1+ZEUS} & 268.1/259 & 263.0/259\\
  HERA $e^- p$ ${\rm NC}~920$~GeV \cite{H1+ZEUS} & 252.3/159 & 244.4/159\\  
  HERA $e p$ $F_{2}^{\text{charm}}$ \cite{HERAhf} & 125.6/79 & 132.3/79\\  
  D{\O} II $p\bar{p}$ incl. jets \cite{D0jet} &  117.2/110 & 120.2/110\\
  CDF II $p\bar{p}$ incl. jets \cite{CDFjet} &  70.4/76 & 60.4/76\\
  CDF II $W$ asym. \cite{CDF-Wasym} &  19.1/13 & 19.0/13\\
  D{\O} II $W\rightarrow \nu e$ asym. \cite{D0Wnue} &  44.4/12 & 33.9/12\\
  D{\O} II $W \rightarrow \nu \mu$ asym. \cite{D0Wnumu} &  13.9/10 & 17.3/10\\
  D{\O} II $Z$ rap. \cite{D0Zrap}   & 15.9/28 & 16.4/28\\
  CDF II $Z$ rap. \cite{CDFZrap}  & 36.9/28 & 37.1/28\\
  D{\O} $W$ asym. \cite{D0Wasym}  & 13.1/14 & 12.0/14 \\ \hline
 \hline
\end{tabular}
\end{center}
\caption{\sf The values of $\chi^2/N$pts. for the non-LHC data sets included in the global fit at NLO and NNLO.}
\label{tab:chisqtable}
\end{table}

\begin{table} 
\begin{center}
\begin{tabular}{l|c|c}\hline \hline
  Data set & NLO & NNLO \\ \hline
  ATLAS $W^+$, $W^-$, $Z$ \cite{ATLASWZ} &  34.7/30 &  29.9/30\\ 
  CMS $W$ asym. $p_T > 35$~GeV \cite{CMS-easym} &  11.8/11 &  7.8/11\\ 
  CMS asym. $p_T > 25, 30$~GeV \cite{CMS-Wasymm} &  11.8/24 &  7.4/24\\ 
  LHCb $Z\rightarrow e^+e^-$ \cite{LHCb-Zee} &  14.1/9 &  22.7/9\\ 
  LHCb $W$ asym. $p_T > 20$~GeV \cite{LHCb-WZ} &  10.5/10 &  12.5/10\\ 
  CMS $Z\rightarrow e^+e^-$ \cite{CMS-Zee} &  18.9/35 &  17.9/35\\ 
  ATLAS High-mass Drell-Yan \cite{ATLAShighmass} &  20.7/13 &  18.9/13\\   
  CMS double diff. Drell-Yan \cite{CMS-ddDY} &  222.2/132 &  144.5/132\\   
  Tevatron, ATLAS, CMS $\sigma_{t\bar{t}}$ \cite{Tevatron-top}-\cite{CMS-top8} &  22.8/17 &  14.5/17\\     
  LHCb 2015 $W$, $Z$ \cite{LHCbZ7,LHCbWZ8} &  114.4/67 &  99.4/67\\
  LHCb 8~TeV $Z\rightarrow ee$ \cite{LHCbZ8}  & 39.0/17 & 26.2/17\\
  CMS 8~TeV $W$ \cite{CMSW8} &  23.2/22 & 12.7/22\\
  ATLAS 7~TeV jets \cite{ATLAS7jets} & 226.2/140 & 221.6/140\\
  CMS 7~TeV $W+c$ \cite{CMS7Wpc} & 8.2/10 & 8.6/10\\
  ATLAS 7~TeV high precision $W$, $Z$ \cite{ATLASWZ7f} & 304.7/61 & 116.6/61\\
  CMS 7~TeV jets \cite{CMS7jetsfinal} &  200.6/158 & 175.8/158\\
  CMS 8~TeV jets \cite{CMS8jets} &  285.7/174 & 261.3/174\\
  CMS 2.76~TeV jet \cite{CMS276jets} &  124.2/81 & 102.9/81\\
  ATLAS 8~TeV $Z$ $p_T$ \cite{ATLASZpT} &  235.0/104 & 188.5/104\\
  ATLAS 8~TeV single diff $t\bar{t}$ \cite{ATLASsdtop} &  39.1/25 & 25.6/25\\
  ATLAS 8~TeV single diff $t\bar{t}$ dilepton \cite{ATLASttbarDilep08_ytt} &  4.7/5 & 3.4/5\\
  CMS 8~TeV double differential $t\bar{t}$ \cite{CMS8ttDD} &  32.8/15 & 22.5/15\\
  CMS 8~TeV single differential $t\bar{t}$ \cite{CMSttbar08_ytt} &  12.9/9 & 13.2/9\\
  ATLAS 8~TeV High-mass Drell-Yan \cite{ATLASHMDY8} & 85.8/48 & 56.7/48\\
  ATLAS 8~TeV $$W$$ \cite{ATLASW8} &  84.6/22 & 57.4/22\\
  ATLAS 8~TeV $W+\text{jets}$ \cite{ATLASWjet}  & 33.9/30 & 18.1/30\\
  ATLAS 8~TeV double differential $Z$ \cite{ATLAS8Z3D} & 157.4/59 & 85.6/59\\ \hline
  Total & 5822.0/4363 & 5121.9/4363 \\ 
 \hline
\end{tabular}
\end{center}
\caption{\sf The values of $\chi^2/N$pts. for the LHC data sets included in the global fit and the overall global fit $\chi^2/N$  at NLO and NNLO. The corresponding values for the non-LHC data sets are shown in Table~\ref{tab:chisqtable}, and the total value corresponds to the sum over both tables.}
\label{tab:LHCchisqtable}
\end{table}

The fit quality is shown at NLO and NNLO in Tables~\ref{tab:chisqtable} 
and \ref{tab:LHCchisqtable}. Overall this is very good, but there are individual 
cases where the data set is less well described. Sometimes this is due to an inherent
difficultly in finding a good quality fit to the data due to large fluctuations 
compared to the data precision, sometimes even in the absence of fluctuations the data is poorly described by fixed--order QCD perturbation theory at the considered order, and sometimes the fit quality of a data set is 
clearly affected by tensions with other data in the fit. Often 
these tensions are between different types of data, so it may be a sign of missing 
theoretical corrections, e.g. being much larger for one type of process than another, 
but sometimes tensions exist between very similar types of data from 
different  experiments, in which case it is clear that it is the data that are in tension. 
In some cases the poor fit quality can be traced to constraints due a small number of correlated 
systematic uncertainties, and improvements are made by relaxing this correlation. 

The overall fit quality is now seen to be considerably better at NNLO than at NLO. 
We note that the fit quality is shown for $\alpha_S(M_Z^2)=0.118$ at both NLO and NNLO. As discussed 
above, this makes almost no difference at NNLO, but the best fit at $\alpha_S(M_Z^2)=0.1203$ at NLO
has a $\chi^2$ that is 50 points lower. However, this makes little impact on the conclusion that 
the NNLO fit quality is far superior. The lower fit quality at $\alpha_S(M_Z^2)=0.1203$ for NLO is essentially 
all due to an improvement in the description of the precision ATLAS and LHCb $W$, $Z$ data, which 
contributes 
$\Delta \chi^2 =-68$. This is partially balanced by the ATLAS $Z$ $p_T$ data, where $\Delta \chi^2=14$ 
is seen. For structure functions, jet data and top quark data individual sets pull either in the direction
of higher $\alpha_S(M_Z^2)$, e.g. NMC structure function data, Tevatron and 2.76~TeV CMS jet data and
inclusive top antitop cross sections, or  in the direction
of lower $\alpha_S(M_Z^2)$, e.g. BCDMS structure function data, most ATLAS and CMS jet data and CMS
single differential top data. However, for each data type the net pull between $\alpha_S(M_Z^2)=0.118$
and $\alpha_S(M_Z^2)=0.1203$ is small. 
The improvement in the fit quality at NNLO relative to NLO for the whole global fit
 is 700 units in $\chi^2$. For the common data points in the different fit orders in 
the MMHT14 analysis \cite{MMHT14} the NNLO fit quality was only 200 units better (most of this due to the 7~TeV double differential
Drell-Yan data, where the lowest mass bin is effectively calculated at one lower order than other mass bins), and
in earlier fits there has been little difference in fit quality at NLO and NNLO, with NLO sometimes lower.

This increase in the difference between NNLO and NLO is due to the fact that perturbative corrections are often larger for hadron-collider processes than for deep inelastic scattering, 
and because the quantity and precision of the LHC data is now such as to make the superiority of NNLO calculations 
completely clear. Indeed, we note that overall there is little to choose between the NLO and NNLO quality in Table
\ref{tab:chisqtable} (non-LHC data sets in the fit), with some data sets preferring NLO and some NNLO, but never that strongly, whereas in 
Table \ref{tab:LHCchisqtable} (LHC data sets in the fit) most data sets are fit better at NNLO, and many of them very significantly so. 
This is particularly clear for the precision electroweak boson data, as highlighted earlier, but also quite clearly true 
for LHC jet data and the top quark data, both inclusive and differential.

\subsection{Central PDF sets and uncertainties} \label{centralpdfs}

The parameters for the central PDF sets at NLO and NNLO are shown in 
Table~\ref{tab:PDFParameterstable}. In order to describe the uncertainties 
on the PDFs we apply
the same general procedure as in \cite{MSTW} (originally presented in a 
similar, but not identical form in 
\cite{Hessian}), i.e. we use the Hessian approach
with a dynamical tolerance, and hence obtain a set of PDF eigenvector
sets, each corresponding to $68\%$ confidence level uncertainty and being 
orthogonal to each other. 

\begin{center}
\begin{table} 
\begin{minipage}[t]{0.5\linewidth}
\begin{tabular}[t]{c|l|l}\hline \hline
 Parameter &  NLO & NNLO \\ \hline
 $\alpha_S(M_Z^2)$ & 0.118 & 0.118 \\ \hline
 $A_u$ & 3.32065 & 3.75143 \\
 $\delta_u$ & 0.60550 & 0.68989 \\
 $\eta_u$ & 2.8039 & 2.7360 \\
 $a_{u,1}$ & -0.14968 & 0.081549 \\
 $a_{u,2}$ & -0.024783 & -0.042876 \\
 $a_{u,3}$ & 0.22001 & 0.23606 \\
 $a_{u,4}$ & 0.11837 & 0.12551 \\
 $a_{u,5}$ & 0.0078489 & 0.0032555 \\
 $a_{u,6}$ & 0.046382 & 0.035996 \\ \hline
 $A_d$ & 1.19171 & 1.52273 \\
 $\delta_d$ & 0.42331 & 0.45142 \\
 $\eta_d-\eta_u$ & 0.81395 & 0.64482 \\
 $a_{d,1}$ & -0.38975 & -0.46093 \\
 $a_{d,2}$ & -0.24774 & -0.0072694 \\
 $a_{d,3}$ & 0.15748 & -0.081052 \\
 $a_{d,4}$ & 0.33173 & 0.47975 \\
 $a_{d,5}$ & -0.20815 & -0.26145 \\
 $a_{d,6}$ & 0.18960 & 0.20334 \\ \hline
 $A_S$ & 48.446 & 56.426 \\
 $\delta_S$ & -0.083176 & -0.0099513 \\
 $\eta_S$ & 12.743 & 12.816 \\
 $a_{S,1}$ & -1.6278 & -1.6176 \\
 $a_{S,2}$ & 0.91925 & 0.92118 \\
 $a_{S,3}$ & -0.35012 & -0.34753 \\
 $a_{S,4}$ & 0.089609 & 0.089414 \\
 $a_{S,5}$ & -0.015122 & -0.013206 \\
 $a_{S,6}$ & 0.0046141 & 0.0045178 \\ \hline
 $A_{\rho}$ & 1.5965 & 3.4499 \\
 $\eta_{\rho}$ & 4.4446 & 5.7130 \\
 $a_{{\rho},1}$ & 0.57502 & -0.28315 \\
 $a_{{\rho},2}$ & -1.8328 & -1.0735 \\
 $a_{{\rho},3}$ & 0.38974 & 0.51791 \\
 $a_{{\rho},4}$ & 0.95378 & 0.52812 \\
 $a_{{\rho},5}$ & -0.92560 & -0.66435 \\
 $a_{{\rho},6}$ & 0.18844 & 0.25281 \\ \hline
\end{tabular}
\end{minipage}
\begin{minipage}[t]{0.5\linewidth}
\begin{tabular}[t]{c|l|l}\hline \hline
 Parameter & NLO & NNLO \\ \hline
  $A_g$ & 1.1807 & 0.95893 \\
 $\delta_g$ & -0.51949 & -0.46745 \\
 $\eta_g$ & 2.9762 & 2.1278 \\
 $a_{g,1}$ & -1.1680 & -1.1361 \\
 $a_{g,2}$ & 0.63137 & 0.73761 \\
 $a_{g,3}$ & -0.35736 & -0.53190 \\
 $a_{g,4}$ & 0.17994 & 0.29747 \\
 $A_{g-}$ & -0.39226 & -0.33211 \\
 $\delta_{g-}$ & -0.50084 & -0.47700 \\
 $\eta_{g-}$ & 75.058 & 63.519 \\ \hline 
  $A_{s+}$ & 99.573 & 93.217 \\
 $\delta_{s+}$ & -0.083176 & -0.0099513 \\ 
 $\eta_{s+}$ & 19.730 & 17.993 \\
 $a_{{s+},1}$ & -1.6720 & -1.6685 \\
 $a_{{s+},2}$ & 0.95463 & 0.95522 \\
 $a_{{s+},3}$ & -0.32818 & -0.32585 \\
 $a_{{s+},4}$ & 0.029202 & 0.024497 \\
 $a_{{s+},5}$ & 0.026356 & 0.031126 \\
 $a_{{s+},6}$ & -0.0096678 & -0.011852 \\ \hline
   $A_{s-}$ & -0.0091729 & -0.017013 \\
 $\delta_{s-}$ & 0.22208 & 0.22208 \\
 $\eta_{s-}$ & 5.5946 & 6.2655 \\
 $x_0$ & 0.034682 & 0.030230 \\ \hline
\end{tabular}
\end{minipage}
\caption{\sf \sf The  optimal  values  of  the  input  PDF  parameters  (as  defined  in  Section~\ref{sec:inputPDF})  at $Q_0^2 = 1~\GeV^2$ determined from the global analyses for our default sets with $\alpha_S(M_Z^2) = 0.118$. $A_u$, $A_d$, $A_g$ and $x_0$, are determined from sum rules and are not fitted parameters.}
\label{tab:PDFParameterstable}
\end{table}
\end{center}

\subsubsection{Procedure to determine PDF uncertainties}

If we have input 
parameters $\{a_i^0\}=\{a_1^0,\ldots,a_n^0\}$,  then we write
\begin{equation} \label{eq:hessian}
  \Delta\chi^2_{\rm global} \equiv \chi^2_{\rm global} - \chi_{\rm min}^2 = \sum_{i,j=1}^n H_{ij}(a_i-a_i^0)(a_j-a_j^0),
\end{equation}
where the Hessian matrix $H$ has components
\begin{equation}
  H_{ij} = \left.\frac{1}{2}\frac{\partial^2\,\chi^2_{\rm global}}{\partial a_i\partial a_j}\right|_{\rm min}.
\end{equation}
The uncertainty on a quantity $F(\{a_i\})$ is then obtained from 
standard linear error propagation:
\begin{equation} \label{eq:heserror}
  \Delta F = T \sqrt{\sum_{i,j=1}^n\frac{\partial F}{\partial a_i}C_{ij}\frac{\partial F}{\partial a_j}},
\end{equation}
where $C\equiv H^{-1}$ is the covariance matrix, and $T = \sqrt{\Delta\chi^2_{\rm global}}$ is the ``tolerance'' for the required confidence interval,
usually defined to be $T=1$ for $68\%$ confidence level. 
We diagonalise the covariance (or Hessian) matrix 
\cite{Hessian}, and work in terms of the eigenvectors.  The covariance matrix 
has a set of normalised {\it orthonormal} eigenvectors $v_k$ defined by
\begin{equation} \label{eq:eigeq}
  \sum_{j=1}^n C_{ij} v_{jk} = \lambda_k v_{ik},
\end{equation}
where $\lambda_k$ is the $k^{\rm th}$ eigenvalue and $v_{ik}$ is the $i^{\rm th}$ 
component of the $k^{\rm th}$ orthonormal eigenvector ($k = 1,\ldots,n$).  
The parameter displacements from the global minimum  are expanded 
in terms of rescaled eigenvectors $e_{ik}  \equiv \sqrt{\lambda_k}v_{ik}$:
\begin{equation} \label{eq:component}
  \Delta a_i\equiv a_i - a_i^0 = \sum_k e_{ik} z_k,
\end{equation}
i.e. the $z_k$ are the coefficients when we express a change in parameters away from 
their best fit values in terms of the rescaled eigenvectors, and a change 
in parameters corresponding to $\Delta \chi^2_{\rm global}=1$ corresponds to $z_k=1$.
This results in the simplification
\begin{equation} \label{eq:hessiandiag}
  \chi^2_{\rm global} = \chi^2_{\rm min} + \sum_k z_k^2.
\end{equation}
Eigenvector PDF sets $S_k^\pm$ are then produced with parameters given by
\begin{equation}
  a_i(S_k^\pm) = a_i^0 \pm t\,e_{ik},
\end{equation}
with $t$ adjusted to give the desired tolerance $T = \sqrt{\Delta\chi^2_{\rm global}}$.

As in MSTW08 and MMHT14 we 
do not determine the size of the eigenvectors using the standard
$\Delta \chi^2=1$ or $T=1$ rule, but allow $T \ne 1$ to account, primarily, 
for the  
tensions in fitting the different data sets within fixed order perturbative 
QCD. Rather than use a fixed value of $T$, we use the ``dynamical 
tolerance'' procedure devised in \cite{MSTW}.
In brief, we define the 68\%  confidence level region for each data set $n$ (comprising $N$ data points) by the condition that
\begin{equation} \label{eq:68percentCL}
  \chi_n^2 < \left(\frac{\chi_{n,0}^2}{\xi_{50}}\right)\xi_{68},
\end{equation}
where $\xi_{68}$ is the 68th percentile of the $\chi^2$-distribution with 
$N$ degrees of freedom, and $\xi_{50}\simeq N$ is the most probable value.  
For each eigenvector (in each of the two directions) we then determine the 
values of $t$ and $T$ for which the $\chi_n^2$ for 
each data set $n$ are minimised, together with $68\%$ confidence level 
limits defined by values at which Eq.~(\ref{eq:68percentCL}) ceases
to be satisfied.  For full details of the 
``dynamical tolerance'' procedure see Section 6.2 of \cite{MSTW}.  

In the limit that  Eq.~(\ref{eq:hessian}) is exact, i.e. there are no 
significant corrections to quadratic behaviour, $t\equiv T$. 
We limit our number of eigenvectors so that this is true to a reasonable 
approximation. This results in the PDF eigenvector sets being 
obtained by fixing a number of the parameters at their best-fit values. 
Letting these additional parameters go free when determining the eigenvectors 
would result in a large degree of correlation between some parameters and to
significant violations in $t\approx T$. In practice, for the resulting additional
eigenvectors the $\chi^2$ would remain flat while PDF parameters almost exactly 
compensate each other, then rise extremely rapidly when this compensation starts to fail. 
This generally results in the required $T$ being reached for rather small $t$, and small changes in the PDFs,
and the resulting extra uncertainty in the PDFs would be minimal. 

\subsubsection{Uncertainties of the MSHT20 PDFs}\label{PDFuncertainties}

The increase in the parameterisation flexibility in the present analysis 
leads to an increase in the number of parameters left free in the 
determination of the PDF uncertainties, as compared to the MMHT14 analysis.  
These additional parameters can then be constrained due to the extra constraints from the new data in the fit. We now have 32 eigenvector pairs, 
as compared to the 20 in MSTW08~\cite{MSTW} or the 25 in MMHT14~\cite{MMHT14}. 
The 32 parameters left free for the determination of the eigenvectors 
consist of: 

$\delta_u$, $\eta_u$, $\delta_d$, $\eta_d - \eta_u$, $A_S$, $\eta_S$, $\eta_g$, $a_{u,2}$, $a_{g,2}$, $\delta_g$, $a_{d,2}$, $\delta_S$, $a_{S,2}$, $\delta_{g'}$, $A_{s+}$, $\eta_{s+}$, $a_{s+,2}$, $A_{s-}$, $\eta_{s-}$, $a_{u,3}$, $a_{d,3}$, $a_{S,3}$, $a_{s+,3}$, $a_{g,3}$, $a_{u,6}$, $a_{d,6}$, $a_{S,6}$, $a_{s+,5}$, $A_{\rho}$, $a_{\rho,1}$, $a_{\rho,3}$, $a_{\rho,6}$. 

In summary, that is the small $x$ power $\delta$; high $x$ power $\eta$; normalisation $A$ (where not already set by sum rules); second, third and sixth  Chebyshev coefficients $a_{2,3,6}$ for the up valence, down valence, sea, gluon, $\bar{d}/\bar{u}$ first generation antiquark ratio, strange sea sum and strange sea asymmetry with several notable exceptions and alterations. The up valence, down valence and gluon normalisations, $A_u$, $A_d$, $A_g$ are not free as they are set by sum rules and momentum conservation, whilst for the gluon there is no sixth Chebyshev coefficient so this is replaced by the low $x$ power of the additional term $\delta_{g'}$. For the strangeness, $s+\bar{s} \equiv s_{+}$, the parameter $\delta_{s+}$ is set to $\delta_{S}$, while for the asymmetry, $s_{-} \equiv s-\bar{s}$, the parameter $\delta_{s-}$ is not free, as this distribution is less constrained by data. In addition, the fifth Chebyshev coefficient rather than the sixth is used for $s_{+}$, and there are no Chebyshev coefficients at all for the $s_{-}$. Finally, there is also no small $x$ power for the $\bar{d}/\bar{u} \equiv \rho$ in our parameterisation so it cannot be free in the eigenvector uncertainty scans, whilst its high $x$ power is not set free as it is relatively unconstrained and its first Chebyshev coefficient rather than second is used. This results in 32 free parameters out of which to form the eigenvectors for the uncertainty determination. This corresponds to one more than in MMHT14
for all PDFs other than those for the strange and antistrange quarks. For $s_{+}$ there are now two more 
parameters left free in the eigenvector determination, but for $s_{-}$ there are no more. 

During the determination of the eigenvectors, all deuteron parameters, 
free coefficients for nuclear corrections and all parameters associated 
with correlated uncertainties, including normalisations, are allowed 
to vary (with appropriate $\chi^2$ penalty). Hence, the true best fit is obtained at all times. 

The most constraining data set for each eigenvector direction, and also 
the values of $t$ and $T$ are shown in Tables~\ref{tab:NLOevecdatasetstable} and~\ref{tab:NNLOevecdatasetstable} for the NLO and NNLO cases, respectively. The eigenvectors here are ordered by size of the eigenvalue and so the ordering is similar, but not identical, between NLO and NNLO due to the underlying similarities in the fits. The final column of the tables also provides the primary parameter for that eigenvector, i.e. the parameter in the input parameterisation with the largest numerical contribution to each eigenvector. This can be useful in understanding the most constraining data sets, however in many cases this parameter is not really dominant as there are often several parameters which form significant parts of  an eigenvector, with only small differences in the magnitude of their contributions. Moreover, it is often the case that several data sets provide similar constraints on an eigenvector direction and so, whilst the information in these tables provides a useful insight, it should not be over-interpreted.

\begin{table} 
\small
\begin{center}
\def\arraystretch{0.92}
\begin{tabular}{|>{\arraybackslash}m{1.035cm}|>{\arraybackslash}m{0.525cm}|>{\arraybackslash}m{0.525cm}|>{\arraybackslash}m{5.05cm}|>{\arraybackslash}m{0.525cm}|>{\arraybackslash}m{0.525cm}|>{\arraybackslash}m{5.05cm}|>{\arraybackslash}m{1.76cm}|}\hline
e- ~~~~~ vector& + \hspace{0.1cm} t & + \hspace{0.1cm} T & most constraining data set & - \hspace{0.2cm} t & - \hspace{0.2cm} T & most constraining data set & primary parameter\\ \hline
1 & 6.88 & 6.86 & ATLAS 7 TeV high prec. $W$,$Z$ & 4.71 & 4.71 & ATLAS 8 TeV $Z$ & $a_{S,3}$ \\
2 & 4.33 & 3.99 & NMC $\mu d$ $F_2$ & 5.52 & 5.86 & NuTeV $\nu N \rightarrow \mu \mu X$ & $a_{s+,5}$ \\
3 & 3.61 & 3.17 & NMC $\mu d$ $F_2$ & 3.37 & 3.69 & NuTeV $\nu N \rightarrow \mu \mu X$ & $a_{S,6}$ \\
4 & 2.80 & 2.77 & CCFR $\nu N \rightarrow \mu \mu X$ & 4.19 & 4.18 & NuTeV $\nu N \rightarrow \mu \mu X$ & $a_{S,2}$ \\ 
5 & 2.87 & 3.21 & CCFR $\nu N \rightarrow \mu \mu X$ & 2.52 & 2.08 & NuTeV $\nu N \rightarrow \mu \mu X$ & $a_{S,3}$ \\ 
6 & 4.77 & 4.41 & BCDMS $\mu p$ $F_2$ & 4.19 & 4.53 & ATLAS 8 TeV $Z$ & $a_{s+,2}$ \\
7 & 3.10 & 3.60 & NMC/... $F_L$ & 6.24 & 5.69 & NMC $\mu d$ $F_2$ & $\delta_{g'}$ \\ 
8 & 3.31 & 2.94 & D{\O} II $W\rightarrow \nu e$ asym. & 1.52 & 1.83 & D{\O} $W$ asym. & $\delta_{u}$ \\ 
9 & 4.84 & 4.59 & NuTeV $\nu N$ $x F_3$ & 2.14 & 2.38 & NuTeV $\nu N \rightarrow \mu \mu X$ & $A_{s-}$ \\ 
10 & 5.00 & 4.90 & D{\O} II $W \rightarrow \nu \mu$ asym. & 2.96 & 3.05 & BCDMS $\mu p$ $F_2$ & $a_{u,6}$ \\ 
11 & 3.78 & 3.72 & ATLAS 8 TeV $Z$ $p_T$ & 1.42 & 1.49 & NuTeV $\nu N \rightarrow \mu \mu X$ & $\delta_{S}$ \\ 
12 & 1.70 & 1.76 & D{\O} $W$ asym. & 4.82 & 4.77 & ATLAS 8 TeV $W$& $\delta_{S}$ \\ 
13 & 2.98 & 3.31 & CMS asym. $p_T > 25, 30$ GeV & 2.79 & 2.42 & ATLAS 8 TeV $W$& $a_{u,3}$ \\ 
14 & 3.91 & 5.15 & CMS 8 TeV jets & 3.59 & 2.23 & ATLAS 8 TeV $Z$ $p_T$ & $a_{g,3}$ \\ 
15 & 2.88 & 3.01 & BCDMS $\mu p$ $F_2$ & 2.59 & 2.46 & D{\O} $W$ asym. & $a_{d,6}$ \\
16 & 3.18 & 2.65 & E866/NuSea $pd/pp$ DY & 2.18 & 2.63 & BCDMS $\mu p$ $F_2$ & $a_{d,6}$ \\ 
17 & 4.42 & 4.26 & E866/NuSea $pd/pp$ DY & 1.53 & 1.70 & E866/NuSea $pd/pp$ DY & $a_{\rho,3}$ \\ 
18 & 2.30 & 2.11 & E866/NuSea $pd/pp$ DY & 5.20 & 5.86 & HERA $e^+ p$ NC 920 GeV & $\delta_{g}$ \\
19 & 3.18 & 2.68 & NuTeV $\nu N \rightarrow \mu \mu X$ & 1.44 & 1.82 & D{\O} $W$ asym. & $a_{d,3}$ \\
20 & 2.61 & 2.21 & CMS 8 TeV $W$ & 1.59 & 1.90 & D{\O} $W$ asym. & $a_{\rho,6}$ \\
21 & 1.61 & 1.42 & D{\O} $W$ asym. & 4.14 & 4.30 & D{\O} II $W\rightarrow \nu e$ asym. & $a_{\rho,6}$ \\ 
22 & 4.02 & 4.31 & CMS $W$ asym. $p_T > 35$ GeV & 1.90 & 1.71 & E866/NuSea $pd/pp$ DY & $A_{\rho}$ \\ 
23 & 2.82 & 4.10 & CMS 8 TeV single diff. $t\bar{t}$ & 4.67 & 5.32 & CDF II $p\bar{p}$ jets & $\delta_{g'}$ \\ 
24 & 1.17 & 1.57 & E866/NuSea $pd/pp$ DY & 4.03 & 3.56 & D{\O} $W$ asym. & $a_{\rho,1}$ \\ 
25 & 4.20 & 3.82 & E866/NuSea $pd/pp$ DY & 1.74 & 1.97 & NuTeV $\nu N$ $x F_3$ & $\eta_{S}$ \\ 
26 & 4.12 & 4.07 & D{\O} $W$ asym. & 4.34 & 4.88 & ATLAS 8 TeV $Z$ $p_T$ & $a_{d,2}$ \\ 
27 & 2.96 & 2.58 & D{\O} $W$ asym. & 2.20 & 1.76 & D{\O} $W$ asym. & $\eta_{d}$ \\ 
28 & 2.66 & 4.92 & NMC/... $F_L$ & 2.21 & 7.14 & HERA $e^+ p$ NC 920 GeV &  $\eta_{g}$ \\ 
29 & 2.59 & 2.78 & CCFR $\nu N \rightarrow \mu \mu X$ & 2.47 & 2.29 & NuTeV $\nu N \rightarrow \mu \mu X$ & $\eta_{s+}$ \\
30 & 4.86 & 4.05 & LHCb 2015 $W$,$Z$ & 1.21 & 1.59 & NuTeV $\nu N \rightarrow \mu \mu X$ & $\eta_{s-}$ \\
31 & 3.61 & 5.53 & LHCb 2015 $W$,$Z$ & 3.92 & 6.35 & ATLAS 7 TeV high prec. $W$,$Z$ & $A_{S}$ \\
32 & 1.91 & 2.44 & CCFR $\nu N \rightarrow \mu \mu X$ & 2.45 & 3.40 & CMS 8 TeV $W$ & $A_{s+}$ \\ \hline
\end{tabular}
\end{center}
\caption{\sf The NLO eigenvectors (in both directions) for the default MSHT20 $\alpha_S(M_Z^2)=0.118$ global NLO fit, including the tolerances for each eigenvector, the data set which most constrains that eigenvector in each direction along it and the parameter with the largest contribution to that eigenvector.}
\label{tab:NLOevecdatasetstable}
\end{table}

\begin{table} 
\small
\begin{center}
\def\arraystretch{0.92}
\begin{tabular}{|>{\arraybackslash}m{1.035cm}|>{\arraybackslash}m{0.525cm}|>{\arraybackslash}m{0.525cm}|>{\arraybackslash}m{5.05cm}|>{\arraybackslash}m{0.525cm}|>{\arraybackslash}m{0.525cm}|>{\arraybackslash}m{5.05cm}|>{\arraybackslash}m{1.76cm}|}\hline
e- ~~~~~ vector& + \hspace{0.1cm} t & + \hspace{0.1cm} T & most constraining data set & - \hspace{0.2cm} t & - \hspace{0.2cm} T & most constraining data set & primary parameter\\ \hline
1 & 3.71 & 3.75 & ATLAS 7 TeV high prec. $W$,$Z$ & 4.76 & 4.75 & CMS 8 TeV $W$ & $a_{S,3}$ \\
2 & 3.12 & 3.33 & NuTeV $\nu N \rightarrow \mu \mu X$ & 2.85 & 2.56 & NMC $\mu d$ $F_2$ & $a_{s+,5}$  \\
3 & 2.48 & 2.58 & NuTeV $\nu N \rightarrow \mu \mu X$ & 4.07 & 3.88 & NMC $\mu d$ $F_2$ & $a_{S,6}$ \\
4 & 3.61 & 3.60 & CMS 8 TeV $W$ & 2.93 & 2.90 & NuTeV $\nu N \rightarrow \mu \mu X$ & $a_{S,2}$ \\
5 & 2.64 & 3.00 & ATLAS 7 TeV high prec. $W$,$Z$ & 2.72 & 2.26 & NuTeV $\nu N \rightarrow \mu \mu X$ & $a_{S,3}$ \\
6 & 5.22 & 5.46 & ATLAS 8 TeV double dif $Z$ & 5.01 & 4.79 & D{\O} $W$ asym. & $a_{s+,2}$\\
7 & 4.07 & 4.37 & NMC/... $F_L$ & 2.90 & 2.58 & D{\O} $W$ asym. & $\delta_{g'}$\\
8 & 3.90 & 3.50 & LHCb 2015 $W$,$Z$ & 3.90 & 3.50 & LHCb 2015 $W$,$Z$ & $\delta_{g'}$\\
9 & 5.48 & 5.59 & LHCb 2015 $W$,$Z$ & 3.73 & 3.64 & BCDMS $\mu p$ $F_2$ & $A_{s-}$\\
10 & 3.55 & 3.58 & BCDMS $\mu p$ $F_2$ & 4.87 & 4.84 & NMC $\mu d$ $F_2$ & $a_{u,6}$\\
11 & 3.06 & 2.91 & D{\O} $W$ asym. & 4.83 & 4.94 & ATLAS 8 TeV $W$ & $\delta_{S}$\\
12 & 1.42 & 1.71 & D{\O} $W$ asym. & 3.40 & 3.06 & CCFR $\nu N \rightarrow \mu \mu X$ & $A_{s-}$\\
13 & 3.87 & 4.10 & CMS asym. $p_T > 25, 30$ GeV & 4.38 & 4.14 & ATLAS 8 TeV $W$& $a_{u,3}$\\
14 & 1.36 & 1.50 & E866/NuSea $pd/pp$ DY & 3.67 & 3.53 & D{\O} $W$ asym. & $a_{u,2}$\\
15 & 5.53 & 5.89 & E866/NuSea $pd/pp$ DY & 3.17 & 2.83 & D{\O} $W$ asym. & $a_{g,3}$\\
16 & 1.89 & 0.52 & E866/NuSea $pd/pp$ DY & 5.64 & 6.65 & ATLAS 8 TeV double dif $Z$ & $a_{d,6}$\\
17 & 2.51 & 2.54 & E866/NuSea $pd/pp$ DY & 2.69 & 2.65 & D{\O} $W$ asym. & $a_{\rho,6}$\\
18 & 1.80 & 1.88 & D{\O} $W$ asym. & 2.47 & 2.38 & CMS 8 TeV $W$ & $a_{\rho,6}$\\
19 & 2.47 & 2.18 & CMS 8 TeV $W$ & 1.37 & 1.63 & D{\O} $W$ asym. & $a_{\rho,1}$\\
20 & 1.82 & 2.22 & D{\O} $W$ asym. & 4.69 & 3.97 & NMC $\mu d$ $F_2$ & $a_{d,2}$\\
21 & 4.41 & 5.36 & ATLAS 8 TeV $Z$ $p_T$ & 4.68 & 4.08 & ATLAS 8TeV sing dif $t\bar{t}$ dilep & $\delta_{g}$\\
22 & 3.49 & 3.23 & D{\O} $W$ asym. & 3.04 & 2.97 & CMS $W$ asym. $p_T > 35$ GeV & $a_{d,3}$\\
23 & 1.84 & 2.43 & ATLAS 8TeV sing dif $t\bar{t}$ dilep & 4.96 & 5.22 & NuTeV $\nu N \rightarrow \mu \mu X$ & $a_{g,2}$\\
24 & 0.99 & 1.23 & E866/NuSea $pd/pp$ DY & 4.61 & 4.43 & E866/NuSea $pd/pp$ DY & $A_{\rho}$\\
25 & 2.01 & 1.35 & D{\O} $W$ asym. & 2.77 & 3.11 & E866/NuSea $pd/pp$ DY & $\eta_{d}$\\
26 & 2.25 & 2.51 & NuTeV $\nu N$ $x F_3$ & 2.06 & 1.94 & E866/NuSea $pd/pp$ DY & $\eta_{S}$\\
27 & 2.83 & 3.65 & ATLAS 8 TeV $t\bar{t}$, dilepton & 2.64 & 5.51 & ATLAS 7 TeV high prec. $W$,$Z$ & $\eta_{g}$\\
28 & 1.74 & 1.92 & D{\O} $W$ asym. & 2.65 & 3.43 & CMS 8 TeV $W$ & $\eta_{d}$\\
29 & 2.57 & 2.85 & CMS 7 TeV $W+c$ & 1.79 & 1.83 & NuTeV $\nu N \rightarrow \mu \mu X$ & $\eta_{s+}$\\
30 & 4.76 & 3.92 & CCFR $\nu N \rightarrow \mu \mu X$ & 2.25 & 2.64 & NuTeV $\nu N \rightarrow \mu \mu X$  & $\eta_{s-}$\\
31 & 2.79 & 4.81 & ATLAS 7TeV high prec $W$,$Z$ & 2.07 & 3.62 & ATLAS 7 TeV high prec. $W$,$Z$ & $A_{S}$\\
32 & 2.57 & 4.27 & CCFR $\nu N \rightarrow \mu \mu X$ & 2.58 & 3.47 & ATLAS 7 TeV high prec. $W$,$Z$ & $A_{s+}$\\
 \hline
\end{tabular}
\end{center}
\caption{\sf The NNLO eigenvectors (in both directions) for the default MSHT20 $\alpha_S(M_Z^2)=0.118$ global fit, including the tolerances for each eigenvector, the data set which most constrains that eigenvector in each direction along it and the parameter with the largest contribution to that eigenvector.}
\label{tab:NNLOevecdatasetstable}
\end{table}

\begin{table} [t]
  \centering
{\footnotesize
  \begin{tabular}{c|c|c|c|c|c|c|c}
    \hline \hline
    evector & $g$ & $u_V$ & $d_V$ & $S({\rm ea})$ & $\bar{d}-\bar{u}$ & $s+\bar{s}$ & $s-\bar{s}$ \\ \hline
    1  & --&  -- &  -- &  0.0 0.03 0.0 & -- &  -- &  -- \\
    2  & -- &  -- &  -- &  -- &  -- &  0.0 0.1 0.0 &  -- \\
    3  &  -- &  -- &  -- &0.1 0.1 0.0&  -- &  0.1 0.2 0.1  &  -- \\
    4  & -- & -- &  -- & 0.0 0.0 0.1 & -- & 0.0 0.2 0.1 &  -- \\
    5  &  -- &  -- &  -- &  0.0 0.1 0.0 & -- &  0.0 0.2 0.0 & --\\
    6  & -- & -- & --& 0.1 0.1 0.1 & --  & 0.2 0.0 0.0 & --\\
    7  & 0.3 0.0 0.0 & 0.0 0.0 0.1 &  -- &  -- & -- &  -- &  -- \\
    8  & 0.1 0.0 0.0 &  -- & --&  0.0 0.1 0.0 &  -- &  -- & --\\
    9  & 0.1 0.1 0.1 & 0.0 0.0 0.3 & -- &0.0 0.1 0.0&  -- & --& 0.1 0.1 0.1\\
    10 & 0.0 0.1 0.1 & 0.0 0.1 0.3 & -- & 0.0 0.2 0.1& -- & -- & 0.0 0.1 0.0 \\
    11 &  -- &  -- & -- & 0.2 0.0 0.1&0.0 0.0 0.1&0.3 0.0 0.3& --\\
    12 & -- & -- & 0.0 0.0 0.1 & -- & -- &  -- & 0.1 0.1 0.2 \\
    13 & 0.0 0.2 0.2 & 0.1 0.1 0.1 & -- & 0.1 0.2 0.0 &0.0 0.1 0.0&  -- & 0.1 0.1 0.1 \\
    14 &  -- &  -- & -- &  -- & 0.0 0.3 0.0 & 0.0 0.1 0.0&  -- \\
    15 & 0.0 0.1 0.2 & 0.2 0.1 0.3 & 0.0 0.0 0.2& 0.1 0.0 0.0 & 0.0 0.1 0.0 & 0.1 0.0 0.0 &  -- \\
    16 & 0.1 0.3 0.5 & 0.0 0.0 0.2 & 0.0 0.0 0.1 &0.0 0.2 0.1 & 0.0 0.2 0.1 &  -- &  -- \\
    17 &  -- & 0.0 0.0 0.2 & 0.0 0.0 0.1 &  -- & 0.0 0.1 0.0&  -- &  -- \\
    18 & -- & 0.1 0.1 0.1 & -- & -- & 0.2 0.1 0.0 &  -- &  -- \\
    19 & -- & 0.1 0.1 0.1 & -- &  -- & 0.1 0.0 0.0 &  -- &  -- \\
    20 & 0.0 0.1 0.1 & 0.1 0.1 0.1 & 0.4 0.5 0.5 & 0.2 0.1 0.1 & 0.2 0.0 0.0 & 0.1 0.0 0.0 & --\\
    21 & 0.6 0.3 0.2 & -- & -- & 0.2 0.1 0.0& 0.4 0.0 0.0 & -- & --\\
    22 &  -- & 0.1 0.1 0.2 & 0.4 0.2 0.3 & 0.0 0.1 0.1 & 0.5 0.4 0.0 &  -- &  -- \\
    23 & 0.1 0.3 0.3 & 0.0 0.0 0.1 & 0.0 0.0 0.1 & -- &  -- & 0.1 0.0 0.0 &  -- \\
    24 & -- &  -- & -- & 0.0 0.0 0.1 & 0.0 0.3 0.6 & -- & -- \\
    25 & -- & 0.0 0.1 0.0 & 0.1 0.1 0.1 & -- & 0.0 0.0 0.1 &  -- &  -- \\
    26 & -- &  -- & 0.0 0.0 0.1 & 0.0 0.0 0.2 & 0.0 0.0 0.3 &  -- &  -- \\
    27 & 0.2 0 0.4 & 0.1 0.1 0.0 & 0.1 0.1 0.1 & 0.0 0.0 0.2 & 0.0 0.0 0.2 & -- & -- \\
    28 & -- & 0.6 0.5 0.1 & 0.2 0.5 0.4 & 0.0 0.1 0.1 & 0.0 0.0 0.1 & -- & --\\
    29 &  -- & -- & 0.0 0.0 0.1 & 0.0 0.0 0.3 & 0.0 0.1 0.1 & 0.0 0.2 0.8 & 0.2 0.2 0.2\\
    30 & -- &  -- & 0.1 0.1 0.1 & -- &  -- & -- & 0.5 1.0 1.0 \\
    31 & 0.1 0.0 0.1 &  -- & 0.0 0.0 0.1 & 0.1 0.0 0.2 & 0.0 0.0 0.2 & 0.0 0.0 0.1 & -- \\
    32 & -- & 0.1 0.1 0.0 &  -- & 0.1 0.0 0.2 & 0.0 0.0 0.1 & 0.2 0.2 0.4 &  -- \\
    \hline \hline
  \end{tabular}
}
\caption{\sf The fractional 
contribution to the total uncertainty  
for the $g,u_V,\ldots$ input distributions in the small $x$ ($x<0.01$), 
medium $x$ ($0.01<x<0.1$) and high $x$ ($x>0.1$) regions, respectively, 
arising from eigenvector $k$ in the NNLO global fit.}
\label{tab:NNLOevecparamstable}
\end{table}

One can see that for the vast majority of cases there is good agreement 
between $t$ and $T$ at both NLO and NNLO. Hence, within the region of 
$68\%$ uncertainty confidence levels for the PDFs, the $\chi^2$ distribution
is quite accurately a quadratic function of the parameters. 
One can see that for some eigenvectors, most notably 27 and 31 in the NNLO 
fit and 28 and 31 in the NLO fit, there is a larger deviation between $t$ and $T$, 
up to a factor of 2 at NNLO in one of the directions and a little worse for the minus direction of eigenvector 28 at NLO. These 
eigenvectors are the ones in which the correlation between some parameters is 
starting to make a significant impact on the quadratic behaviour of the $\chi^2$. We determine our maximum 
number of parameters by insisting at worst a factor of about 2 between 
$t$ and $T$, so these are the final eigenvectors making this cut (the number 32 is 
set using the NNLO fit, so we allow one marginally worse eigenvector in the NLO fit).
Extending beyond 32 eigenvectors would quickly result in mismatches between $t$ and 
$T$ of much more than a factor of 2.

Even when $t =T$ to a good approximation, there is, however,
a reasonable degree of asymmetry between the $t$ and $T$ values in the two 
directions for a single eigenvector, and it is nearly always the case that
it is a different data set which is the main constraint in the two directions. 
In fact, the data set which has the most rapid deterioration in fit quality
in one direction is often improving in fit quality until quite a 
high value of $t$ along the other direction. This is an indication of the 
tension between data sets, with nearly all eigenvectors having some data
sets which pull in opposite directions. 

As alternative information for the NNLO fit we also present the contribution of each of the different eigenvectors to the different PDF error bands in different $x$ regions. This is shown in 
Table \ref{tab:NNLOevecparamstable}. For each eigenvector and for seven independent PDF 
combinations, $g$, $u_V$, $d_V$, $S({\rm ea})$, $\bar{d}-\bar{u}$, $s+\bar{s}$ and $s-\bar{s}$,
we show the fractional contribution each eigenvector makes to the total uncertainty. This is done
in the small $x$ ($x<0.01$), medium $x$ ($0.01<x<0.1$) and high $x$ ($x>0.1$) regions, 
respectively. For each of these regions the maximum contribution to the total PDF
uncertainty coming from this eigenvector is shown, rounded to the nearest 0.1 (corresponding to 10\%). Hence, if the 
$x$ range were very narrow and there were no rounding the total in each vertical column would sum 
to $1$. However, both the rounding and fact that different eigenvectors maximise their 
contributions to the uncertainty at different $x$ values within the three regions mean this 
summation results in a number often slightly in excess of 1. 

Let us now consider some particular eigenvectors of the NNLO fit as examples. We first consider eigenvector 1. As shown in 
Table~\ref{tab:NNLOevecdatasetstable} this eigenvector's largest component is the third Chebyshev coefficient of
 the light sea. In this unique case the largest value in Table~\ref{tab:NNLOevecparamstable} is 
0.03, and we have not rounded to nearest 0.1, because the eigenvector contributes less than
$10\%$ of the uncertainty to all PDFs. As the largest contributing parameter implies though, the 
uncertainty contribution is largest in the light sea. The most constraining data sets are ATLAS 7~TeV $W$, $Z$ data and CMS 8~TeV $W$ data, clearly both sensitive to the light sea. Although the 
absolute change in PDFs for eigenvector 1 is small, it must produce changes in shape sufficient to 
cause the fit quality for these two data sets to deteriorate at a sufficient level for the 
tolerance criterion to be satisfied. 

As a second example, let us consider eigenvector 4 for the default NNLO fit. As indicated in Table~\ref{tab:NNLOevecdatasetstable} the largest component of this eigenvector is the second Chebyshev coefficient of the light sea, therefore it is unsurprising that it is most constrained in the positive direction by a Drell-Yan data set, the CMS 8~TeV $W$ data. On the other hand that it is most constrained by the NuTeV dimuon data in the negative direction is less obvious. However, the eigenvector also has a sizeable contribution from the third Chebyshev coefficient of the total strangeness $s + \bar{s}$, and in Table~\ref{tab:NNLOevecparamstable} we see that the contribution to the uncertainty from this eigenvector is actually in the light sea and the total strangeness, and is a little larger in the latter. The pulls of different data sets on this eigenvector and the consequent tensions for this direction in the parameter space are revealed by the changes in $\chi^2/N$ as one moves along the eigenvector. In the positive direction, not only does the $\chi^2/N$ of the CMS 8~TeV $W$ data increase markedly, thereby constraining this eigenvector in this direction, but the ATLAS 7~TeV $W$, $Z$ data set is also nearly as constraining. This reflects its known sensitivity to the total strangeness and also it is interesting to note that this constrains the eigenvector in the opposite direction to the NuTeV dimuon data, indicative of the known opposite pulls on the strangeness of these data sets. Meanwhile in the positive direction further constraints are provided, albeit further along the eigenvector direction, by the NMC $d$ data and the CCFR dimuon data, the latter having been also previously observed to pull slightly differently to the corresponding NuTeV data \cite{MMHT14}. In the negative direction, the CMS 7~TeV $W+c$ data set is also very constraining, as one might expect given it pulls on the strangeness similar to the NuTeV data, whilst the CDF $W$ asymmetry and the ATLAS 8~TeV $W^{\pm}$ are also constraining, albeit less so. All of these pulls are sensible given the eigenvector is composed largely of sea and total strangeness parameters.

The final eigenvector, number 32, is also interesting in regard to the total strangeness, with it overwhelmingly being made up of the overall normalisation of the total strangeness $A_{s+}$. The fact that it contributes mainly to the 
total strangeness uncertainty is confirmed from Table~\ref{tab:NNLOevecparamstable}.
Therefore it is constrained by the CCFR dimuon data in the positive (increased strangeness) direction and the ATLAS 7~TeV $W$, $Z$ data in the negative (reduced strangeness direction), as one might naively expect. If the eigenvector is examined in more detail however further subtleties are revealed, with the NuTeV dimuon data improving slightly in both directions along this eigenvector, although more along the negative direction, perhaps indicating its tensions with both the CCFR dimuon and the ATLAS 7~TeV $W$, $Z$ data and supporting it favouring reduced strangeness. It is also interesting to note that the CMS 7~TeV $W+c$ data set improves in $\chi^2/N$ along the positive direction, perhaps therefore favouring a marginally increased strangeness relative to the CCFR dimuon, albeit still much less than the ATLAS 7~TeV $W$, $Z$. Furthermore, the ATLAS 7~TeV $W$, $Z$ data in actuality, not only constrains the negative direction (reduced strangeness) in our fit, where it is the most constraining data set, it also worsens in $\chi^2$ significantly along the positive (increased strangeness) direction away from the central fit, providing a significant constraint in this direction as well. This indicates that in our global fit the ATLAS 7~TeV $W$, $Z$ data seem to favour a specific magnitude of the strangeness normalisation and apparently worsens in fit quality if this is increased or decreased away from the central value. However, it should be noted at this stage, that the magnitude of the total strangeness in any individual $x$ range is determined not only by the normalisation $A_{s+}$, but also by the Chebyshev coefficients $a_{s+,i}$, and this variation in different $x$ ranges is indeed seen  in Table~\ref{tab:NNLOevecparamstable}. Finally, it is also worth noting that the ATLAS 8~TeV double differential $Z$ data also worsens in fit quality along both directions of this eigenvector, albeit more along the positive direction, indicating a preference perhaps for a reduced $A_{s+}$, whilst the ATLAS 8~TeV $W^{\pm}$ indicates the opposite preference. Finally, the ATLAS 8~TeV $Z$ $p_T$ data set worsens in $\chi^2$ along both directions, indicative of its tensions with the Drell-Yan data sets.

From the point of view of the strangeness, eigenvector 30 is also worth commenting on as it is dominantly the high $x$ power of the strangeness asymmetry, $\eta_{s-}$, and this is clearly confirmed by Table~\ref{tab:NNLOevecparamstable}. Therefore, the observation that it is constrained most by the CCFR dimuon data in one direction (increasing the power and so suppressing the asymmetry at high $x$) and by the NuTeV dimuon data (reducing the power and so increasing the asymmetry at high $x$) in the other direction is as one would expect. In addition there is perhaps some tension at the level of this eigenvector between the LHCb 2015 $W$, $Z$ data at 7 and 8~TeV and the ATLAS 8~TeV $W^{\pm}$ data, with the former showing a worsening $\chi^2/N$ in both directions along this eigenvector, whilst the latter improves in the positive direction and is flat in the negative direction.

Away from the quark and antiquark decomposition we look at eigenvector 21. This is largely associated with the small $x$ parameter $\delta_g$, and indeed we see its major contribution is to the small $x$ gluon uncertainty. 
However, from the momentum sum rule this also affects the high-$x$ gluon, and indeed the eigenvector is mainly constrained by the ATLAS $8~\TeV$ $Z$ $p_T$ data in one direction and ATLAS single differential top pair production data in the other direction, both of which are primarily sensitive to the higher $x$ gluon. The clear tension between these two data sets is discussed in detail in Section~\ref{MSHT20gluonlightquark}.

Finally, let us comment on eigenvector 24, which is related to the difference between the up and down antiquarks, as it is largely made up of the normalisation of the ratio of the down antiquarks to the up antiquarks $\bar{d}/\bar{u}$, $A_{\rho}$. As will be described further in Section~\ref{newparameterisationeffects}, the E866 Drell-Yan ratio data are particularly sensitive to this ratio and so this data set provides the greatest constraints in both directions on this eigenvector. There are nonetheless other constraining data sets with the D{\O} $W$ asymmetry data, LHCb 2015 $W$, $Z$ 7 and 8~TeV data, and ATLAS 8~TeV $W^{\pm}$ all constraining this eigenvector strongly along the negative direction. As we see from Table~\ref{tab:NNLOevecparamstable}, this eigenvector also 
contributes a little to the sea uncertainty, consistent with these observations. 

We can also comment at this stage on more general constraints offered by different data sets on the error bands in the global fit. First, in addition to the most constraining data sets and primary parameters for each eigenvector, the values of $t$ and $T$ for the
$68\%$ confidence levels are also given. On average across the eigenvectors we achieve $t\approx T \approx 3$, i.e. 
$\Delta \chi^2_{\rm global} \approx 10$, though $T^2$ does vary between about
1 unit and at most $T^2\approx 40$ in the NNLO case. This is similar to that achieved in MMHT14, which confirms that the additional eigenvectors and parameters added for MSHT20 are adequately constrained by the enlarged data set included. For the NLO fit actually $T^2 \approx 50$ is reached, perhaps indicating that given the worse fit obtained at NLO (particularly in the high precision LHC data sets), there are fewer constraints on the eigenvectors as a result.

Secondly, it is reassuring to observe that the eigenvectors are largely constrained most by data sets obviously related to their dominant parameters; for example the ATLAS 7~TeV $W$, $Z$ data constrain eigenvectors comprising of sea and total strangeness parameters, the NuTeV dimuon data constrain eigenvectors associated with total strangeness and sea parameters, the D{\O} $W$ asymmetry data constrain eigenvectors largely related to the up and down quarks and antiquarks, the E866 Drell-Yan ratio data do similarly, and this applies to many data sets. 

Beyond this we can also remark on the variety of different data sets at play in a full global fit. It is very clear from Tables \ref{tab:NLOevecdatasetstable} and \ref{tab:NNLOevecdatasetstable} that a wide
variety of different data types are responsible for constraining the PDFs. This indicates the clear need for both older structure function data, dimuon data and the like, and the newer LHC data sets, with them clearly offering complementary constraints on the PDFs. At a simplified level we can consider the number of eigenvector directions constrained by each data set in the NNLO global fit in Table~\ref{tab:NNLOevecdatasetstable}. This reveals that the data sets which constrain 4 or more eigenvector directions are the D{\O} $W$ asymmetry (13), E866 Drell-Yan ratio (8), NuTeV dimuon (7), ATLAS 7~TeV $W$, $Z$ (6), CMS 8~TeV $W$ (5) and NMC $d$ (4). Moreover, while 39 of the 64 eigenvector directions are constrained by non-LHC data sets, including the three most constraining,  the LHC data sets, particularly high precision measurements, are increasingly playing an important role. 

It should be noted however that this analysis is over-simplified because, as illustrated by the discussion of a selection of eigenvectors above, many of the eigenvector directions are constrained by several data sets, with the second and even third most constraining data sets often offering comparable constraints. For example, this is particularly true of the HERA data (see Section~\ref{noHERA}), which offer significant constraints on many of the eigenvectors but are not the most constraining in any case. This in part reflects the large number of data points in the combined HERA data set, which means it can have a large global effect on the PDFs without necessarily being the most constraining for any given eigenvector direction. Indeed, for data sets with larger pulls on the central fit PDFs, either through their sensitivity to the PDFs or through their large numbers of data points (such as the HERA data), we would expect the central fit PDFs to be close to their minimum in $\chi^2$ space for eigenvectors to which they are relevant. As a result, their changes in $\chi^2$ away from the central fit will naturally be smaller, particularly once the number of data points is factored in via the tolerance procedure. In contrast, smaller data sets are more likely to have little effect on the overall central fit PDFs and so sit away from their minimum in $\chi^2$ in a particular direction. Consequently their $\chi^2$ will change more rapidly as we move along the eigenvector, thereby having a more significant effect in constraining the error bands. This perhaps also partly explains why the D{\O} $W$ asymmetry data set is one of the most constraining despite only affecting the central PDFs at high $x$, and doing so mainly for the valence quarks.  This reflects that the simplified analysis of counting the number of eigenvector directions constrained most by a given data set deals with which data sets are relevant for the error bands only, rather than an overall analysis of their pull or effect on the PDF central values, which is a different (albeit related) question. Moreover, data sets which have a significant effect on the central values of the PDFs also have an indirect effect on the uncertainties by pulling the global fit into a different minimum position in the PDF parameter space.

Finally, we recall that in the MMHT14 fit~\cite{MMHT14} the lower number of  degrees of freedom (40), rather than the number of data points (84), was used in the case of the NuTeV dimuon data set, to account for the high degree of correlation between nearby data points. In reality, this is only an approximate way to deal with this issue, and in modern data sets this would be accounted for by providing full statistical correlations between data points. In addition, both the CT18~\cite{CT18} and NNPDF3.1~\cite{NNPDF3.1} sets quote the number of data points. Therefore, for simplicity and for ease of comparison we now use the number of data points, both in quoting the $\chi^2/ N$, see Table~\ref{tab:chisqtable}, and in determining the confidence level as in \eqref{eq:68percentCL}. If we instead use the lower number of degrees freedom, for the NNLO fit we find that 3 of the 7 eigenvector directions constrained by  NuTeV are instead constrained by other data sets, while the value of $t$ for the eigenvector directions that are still constrained by NuTeV are higher. In particular, we find that the relevant directions of eigenvectors 5, 23 and 30 would in this case be constrained by the ATLAS 8~TeV $Z$ $p_\perp$, CMS 8~TeV inclusive jet and D{\O} $W$ asymmetry, respectively. The impact on the PDF uncertainties is on the other hand very mild, with by far the largest difference being in the strangeness, which is $\sim 1\%$ larger in the $x \sim 0.01-0.4$ region.

\section{Comparison of MSHT20 with MMHT14 PDFs}\label{sec:6}

We now show the change in both the central values and the uncertainties of 
the NNLO PDFs at $Q^2=10^4~\GeV^2$ (unless otherwise specified)
in going from the NNLO MMHT14 analysis. By default we choose to show the ratio of the 
MSHT20 PDFs to those of MMHT14 as this allows any differences to be most easily identified. In addition we also provide plots of the percentage uncertainties of both MSHT20 and MMHT14 together in order to observe changes in the error bands with this new PDF set and accompanying enlarged collection of data sets. We then repeat this process in comparing the NLO MSHT20 PDFs with those of MMHT14. Throughout the plots MMHT14 at NNLO is shown in red, MSHT20 at NNLO in green, MMHT14 at NLO is blue and MSHT20 at NLO is in purple.
 
\subsection{Gluon and light quark distributions} \label{MSHT20gluonlightquark}

\begin{figure}
\begin{center}
\includegraphics[scale=0.24, trim = 50 0 0 0 , clip]{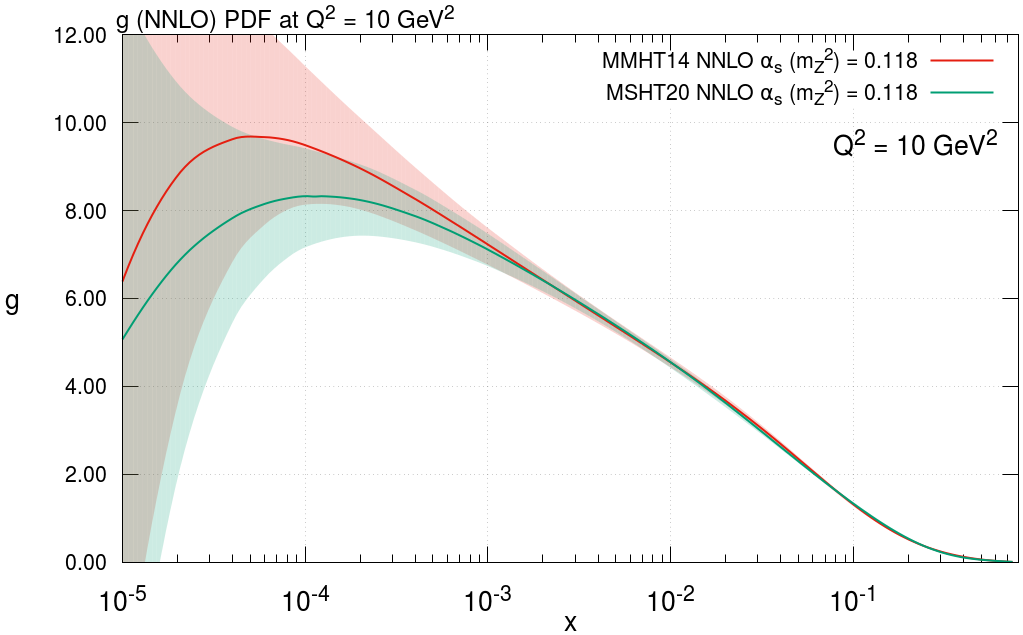}
\includegraphics[scale=0.24, trim = 50 0 0 0 , clip]{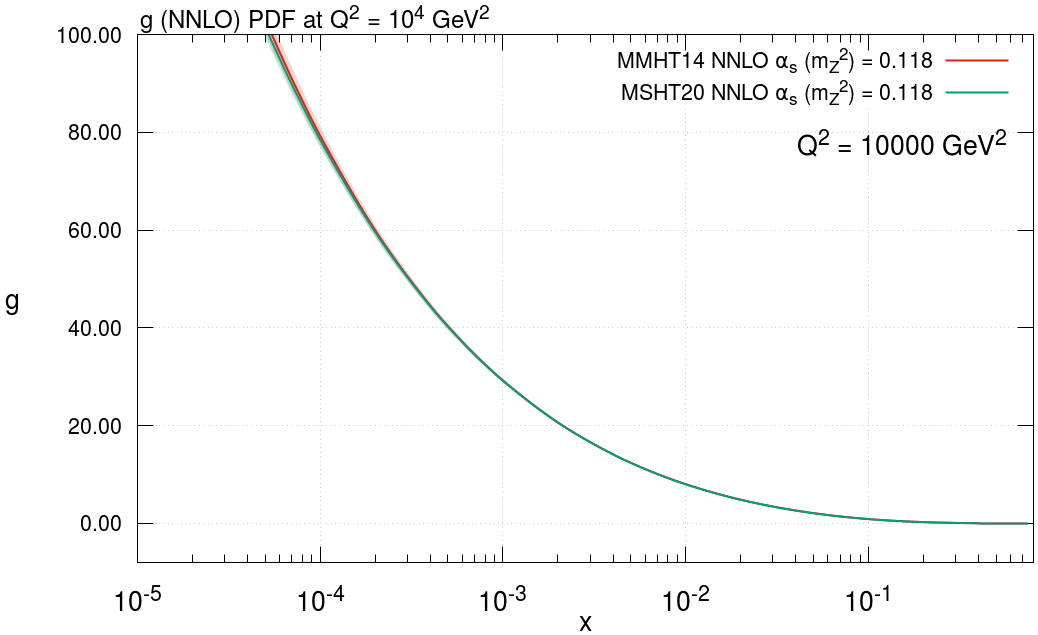}
\caption{\sf Gluon PDF for MSHT20 and MMHT14, at NNLO and (left) $Q^2=10~\GeV^2$ and (right) $Q^2=10^4~\GeV^2$.}
\label{gluon_q210_q210000_NNLO}
\end{center}
\end{figure} 

\begin{figure}
\begin{center}
\includegraphics[scale=0.24, trim = 50 0 0 0 , clip]{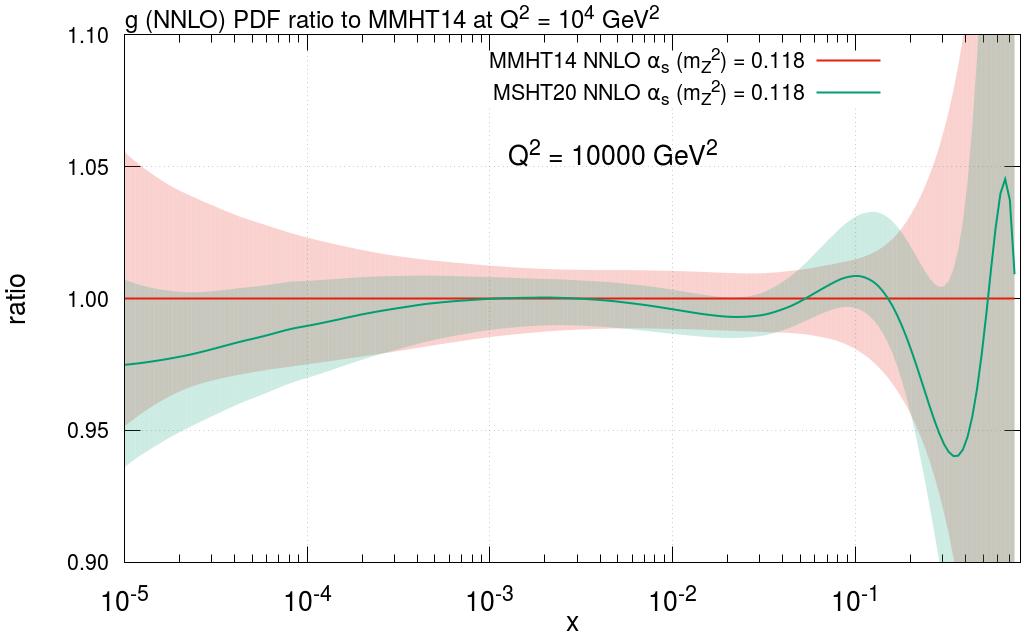}
\includegraphics[scale=0.24, trim = 50 0 0 0 , clip]{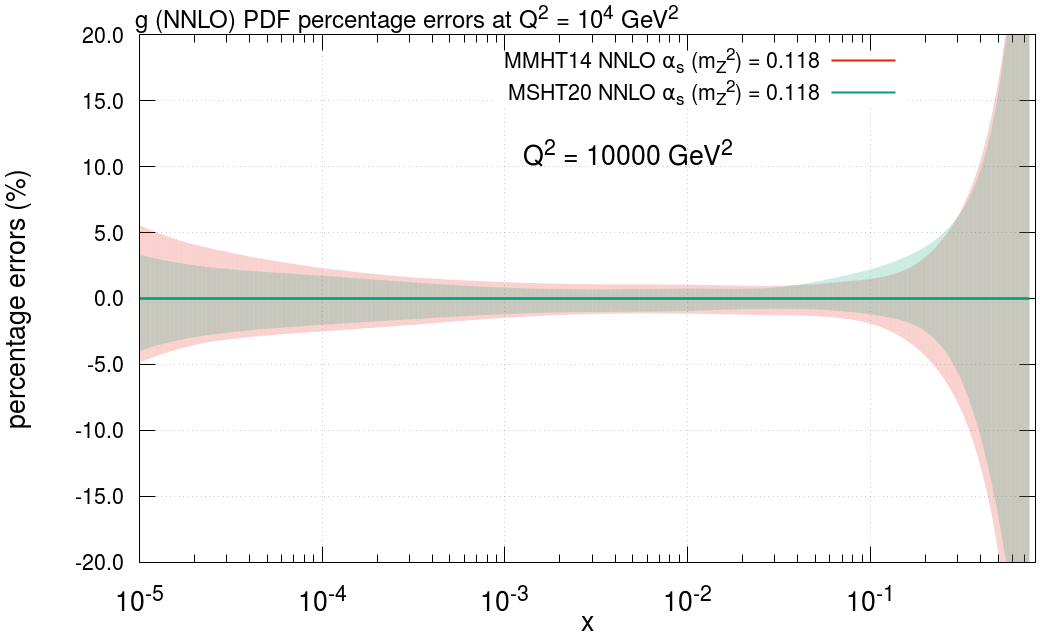}
\caption{\sf Gluon PDF for MSHT20 and MMHT14, at NNLO and $Q^2=10^4~\GeV^2$.  (Left) ratio to MMHT14. (Right) percentage uncertainties.}
\label{gluon_q210000_ratio__percentageerrors_NNLO}
\end{center}
\end{figure} 

\begin{figure}
\begin{center}
\includegraphics[scale=0.27, trim = 50 0 0 0 , clip]{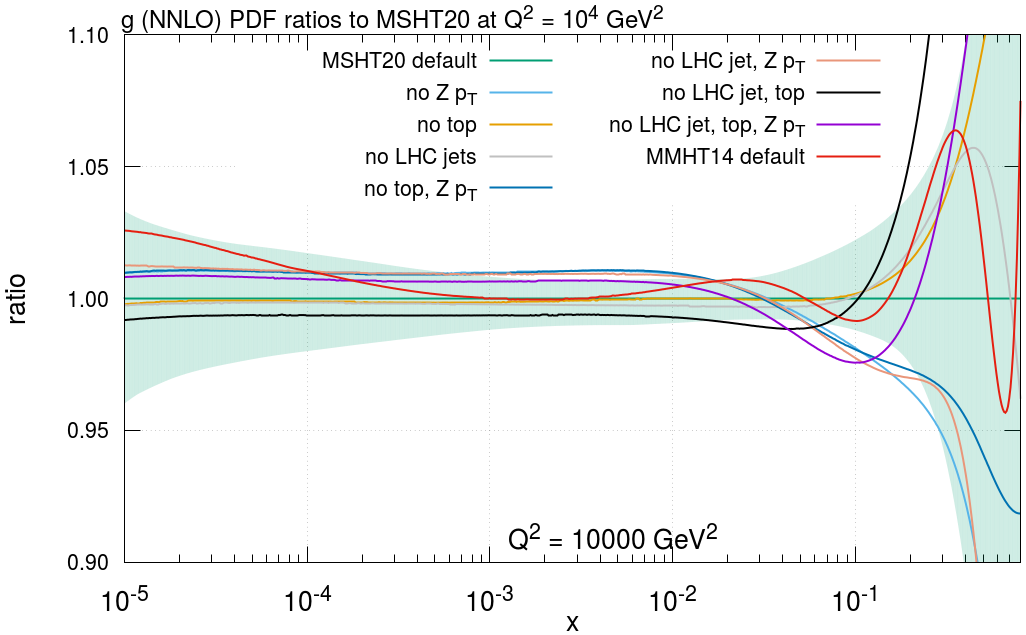}
\caption{\sf The effects of removing the LHC jet, $Z$ $p_T$ and top data sets on the high $x$ gluon, shown in ratios to the default MSHT20 PDFs. }
\label{MSHT20_NNLO_gluonpulls_q210000_NNLO}
\end{center}
\end{figure} 

\begin{figure}
\begin{center}
\includegraphics[scale=0.24, trim = 50 0 0 0 , clip]{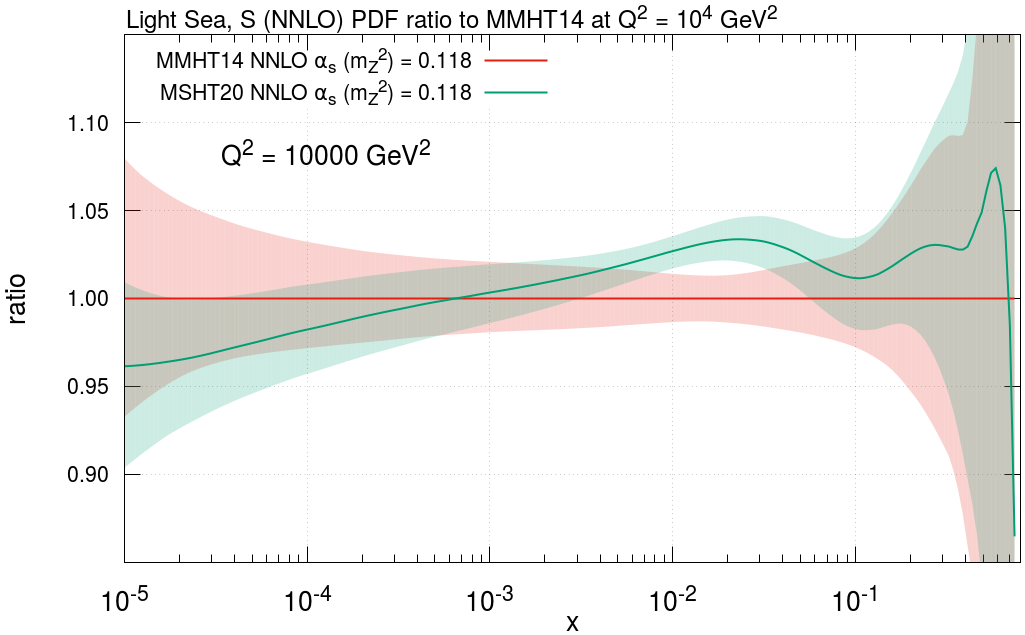}
\includegraphics[scale=0.24, trim = 50 0 0 0 , clip]{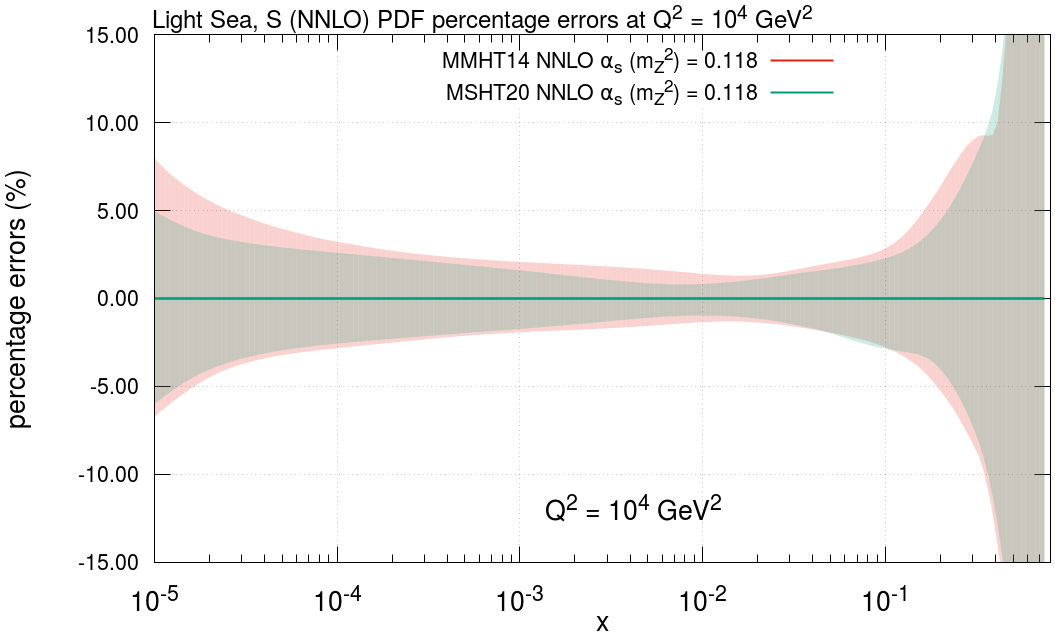}
\caption{\sf Light sea (S) PDFs, for MSHT20 and MMHT14 at NNLO at $Q^2=10^4~\GeV^2$.  (Left) ratio to MMHT14. (Right) percentage uncertainties.}
\label{lightseafigures}
\end{center}
\end{figure}

The central values and uncertainties of the gluon at both the low scale $Q^2 = 10~\GeV^2$ and a high scale $Q^2 = 10^4~\GeV^2$ are given in Fig.~\ref{gluon_q210_q210000_NNLO}, whilst the ratio to MMHT14 is shown in Fig.~\ref{gluon_q210000_ratio__percentageerrors_NNLO} (left) along with the percentage uncertainties in Fig.~\ref{gluon_q210000_ratio__percentageerrors_NNLO} (right). In Fig.~\ref{gluon_q210000_ratio__percentageerrors_NNLO} (left) the MSHT20 gluon is seen to be consistent within uncertainties across the whole $x$ range with that of MMHT14. 
There is a small reduction at very low $x$ reflecting the slightly larger difference seen in this region at lower $Q^2$, as seen in Fig.~\ref{gluon_q210_q210000_NNLO} (left).
The main details of the shape change in the intermediate to low $x$ gluon appeared upon extending the input parameterisation and resulted in  small improvements in the $\chi^2$ of several data sets, including the HERA cross section data, CMS jets and the ATLAS $W$, $Z$ data. For the $x$ range from $0.001-0.01$ there is very little change. At high $x>0.1$, there is a significant decrease near $x=0.3-0.4$ and then some oscillation at very high $x$, with the central value  consistent with the uncertainties, which are very large in this region due to the limited data constraints here. The LHC jet, $Z$ $p_T$ and $t\bar{t}$ data all have effects on the gluon in the high $x$ region, with the net result of the reduced gluon noted, although the details of their interplay are complicated.
The result of dropping each of these three types of data, either individually, or in combination, 
is shown Fig.~\ref{MSHT20_NNLO_gluonpulls_q210000_NNLO}. It can be seen that omitting either jet 
or  top data results in an increase in the high-$x$ gluon, and for both at the same time this is 
well outside the error band. Omitting only the ATLAS $Z$ $p_T$ data results in a reduction of
the gluon near $x\sim 0.1-0.2$ - indeed all combinations which omit the $Z$ $p_T$
data lead to a reduction in this $x$ region. Omitting all three types of distribution result in 
a gluon roughly similar to that of MMHT14 (except for $x>0.4$, where the uncertainty is large), but 
clearly the detailed shape is sensitive to other new data and the parameterisation change. Of these 
new data which constrain the high-$x$ gluon, the ATLAS $Z$ $p_T$ data clearly prefer a very large 
high-$x$ gluon, the top data a suppressed high-$x$ gluon and the jets more weakly a reduced 
high-$x$ gluon (though different jet data sets have some difference in pull). 
It can be seen in Fig.~\ref{gluon_q210000_ratio__percentageerrors_NNLO} (right) that the uncertainties are slightly reduced in MSHT20 relative to MMHT14 across the vast majority of the $x$ range, due to the impact of new data. The final HERA data, for example, reduce the uncertainty at very small $x$. The changes are not substantial at intermediate $x$, where the gluon is already very well constrained. The uncertainty at higher $x$ values is systematically reduced by up to $20\%$, reflecting the impact of additional LHC data sets in this region.

For the light quark sea $S(x)$, Fig.~\ref{lightseafigures} provides both the ratios to MMHT14 and the percentage uncertainties. The shape change in the light sea is evident in the ratio in Fig.~\ref{lightseafigures} (left), with the enhancement at $x\sim 10^{-2}$ largely due to an increase in the strangeness in this region resulting from the ATLAS $W$, $Z$ data. There is also a decrease in the valence quarks in this region, due both to the improved flexibility in the parameterisations and to the LHC precision data, which prefer a smaller $d_V$ distribution. Similarly there is a reduction in all of the $\bar{u}$, $\bar{d}$ and $s+\bar{s}$ at very low $x$ which is visible consequently in the light sea (all of these individual changes are discussed in subsequent sections). This is a reflection of the reduced gluon and the evolution, the light sea being driven almost entirely by gluon splitting in this region. The light sea itself shows lower uncertainties than the individual antiquarks, as it is the combinations of the antiquarks (albeit with process-dependent charge weightings) which are often constrained directly via data. As a result, in the light sea the enhancement around $x \sim 0.02$ is not consistent with MMHT14 within its error bands whilst the $s+\bar{s}$ and $R_s$ distributions, discussed later in Section~\ref{strangeness} (and defined in equation \eqref{eq:Rs}) and shown in Figs.~\ref{splussbarfigs} and \ref{Rs_q21p9_NNLO}, are actually consistent with MMHT14. As noted earlier though, the light sea enhancement is also related to the valence quark suppression in the same region, and there is no consistency between MSHT20 and MMHT14 for $d_V$.

\subsection{Up and down quark and antiquarks distributions} \label{updown}

\begin{figure}
\begin{center}
\includegraphics[scale=0.24, trim = 50 0 0 0 , clip]{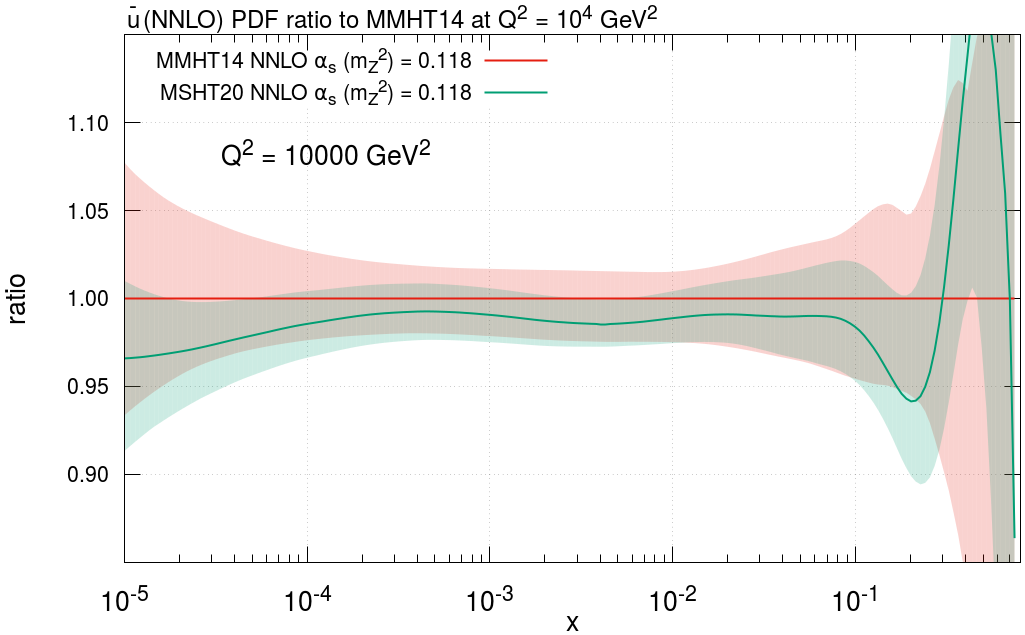}
\includegraphics[scale=0.24, trim = 50 0 0 0 , clip]{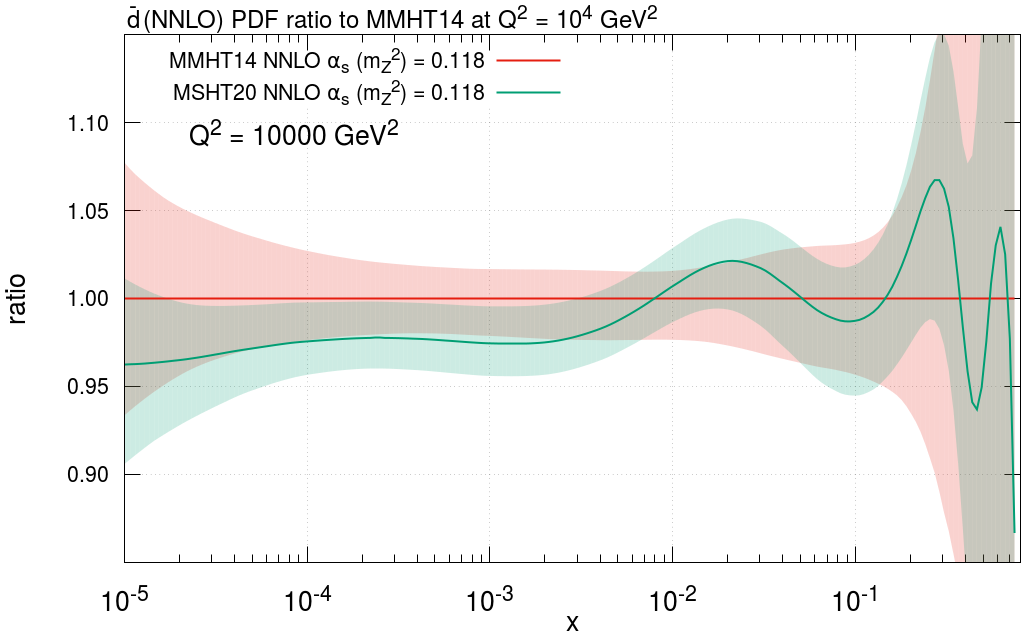}
\caption{\sf Ratio of (left) Up antiquark and (right) Down antiquark PDFs to MMHT14, at NNLO and $Q^2=10^4~\GeV^2$.}
 \label{updownbarratios}
\end{center}
\end{figure}

\begin{figure}
\begin{center}
\includegraphics[scale=0.24, trim = 50 0 0 0 , clip]{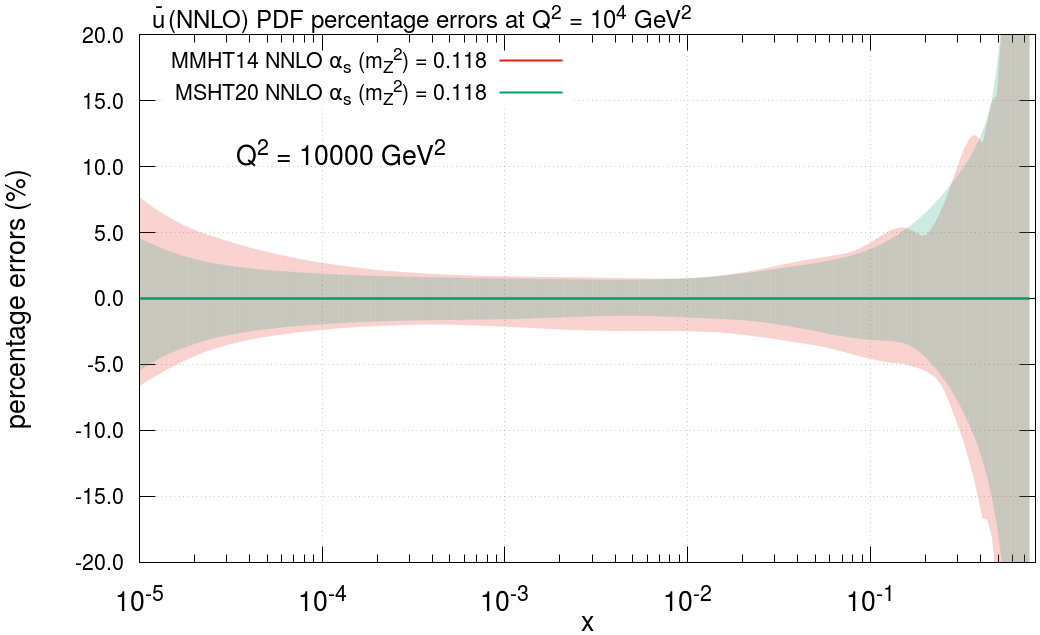}
\includegraphics[scale=0.24, trim = 50 0 0 0 , clip]{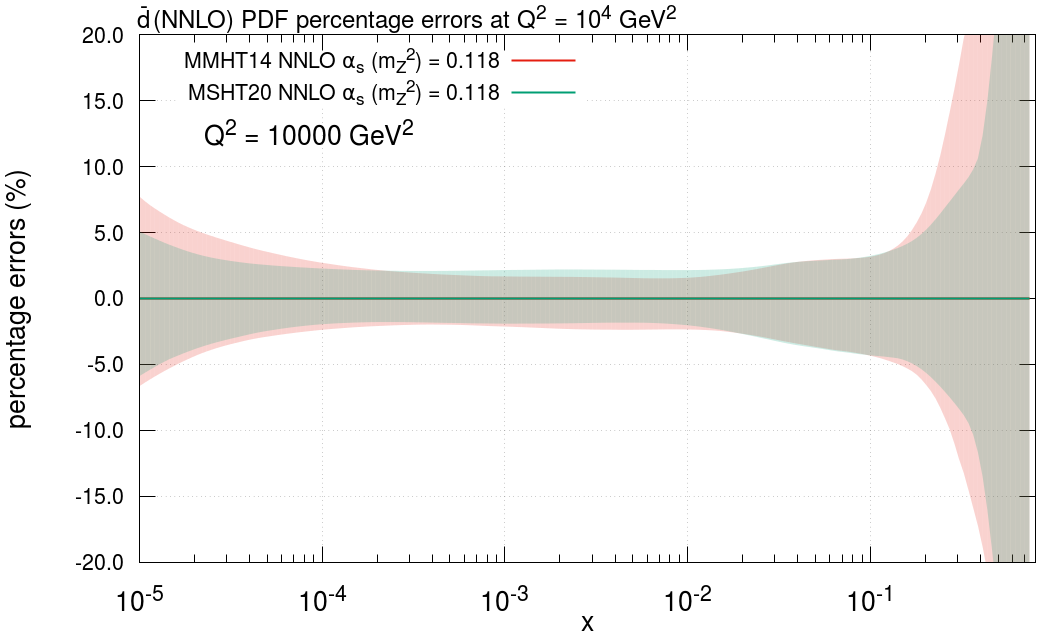}
\caption{\sf Percentage uncertainties of (left) Up antiquark and (right) Down antiquark PDFs, for MSHT20 and MMHT14 at NNLO and $Q^2=10^4~\GeV^2$}
\label{updownbarerrs}
\end{center}
\end{figure}

First we focus on the antiquarks. The ratios to MMHT14 and the percentage uncertainties of the up and down antiquarks are given in Figs.~\ref{updownbarratios} and \ref{updownbarerrs} respectively. The up and down antiquark PDFs are lowered over the majority of the $x$ range relative to MMHT14. This occurs very largely to compensate for the increase in the strangeness due to the ATLAS 7~TeV $W$, $Z$ data whilst keeping the combinations of antiquarks needed in structure function and other data unchanged. In particular this explains the larger reduction in the down as the charge weighted combinations of up, down and strange must remain the same for the HERA data. The down antiquark shows a small bump at $x\sim 0.02$ which is likely related to the ATLAS 7~TeV $W$, $Z$ data set, which enhances $\bar{s}$ in this region, since it can also enhance the $\bar{d}$ through Cabbibo mixing. The enhancement in $\bar d$ here is also likely related to the decrease in $d_V$, leaving the total $d$ distribution less altered.  
The percentage uncertainties for the up and down antiquarks are reduced relative to MMHT14, particularly at low and high $x$. At very small $x$ this is related to the corresponding 
decrease in the gluon uncertainty, but at higher $x$, particularly very high $x$, it 
is clearly due to improved data constraints, as the parameterisation is now more flexible. 

\begin{figure}
\begin{center}
\includegraphics[scale=0.24, trim = 50 0 0 0 , clip]{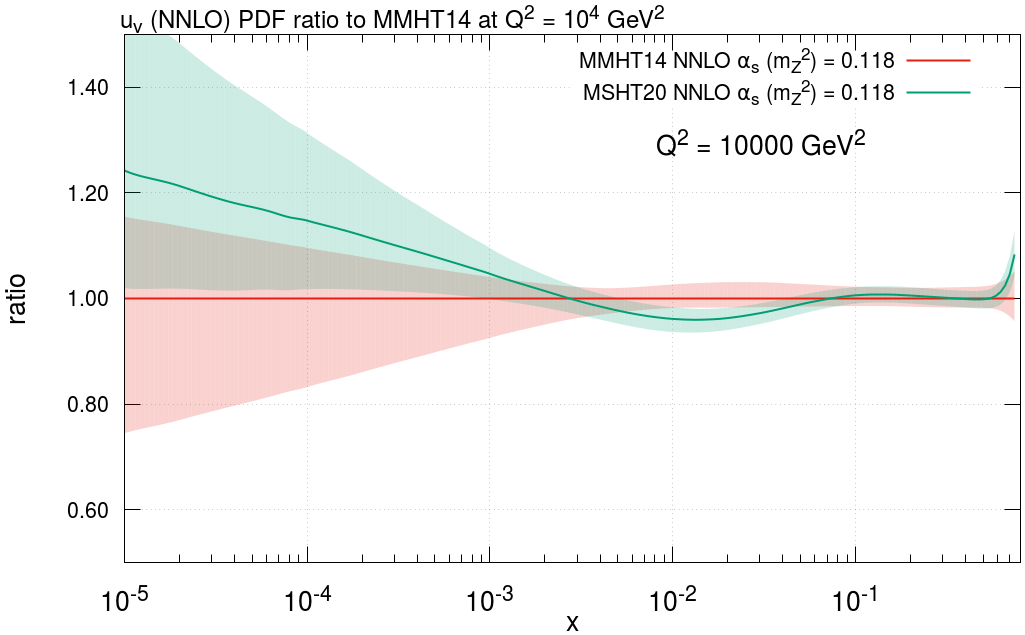}
\includegraphics[scale=0.24, trim = 50 0 0 0 , clip]{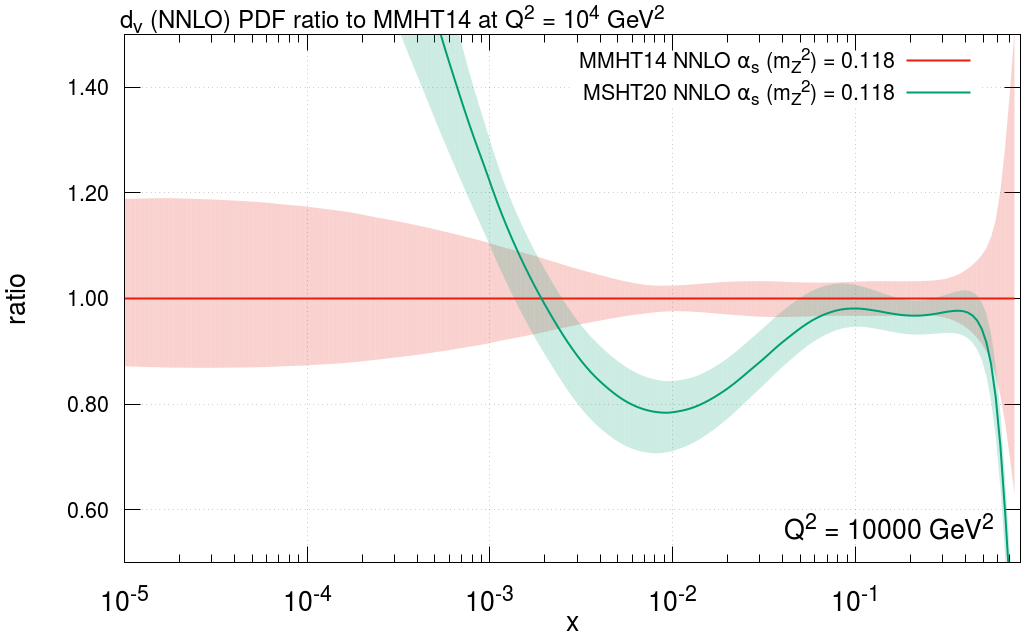}
\caption{\sf Ratio of (left) Up valence and (right) Down valence PDFs to MMHT14 at NNLO and $Q^2=10^4~\GeV^2$. }
\label{uvdvratios}
\end{center}
\end{figure} 

\begin{figure}
\begin{center}
\includegraphics[scale=0.24, trim = 50 0 0 0 , clip]{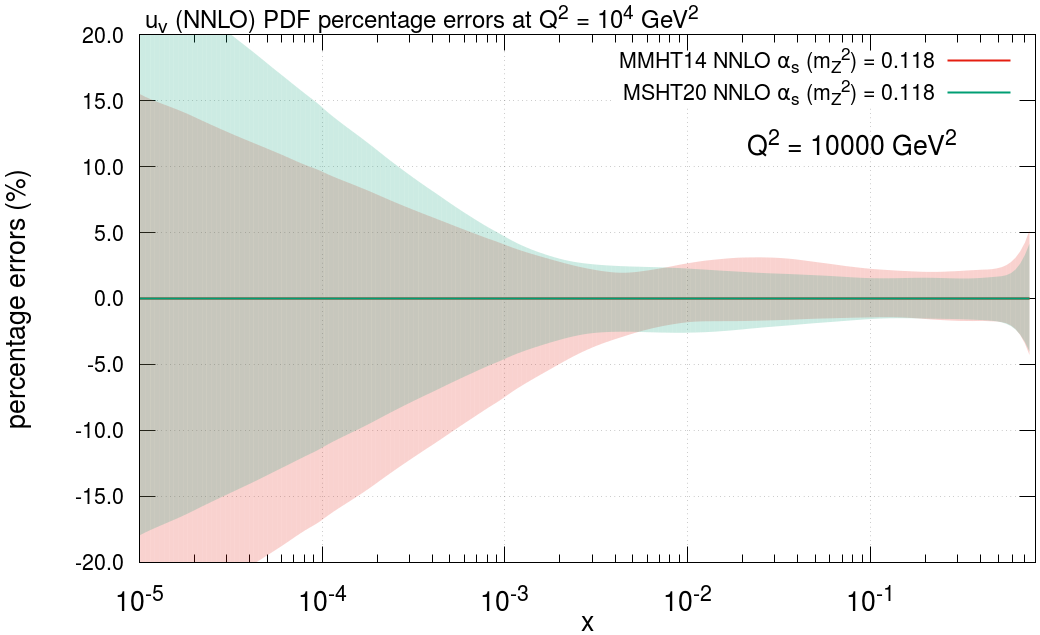}
\includegraphics[scale=0.24, trim = 50 0 0 0 , clip]{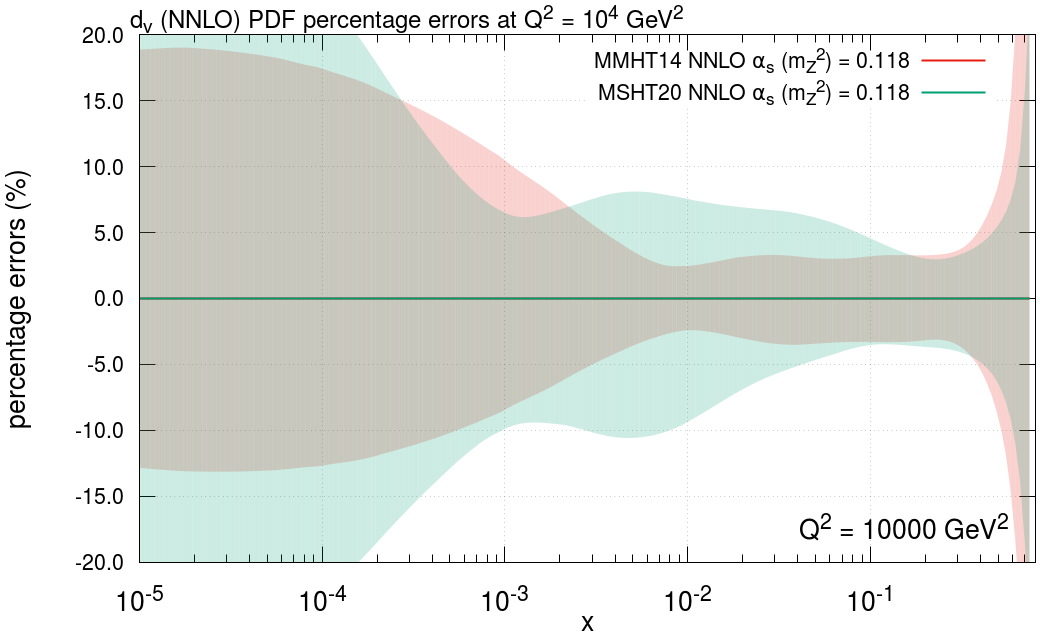}
\caption{\sf Percentage uncertainties of (left) Up valence and (right) Down valence PDFs for MSHT20 and MMHT14 at NNLO and $Q^2=10^4~\GeV^2$ }
\label{uvdvpercentageuncertainties}
\end{center}
\end{figure} 

However, whilst the up and down antiquarks show little change at high $Q^2$, the valence quarks show significant changes in shape, as evidenced in Fig.~\ref{uvdvratios} (left) and \ref{uvdvratios} (right). This results from the wide variety of precise new LHC electroweak data added, and also the improved parameterisation flexibility alleviating tensions between LHC and previous data, with this being particularly the case for the down valence PDF. More details on the parameterisation effects are given  in Section~\ref{newparameterisationeffects}. The D{\O} $W$ asymmetry data has an effect on both the central values and uncertainties of the valence quark PDFs at high $x$, bringing the $d_V$ down in this region and reducing the uncertainties of both the $u_V$ and $d_V$. The effects of all of these changes are also evident in the percentage uncertainties in Fig.~\ref{uvdvpercentageuncertainties}. Whilst the error bands are mildly reduced for $u_V$ at intermediate to high $x$, there are significant differences at low $x$,  with the error bands nonetheless being of similar magnitude here to MMHT14. The $d_V$ on the other hand has enlarged error bands relative to MMHT14 over most of the $x$ range, with the exception of high $x$. This is a result of the extended parameterisation for $d_V$ itself, but also of the very much improved flexibility in the parameterisation for the difference between $\bar u$ and $\bar d$.

\subsection{$u_V-d_V$ distribution}

\begin{figure}
\begin{center}
\includegraphics[scale=0.24, trim = 50 0 0 0 , clip]{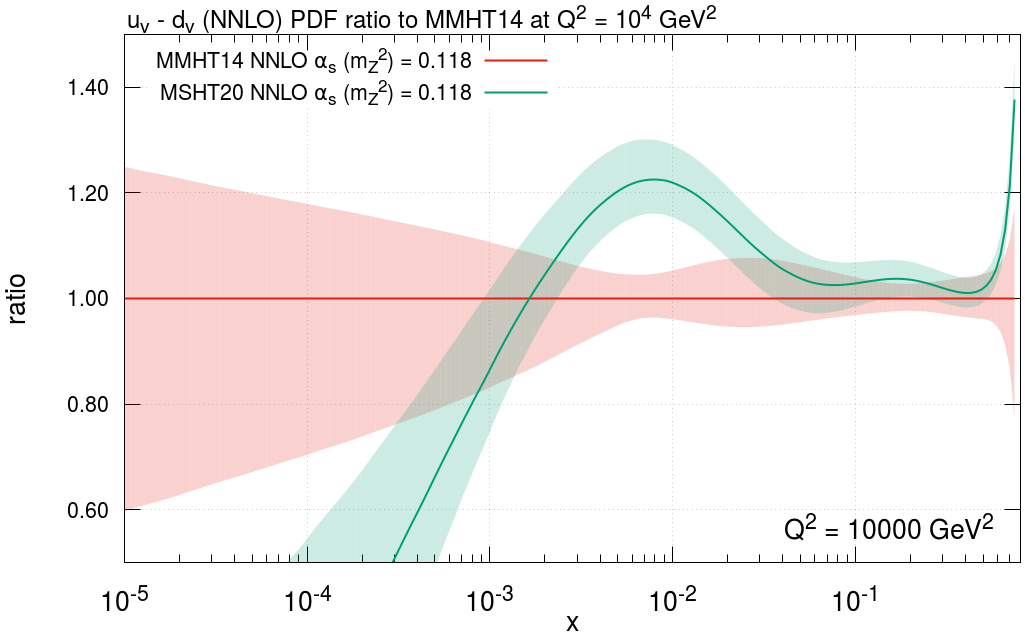}
\includegraphics[scale=0.24, trim = 50 0 0 0 , clip]{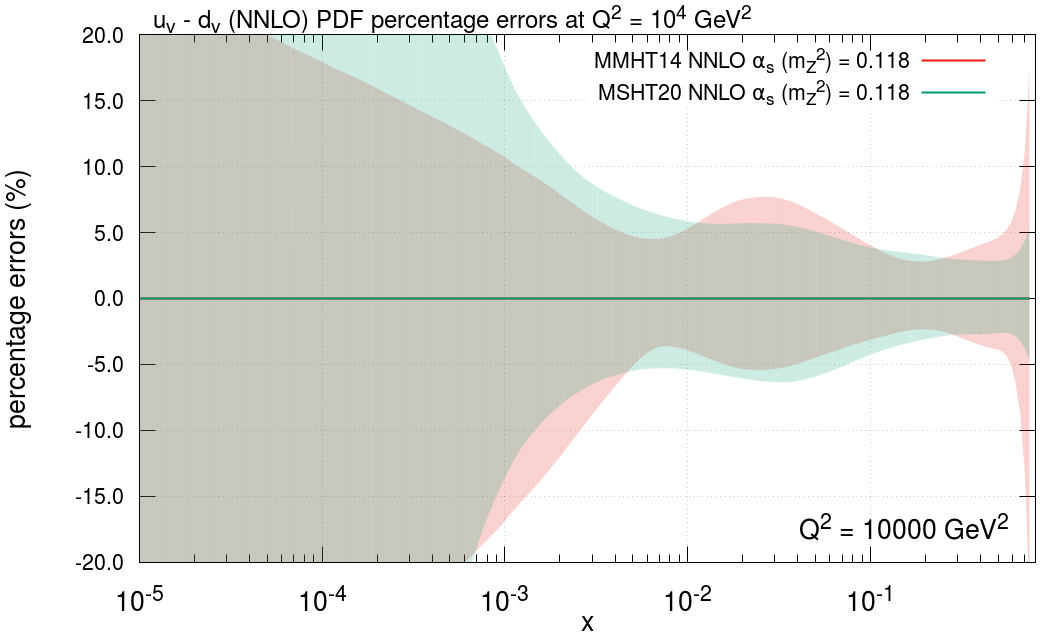}
\caption{\sf $u_V$ - $d_V$ PDF, for MSHT20 and MMHT14 at NNLO and $Q^2=10^4~\GeV^2$.  (Left) ratio to MMHT14. (Right) percentage uncertainties.}
\label{uvminusdv}
\end{center}\end{figure} 

The changes in the individual $u_V$ and $d_V$ PDFs are also reflected in their difference $u_V-d_V$ in Fig.~\ref{uvminusdv}. The substantial shape change is evident and driven by precise LHC $W^+, W^-$ data, including the asymmetry. This asymmetry at low rapidity is a very direct constraint 
on $u_V-d_V$, see e.g. Section 5 of \cite{MMSTWW}.   
We observe an increase in the uncertainty at very small $x$ reflecting the extended parameterisation here. The new PDFs are therefore more representative of the lack of experimental constraints in this region. Meanwhile, at high $x$ the $u_V-d_V$ error bands are reduced, largely down to the D{\O} $W$ asymmetry data constraints.

\subsection{$\bar d - \bar u$ or $\bar d /\bar u$ distributions} \label{dbarminusubarMSHT20}

\begin{figure}[t]
\begin{center}
\includegraphics[scale=0.24, trim = 50 0 0 0 , clip]{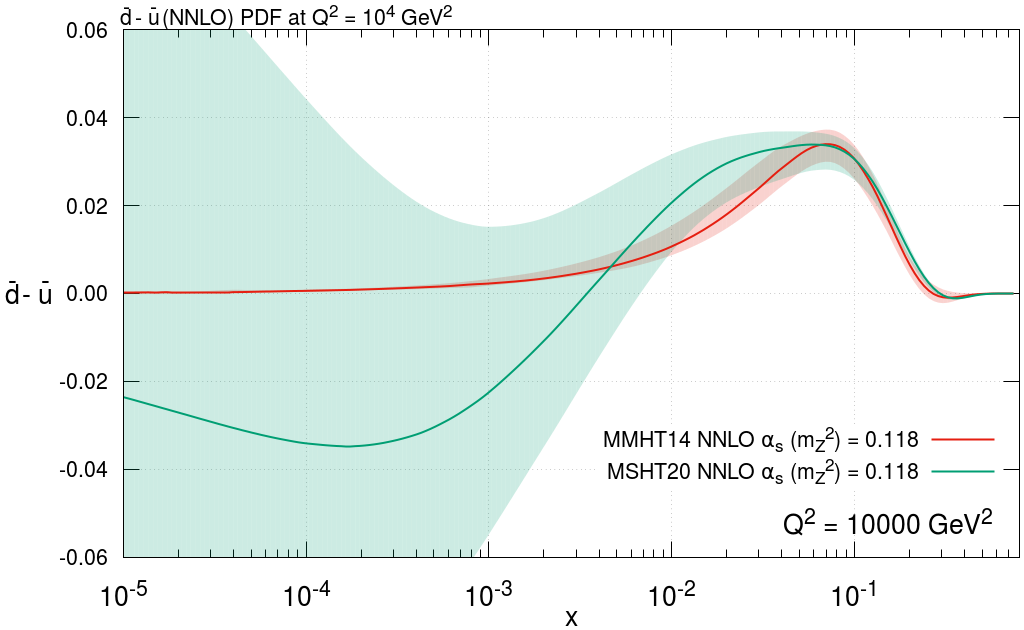}
\includegraphics[scale=0.24, trim = 50 0 0 0 , clip]{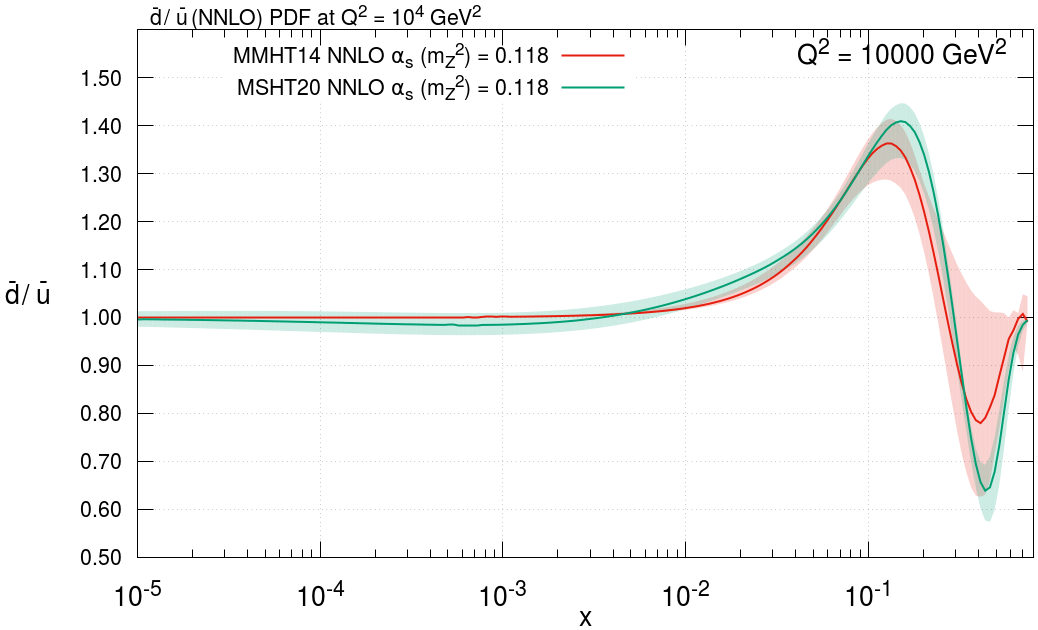}
\caption{\sf (Left) $\bar{d} - \bar{u}$ and (right) $\bar{d} /\bar{u}$ PDFs, for MSHT20 and MMHT14 at NNLO and $Q^2=10^4~\GeV^2$. In MMHT14 the input PDFs parameterised $\bar{d} - \bar{u}$, whereas now in MSHT20 we parametrise $\bar{d}/\bar{u}$. }\label{dbarminusubar}
\end{center}
\end{figure} 

\begin{figure}[t]
\begin{center}
\includegraphics[scale=0.24, trim = 50 0 0 0 , clip]{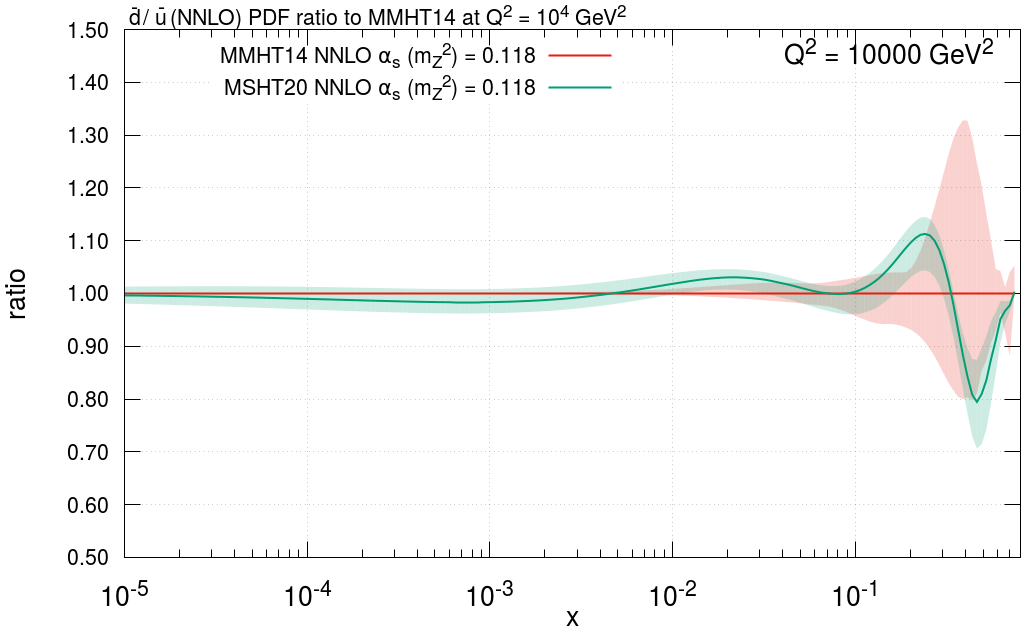}
\includegraphics[scale=0.24, trim = 50 0 0 0 , clip]{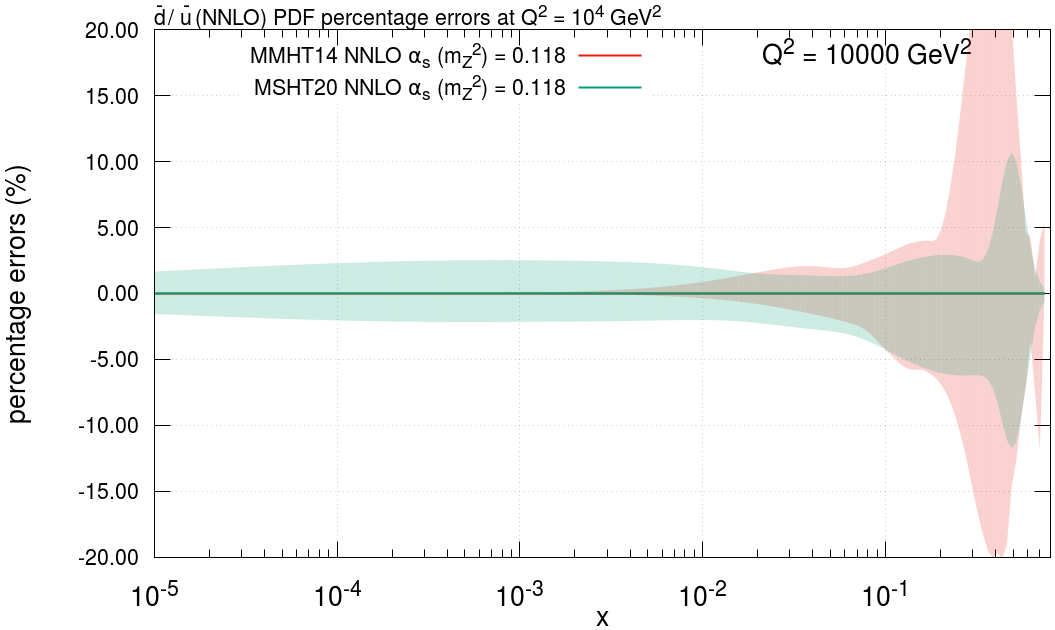}
\caption{\sf $\bar{d}/\bar{u}$ PDFs, for MSHT20 and MMHT14 at NNLO and $Q^2=10^4~\GeV^2$.   (Left) ratio to MMHT14. (Right) percentage uncertainties.}
\label{dbaroverubar}
\end{center}
\end{figure}

The $\bar{d}$ - $\bar{u}$ difference is where one of the most substantial changes relative to MMHT14 are present in our new MSHT20 PDF set. As outlined in Section~\ref{sec:inputPDF}, and to be detailed further in Section~\ref{newparameterisationeffects}, we have considerably improved our description of this PDF, not only changing to a parameterisation in Chebyshevs, but also now parameterising the isospin ratio $\bar{d} / \bar{u}$ rather than the difference $\bar{d} -\bar{u}$. The effects of these changes are numerous, and will be elucidated later. Nonetheless for now we present their effects on the central values and uncertainties. In Fig.~\ref{dbarminusubar} the MSHT20 and MMHT14 PDFs at NNLO are shown both in terms of the difference and the ratio. The shape difference at intermediate $x$ is clear, with the extended parameterisation allowing for an enhanced, broadened peak structure. This alleviates tensions between new LHC data and older data, including between the ATLAS $W$, $Z$ data and the E866 Drell-Yan ratio data. 
There is also now a preference for a very small negative $\bar{d}-\bar{u}$ at high $x$. The effects at low $x$ are even more dramatic than the shape change. In the old parameterisation used by MMHT14 the isospin asymmetry, $\bar{d} - \bar{u}$, was forced to go to $0$ at very low $x$. By now parameterising the ratio (and having no low $x$ power in the parameterisation  in eq.(\ref{eq:dbaroverubar})) we do not enforce this, rather we insist that the ratio $\bar{d}/\bar{u}$ tends to a constant, but not that this be 1. In Fig.~\ref{dbarminusubar} (right) it is clear that the fit naturally chooses the ratio to tend to 1, to rather high precision, at low $x$, an interesting effect of the data, and theoretically consistent with our expectations. Moreover, the effect of the MMHT14 parameterisation was not only to ensure the difference tended to 0 at very low $x$, but also as a consequence to suppress the uncertainties in this region, resulting in the very small error band, clearest in Fig.~\ref{dbarminusubar} (left) at very low $x$. This was not a reflection of the uncertainty in the data in this region but instead a feature of the input parameterisation used. By using the ratio without a low $x$ power we prevent this suppression of the uncertainties in this region and allow the substantially increased error bands present in MSHT20, which much more accurately mirror the lack of data constraints at low $x$. The ratio of the $\bar{d}/\bar{u}$ in MSHT20 relative to MMHT14 and the percentage uncertainties on $\bar{d}/\bar{u}$ in both MSHT20 and MMHT14 (in the latter this ratio is not parameterised but can still be plotted of course) are given in Fig.~\ref{dbaroverubar}. The large increase in the error bands at very low $x$ is clear. In addition, the reduction in the uncertainty bands in the data region at intermediate to high $x$, resulting from the larger collection of data sets used in MSHT20, is obvious.

\subsection{$s+\bar s$ and $s-\bar s$ distributions} \label{strangeness}

\begin{figure}
\begin{center}
\includegraphics[scale=0.24, trim = 50 0 0 0 , clip]{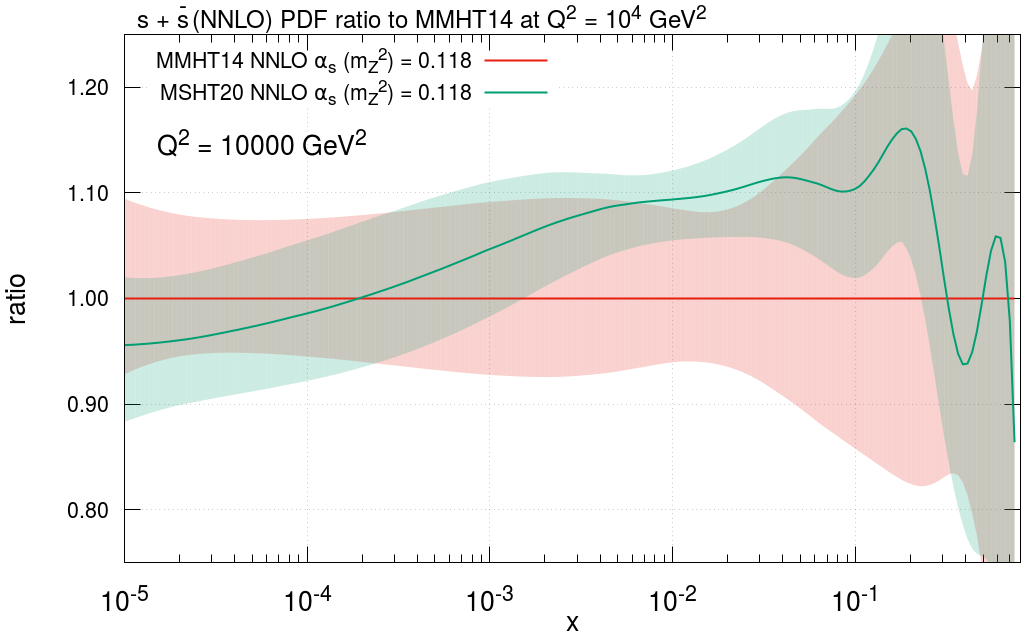}
\includegraphics[scale=0.24, trim = 50 0 0 0 , clip]{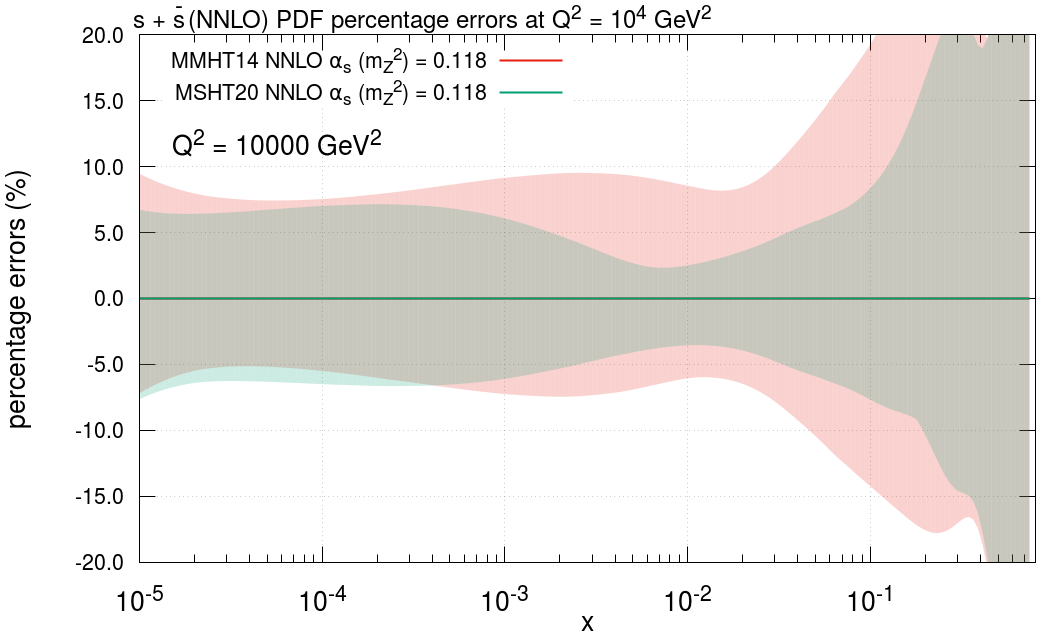}
\caption{\sf $s+\bar{s}$ PDFs, for MSHT20 and MMHT14 at NNLO and $Q^2=10^4~\GeV^2$. (Left) ratio to MMHT14. (Right) percentage uncertainties.}
\label{splussbarfigs}
\end{center}
\end{figure} 

\begin{figure}
\begin{center}
\includegraphics[scale=0.24, trim = 60 0 0 0 , clip]{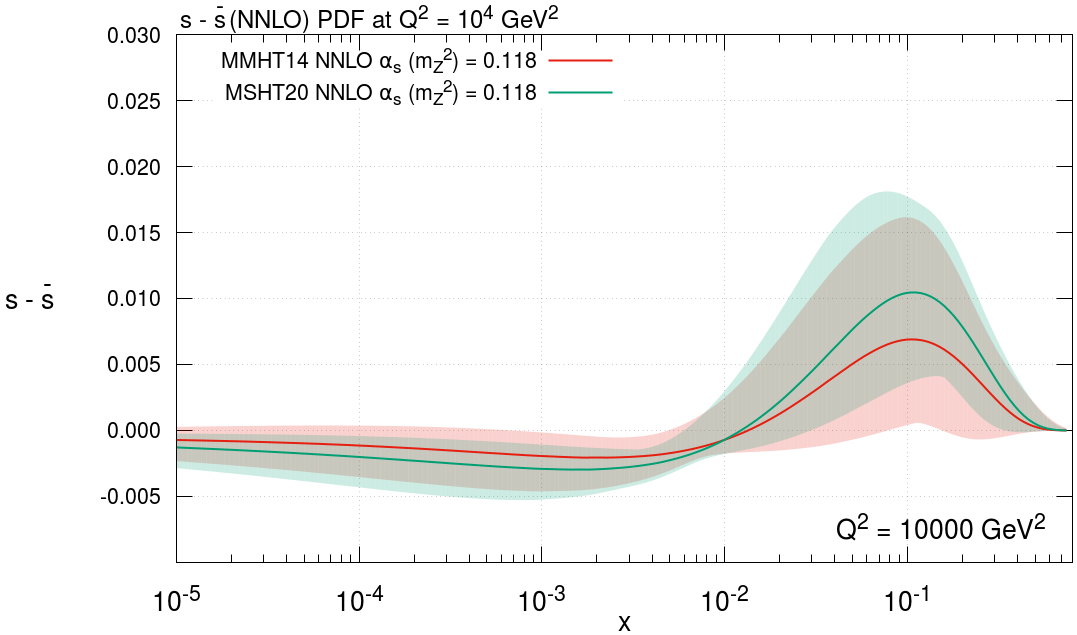}
\includegraphics[scale=0.24, trim = 50 0 0 0 , clip]{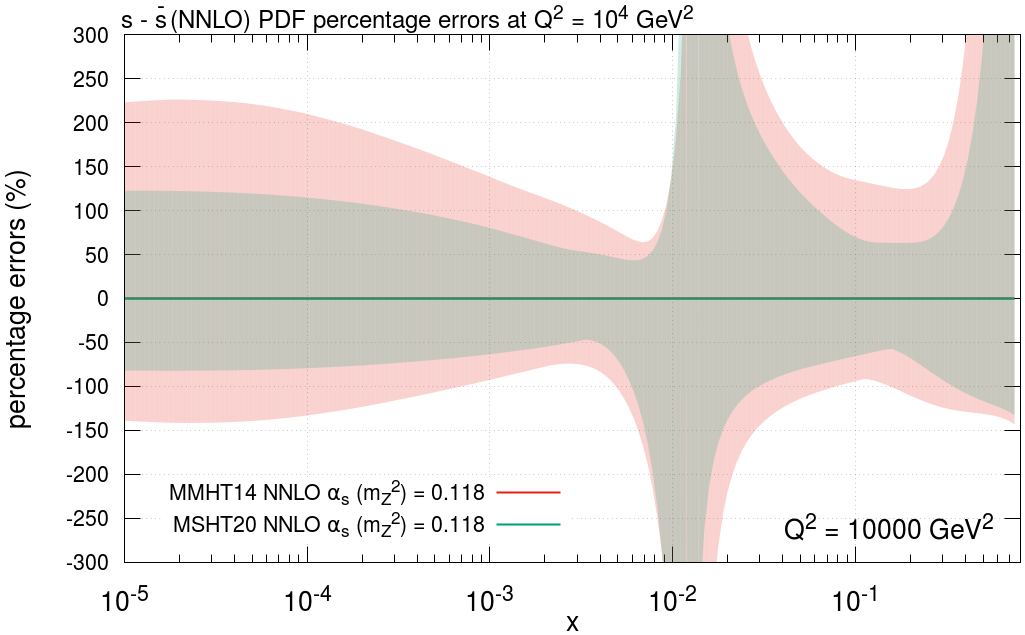}
\caption{\sf $s-\bar{s}$ PDFs, for MSHT20 and MMHT14 at NNLO and $Q^2=10^4~\GeV^2$. In the (left) plot the absolute values of the PDFs are shown rather than the ratio as it passes through zero. In the (right) plot the percentage uncertainties are shown, however they become large in the regions where the asymmetry passes through or tends zero.}
\label{sminussbarfigs}
\end{center}
\end{figure} 

\begin{figure}
\begin{center}
\includegraphics[scale=0.24, trim = 50 0 0 0 , clip]{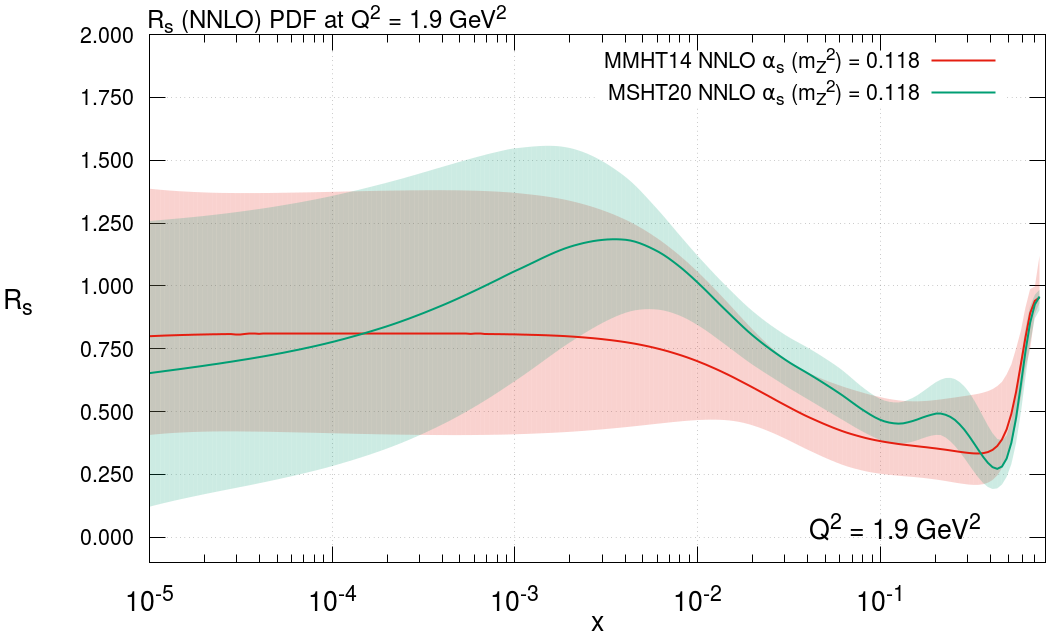}
\caption{\sf $R_s$ PDF,  for MSHT20 and MMHT14 at NNLO and $Q^2=1.9~\GeV^2$. }
\label{Rs_q21p9_NNLO}
\end{center}
\end{figure} 

\begin{figure} 
\begin{center}
\includegraphics[scale=0.6]{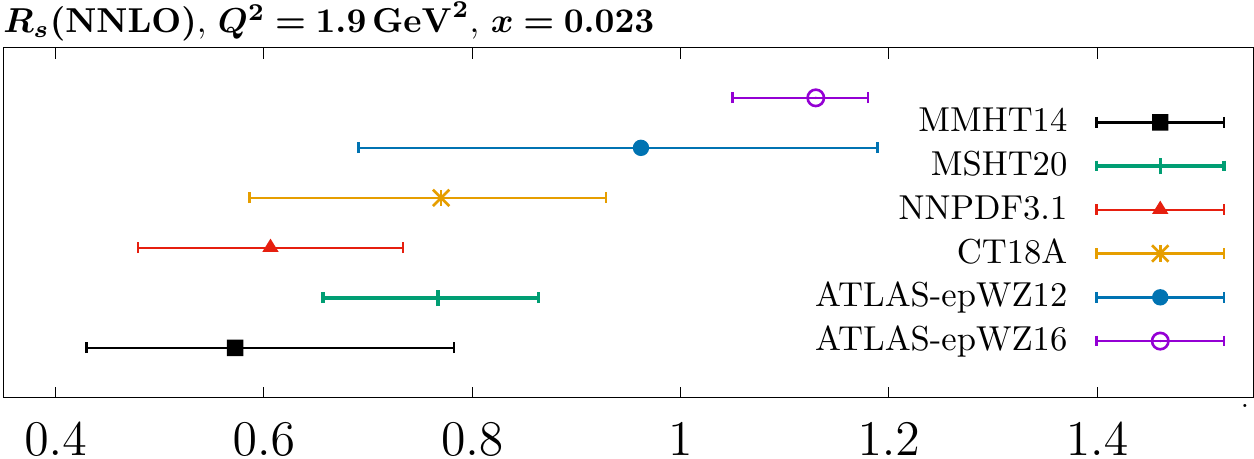}
\caption{\sf Comparison of the value of $R_s$ at $x=0.023$ and $Q^2=1.9 \,{\rm GeV}^2$ for a range of NNLO PDF sets.}
\label{fig:rs}
\end{center}
\end{figure}

The total strangeness and strangeness asymmetry are shown in Figs.~\ref{splussbarfigs} and \ref{sminussbarfigs}. There is a definite increase in $s+\bar s$ below $x \sim 0.1$, driven by ATLAS $W$, $Z$ data (both 7 and 8~TeV)
mainly. This is the so-called ``strangeness unsuppression'' reported in \cite{ATLASWZ7f}, which is often shown in terms of the variable $R_s$:
\begin{equation} \label{eq:Rs}
R_s = \frac{s+\bar{s}}{\bar{u}+\bar{d}}.
\end{equation}
As a result of the precise ATLAS Drell-Yan data here there is also a consequent large reduction in the uncertainties in this region. The other main data sets which constrain the total strangeness are the CMS $W+c$ cross section at 7~TeV and the dimuon cross sections. The latter prefers a rather lower
$R_s$ value and the former a more intermediate value. As a result the MSHT20 default strangeness is 
influenced by all these pulls and $R_s$ is still considerably smaller than 1, outside of the error bands, in the high $x$ region (see Figure~\ref{Rs_q21p9_NNLO}). We recall that we now include a NNLO calculation of the dimuon production cross section; whilst this has a limited impact on the PDFs, it does shift the preferred value of the $D\to \mu$ branching ratio  to be more consistent with experimental determinations, see Section~\ref{NNLOdimuoneffects} for more detail. In Fig.~\ref{Rs_q21p9_NNLO} we show $R_s$ at $Q^2=1.9~\GeV^2$, while in Fig.~\ref{fig:rs} we show results for $R_s$ at a fixed value of  $x=0.023$ and $Q^2=1.9 \,{\rm GeV}^2$, for a range of PDF sets. We can see that the MSHT20 value is rather higher than MMHT14, and with smaller uncertainties, due in large part to the inclusion of the ATLAS data. On the other hand, the dedicated ATLAS-epWZ16~\cite{ATLASWZ7f} fit to this data set, in addition to HERA data, is significantly higher and inconsistent with both MMHT14 and MSHT20 at the $\sim 2 \sigma$ level; for the  ATLAS-epWZ12~\cite{Aad:2012sb}  fit to earlier lower statistics data a similar, but milder, trend can also be seen. The size of this effect however appears to be driven by the rather reduced dataset considered in the ATLAS fits. In the MSHT20 case, the value of $R_s=0.77^{+0.10}_{-0.13}$ is a compromise between the dimuon and, to a lesser extent, $W+c$ data, which favour a lower $R_s$ value, and the ATLAS precision $W$ and $Z$ data, which prefer a larger value. The NNPDF3.1 value lies slightly lower than MSHT20 (see~\cite{Faura:2020oom} for updated values), while CT18A is consistent, and with rather larger errors.

%

The impact of the extended parameterisation on the $s+\bar{s}$ is also evident in these plots with the shape change causing a smaller total strangeness at very low $x$, with $R_s < 1$ although consistent with 1 within errors. The uncertainty band on $R_s$ has grown slightly at very low $x$ due to the extended parameterisation, thereby reflecting the lack of data constraints at very low $x$. Nonetheless in Fig.~\ref{splussbarfigs} (right) it is clear that the total strangeness has reduced errors relative to MMHT14 over nearly the entire $x$ range. 

The strangeness asymmetry, which we recall is parameterised such that its integral is zero, is shown in Fig.~\ref{sminussbarfigs}. This remains relatively unconstrained by the data in the fit. As a result its parameterisation is unchanged relative to MMHT14. Nonetheless the impact of the new LHC data added, particularly the combined high precision ATLAS $W$, $Z$ data at 7~TeV and the individual $W$ and $Z$ data at 8~TeV, is to enlarge the strangeness asymmetry, with it now clearly non-zero at $x\approx 0.1$ outside of the uncertainty band. Fig.~\ref{sminussbarfigs} (right) also shows the impact this additional data has had on its uncertainty bands (with the region $x \sim 0.01$ showing a large uncertainty as the asymmetry passes through zero), with a clear reduction in the error bands over the entire $x$ range.

\subsection{Comparison at NLO}

The changes in the NLO PDFs are generally very similar 
to those at NNLO. However, there are nonetheless some noteworthy features, resulting from both the different order of evolution and theoretical predictions. We detail these here although some of the differences resulting from the order will be discussed also in Section~\ref{sec:7}.

\begin{figure}
\begin{center}
\includegraphics[scale=0.24, trim = 50 0 0 0 , clip]{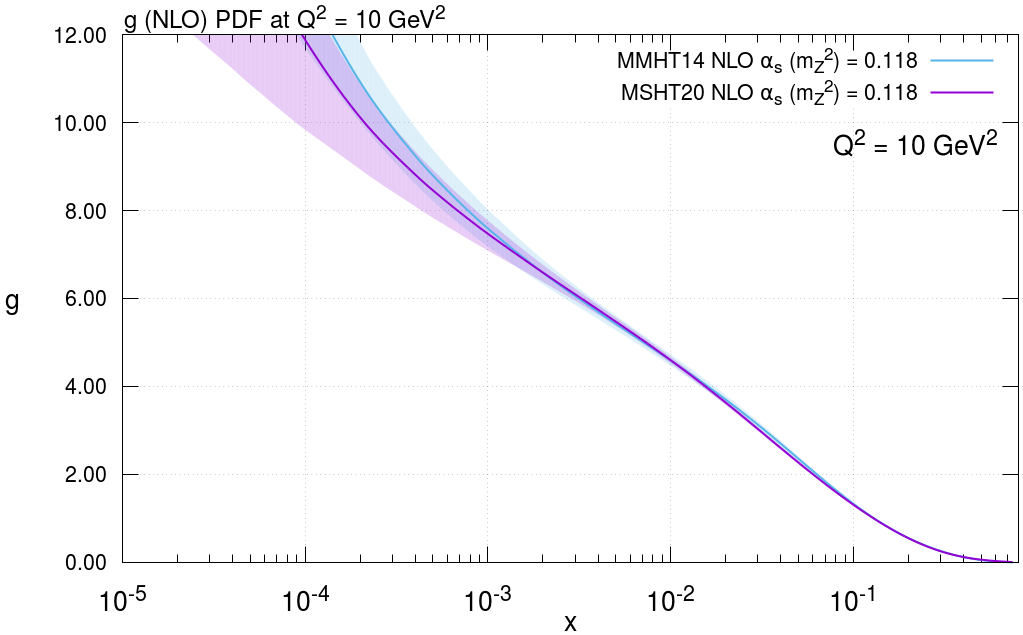}
\includegraphics[scale=0.24, trim = 50 0 0 0 , clip]{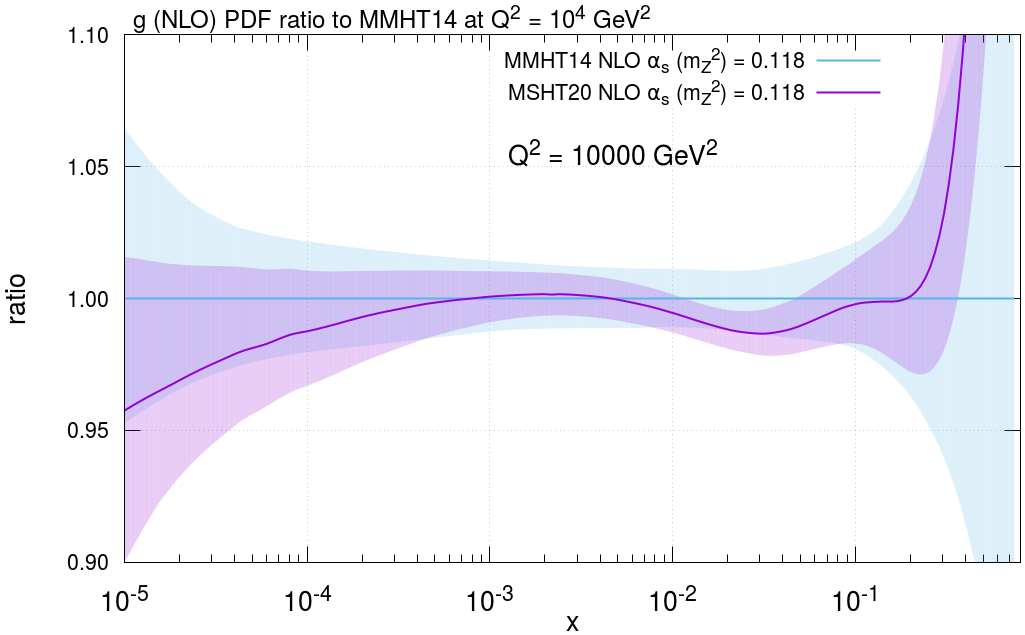}
\caption{\sf Gluon PDFs, for MSHT20 and MMHT14 at NLO. (Left) absolute PDFs at $Q^2=10~\GeV^2$. (Right)  the ratio to MMHT14 at $Q^2=10^4~\GeV^2$ .}\label{gluon_NLO_MMHT14comp}
\end{center}
\end{figure}

The comparison of the gluon PDF in MSHT20 at NLO with that of MMHT14 shows some differences relative to those seen at NNLO. Fig.~\ref{gluon_NLO_MMHT14comp} provides the absolute gluon PDF at NLO at $Q^2=10~\GeV^2$ on the left and the ratio at $Q^2=10^4~\GeV^2$ to MMHT14 at NLO on the right.  At NLO the MSHT20 gluon is a little lower relative to MMHT14 than at NNLO and this feeds through into the quarks at low $x$. Meanwhile there are also differences at high $x>0.2$, where the gluon at NLO is now larger than MMHT14. This results from the sum rule causing the gluon to compensate for being lower at low and intermediate $x$, i.e. $x \sim 0.05$, by becoming larger at high $x$. The change near $x=0.05$ is due to different pulls 
on the gluon from new data, e.g. $Z$ $p_T$ data, at NLO compared to NNLO, while that at very 
high-$x$ is also due to e.g. LHC jet data. 

\begin{figure}
\begin{center}
\includegraphics[scale=0.24, trim = 50 0 0 0 , clip]{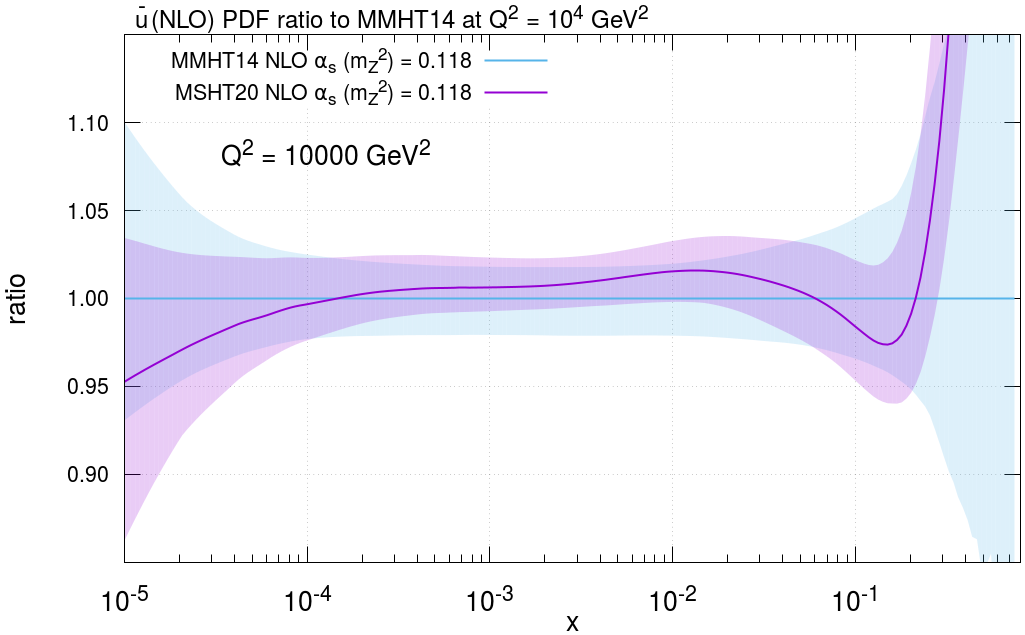}
\includegraphics[scale=0.24, trim = 50 0 0 0 , clip]{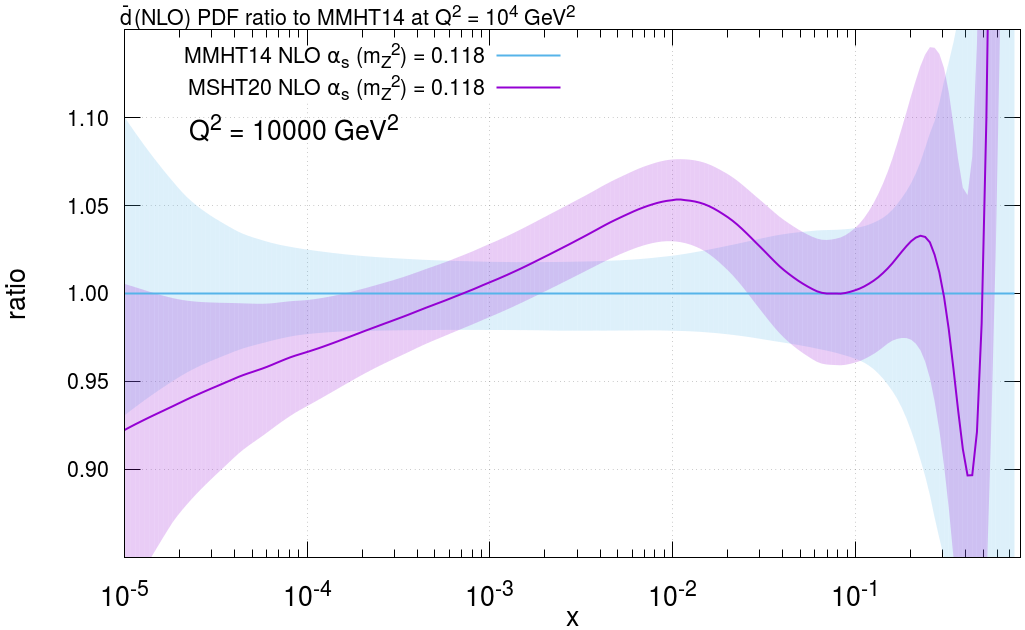}
\caption{\sf (Left) Up antiquark, $\bar{u}$, and (right) Down antiquark, $\bar{d}$, PDF ratios to MMHT14 at NLO and $Q^2=10^4~\GeV^2$. }\label{dbarubar_q210000_NLOratio}
\end{center}
\end{figure} 

There are also some changes in the $\bar{d}$ and $\bar{u}$ both in their absolute values and in their difference and ratio to one another. In Fig.~\ref{dbarubar_q210000_NLOratio} their ratios to MMHT14 at NLO at $Q^2=10^4~\GeV^2$ are given. It is noticeable that at NLO the up and down antiquarks are not reduced relative to MMHT14 in the way they are at NNLO. This is as a result of the difference in the total strangeness between NLO and NNLO. As explained later, at NLO the total strangeness is not enhanced relative to MMHT14 and so consequently $\bar{u}$ and $\bar{d}$ are not reduced to compensate for this in the fit. There is also a noticeable peak in $\bar{d}$ around $x \approx 10^{-2}$ which is not present at NNLO. This appears to allow for the fit to the ATLAS 7~TeV $W$, $Z$ data, without needing the same increase in strangeness in this region as occurs at NNLO. However,  it is noteworthy that the quality of the fit at NLO for this data set is much worse than at NNLO (as seen in Table~\ref{tab:LHCchisqtable}). The reduction in the gluon at low $x$ has also translated to a reduction to the antiquarks in this region.

\begin{figure}
\begin{center}
\includegraphics[scale=0.24, trim = 50 0 0 0 , clip]{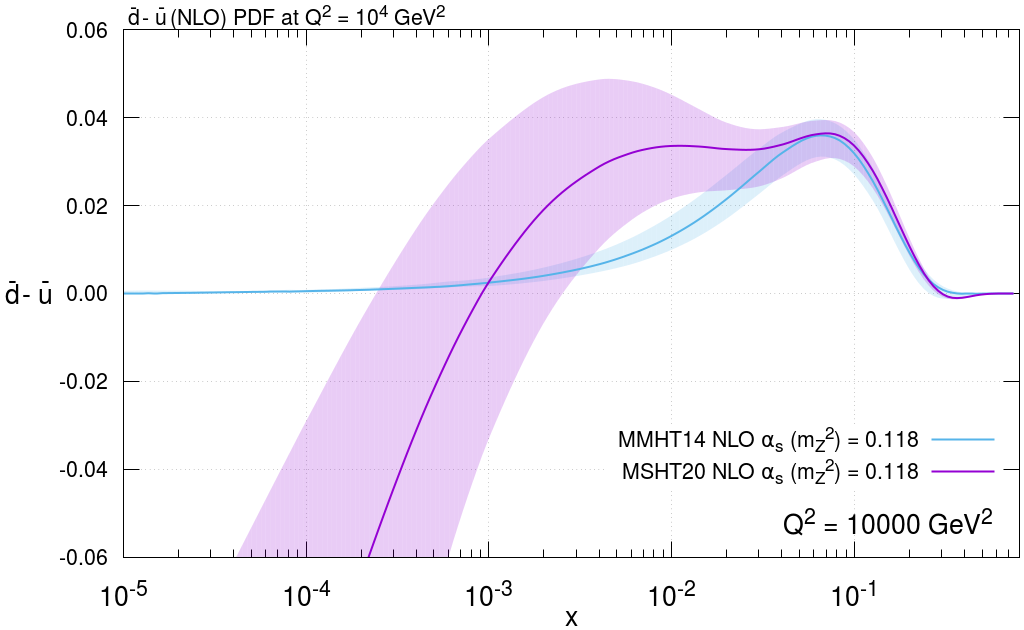}
\includegraphics[scale=0.24, trim = 50 0 0 0 , clip]{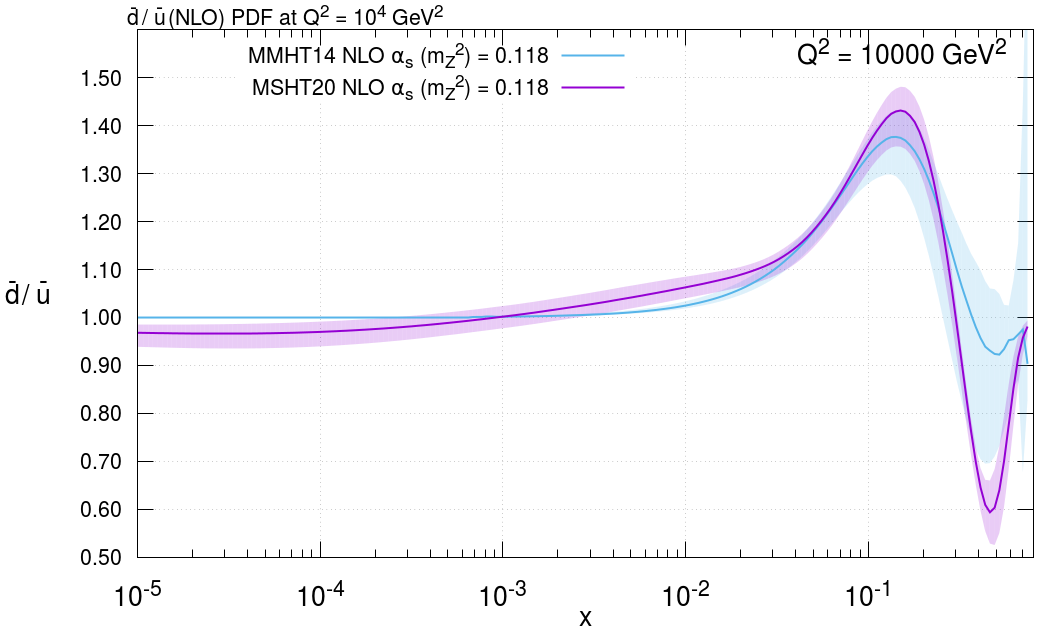}
\caption{\sf (Left) $\bar{d} - \bar{u}$ and (right) $\bar{d} / \bar{u}$  PDFs at NLO and $Q^2=10^4~\GeV^2$.}\label{dbartoubar_q210000_NLO_absolute_plots}
\end{center}
\end{figure} 

\begin{figure}
\begin{center}
\includegraphics[scale=0.24, trim = 50 0 0 0 , clip]{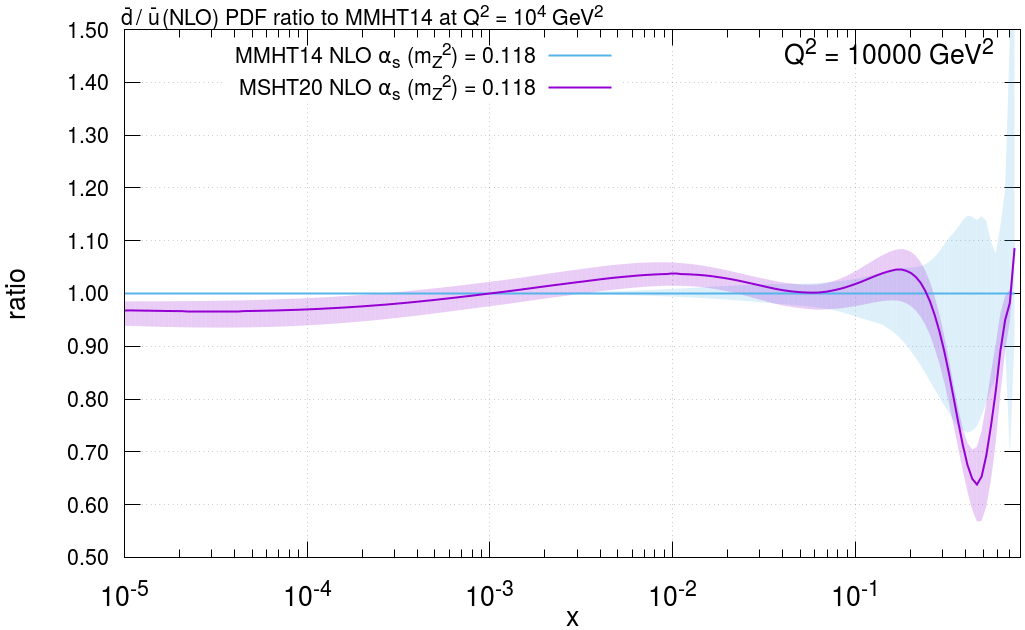}
\includegraphics[scale=0.24, trim = 50 0 0 0 , clip]{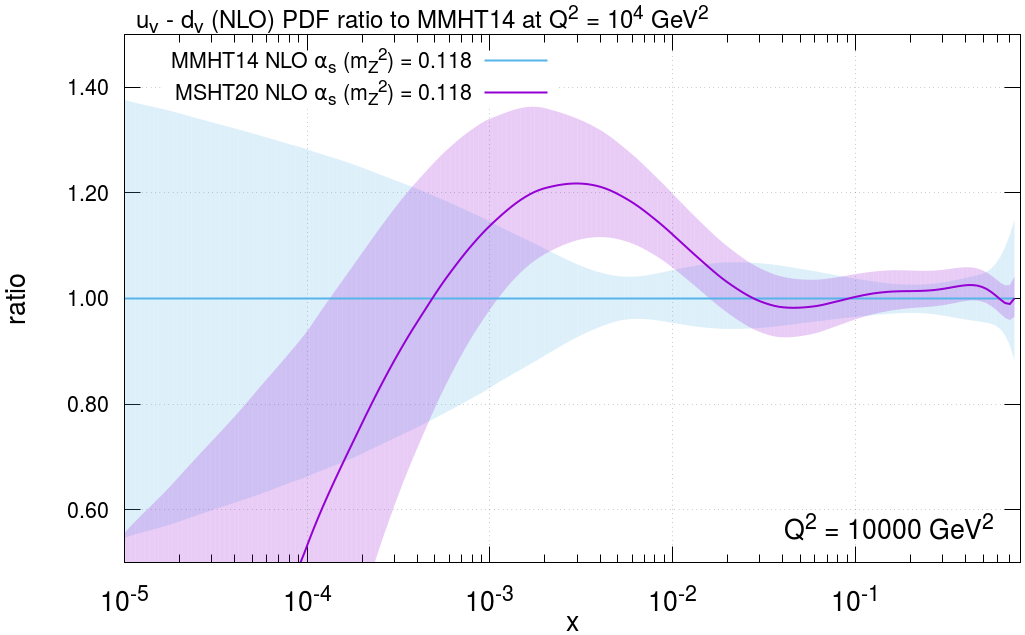}
\caption{\sf (Left) $\bar{d} / \bar{u}$ and (right) $u_V - d_V$ PDFs ratios to MMHT14 at NLO and $Q^2=10^4~\GeV^2$.}\label{uvminusdv_NLO}
\end{center}
\end{figure} 

\begin{figure}[t]
\begin{center}
\includegraphics[scale=0.24, trim = 50 0 0 0 , clip]{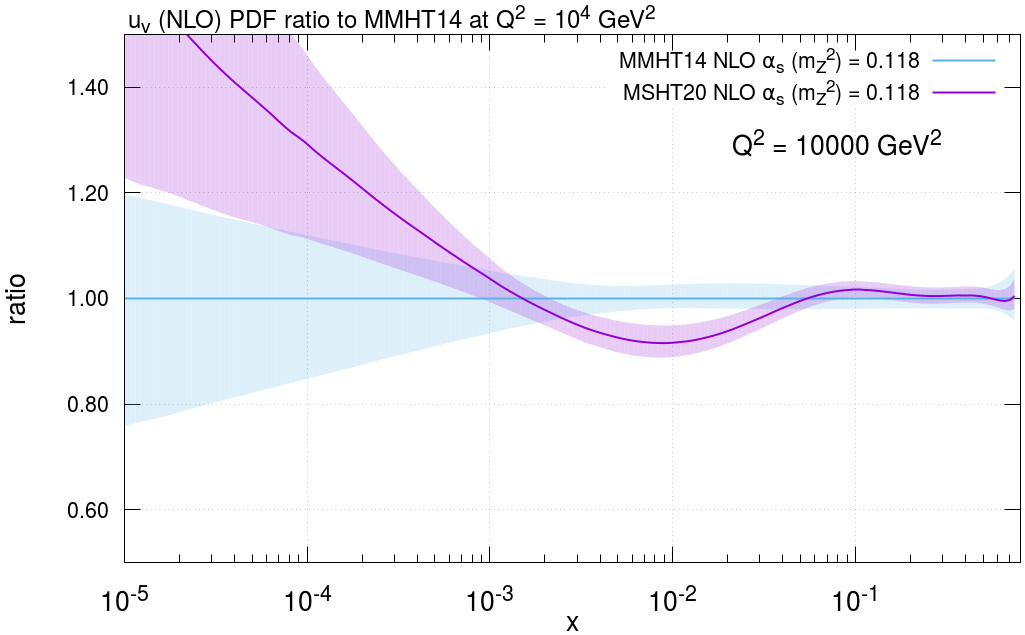}
\includegraphics[scale=0.24, trim = 50 0 0 0 , clip]{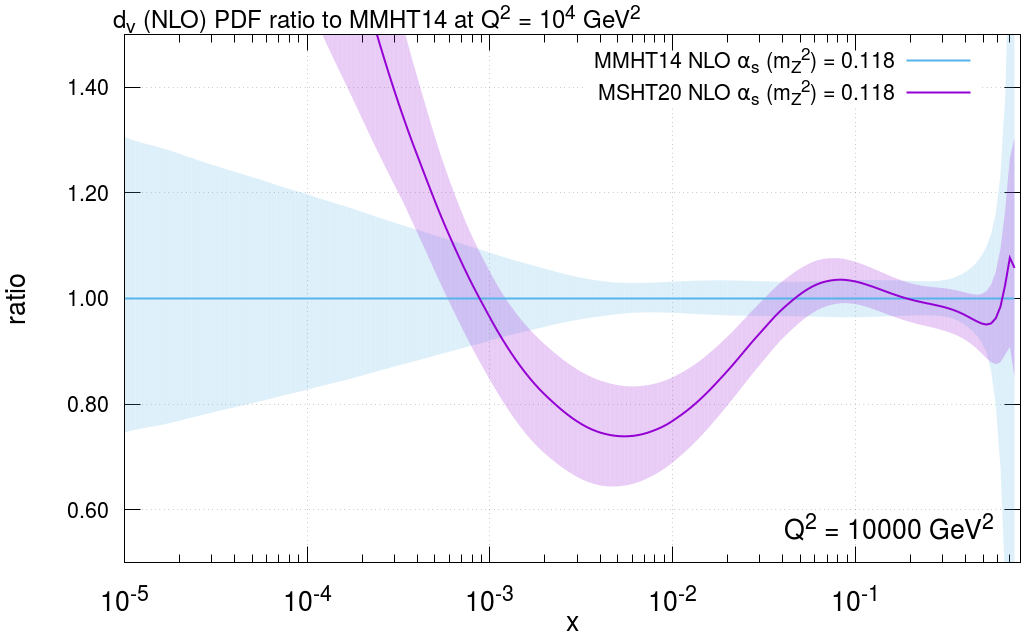}
\caption{\sf (Left) $u_V$ and (right) $d_V$ PDF ratio to MMHT14 at NLO and $Q^2=10^4~\GeV^2$.}\label{uv_dv_NLO}
\end{center}
\end{figure} 

\begin{figure}[t]
\begin{center}
\includegraphics[scale=0.24, trim = 50 0 0 0 , clip]{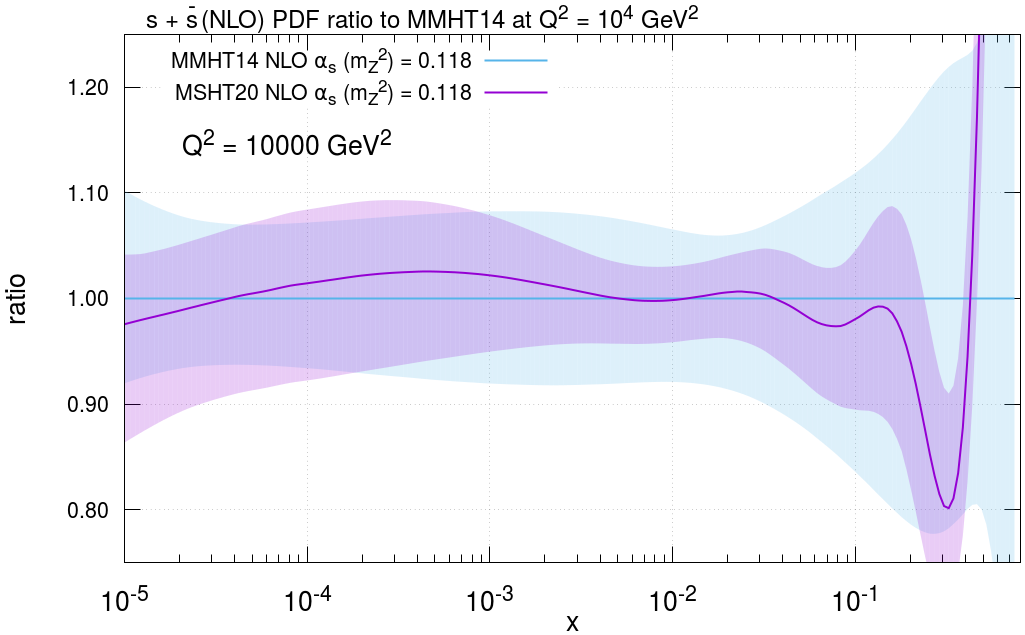}
\includegraphics[scale=0.24, trim = 50 0 0 0 , clip]{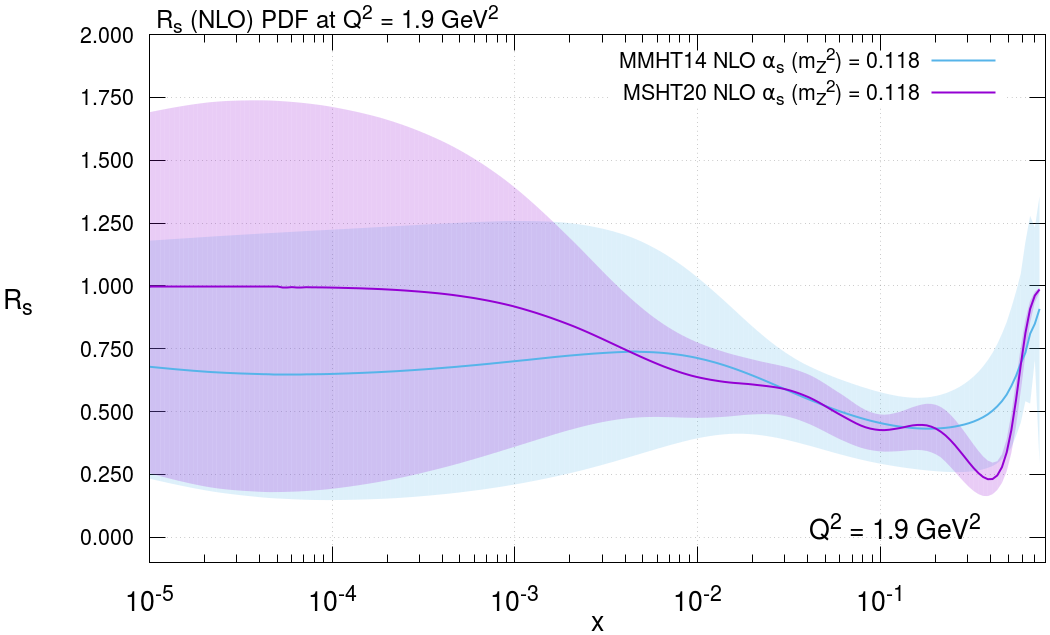}
\caption{\sf (Left) $s+\bar{s}$ at $Q^2=10^4~\GeV^2$ and (right) $R_s$ at $Q^2=1.9~\GeV^2$ PDF ratio to MMHT14 at NLO.}\label{totalstrangeness_NLOratios}
\end{center}
\end{figure} 

Next we focus on the comparison to MMHT14 at NLO in the asymmetry and the ratio of $\bar{d}$ and $\bar{u}$. In Fig.~\ref{dbartoubar_q210000_NLO_absolute_plots} the central values of the asymmetry $\bar{d}-\bar{u}$ and ratio $\bar{d}/\bar{u}$ are given. Comparing this with the behaviour at NNLO in the corresponding plots in Fig.~\ref{dbarminusubar}, clear differences at both low and high $x$ are visible. Whilst at high $x$ the same broadened peak is visible, the increase in the amplitude of the ratio of $\bar{d}/\bar{u}$ between MMHT14 and MSHT20 is enhanced at NLO relative to NNLO. This is a reflection of the increased $\bar d$ distribution at $x\sim 0.01$ already noted. 
We can see that, in contrast to the NNLO case, the ratio of $\bar{d}/\bar{u}$ at very low $x$ is no longer consistent with 1 within errors.
 This is clearer in the asymmetry in Fig.~\ref{dbartoubar_q210000_NLO_absolute_plots} (left) where at NLO the MSHT20 $\bar d -\bar u$ becomes negative and does not turn back up towards zero at very low $x$. These points can also be seen in the ratio to MMHT14 for $\bar{d}/\bar{u}$ in Fig.~\ref{uvminusdv_NLO} (left). This emphasises the fact that we see the ratio of $\bar{d}$ and $\bar{u}$ tend to 1 at low $x$ at NNLO in MSHT20 is not fixed by the parameterisation. 

Fig.~\ref{uvminusdv_NLO} (right) presents the ratio to MMHT14 of the difference of the valence PDFs $u_V-d_V$ at NLO, and is to be compared with the NNLO plot in Fig.~\ref{uvminusdv} (left). The general shape of the difference between MSHT20 and MMHT14 is consistent between NLO and NNLO, emphasising that this difference is driven by the data, albeit aided by the extra flexibility in the MSHT20 parameterisation. Nevertheless the details of the ratios are different, with the ratio to MMHT14 at NLO not becoming large at very high $x$ in the way it did at NNLO. In addition, the differences at low $x$ between MMHT14 and MSHT20 are reduced, with the latter becoming negative at lower $x$ at NLO. Also, partly down to the increased error bands at NLO, MSHT20 is consistent within uncertainties with the MMHT14 $u_V-d_V$ at NLO at low $x$, unlike at NNLO, despite its drastically different shape in this region. This greater similarity between MSHT20 and MMHT14 at NLO than NNLO in $u_V-d_V$ is not due to more similar individual valence quarks. For the down valence, shown in Fig.~\ref{uv_dv_NLO} (right), the difference is quite similar to the NNLO version in Fig.~\ref{uvdvratios} (right), but the up valence shown in Fig.~\ref{uv_dv_NLO} (left) actually gets further away from MMHT14 at NLO than at NNLO (Fig.~\ref{uvdvratios} (right)). Hence, at NLO a similar change in shape of both valence quarks cancels to some extent in their difference. 

Finally, we compare the total strangeness and $R_s$. Comparing Fig.~\ref{totalstrangeness_NLOratios} (left) with the comparable NNLO in Fig.~\ref{splussbarfigs} (left) there are significant differences, with MSHT20 much closer to MMHT14 at NLO than at NNLO. In particular, the strangeness enhancement at $x \approx 0.02$ resulting from the high precision ATLAS $W$, $Z$ data at 7~TeV at NNLO (and also supported by the separate ATLAS 8~TeV $W$ and $Z$ data sets) is no longer present at NLO. This feature is reflected in the comparisons of $R_s$ in Fig.~\ref{totalstrangeness_NLOratios} (right) and \ref{Rs_q21p9_NNLO}. In this comparison it is also clear that the form of $R_s$ at low $x$ is also altered, albeit well within uncertainties, with MSHT20 at NLO tending to 1, and being higher than MMHT14, which is not true at NNLO. In addition, the errors on $R_s$ in this region in MSHT20 at NLO are also notably larger than at NNLO.

\section{Dependence on the perturbative order} \label{sec:7}

\subsection{Comparison of NLO and NNLO}

\begin{figure}
\begin{center}
\includegraphics[scale=0.24, trim = 50 0 0 0 , clip]{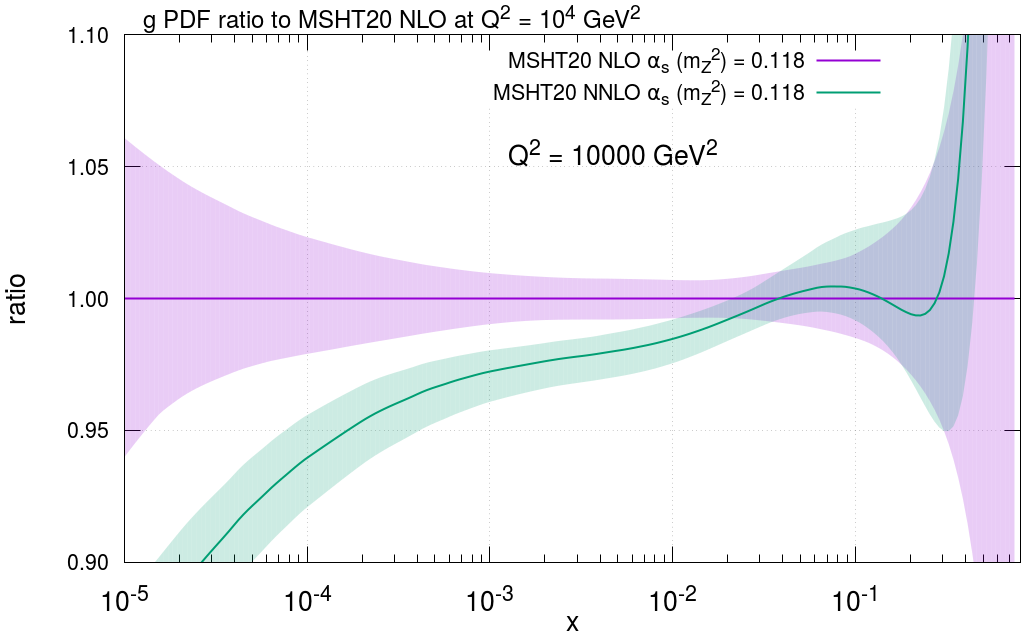}
\includegraphics[scale=0.24, trim = 50 0 0 0 , clip]{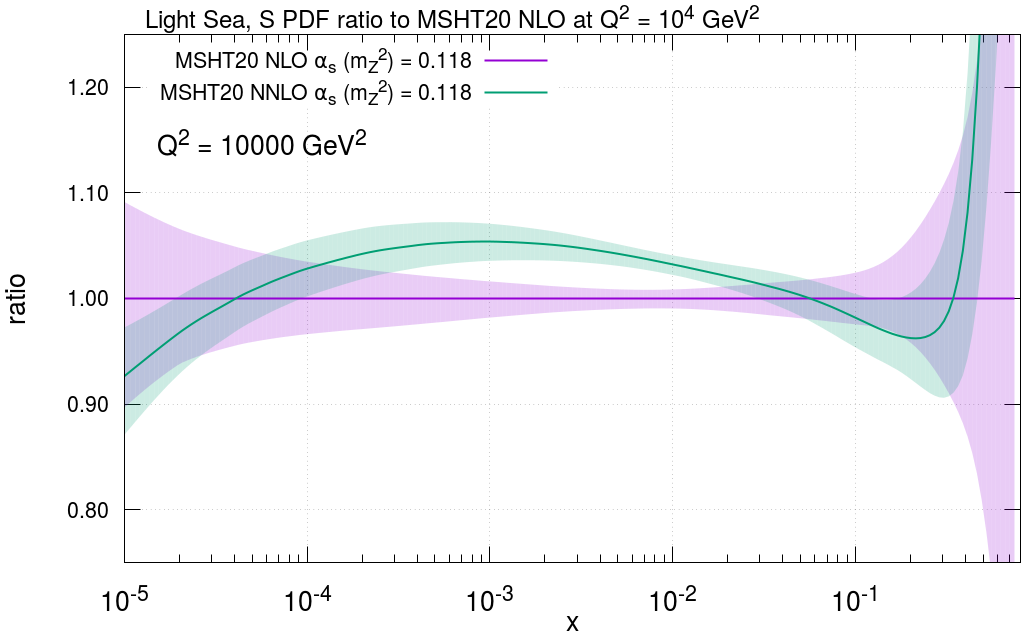}
\caption{\sf (Left) Gluon and (right) Light sea PDFs showing the ratio of the MSHT20 PDFs to their NLO values, at $Q^2=10^4~\GeV^2$.}\label{gluonlightseaMSHT20_NLONNLOcomp}
\end{center}
\end{figure} 

\begin{figure}[t]
\begin{center}
\includegraphics[scale=0.24, trim = 50 0 0 0 , clip]{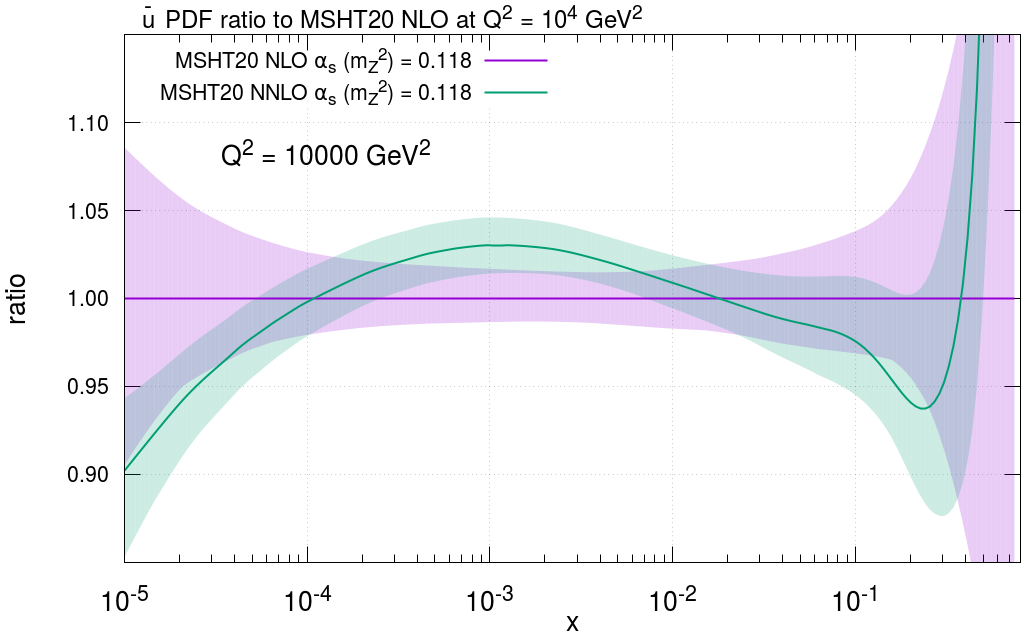}
\includegraphics[scale=0.24, trim = 50 0 0 0 , clip]{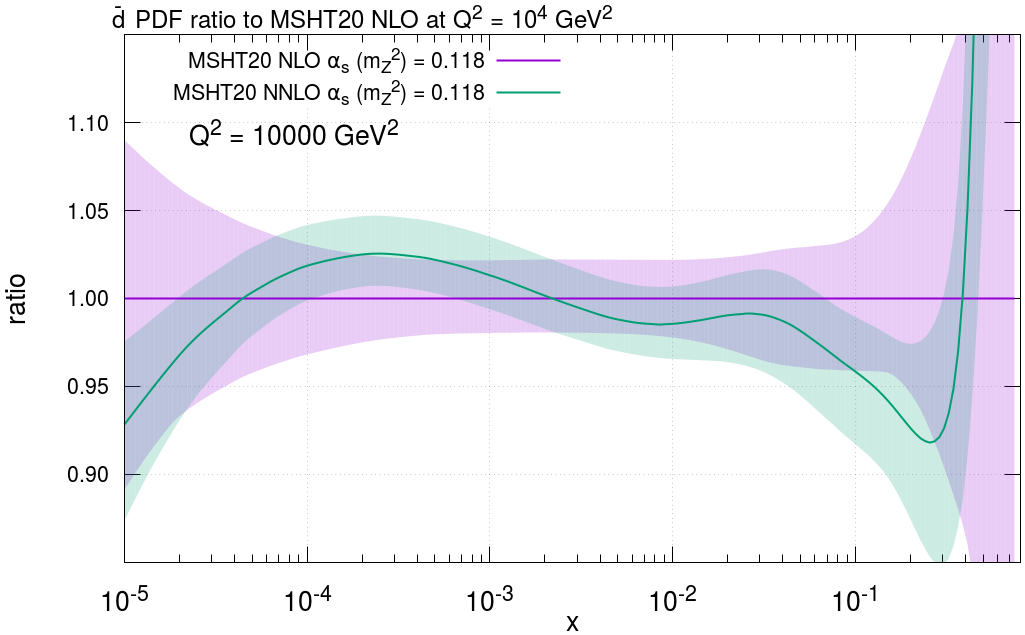}
\caption{\sf (Left) $\bar{u}$ and (right) $\bar{d}$ PDFs showing the ratio of the MSHT20 PDFs to their NLO values, at $Q^2=10^4~\GeV^2$.}\label{lightantiquarksMSHT20_NLONNLOcomp}
\end{center}
\end{figure} 

\begin{figure}[t]
\begin{center}
\includegraphics[scale=0.24, trim = 50 0 0 0 , clip]{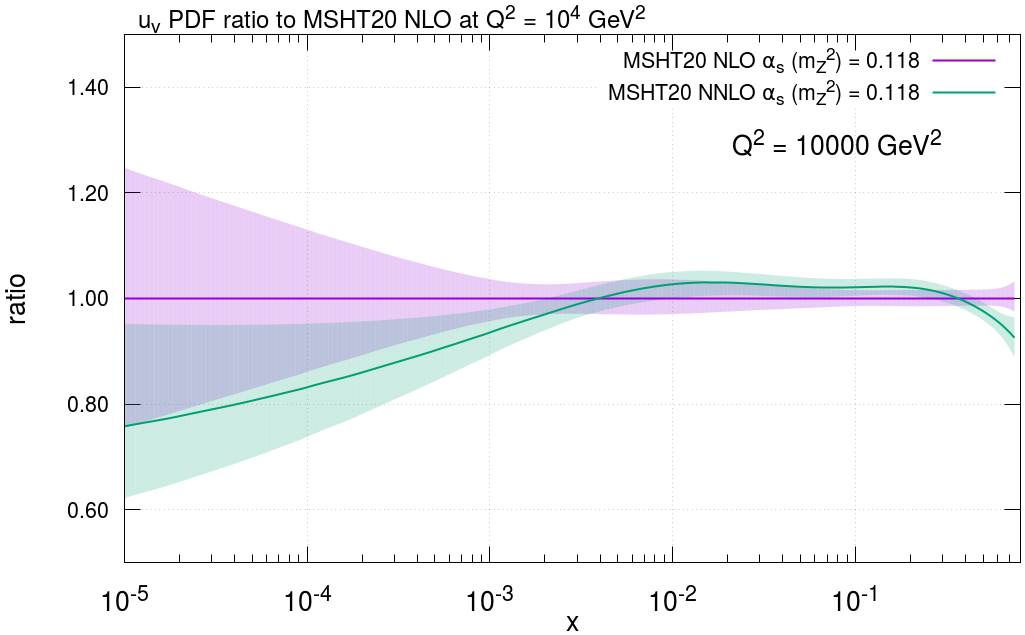}
\includegraphics[scale=0.24, trim = 50 0 0 0 , clip]{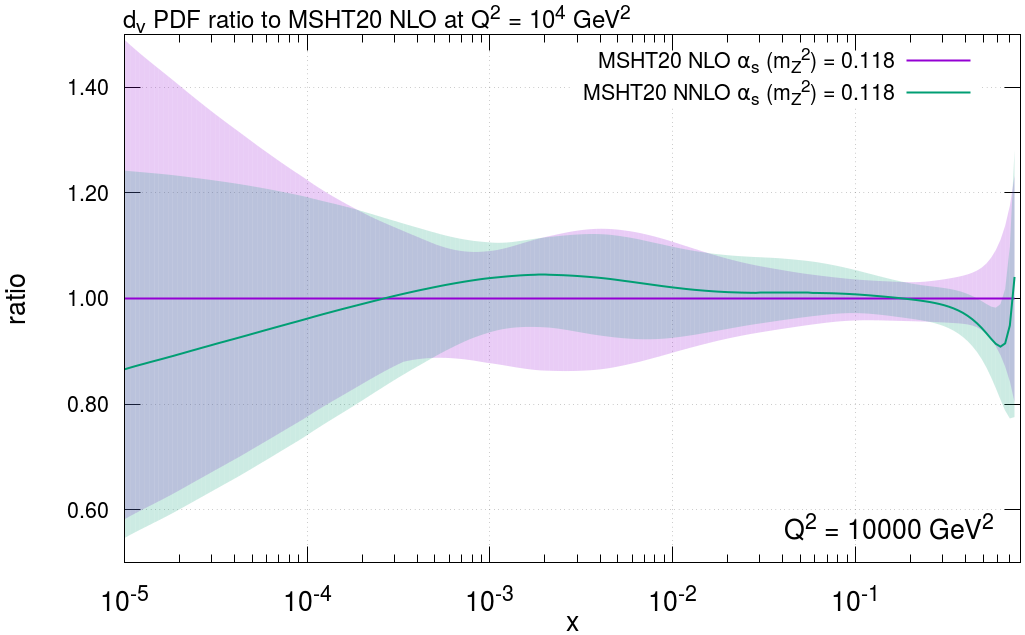}
\caption{\sf (Left) $u_V$ and (right) $d_V$ PDFs showing the ratio of the MSHT20 PDFs to their NLO values, at $Q^2=10^4~\GeV^2$.}\label{uv_dv_q210000_ratioNLONNLO}
\end{center}
\end{figure}

In this section we directly compare MSHT20 at NNLO with the MSHT20 NLO fit. This reveals more straightforwardly some of the features present in comparing the NLO results and the NNLO results against MMHT14 in the previous section.

First, in Fig.~\ref{gluonlightseaMSHT20_NLONNLOcomp} we show the changes in the gluon and the light quark sea, which are some of the more significant changes between NLO and NNLO. In particular the gluon is affected at low $x$ by the change in the order of the evolution, with the NNLO splitting function $P_{qg}$ having an additional small $\ln (x)/x$ enhancement which is therefore compensated for in the fit by a reduced gluon at NNLO, similar to the standard reduction in $\alpha_S$ as the order is increased. The gluon at high $x$ is then increased as a result of the momentum sum rule. The light sea shows a different shape in the intermediate $x$ region with it being enhanced at NNLO below $x \approx 0.1$. This is a consequence of the negative NNLO structure function coefficient function in this $x$ region, which in turn causes more sea quarks to be needed at NNLO in order to fit the data.

Next, in Fig.~\ref{lightantiquarksMSHT20_NLONNLOcomp} we show results for the $\bar{u}$ and $\bar{d}$. As for the light quark distribution, these are enhanced at intermediate $x$, however in the latter case this effect is complicated by the competing increase in the $\bar{d}$ around $x \approx 0.02$ at NLO. This arises from the impact of the strangeness no longer being enhanced at NLO relative to MMHT14 and the need to fit the ATLAS 7~TeV $W$, $Z$ data.

\begin{figure}
\begin{center}
\includegraphics[scale=0.24, trim = 50 0 0 0 , clip]{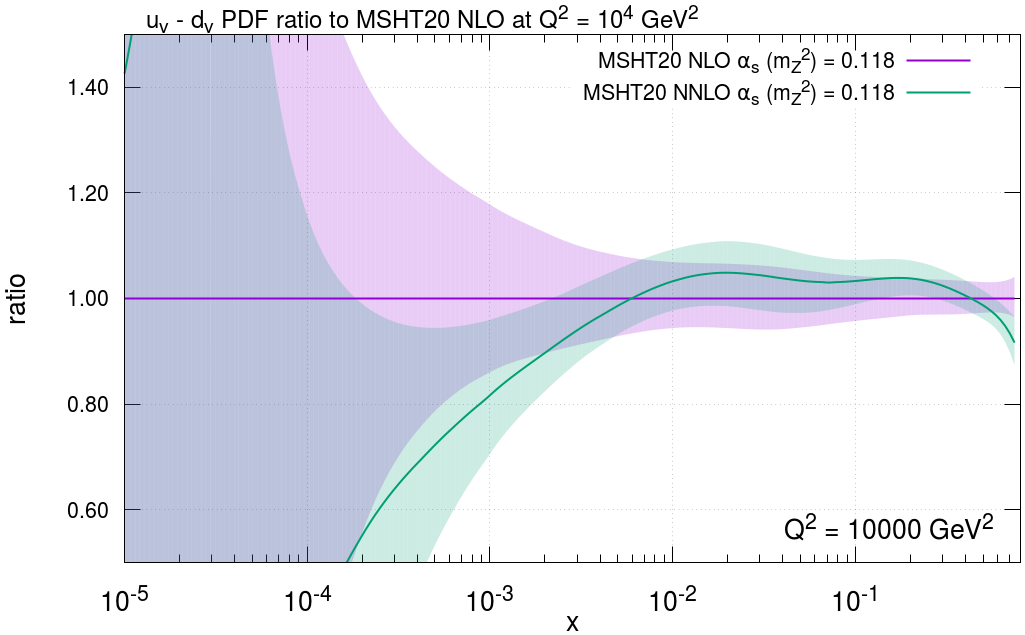}
\includegraphics[scale=0.24, trim = 60 0 0 0 , clip]{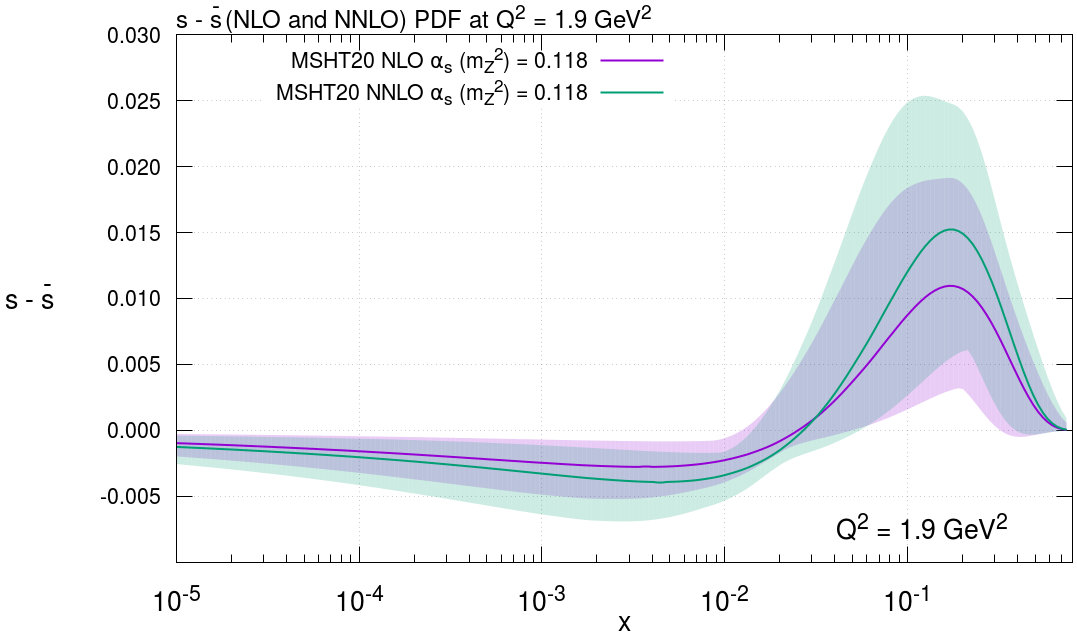}
\caption{\sf (Left) $u_V-d_V$ PDF showing the ratio of the MSHT20 PDFs to their NLO value at $Q^2=10^4~\GeV^2$. (right) $s-\bar{s}$ PDFs showing the absolute values of the MSHT20 PDFs at NLO and NNLO at $Q^2=1.9~\GeV^2$.}\label{uvmdv_sminussbar_NLO}
\end{center}
\end{figure}

The valence quarks and their difference also have appreciable differences between NLO and NNLO as shown in Figs.~\ref{uv_dv_q210000_ratioNLONNLO} and \ref{uvmdv_sminussbar_NLO} (left). The up valence quark is reduced at low and very high $x$ and marginally enlarged at intermediate $x$. The reduction at low $x$ is likely a consequence of the valence sum rule, whilst at high $x$ this is due to NNLO corrections to cross sections enlarging the contribution of the quarks, and in turn therefore requiring less quark than at NLO. The small increase at intermediate $x$ results from the opposite effect, i.e. the 
NNLO coefficient functions giving a negative correction, resulting in larger up valence at 
NNLO (the $F_2$ coefficient functions, as well as the valence quarks, satisfy the valence sum rule). In contrast, whilst the down valence altered drastically between MMHT14 and MSHT20, with a significant change in shape outside the MMHT14 error bands, it is consistent between NLO and NNLO in MSHT20. This emphasises that the change is a consistent effect arising from the extended parameterisation and the additional data in the fit. Note, however, that the general trend of the shape difference between NLO and NNLO is rather similar to the up valence, but is less obvious due to larger uncertainties for $d_V$. 
The difference $u_V - d_V$ therefore alters between NLO and NNLO, mainly reflecting the difference in the up valence (which is twice the size, on average, of the down valence), falling sharply below intermediate $x$, and for part of the region it is outside the $u_V-d_V$ MMHT14 error bands.

\begin{figure}
\begin{center}
\includegraphics[scale=0.24, trim = 50 0 0 0 , clip]{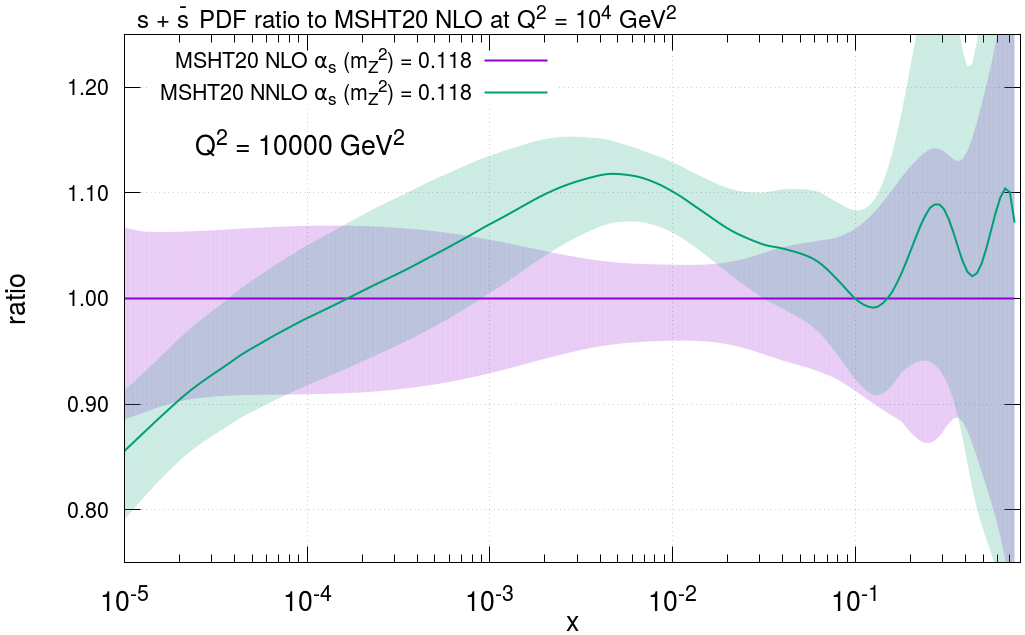}
\includegraphics[scale=0.24, trim = 50 0 0 0 , clip]{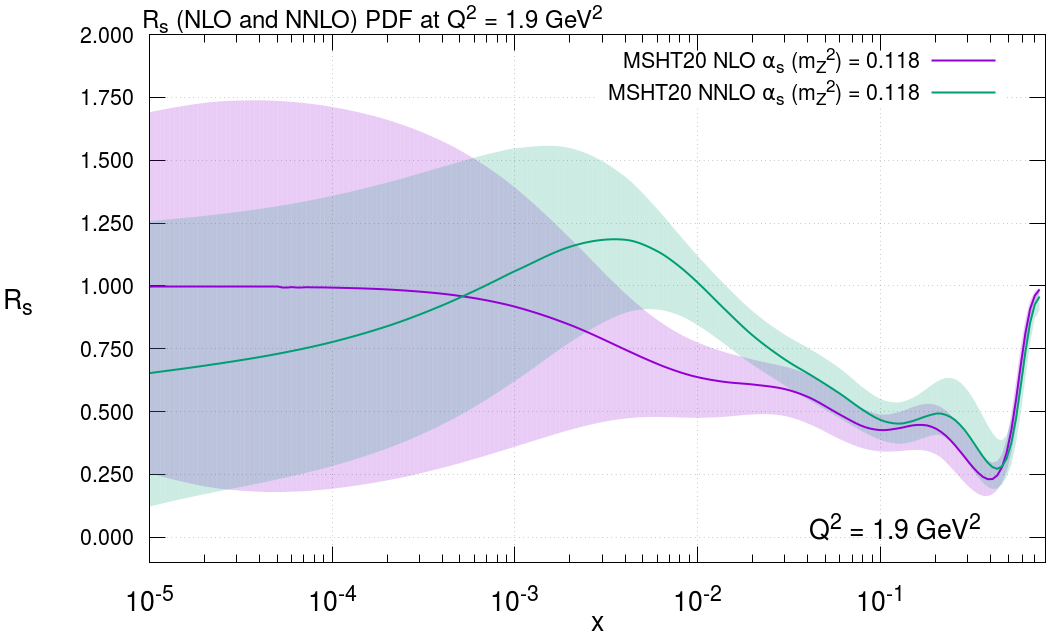}
\caption{\sf (Left) $s+\bar{s}$ PDF showing the ratio of the MSHT20 PDFs to their NLO value at $Q^2=10^4~\GeV^2$. (right) $R_s$ PDFs showing the absolute values of the MSHT20 PDFs at NLO and NNLO at $Q^2=1.9~\GeV^2$.}\label{totalstrangenessMSHT20_NLONNLOcomp}
\end{center}
\end{figure}

Finally, we consider the total strangeness and strangeness asymmetry. In the previous two sections when comparing with MMHT14 we found that the MSHT20 $s - \bar{s}$ was larger in amplitude than MMHT14 (Fig.~\ref{sminussbarfigs} (left)). However, not only is the asymmetry larger in MSHT20, it is also larger at NNLO than NLO, as observed in Fig.~\ref{uvmdv_sminussbar_NLO} (right). As described in Section~\ref{strangeness}, this is also the result of the impact of the ATLAS 7~TeV $W$, $Z$ data. As shown in Fig.~\ref{totalstrangenessMSHT20_NLONNLOcomp}, at NNLO this data set (and partly the corresponding 8~TeV data sets) pulls the total strangeness up, leading to the well known ``strangeness unsuppression'' at $x \approx 0.02$. This consequently lowers the up and down antiquarks to compensate for this effect, to fit HERA data, and also seems to enhance the strangeness asymmetry. However, at NLO the ATLAS 7~TeV $W$, $Z$ is much more poorly fit and rather than cause an enhancement in the total strangeness, the best global fit results in an enlargement in the $\bar{d}$ in this region. As a consequence the strangeness asymmetry also appears smaller at NLO than NNLO.

In some senses we might regard the difference between the NLO and NNLO PDFs as a representation of the theory
uncertainty of the PDFs, that is due to the approximate nature of the pQCD theory predictions entering the fit. However, this observation is misleading and complicated by a number of issues. First, in a given scheme the NLO and NNLO PDFs
are not strictly the same object - for any physical quantity the change in the PDFs from one order to the next may be 
compensated for, or exacerbated by, the change in the QCD cross section. Indeed, we see this in the change from NLO to 
NNLO in the light sea quarks for $x \sim 0.001$, compensating for the change in structure function coefficient 
functions. These cross section corrections will be different from process to process, but in a global fit may still represent a cumulative effect.
Also, the NNLO PDFs evolve differently to the NLO PDFs, so equivalence at one scale does not imply equivalence at 
another. Nevertheless, overall one may potentially think of the change in the PDF from NLO to NNLO at given $x$ and 
$Q^2$ giving some indication of a theoretical uncertainty on the PDFs and also of the possible change in going from NNLO to 
N${}^3$LO. Again, the scope for such an observation is limited somewhat by the difficulty we see in obtaining a very high quality fit at 
NLO. It seems that relative changes in PDFs are, to some extent, attempting to compensate for distinct limitations in the 
NLO calculations. This is clear in the poor fit to ATLAS $W,Z$ data and the very different strange quark fraction at 
NLO compared to NNLO. Since the NNLO fit does achieve a much better fit to the ATLAS $W,Z$ data we would not obviously 
expect a similar large change in the strange quark when going from NNLO to N${}^3$LO. It is much more meaningful to infer
the size of theoretical uncertainty on specific physical quantities by comparing NLO and NNLO predictions for these quantities.

\subsection{LO fit}

 \begin{figure} [t]
\begin{center}
\includegraphics[scale=0.24]{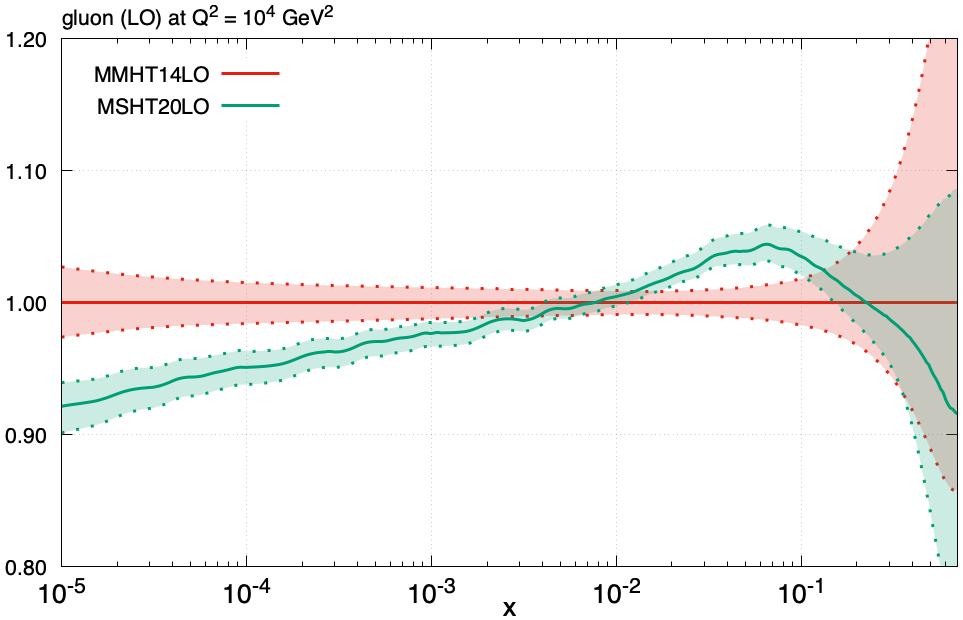}
\includegraphics[scale=0.24]{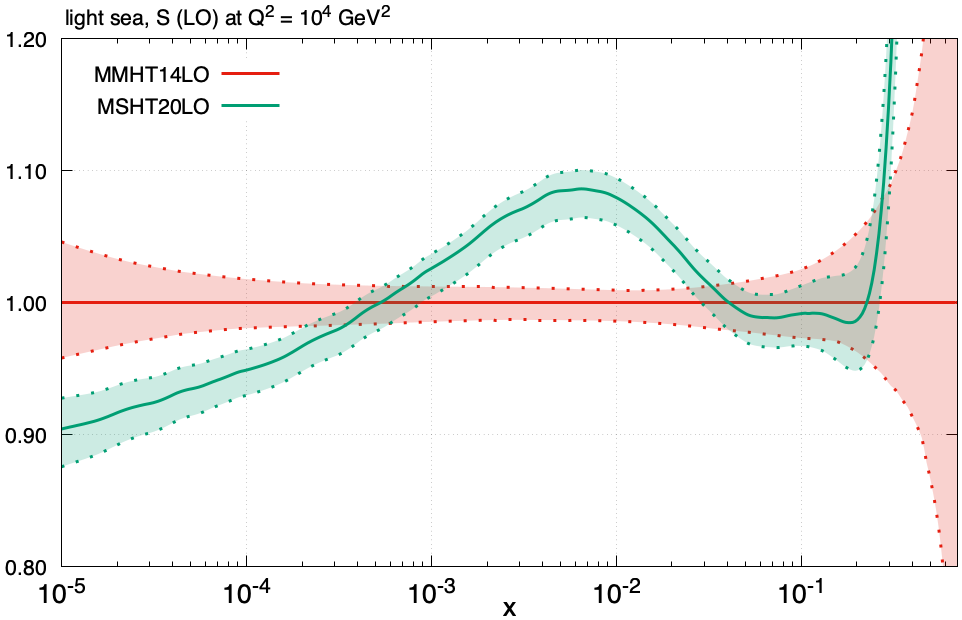}
\includegraphics[scale=0.24]{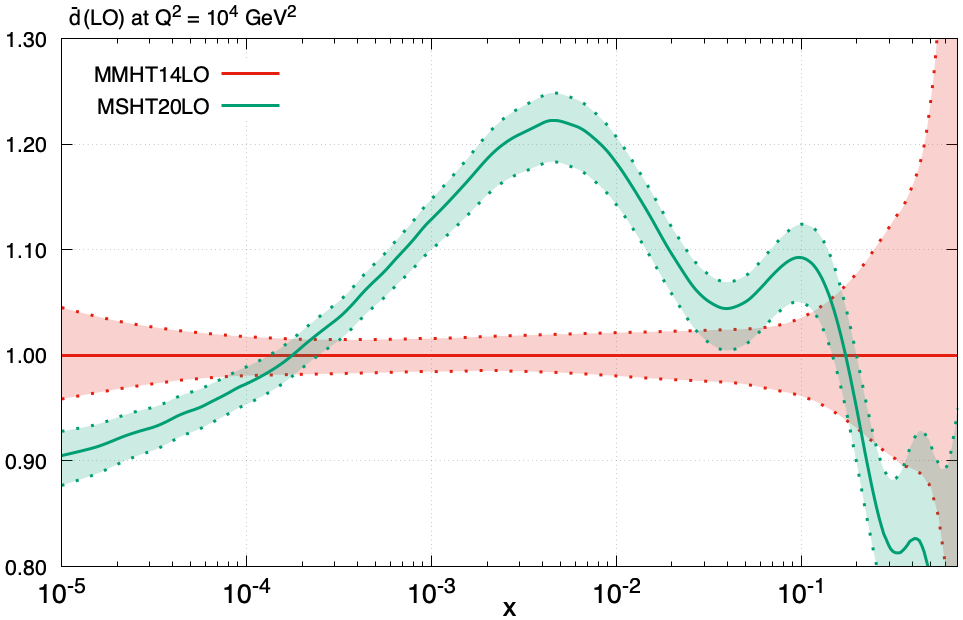}
\includegraphics[scale=0.24]{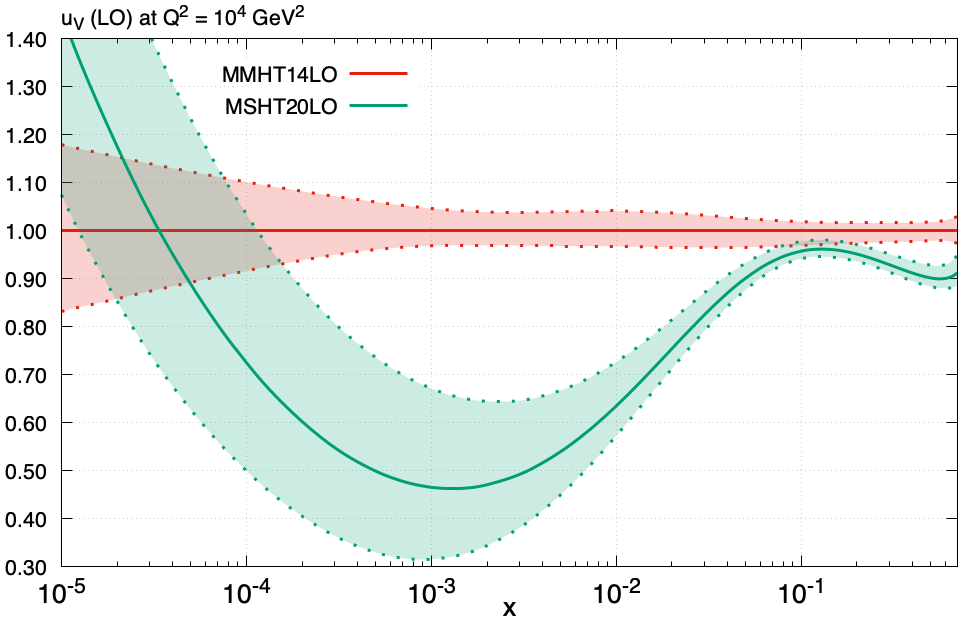}
\caption{\sf Ratio of MSHT20 LO to MMHT14LO PDFs at $Q^2=10^4~\GeV^2$.}
\label{fig:LO}
\end{center}
\end{figure}

In addition to the above sets, we provide a LO fit. As in previous fits, a larger value of the coupling is preferred, and we fix this to $\alpha_S(M_Z^2)=0.130$ for concreteness. We also continue to apply an overall $K$--factor of $1+ \alpha_S(m_{ll}^2)C_F \pi/2$ to DY predictions, in order to account for the large difference in the space--like and time--like regimes relevant for the DIS and DY processes, respectively.

If a fit is attempted with the same parametric freedom as in the NLO/NNLO fits, rather pathological behaviour is observed in the resulting quark distributions; a similar effect was seen in the MMHT14 fit~\cite{MMHT14} with respect to the $s_+$ distribution. We therefore fix certain parameters in order to avoid this. In particular, we apply the following restrictions: fix the normalisation parameter $A_{s_+}$ in order to impose that the value of the strangeness, $s_+$, normalisation, is fixed to that of the sea, $S$, this is particularly relevant at low $x$; fix the three Chebyshev coefficients $a_{s_+,i}$, with $i=1,4,6$ to values of the sea, $S$; set the high $x$ power of the second gluon term, to $\eta_{g_-}\approx 66$ (the precise value is arbitrary); set the sixth Chebyshev coefficient, $a_{\rho,6}$, such that $\overline{d}/\overline{u}\to 1$ as $x\to 0$. $A_{s_+}$ is a free parameter in the calculation of the PDF eigenvectors at NLO and NNLO, and we in addition fix $a_{s_+,3}$ in the LO eigenvector calculation in order to stabilise the corresponding Hessian matrix. Therefore, the resulting PDF set has 30, rather than 32, eigenvector sets.

The fit quality is extremely poor, with $\chi^2/N_{\rm pt} =2.58$. This is a significant deterioration with respect to the MMHT14 case, which found $\chi^2/N_{\rm pt} =1.34$. As we have seen at NLO, the fit quality is now rather worse, with $\chi^2/N_{\rm pt} =1.09$ (1.33) for MMHT14 (MSHT20). The reason for this is due to the new high precision LHC data in the fit, where NNLO theory is essential in order to provide a reasonable description. While NLO theory already gives a rather poor description in these cases, this becomes dramatically worse at LO, as we might expect.

In Fig.~\ref{fig:LO} we compare the LO MSHT20 set with MMHT14 for a selection of PDFs, at $Q^2=10^4\,{\rm GeV}^2$. In general, the difference between the sets is much more significant than that seen at NLO or NNLO, and is often well outside the quoted uncertainty bands. This is perhaps unsurprising, given just how poor the underlying fit quality is. In such a situation, it is far from expected that the uncertainty bands will provide a meaningful estimate.  Nonetheless, some of the effects are qualitatively in line with differences between MMHT and MSHT at NLO and NNLO, such as the reduction in the gluon at lower $x$, enhancement in the light sea at intermediate $x$ and reduction at low $x$, and reduction in $u_V$ at intermediate $x$. It is therefore tempting to conclude that these changes are driven at least in part by the new LHC data in the fit in a meaningful way, albeit within the context that the fit to such LHC data is extremely poor at LO. 

In summary, we provide the updated MSHT20LO set for potential use in LO Monte Carlo generators, though the use of NLO matrix elements is now rather standard. Beyond that, however, we would not strongly advocate for the use of these LO PDFs. Given the significant deterioration in fit quality in the current fit, we would not necessarily consider the MSHT20LO PDFs as preferable to MMHT14LO, but would rather consider that the difference between the two sets might provide some estimate of the underlying uncertainty at LO. Arguably, the most important conclusion from the above results is that as we approach the high precision LHC era, a purely LO PDF set can only be regarded as a qualitative concept.

\section{MSHT20: dedicated studies}\label{sec:8} 

There have been a range of updates, both theoretical and in terms of the data sets included, in the MSHT20 PDFs. In this section we present in detail a range of dedicated studies to highlight and explore further some of the key updates. In particular we consider: the extension of the PDF parameterisation;  the impact of the new LHC data; the inclusion of NNLO dimuon corrections; the use of the D{\O} $W$ (rather than lepton) asymmetry data; the impact and treatment of the ATLAS high precision Drell-Yan data; the result of allowing a non--zero strangeness asymmetry ($s \ne \bar s$) at input; the result of omitting the HERA data from the fit; and a detailed discussion of the tensions between various data sets in the global fit.

\subsection{New parameterisation, in particular for $\bar d /\bar u$.} \label{newparameterisationeffects}

\begin{table} 
\begin{center}
\begin{tabular}{|>{\centering\arraybackslash}m{4.2cm}|>{\centering\arraybackslash}m{2cm}|>{\centering\arraybackslash}m{1.15cm}|>{\centering\arraybackslash}m{1.35cm}|>{\centering\arraybackslash}m{1.65cm}|>{\centering\arraybackslash}m{1.6cm}|>{\centering\arraybackslash}m{1.9cm}|} \hline
  \multirow{2}{*}{Data set} & \multicolumn{6}{c|}{$\Delta \chi^2$ relative to MSHT20} \\ \cline{2-7}
  & Old parameterisation & $\bar{d}/\bar{u}$ 6 & $\bar{d}/\bar{u}$ 6, $d_V$ 6 & $\bar{d}/\bar{u}$ 6, $d_V$ 6, $u_V$ 6 & $\bar{d}/\bar{u}$ 6, $d_V$ 6, $u_V$ 6, $g$ 6 & $\bar{d}/\bar{u}$ 6, $d_V$ 6, $u_V$ 6, $g$ 6, sea 6  \\ \hline
  BCDMS $\mu p$ $F_2$ & -5.8 & -3.6 & -3.9 & -4.0 & -1.5 & -0.6 \\
  BCDMS $\mu d$ $F_2$ & 11.8 & 13.4 & 6.3 & 5.0 & 3.6 & 3.7 \\
  NMC $\mu p$ $F_2$ & 4.4 & 2.5 & 2.6 & 2.4 & 3.5 & 2.5 \\
  NMC $\mu d$ $F_2$ &  3.4 & 1.6 & 0.5 & 0.2 & 4.2 & 3.6 \\  
  E866/NuSea $pd/pp$ DY  & 8.3 & 3.0 & 2.1 & 1.5 & 1.6 & 1.8 \\  
  HERA $e^+ p$ NC 920~GeV & 10.9 & 9.3 & 10.9 & 11.6 & 5.3 & 2.4 \\
  CDF II $W$ asym. & 1.3 & 0.4 & 2.7 & 2.2 & 2.7 & 2.2 \\
  D{\O} II $W\rightarrow \nu e$ asym. & -3.8 & -1.2 & -0.1 & -0.3 & -0.2 & -0.6 \\
  LHCb 2015 $W$, $Z$ & -3.4 & -4.2 & -3.3 & -3.3 & -3.1 & -1.2 \\
  CMS 8~TeV $W$ & -0.8 & -1.2 & -0.4 & 1.2 & 1.1 & 0.9 \\
  ATLAS 7~TeV high precision $W$, $Z$ & 10.7 & 8.1 & 4.3 & 5.4 & 3.5 & 0.6 \\
  CMS 7~TeV jets & 8.8 & 7.3 & 8.2 & 8.3 & 0.3 & -0.2 \\
  D{\O} $W$ asym. & 3.6 & 2.5 & -0.8 & -1.9 & -1.2 & -0.8 \\
  ATLAS 8~TeV $Z$ $p_T$ & 11.9 & 9.6 & 8.6 & 8.1 & 1.7 & 3.2 \\
  CMS 8~TeV jets & 2.6 & 4.9 & 3.2 & 3.4 & 2.4 & 0.9 \\
  ATLAS 8~TeV $W$& -3.9 & -0.1 & 0.9 & 0.9 & 1.0 & 0.9 \\
  CMS 2.76~TeV jets & 3.8 & 3.4 & 3.3 & 3.4 & 2.3 & 2.0 \\
  ATLAS 8~TeV double differential $Z$ & -5.3 & -2.1 & -0.9 & -1.6 & -1.5 & 0.1 \\    \hline
  TOTAL & 73.3 & 54.8 & 40.8 & 40.8 & 22.8 & 16.9 \\ 
 \hline
\end{tabular}
\end{center}
\caption{ \sf The changes in $\chi^2$ relative to MSHT20 as the parameterisation is changed from the old MMHT14 parameterisation to the MSHT20 (default) enhanced parameterisation with $\bar{d}/\bar{u}$ and the extension of the parameterisation of the $d_V$, $u_V$, $g$, sea ($S$) and $s+\bar{s}$. The number 6 in the column title means that the PDF parameterisation was extended to 6 Chebyshevs. Positive $\Delta\chi^2$ means MSHT20 fits the data set better, i.e. it has a lower $\chi^2$ in MSHT20 than in the parameterisation used for that column.}
\label{tab:oldparamdeltachisqtable2}
\end{table}

One of the major changes in MSHT20 has been introduced in the parameterisation of the input distributions. This had been previously outlined in \cite{Thorne:2019mpt}. Moreover, it has been updated and described in more detail in Section~\ref{sec:inputPDF}, where several of the effects of this change have already been set out, whilst it was necessarily discussed alongside the comparison of MSHT20 and MMHT14 PDFs in Section~\ref{sec:6}. Nonetheless, it is informative to give these changes a little more attention. In this section we investigate the specific effects of just the parameterisation changes by performing several further global fits, beginning with the MSHT20 data set and theoretical settings but starting with the old MMHT14 parameterisation. This is then incremented towards the final MSHT20 new parameterisation in order to investigate the effects of each of the changes in isolation.

The changes in the parameterisation are numerous, but can broadly be split into changes made in the $\bar{d}$-$\bar{u}$ difference (or the $\bar{d}/\bar{u}$ ratio in MSHT20) and extensions to the parameterisations of the other PDFs, with the change to six Chebyshevs rather than the four or fewer used in MMHT14. The choice of six Chebyshevs is made 
based on the results in \cite{MMSTWW}, where a fit to pseudodata for PDF-type functional forms was made using 
parameterisations with increasing numbers of Chebyshev polynomials. It was seen that in order to obtain a fit with 
${\cal O}(1\%)$ precision four Chebyshevs are sufficient, but in order to be confident of at worst about $1\%$ 
precision, but generally somewhat better than this, six Chebyshevs are needed. The precision and variety of data 
constraints are now such that PDF uncertainties of ${\cal O}(1\%)$ are now possible, so the bias from parameterisation 
limitation should be comfortably lower than this. Hence, we choose to use six Chebyshevs. When using this number we start to see 
some signs of redundancy in parameters in our best fits, so using even more parameters is currently both unwarranted and 
likely impractical.
In order to understand in more detail these changes between the MSHT20 and MMHT14 parameterisations, the alterations in the PDF parameterisation are made incrementally in the fits listed in Table~\ref{tab:oldparamdeltachisqtable2}.
We list the key data sets which are sensitive to the parameterisation extensions and positive values indicate that the fit there is worse than the MSHT20 default. However, first we can simply compare the overall fit quality of the old parameterisation fit with the default MSHT20 fit, as shown in the first column of Table~\ref{tab:oldparamdeltachisqtable2}. As can be seen, the result of the various improvements in parameterisation between the MMHT14 ``old'' parameterisation and the new MSHT20 parameterisation is a significant improvement in the fit quality of 73.3 points in $\chi^2$. In detail it is clear that much of these improvements result from augmentations in the fit to the BCDMS $d$, E866/NuSea Drell-Yan, HERA $e^+ p$ neutral current, ATLAS 7~TeV $W$, $Z$, CMS 7~TeV jets and ATLAS 8~TeV $Z$ $p_T$ data sets, with these changes partially balanced by more mild reductions in the fit quality to the BCDMS $p$, ATLAS 8~TeV $W^{\pm}$ and ATLAS 8~TeV double differential $Z$ data sets. To further understand these changes we now focus upon them one-by-one.

The most major, and hence first alteration to the parameterisation investigated was in the $\bar{d} - \bar{u}$, where a change to a description in terms of Chebyshevs (previously in MMHT14 the $\bar{d}-\bar{u}$ was fit by a quadratic polynomial) was motivated by the increased data available that would constrain this distribution, particularly including the ATLAS 7 and 8~TeV $W$, $Z$ data. We first discuss the former case where the difference $\bar d -\bar u$ was parameterised.  Then in the following paragraph we describe the present situation where the ratio $\bar d/\bar u$ is parameterised. For the latter case the detailed changes in $\chi^2$ are shown in Table~\ref{tab:oldparamdeltachisqtable2}.

\begin{figure}
\centerline{\includegraphics[scale=0.28,trim = 50 0 0 0,clip]{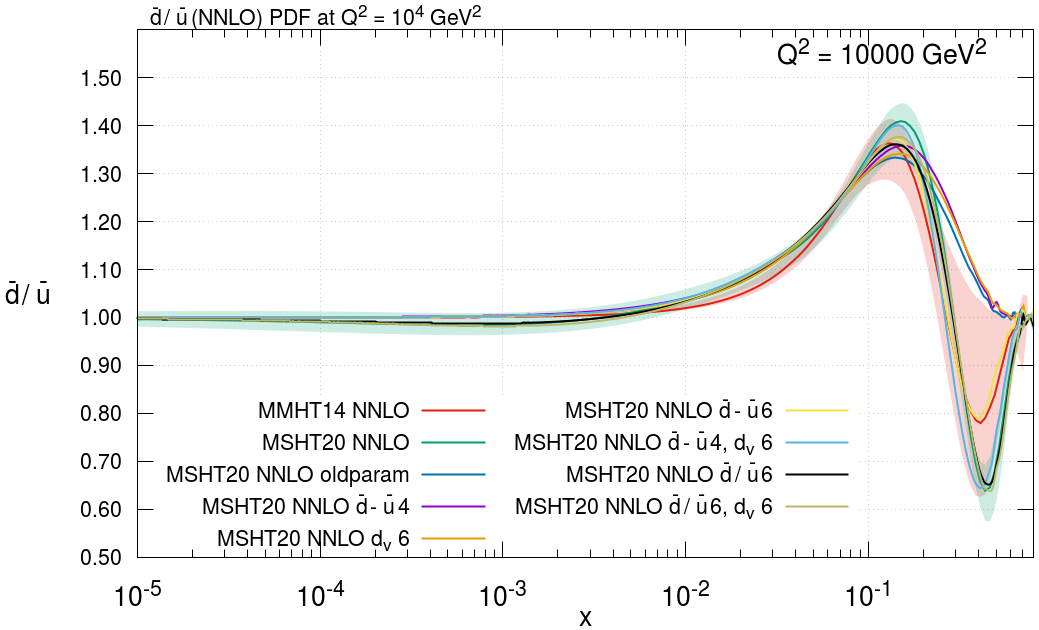} }
\caption{\sf {$\bar{d}/\bar{u}$ PDF at $Q^2=10^4~\GeV^2$ at NNLO showing how incrementing the old parameterisation to the new parameterisation affects the PDFs produced.}\label{dbaroverubar_q210000_NNLOoldparamtonewparam}}
\end{figure}

The effects of the alterations in the SU(2) antiquark asymmetry parameterisation have been outlined previously in \cite{Thorne:2019mpt}, although with a reduced data set relative to the final choice in MSHT20. Various parameterisations of the difference of these antiquarks were investigated and it was found that a conversion (and extension) of the $\bar{d}-\bar{u}$ parameterisation alone to four or even six Chebyshev polynomials made little overall difference to the global fit $\chi^2$ ($\Delta\chi^2=2$) with only a minor shuffling of the individual data set $\chi^2$s within approximately the precision of the fit. This is in slight contrast to \cite{Thorne:2019mpt} where this was found to make more of a difference, albeit with a reduced data set. Similarly, if the down valence parameterisation is extended from four to six Chebyshevs alone it results in only a minor improvement of 5 points in $\chi^2$ across the global fit, centred on the E866 and NuSea Drell-Yan data and the ATLAS 7~TeV $W$, $Z$ data (with improvement also in the BCDMS $d$ fit). Nonetheless the changes to the $\bar{d}-\bar{u}$ parameterisation cannot be viewed in isolation, as they will be impacted by alterations to the up and down valence quarks and it is only when the extension of the latter to six Chebyshev coefficients is made that the full impact of the change in the $\bar{d}-\bar{u}$ becomes apparent. In common with the conclusions in \cite{Thorne:2019mpt}, we find that an enhancement in both the $\bar{d}-\bar{u}$ and the $d_V$ parameterisations together are needed in order improve the overall fit quality significantly; doing just this enables an improvement of $\Delta\chi^2=35$. This improvement in the fit quality results from an easing of the tension between the E866/NuSea Drell-Yan and ATLAS 7~TeV $W$, $Z$ data sets, with the former improving by 8 points whilst the fit to the latter also improves mildly by a few points in $\chi^2$, with some further small improvement in the CMS 8~TeV $W$ data set fit quality. These changes make the benefit of extending the parameterisations of the $\bar{d} - \bar{u}$ and the $d_V$ obvious, and this is reflected in the PDFs themselves, shown later in Fig.~\ref{dbaroverubar_q210000_NNLOoldparamtonewparam}.

In reality however, in  MSHT20, rather than fit $\bar{d}-\bar{u}$, we fit the ratio $\bar{d}/\bar{u}$ in order to allow a better description of the very low $x$ region and, in particular, to ensure that the error bands in this region reflect the lack of data constraints. Therefore, in Table~\ref{tab:oldparamdeltachisqtable2} the changes in the fit quality are indicated in the second column for the sole change of replacing the old
$\bar{d}-\bar{u}$ parameterisation with  $\bar{d}/\bar{u}$ with 6 Chebyshevs. Unlike the case for fitting $\bar{d}-\bar{u}$ with Chebyshevs, there is already a notable improvement in $\chi^2$ of 18.5 points when this alteration is made, in the absence of any additional changes also to the down valence parameterisation. It is therefore clear the fit favours fitting the ratio of these antiquarks rather than the difference. This results from small improvements in the fit to the NMC data, an easing of the tension of the aforementioned E866/NuSea Drell-Yan data and the ATLAS 7~TeV $W$, $Z$ data without the need for the $d_V$ extension and small improvements to various other data sets, most notably the ATLAS 8~TeV $Z$ $p_T$  data. These improvements come at the expense of minor reductions in the fit quality for the ATLAS 8~TeV $W^{\pm}$ and double differential $Z$ data, both of which are in slight tension with the $Z$ $p_T$  data. Nonetheless, even though the amelioration of the tension of the E866/NuSea Drell-Yan data and the ATLAS 7~TeV $W$, $Z$ data is able to occur purely with the fitting of the ratio of the SU(2) antiquarks, there is still further improvement in the global $\chi^2$ with the extension of the down valence parameterisation, as shown in the third column of Table~\ref{tab:oldparamdeltachisqtable2}. However this now comes from improvements to the fit of the BCDMS $d$ data and a further improvement in the ATLAS $W$, $Z$ 7~TeV data as well as marginal improvements in the fit to the D{\O} $W$ asymmetry data. Overall these two changes in the fit parameterisation have improved the overall fit quality at this stage by $\Delta \chi^2 = 73.3 - 40.8 = 32.5$ relative to the MMHT14 old parameterisation fit and with the changes, as might be expected, focused on data sets sensitive to the SU(2) quarks and antiquarks.

These alterations in the parameterisation and consequently the $\chi^2$ of the global fit to the different data sets, also affect the shapes of the PDFs themselves. Fig.~\ref{dbaroverubar_q210000_NNLOoldparamtonewparam} presents the $\bar{d}/\bar{u}$ PDF produced in a variety of different fits as the parameterisation of the $\bar{d}-\bar{u}$ and $d_V$ are altered. This makes the differences in the PDFs, and causes of the improvement in the fit quality, clearer as the shape of the $\bar{d}/\bar{u}$ is significantly altered in the intermediate to high $x$ region. As discussed in Section~\ref{dbarminusubarMSHT20}, MMHT14 has a smaller peak width relative to MSHT20 as a result of the data added to the fit, principally the ATLAS 7~TeV $W$, $Z$ data. Consequently, when the old parameterisation is used to fit all the new data sets, the lack of flexibility in the parameterisation results in the ratio of the antiquarks being greater than 1 throughout the high $x$ region in order to allow for the broadened peak in the $10^{-2} < x < 10^{-1}$ range. As a result the $\bar{d}/\bar{u}$ ratio is rather larger than MSHT20 for $x\sim 0.2-0.3$. This difference at high $x$ is still present when the $\bar{d}-\bar{u}$ is extended to four or six Chebyshevs (although the latter does at least allow it to be negative, but no more than MMHT14) and also when the $d_V$ is extended in the absence of any changes to the $\bar{d}-\bar{u}$. In contrast, however, once both the $\bar{d}-\bar{u}$ and the $d_V$ are extended, the flexibility in the parameterisation allows both the broadened peak structure favoured by the ATLAS data and the negative asymmetry at high $x$ peaking around $x \approx 0.4$ favoured by fixed target experiments (in particular the E866/NuSea Drell-Yan data set). This therefore allows for the improved quality of the global fit and in particular the improvement to the E866/NuSea Drell-Yan $\chi^2$. It is also clear that parameterising this instead as the ratio $\bar{d}/\bar{u}$, with a free $(1-x)$ power, provides the flexibility to allow for these changes without the absolute requirement for the $d_V$ to be extended simultaneously, although the latter does allow the amplitude of the peaks to be larger than in the absence of any change to the down valence. This enables a further improvement in the fit quality, as already discussed. 

Further to these major alterations in the parameterisation of the light quark flavour decomposition, the input distributions of all PDFs (with the exception of the strangeness asymmetry) are also extended to six Chebyshevs in the new parameterisation. We therefore now increment these changes on top of the changes to the $\bar{d}/\bar{u}$ and $d_V$ already discussed. The fourth column of Table~\ref{tab:oldparamdeltachisqtable2} adds the further extension of the up valence quark onto these two changes, however this makes no  difference to the overall fit quality, as the up valence was already well constrained in the MMHT14 fit, and instead there is only minor reshuffling of the $\chi^2$ between data sets.

In the fifth column, the gluon PDF is extended from four to six Chebyshevs\footnote{In actuality it is more complex than this for the gluon, as shown in Section~\ref{sec:inputPDF} the gluon has an additional term focused on low $x$ due to the fact that with our low input scale of $Q_0^2=1~\GeV^2$ the fit to HERA data favours an input gluon with complex shape, and which may become negative at very low $x$. This additional term itself has 3 free parameters, therefore the main term in the gluon parameterisation is extended from two Chebyshevs in MMHT14 to four in MSHT20, which then gives 9 free parameters overall in the gluon - equivalent to seven Chebyshevs in a single polynomial, since the normalisation is fixed by the 
momentum sum rule.}. This leads to significant improvements in the fit to the HERA $e^+p$ neutral current data at 920~GeV, the CMS 7~TeV jets data (and 8~TeV a little as well) and the ATLAS 8~TeV $Z$ $p_T$  data. The latter two of these data sets have been shown in Fig.~\ref{MSHT20_NNLO_gluonpulls_q210000_NNLO} to have different effects on the high $x$ gluon and so the additional parameterisation flexibility allows these tensions to be eased somewhat. As a result of the gluon extension, the overall fit quality improves by approximately a further 18 points in $\chi^2$. 

The penultimate change to the parameterisation is the extension of the sea parameterisation to six Chebyshevs, the effects of this are presented in column six, and this causes a small improvement of $\Delta \chi^2 = 5.9$. This improvement occurs in the HERA $e^+p$ neutral current data and the ATLAS 7~TeV $W$, $Z$ data with a minor improvement in other data sets, such as the CMS 8~TeV jets. This comes again at the expense of a small deterioration in the fit quality to the ATLAS 8~TeV double differential $Z$ data and the 8~TeV $Z p_T$ data by 1-2 points in $\chi^2$.

Finally, the extension of the total strangeness to six Chebyshevs brings the parameterisation all the way to the default MSHT20 set-up. This accounts for the remaining improvement of $\Delta \chi^2 = 17$ with notable improvements in many of the data sets listed, including the BCDMS, NMC, CDF $W$ asymmetry, ATLAS 8~TeV $Z$ $p_T$ data and several others. This therefore seems to ease somewhat the slight tension observed between the ATLAS 7~TeV $W$, $Z$ data and the 8~TeV $Z$ $p_T$  data with the latter improving by 3 points without any reduction in the fit quality of the former. The effect this extension has on several data sets is most likely indicative of the interplay between the 3 light quarks (and in particular the strange and the down quarks) in various DIS data sets, where their combinations are constrained relative to the newer LHC data sets, with the strangeness offering an often small additional (but less constrained) contribution to electroweak processes.

\begin{figure}
\centerline{\includegraphics[scale=0.28,trim = 50 0 0 0, clip]{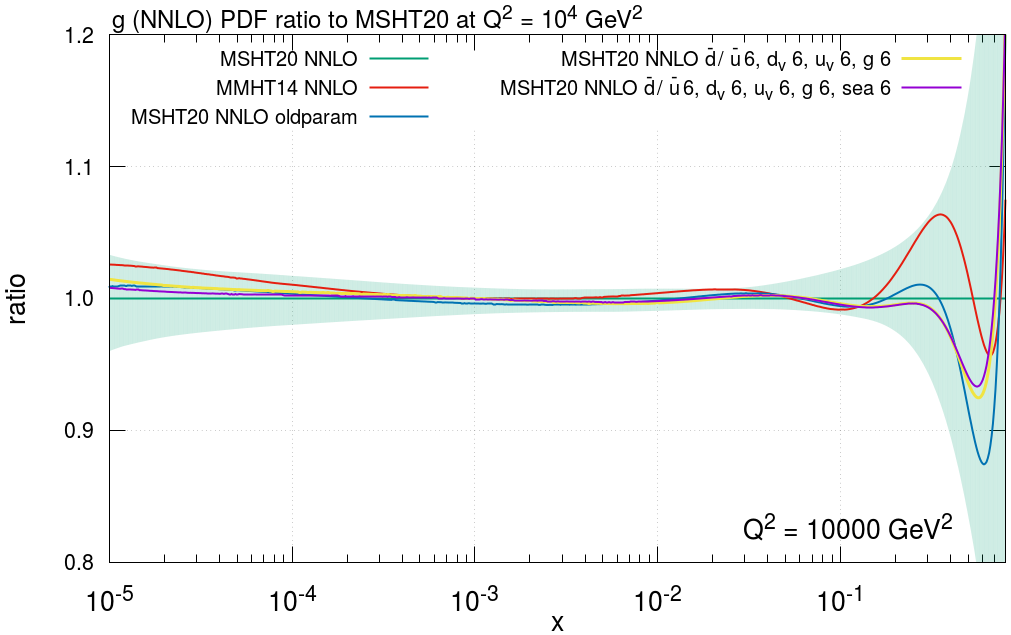} } 
\caption{\sf {Ratio of the gluon PDF to the MSHT20 default at $Q^2=10^4~\GeV^2$ at NNLO showing how incrementing the old parameterisation to the new parameterisation affects the PDFs produced.}\label{gluonratio_q210000_NNLOoldparamtonewparam}}
\end{figure}  

\begin{figure}[t]
\centerline{\includegraphics[scale=0.28,trim = 50 0 0 0, clip]{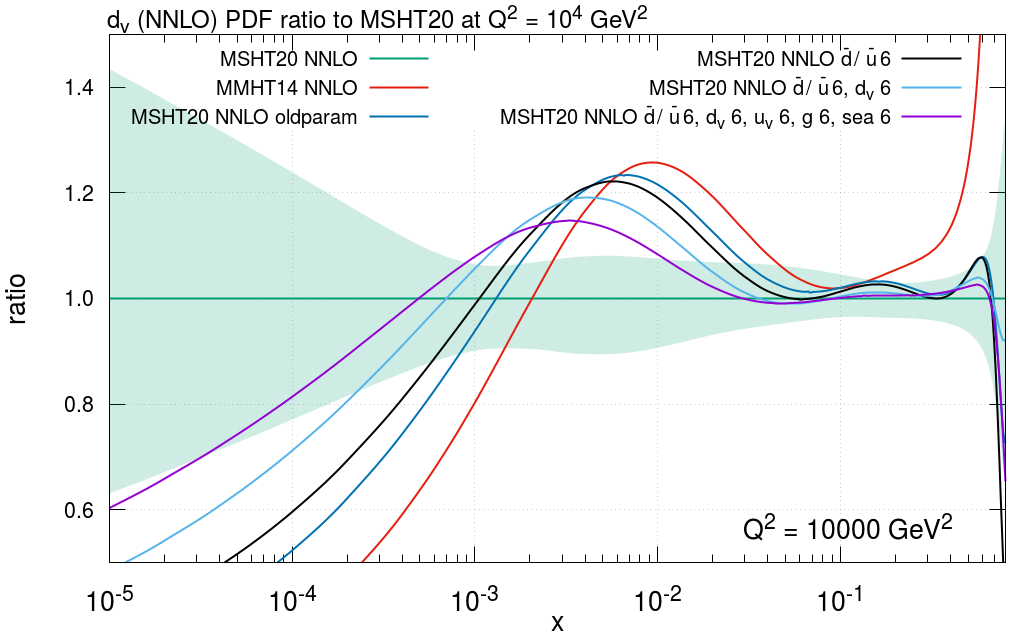} }
\caption{\sf {Ratio of the down valence PDF to the MSHT20 default at $Q^2=10^4~\GeV^2$ at NNLO showing how incrementing the old parameterisation to the new parameterisation affects the PDFs produced.}\label{dnvratio_q210000_NNLOoldparamtonewparam}}
\end{figure}  

\begin{figure}[t]
\centerline{\includegraphics[scale=0.28,trim = 50 0 0 0, clip]{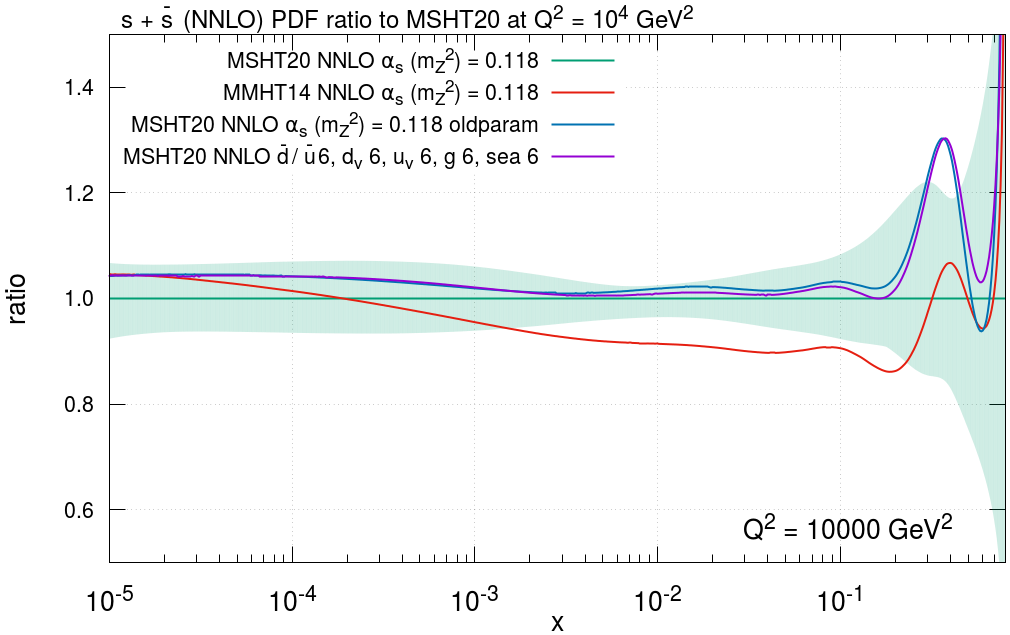} }
\caption{\sf {Ratio of the total strangeness, $s+\bar{s}$, to the MSHT20 default at $Q^2=10^4~\GeV^2$ at NNLO showing how incrementing the old parameterisation to the new parameterisation affects the PDFs produced.}\label{splussbarratio_q210000_NNLOoldparamtonewparam}}
\end{figure}  

All of these further extensions to the parameterisation have notable effects on the PDFs. Figs.~\ref{gluonratio_q210000_NNLOoldparamtonewparam} and \ref{dnvratio_q210000_NNLOoldparamtonewparam} illustrate these changes for the gluon and the down valence PDFs respectively. For the gluon it is clear that the differences that arise are well within the error bands and are also confined to the high $x$ region where there are fewer constraints. Extending the gluon parameterisation enables the gluon to rise in the high $x$ region, and this effect is strengthened when the sea and the $s+\bar{s}$ are also subsequently extended to six Chebyshev coefficients, removing a dip relative to the MSHT20 default fit at $x\approx 0.5$. This rise allowed in the high $x$ gluon presumably contributes to the improvement in the ATLAS 8~TeV $Z p_T$ fit quality seen upon these extensions in Table~\ref{tab:oldparamdeltachisqtable2}. 

The changes in the down valence are much more pronounced. As we have already seen in Section~\ref{updown}, there are large differences between MMHT14 and MSHT20, well outside the error bands, with a clear shape change below $x \approx 0.03$. Fig.~\ref{dnvratio_q210000_NNLOoldparamtonewparam} offers a greater insight into these differences. The MSHT20 fit using the old parameterisation already moves 
significantly in the direction of the default (new parameterisation) MSHT20 fit, showing that the effect is partly a result of new data sets added. However much of the difference still remains and is a result of the extensions of the parameterisation. Altering or extending in sequence the $\bar{d}/\bar{u}$, $d_V$, $g$, sea and finally $s+\bar{s}$ (the latter takes us to the MSHT20 default fit) moves the shape and absolute values of the PDFs in this intermediate to low $x$ range gradually closer to that of MSHT20, with the final extension of the $s+\bar{s}$ having a particularly significant impact. The latter results from the combination of down and strange contributions, which are constrained by many of the data sets. It should also be noted that whilst the differences in the down valence are large, they occur in the intermediate to low $x$ region where there was previously very little constraint on the down valence quark, and where now the, still relatively limited, constraints are provided mainly by new
ATLAS, CMS and (at the lower $x$ range) LHCb data. 

Lastly, Fig.~\ref{splussbarratio_q210000_NNLOoldparamtonewparam} presents the effects of the parameterisation on the total strangeness, $s+\bar{s}$. The MMHT14 strangeness is much lower at intermediate to high $x$ than MSHT20 as a result of the addition of the ATLAS 7~TeV $W$, $Z$ data and the two corresponding separate 8~TeV data sets to the latter. If the MSHT20 data are fit using the MMHT14 old parameterisation this large difference is removed, nonetheless there is still a clear difference in the $s+\bar{s}$ at high $x \approx 0.35$ with a peak above the MSHT20 value (just outside of the MSHT20 error bands) and there are smaller differences between $10^{-2}$ and $10^{-1}$. The total strangeness also is raised at very low $x$. All of this shows the clear impact of the extended parameterisation on the strangeness, which is not as well constrained as some of the other PDFs. If the old parameterisation is then incremented to the new parameterisation, very little changes (as seen in the ``$\bar{d}/\bar{u}$, $d_V$, $u_V$, $g$, sea 6'' line in the figure, in which all changes apart from those to the strangeness parameterisation have been made) until the final step of extending the $s+\bar{s}$ parameterisation to six Chebyshev coefficients is made. At that stage this peak at high $x$ is removed and overall we return to the MSHT20 strangeness shape. These effects at high $x$ are a reflection of the fact that the 5th and 6th Chebyshev polynomials oscillate more rapidly at high $x$ and so allow significant changes of shape in this region. This oscillation is then seen in the ratio of the old parameterisation to the new one.

\subsection{New LHC data} \label{nonewLHCdatacomp}

\begin{figure}
\begin{center}
\includegraphics[scale=0.24, trim = 50 0 0 0 , clip]{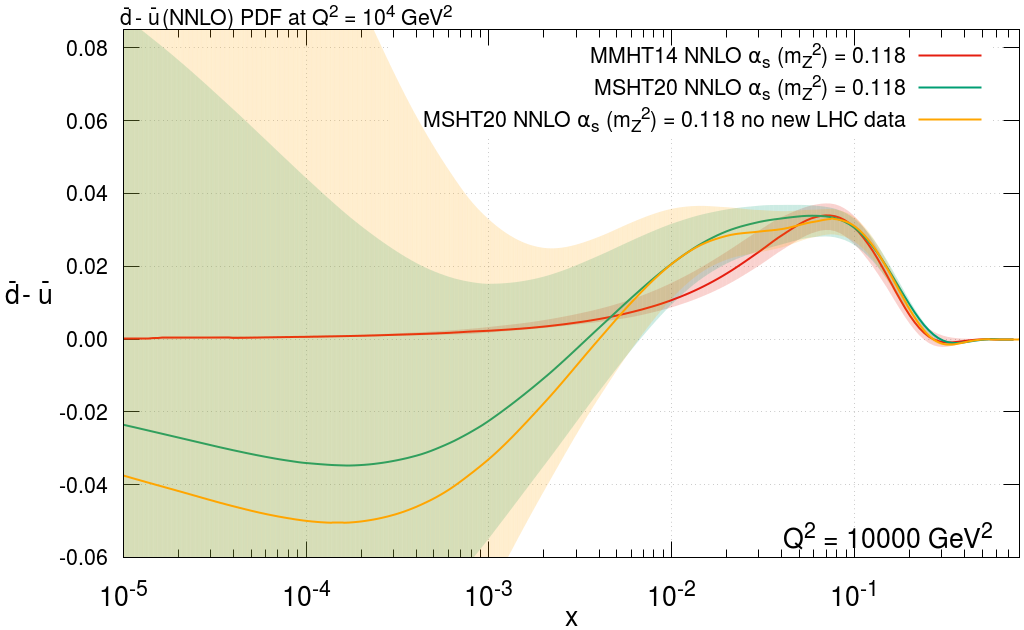}
\includegraphics[scale=0.24, trim = 50 0 0 0 , clip]{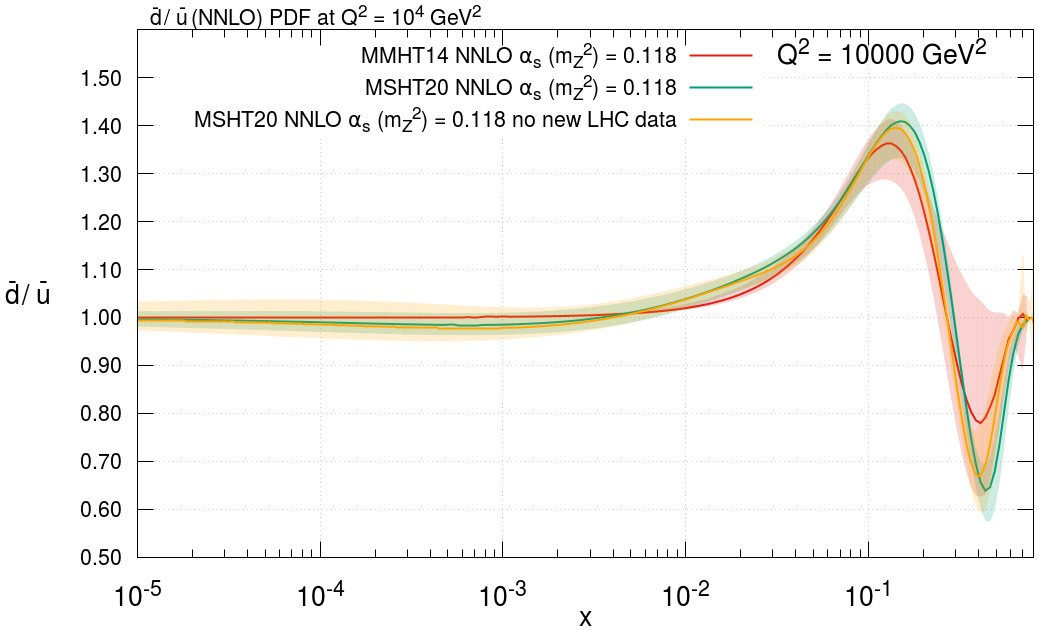}
\caption{\sf (Left) Difference and (right) ratio of the $\bar{d}$ and $\bar{u}$ PDFs at $Q^2=10^4~\GeV^2$ at NNLO showing the effect of removing the new LHC data added since MMHT14 on the PDFs.}\label{MSHT20_NNLO_nonewLHC_dbarubar}
\end{center}
\end{figure}

\begin{figure}
\begin{center}
\includegraphics[scale=0.24, trim = 50 0 0 0 , clip]{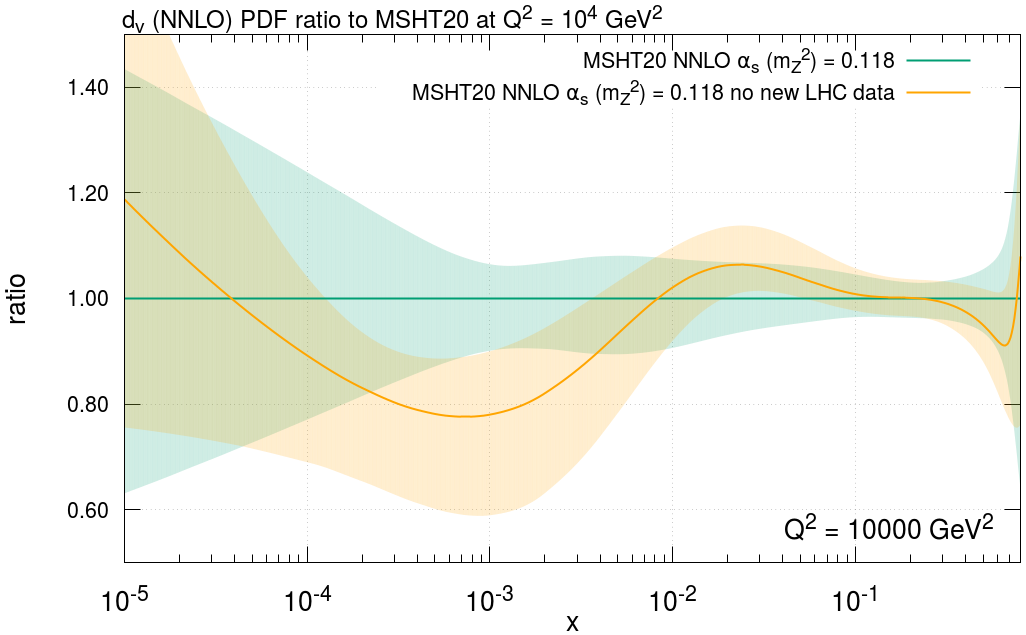}
\includegraphics[scale=0.24, trim = 50 0 0 0 , clip]{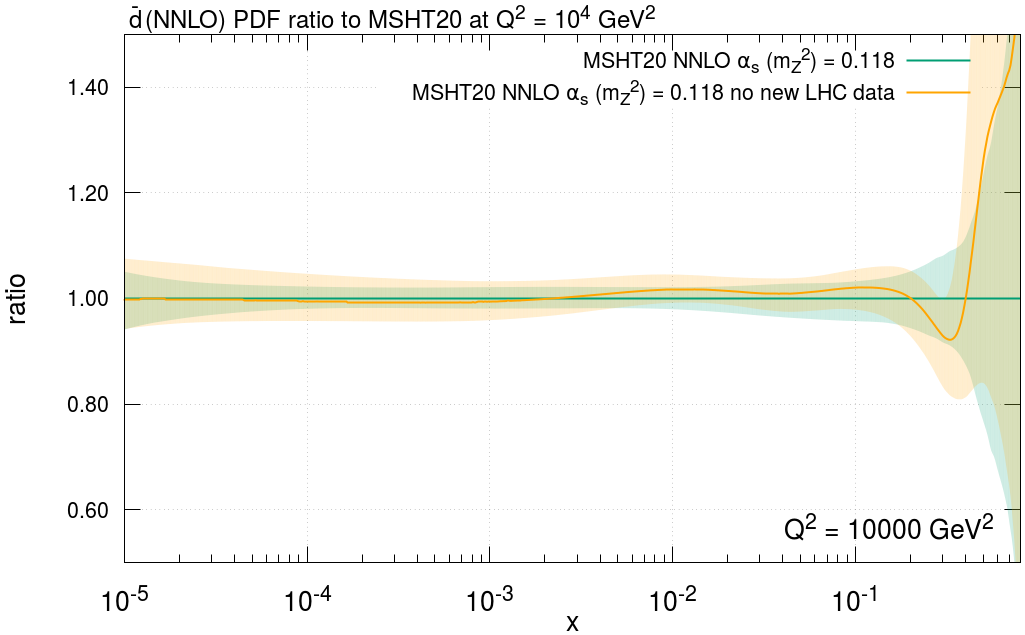}
\caption{\sf (Left) Down valence and (right) down antiquark PDF ratios to the MSHT20 default fit showing the effect of the removal of new LHC data at $Q^2=10^4~\GeV^2$ at NNLO.}\label{MSHT20_NNLO_nonewLHC_downratios}
\end{center}
\end{figure}

\begin{figure}
\begin{center}
\includegraphics[scale=0.24, trim = 50 0 0 0 , clip]{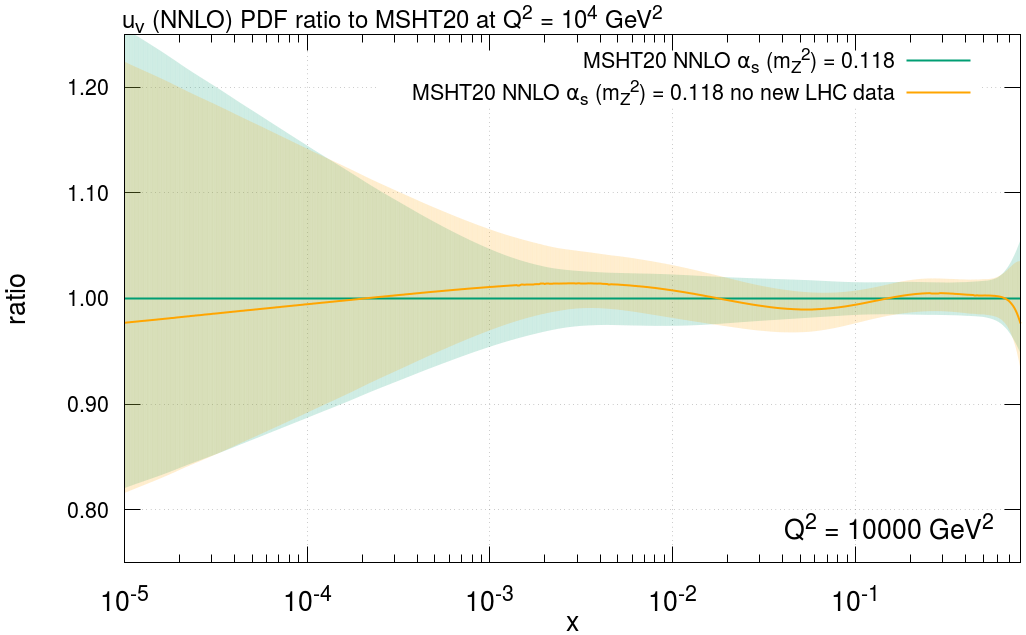}
\includegraphics[scale=0.24, trim = 50 0 0 0 , clip]{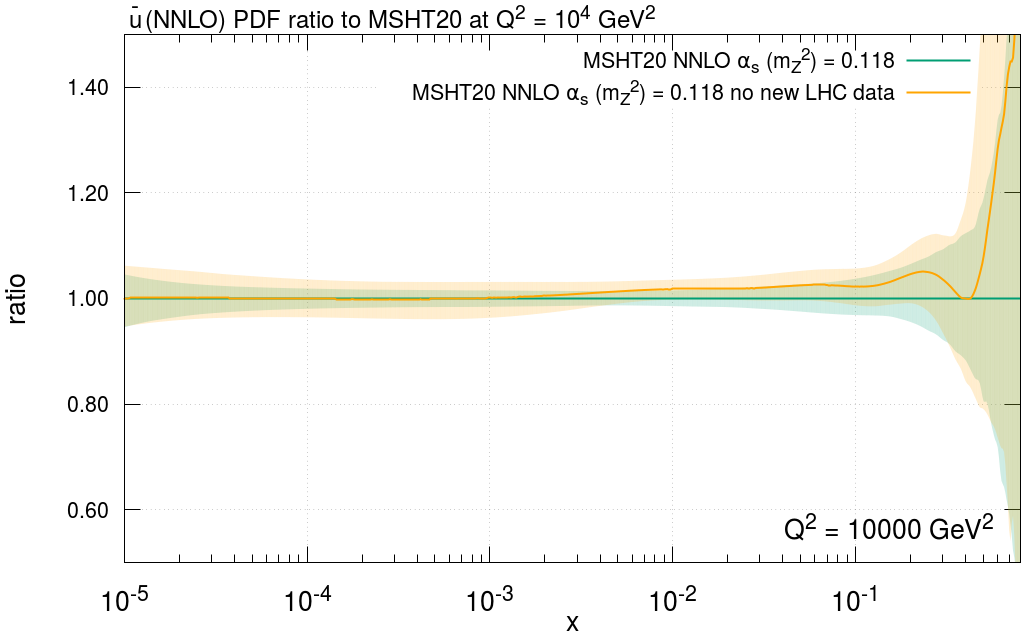}
\caption{\sf (Left) Up valence and (right) up antiquark PDF ratios to the default MSHT20 default fit showing the effect of the removal of new LHC data at $Q^2=10^4~\GeV^2$ at NNLO.}\label{MSHT20_NNLO_nonewLHC_upratios}
\end{center}
\end{figure}

\begin{figure}
\begin{center}
\includegraphics[scale=0.24, trim = 80 0 0 0 , clip]{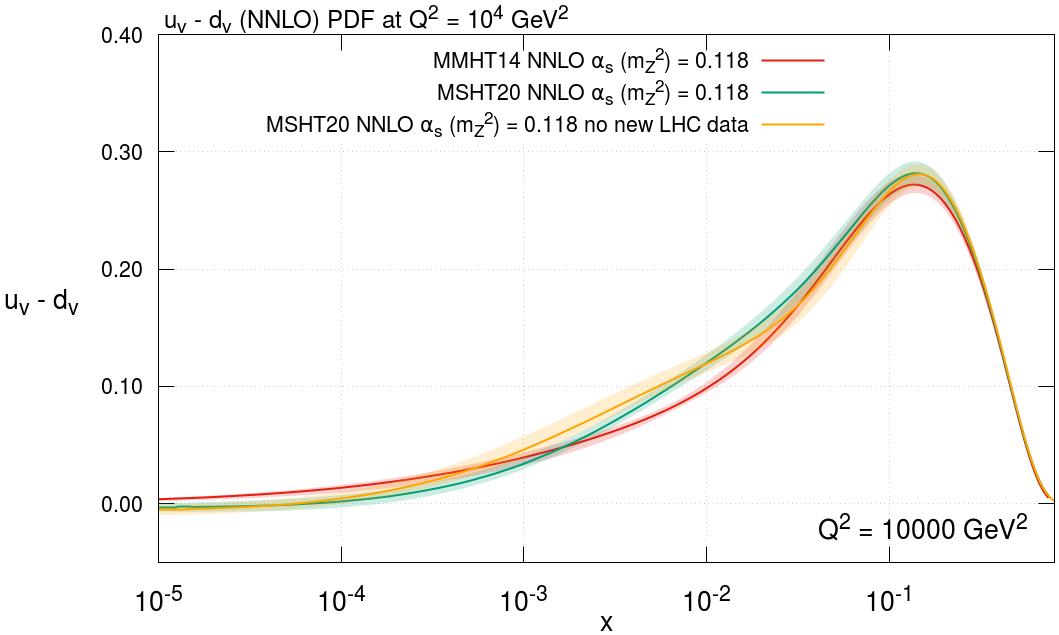}
\includegraphics[scale=0.24, trim = 50 0 0 0 , clip]{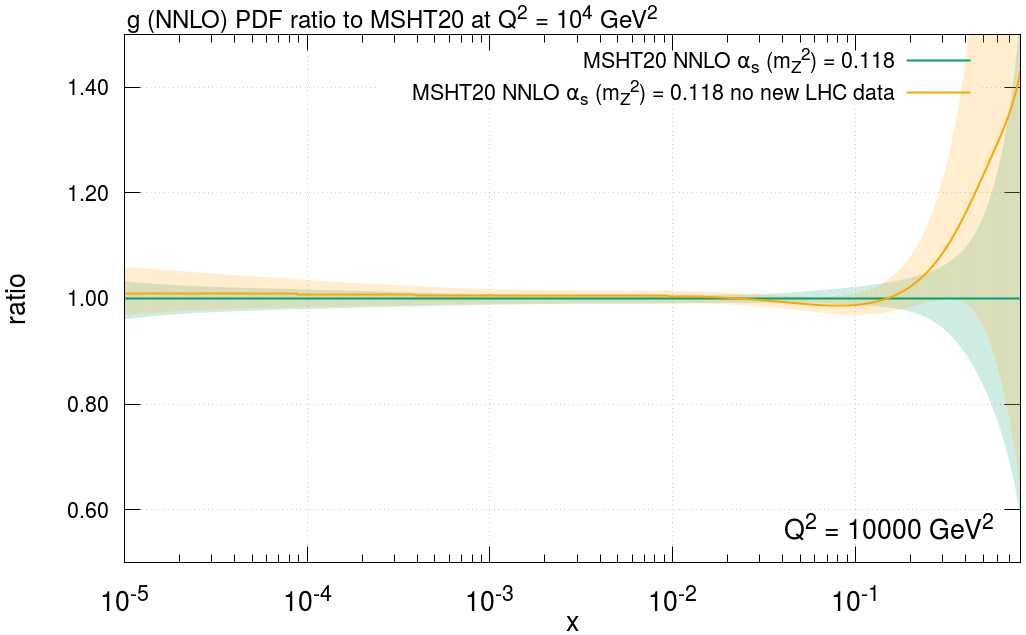}
\caption{\sf (Left) $u_V - d_V$ and (right) $g$ PDFs comparing the MSHT20 default fit with the same upon removal of the new LHC data. The former plot presents the PDFs themselves and the latter plot presents the ratio to the MSHT20 default fit, both  at NNLO at $Q^2=10^4~\GeV^2$.}\label{MSHT20_NNLO_nonewLHC_ratiosetc}
\end{center}
\end{figure}

\begin{figure}
\begin{center}
\includegraphics[scale=0.23, trim = 50 0 0 0 , clip]{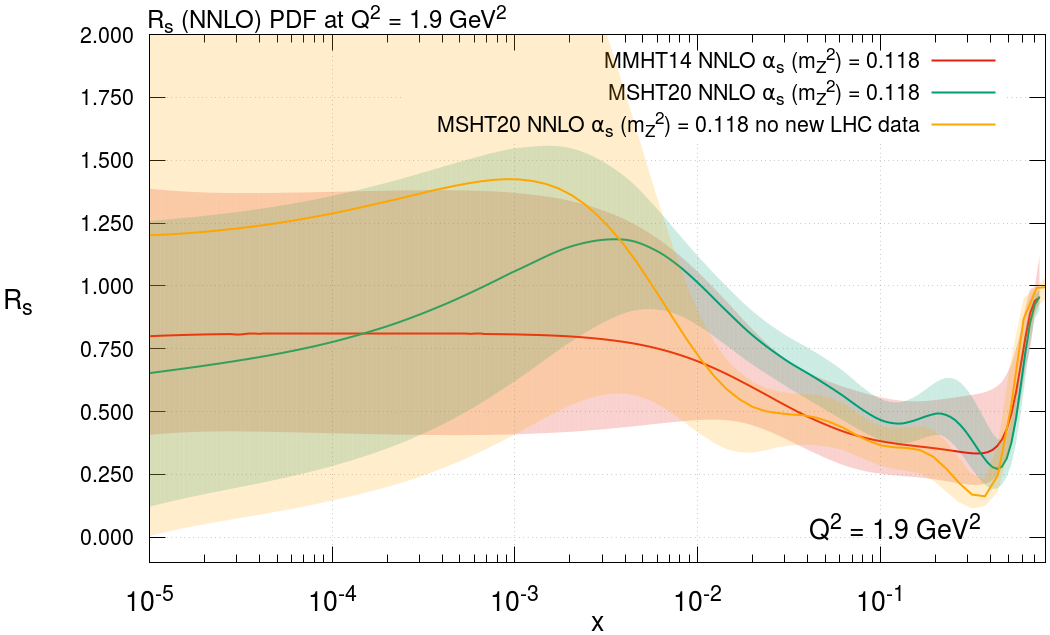}
\includegraphics[scale=0.23, trim = 60 0 0 0 , clip]{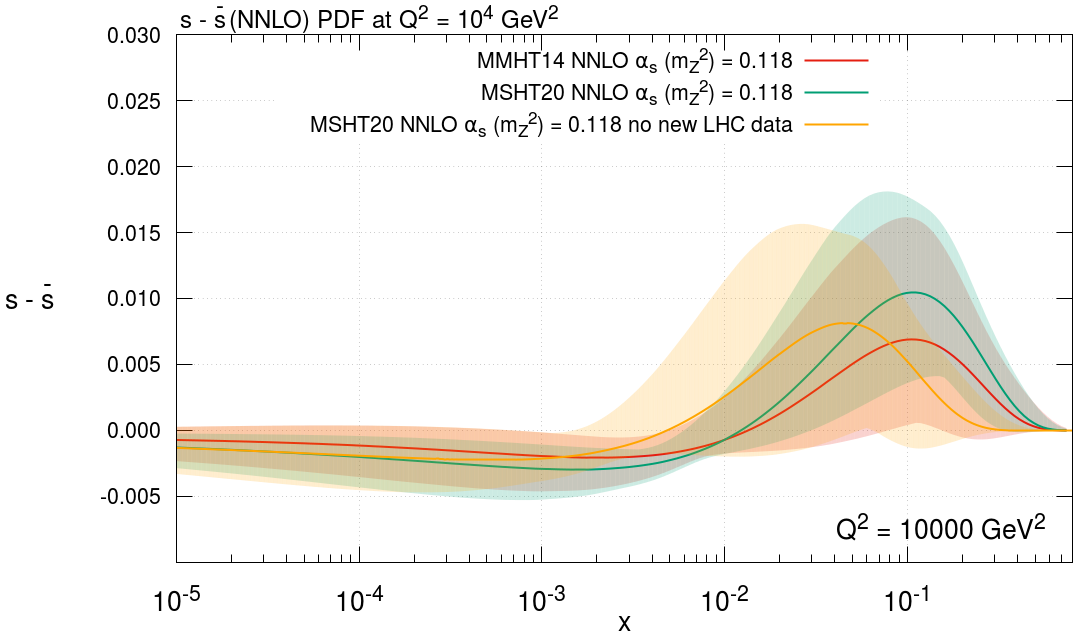}
\caption{\sf (Left) $R_s$ at at $Q^2=1.9\text{GeV}^2$  and (right) $s-\bar{s}$ at $Q^2=10^4~\GeV^2$ at NNLO comparing the PDFs upon removal of the new LHC data  with the default MSHT20 fit and the MMHT14 fit.}\label{MSHT20_NNLO_nonewLHC_Rssminussbar}
\end{center}
\end{figure}

\begin{figure}
\begin{center}
\includegraphics[scale=0.24, trim = 35 0 0 0 , clip]{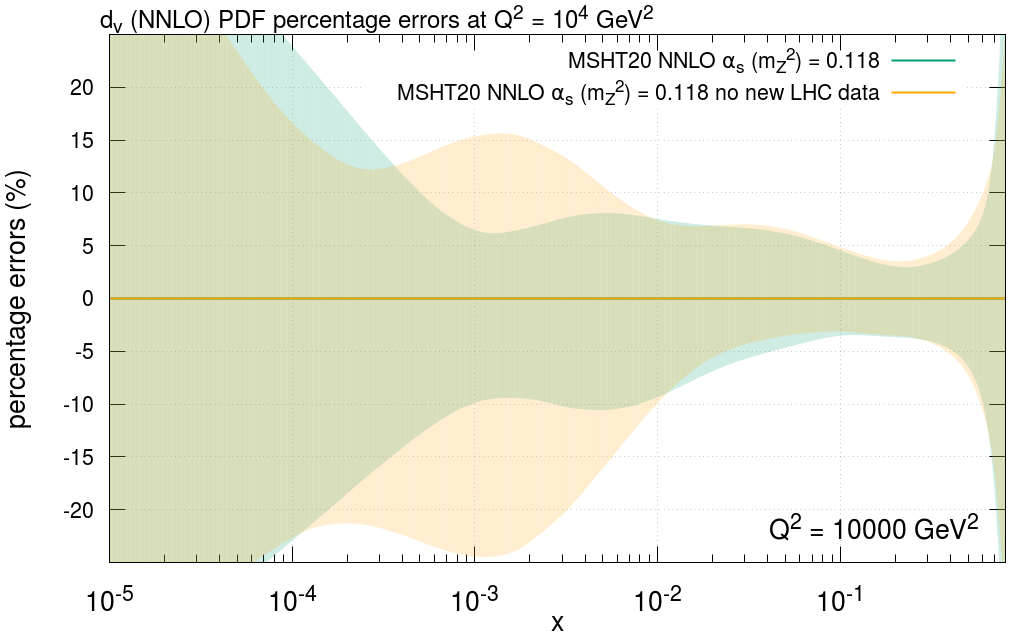}
\includegraphics[scale=0.24, trim = 40 0 0 0 , clip]{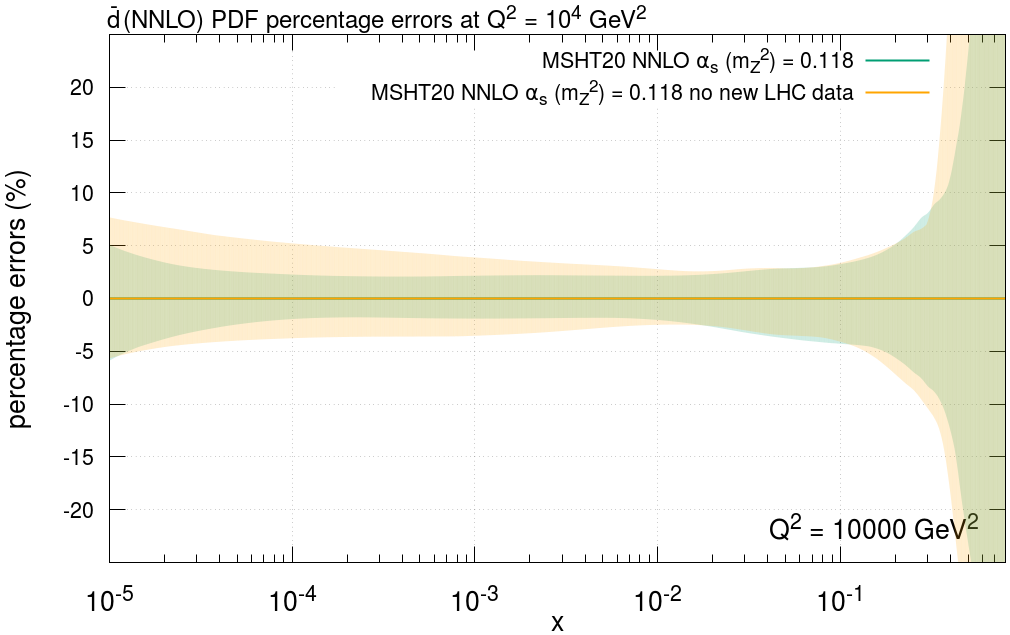}
\caption{\sf (Left) Down valence and (right) down antiquark PDF percentage errors upon removal of the new LHC data compared to MSHT20 at $Q^2=10^4~\GeV^2$ at NNLO.}\label{MSHT20_NNLO_nonewLHC_downpercentageerrors}
\end{center}
\end{figure}

\begin{figure}
\begin{center}
\includegraphics[scale=0.24, trim = 35 0 0 0 , clip]{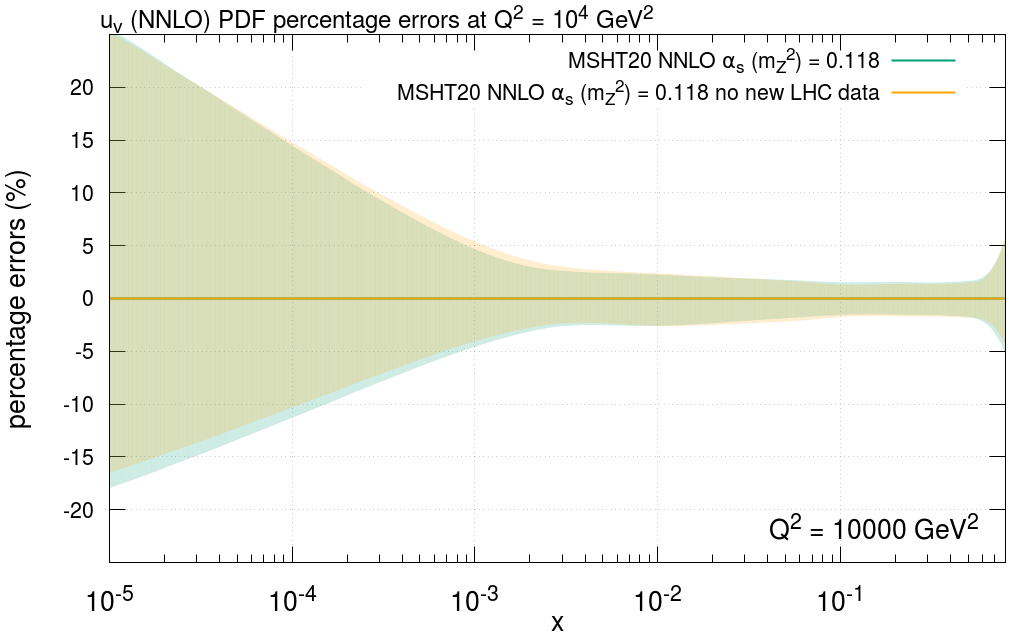}
\includegraphics[scale=0.24, trim = 40 0 0 0 , clip]{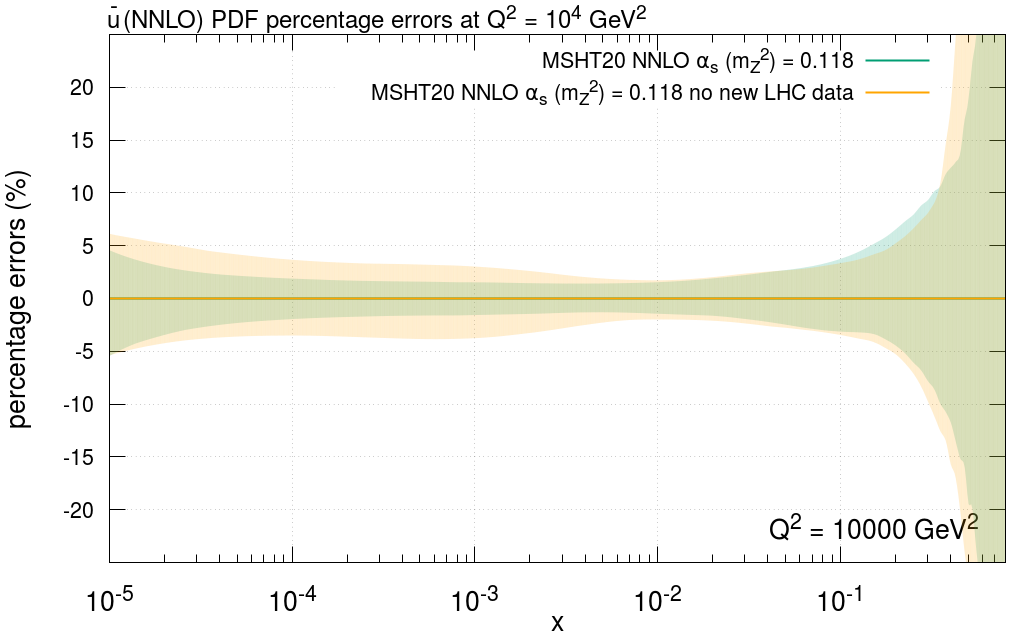}
\caption{\sf (Left) Up valence and (right) up antiquark PDF percentage errors upon removal of the new LHC data compared to MSHT20 at $Q^2=10^4~\GeV^2$ at NNLO.}\label{MSHT20_NNLO_nonewLHC_uppercentageerrors}
\end{center}
\end{figure}

\begin{figure}
\begin{center}
\includegraphics[scale=0.23, trim = 40 0 0 0 , clip]{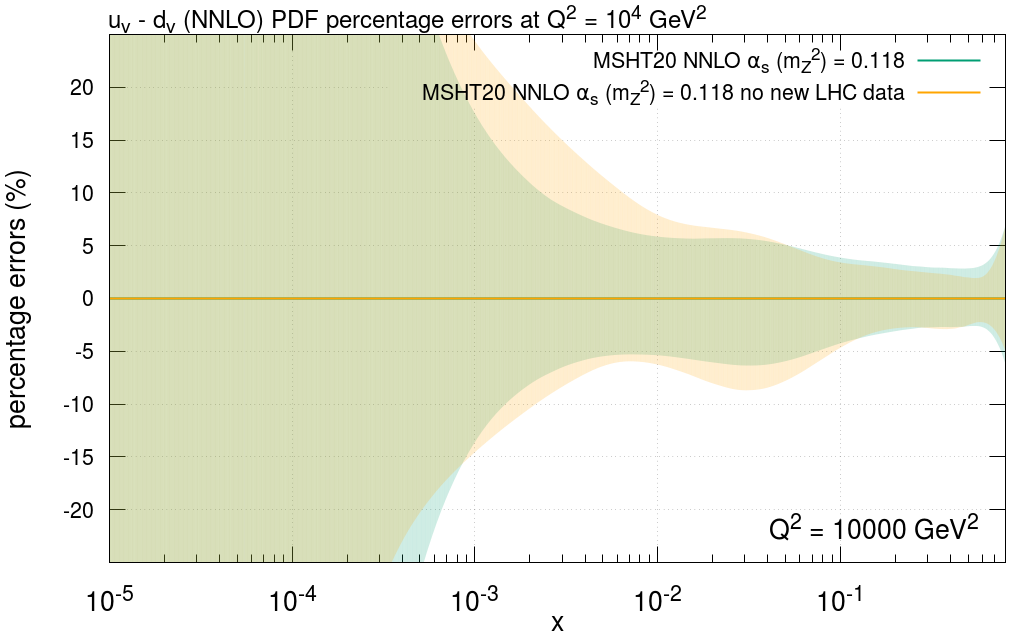}
\includegraphics[scale=0.23, trim = 40 0 0 0 , clip]{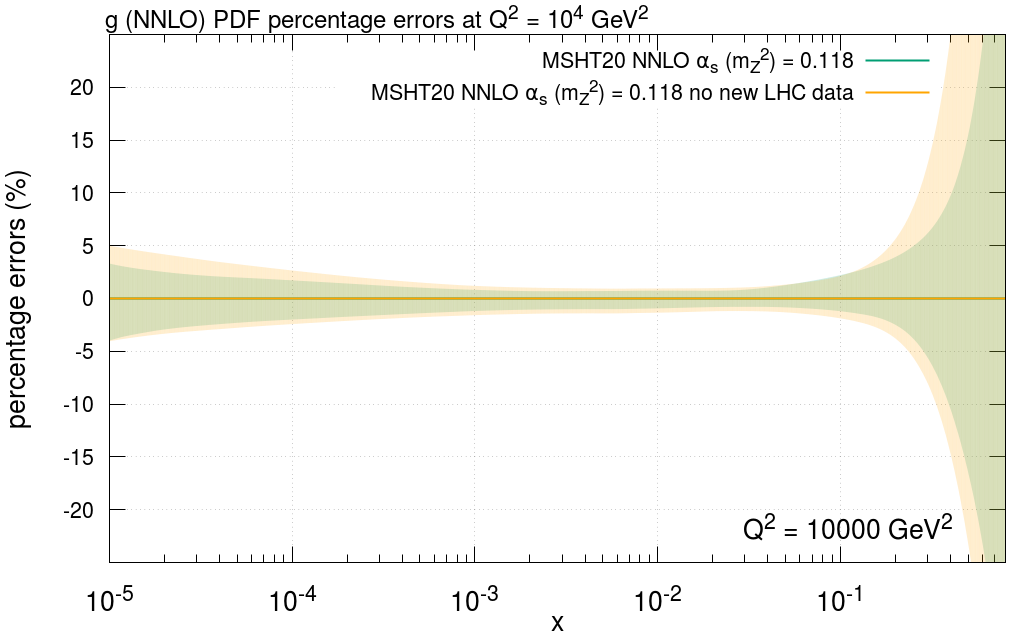}
\caption{\sf (Left) $u_V - d_V$ and (right) gluon PDF percentage errors upon removal of the new LHC data compared to MSHT20 at $Q^2=10^4~\GeV^2$ at NNLO.}\label{MSHT20_NNLO_nonewLHC_percentageerrorsetc}
\end{center}
\end{figure}

The  majority of the new data sets added since MMHT14 have been from the LHC, and it is therefore informative to determine their effect on the MSHT20 PDFs.
In this section we provide such a comparison, presenting the exact same theoretical choices and parameterisation as used in the default MSHT20 fit but with the new LHC data sets removed. The results are shown in Figs.~\ref{MSHT20_NNLO_nonewLHC_dbarubar}-\ref{MSHT20_NNLO_nonewLHC_percentageerrorsetc}. First we present comparisons of the central values through absolute plots or ratios to MSHT20. Later, we show figures illustrating the differences in the error bands more clearly. 

We begin with the the asymmetry of the SU$(2)$ antiquarks, $\bar{d}$ and $\bar{u}$, which is presented as both the difference ($\bar{d}-\bar{u}$) and the ratio ($\bar{d}/\bar{u}$) in Fig.~\ref{MSHT20_NNLO_nonewLHC_dbarubar}. In Fig.~\ref{MSHT20_NNLO_nonewLHC_dbarubar} (left) we can see that both the central values and the uncertainties are affected by the absence of the LHC data.
The central value of the $\bar{d}-\bar{u}$ has been shifted down,
albeit only slightly, in the low $x$ region, most likely due to the ATLAS and the LHCb $W$, $Z$ data. Nonetheless, it remains very consistent with the large uncertainties in this region and also is clearly tending to 0 within errors. The small-$x$ limit can be appreciated more clearly from the result that the ratio is tending to 1 in Fig.~\ref{MSHT20_NNLO_nonewLHC_dbarubar} (right). In addition to the impact on the central values, it is clear that LHC data also constrain the error bands on the PDFs considerably in the low $x$ region. There is also a small difference around the broadened peak at intermediate $x$. This arose in the new parameterisation mainly as a means of alleviating tensions between the older Drell-Yan ratio data and the new ATLAS $W$, $Z$ data. The absence of the latter then enables the small reduction here, but it is notable that a tendency towards a broader peak exists even without the new LHC data. This is due in part just to the parameterisation change.

For the differences in the down quark, we provide ratios of the down valence and down antiquark PDFs relative to MSHT20 in Fig.~\ref{MSHT20_NNLO_nonewLHC_downratios}. The $d_V$ ratio shows a significantly different shape even in the absence of the new LHC data, particularly being 
higher for $x\sim 0.02$, where the constraint of central rapidity LHC $W$ data applies, and then lower around $x\sim 10^{-3}$, with uncertainty bands barely overlapping, before raising at very low $x$. The down valence also showed very large changes between MMHT14 and MSHT20 in Fig.~\ref{uvdvratios} (right), where the differences are much larger than here. This implies that the majority of the differences between MMHT14 and MSHT20 are driven by the parameterisation changes, as discussed in Section~\ref{newparameterisationeffects}, but there remains a significant effect just from the impact of the new LHC data. The pull of the new LHC data on the region around $x\sim 10^{-3}$ is also clear in the comparison of the size of the error bands in Fig.~\ref{MSHT20_NNLO_nonewLHC_downpercentageerrors} (left), which in the absence of the LHC data are more than doubled relative to the MSHT20 default. The ratio of the down antiquark relative to MSHT20 shows far milder differences in Fig.~\ref{MSHT20_NNLO_nonewLHC_downratios} (right), with the $\bar{d}$ slightly enhanced in the $x\sim 10^{-2}$ region. This is a reflection of the reduced strangeness in the absence of the ATLAS $W$, $Z$ data. The $\bar{d}$ is also reduced around $x\approx 0.2$ before rising sharply at very high $x$ relative to MSHT20, being on the edge of the MSHT20 error bands. Again, focusing on the comparison of the size of the error bands in Fig.~\ref{MSHT20_NNLO_nonewLHC_downpercentageerrors} (right), the LHC data have the largest impact on the low $x$ region, with the PDF errors larger than the default MSHT20 fit for $x<0.01$, and indeed nearly doubled in size in the $10^{-4}<x<10^{-3}$ range. This is partly a reflection of the change in the gluon at small $x$, which drives the $\bar d$ 
distribution via evolution, but also an illustration of constraints from LHCb data.
The $\bar d$ distribution rises significantly for $x>0.5$, but this is where the uncertainty becomes very large and there is little direct constraint from data, so the rise is unlikely to be of real significance. 

The up valence quark ratio in Fig.~\ref{MSHT20_NNLO_nonewLHC_upratios} (left) displays a much reduced impact from the new LHC data in the fit, with only a slight change in shape visible, within the small error bands of the MSHT20 default global fit. The reason for this much reduced difference is that the up valence was already well constrained by structure function data. This is also reflected in the error bands in Fig.~\ref{MSHT20_NNLO_nonewLHC_uppercentageerrors} (left) which show no significant difference upon the removal of the LHC data. The $\bar{u}$ ratio in Fig.~\ref{MSHT20_NNLO_nonewLHC_upratios} (right) also shows a reduced impact relative to the differences seen in the $\bar{d}$, although both become large at very high $x$. The $\bar{u}$ is less impacted by the removal of the LHC data because the removal of the ATLAS $W$, $Z$ data and its effect on the strangeness is more closely tied, through Cabbibo mixing, to the $\bar{d}$. Additionally, in order to ensure the charge weighted sum of $u$, $d$ and $s$ quarks is maintained in the fit to proton structure data with the enhanced strangeness caused by the addition of the ATLAS $W$, $Z$ data, larger differences in $s$ (or $d$) can be compensated by smaller differences in the $u$, as a result of the charge weighting. The impact of the LHC data on the $\bar{u}$ errors is very similar to that described for the $\bar{d}$, being predominant at small $x$. 

As a result of the significant change in the down valence and rather more marginal changes in the up valence, the difference of the valence quarks $u_V - d_V$ has a considerable change in shape. This  is lower than MSHT20, and consequently closer to MMHT14, as one would expect, in the range $0.01<x<0.1$ and larger than MSHT20 below this, see Fig.~\ref{MSHT20_NNLO_nonewLHC_ratiosetc} (left). The effects of the differences in the parameterisation are notable in the intermediate $x$ region from $10^{-3}$ to $10^{-2}$, as $u_V - d_V$ is closer to MSHT20 than MMHT14 despite the removal of the majority of the new data. The error bands of $u_V - d_V$ are shown in Fig.~\ref{MSHT20_NNLO_nonewLHC_percentageerrorsetc} (left), and reflect the impacts described above on the central values, with reductions in the error bands from $x\sim 10^{-3}$ to 0.1. This reflects the influence the LHC data have in this $x$ range, whereas at high $x$ the structure function data remain the overwhelming constraint. 

In Fig.~\ref{MSHT20_NNLO_nonewLHC_ratiosetc} (right) we show the effect of the LHC data on the MSHT20 gluon PDF. The removal of the LHC data causes a large increase in the gluon at high $x$, outside of the MSHT20 error bands (although MSHT20 is just within the error bands of the no new LHC data fit). This is the result of the removal of the top, jet and $Z$ $p_T$ data, which in combination lower the gluon at high $x$ as discussed in Section~\ref{MSHT20gluonlightquark} and evident in Fig.~\ref{MSHT20_NNLO_gluonpulls_q210000_NNLO}. The constraints of these LHC data sets on the high $x$ gluon is also clear in the percentage errors in Fig.~\ref{MSHT20_NNLO_nonewLHC_percentageerrorsetc} (right), with the error bands at high $x$ enlarged upon the removal of the LHC data. Indeed the gluon uncertainty bands are reduced over the whole $x$ range by the addition of the LHC data, with the effects greatest at low and high $x$ and more marginal reductions at intermediate $x$ where the gluon was already well constrained predominantly by HERA data.

Finally, probably the largest impact of the LHC data on the PDFs in MSHT20 is on the total strangeness. As is well-known and outlined in Section~\ref{strangeness} (further examined in Section~\ref{ATLASDYeffects}), the ATLAS $W$, $Z$ data enhance the strangeness around $x\approx 0.02$. This effect is visible in the MSHT20 PDFs in an increased value of $R_s$ on the edge of the MMHT14 uncertainty band from just below $x\approx 10^{-2}$ to $x\approx 0.1$. The removal of these ATLAS $W$, $Z$ data sets therefore can be seen to cause the expected reduction in the total strangeness in this region in Fig.~\ref{MSHT20_NNLO_nonewLHC_Rssminussbar} (left), with $R_s$ returning to the levels observed in MMHT14. Nonetheless, the impact of these ATLAS data, as well as additional LHC data such as that from LHCb, also effects the strangeness at lower $x$. Namely, the central value of $R_s$ without 
new LHC data continues to rise for lower $x$ values than in the default MSHT20 fit and consequently has larger values in the very low $x$ region. These differences are well within the large uncertainty bands at low $x$, which themselves are enlarged enormously upon removal of the new LHC data, with 
effectively no constraint at all on $R_s$ below $x\sim 0.005$. There are of course little data at very low $x$ so this effect results from an extrapolation of the behaviour of $R_s$ at low $x$ by the parameterisation (which itself has remained unchanged in the comparison of the two fits). However, the clear impact of LHCb data on imposing 
some constraint at $x\sim 0.001$ in the full MSHT20 fit is seen, with consequent reduction of uncertainty in the lower $x$ region, where extrapolation takes place. 

The strangeness asymmetry in Fig.~\ref{MSHT20_NNLO_nonewLHC_Rssminussbar} (right) additionally exhibits changes upon the removal of the LHC data. The peak of the $s-\bar{s}$ moves to lower $x\approx 0.05$ from $x\approx 0.1$ in MSHT20 and MMHT14. This effect is particularly interesting as the strangeness asymmetry parameterisation is the only one unaltered since MMHT14. Therefore the fact that differences are still seen relative to MMHT14 once all of the new LHC data (which provide the major constraint on the strangeness asymmetry of the new data since MMHT14) are removed  is indicative of the indirect effect of the changes in parametrisation of other PDFs since MMHT14. In particular, alterations in the $s+\bar{s}$ distribution can  then impact upon the strangeness asymmetry. Upon then adding the LHC data the strangeness asymmetry peak is restored to $x \approx 0.1$, back to its MMHT14 $x$ position but with an enhanced amplitude. This enhanced amplitude is also present without the new LHC data (although a little reduced) and so is to some extent an effect enabled by changes in parameterisation.

\subsection{Impact of NNLO dimuon corrections}  \label{NNLOdimuoneffects}

As discussed in Section~\ref{sec:dimuonfit} we now include the full NNLO calculation of~\cite{NNLOdimuon} in the FFNS in our fit to the CCFR and NuTeV neutrino--induced dimuon production data, suitably modified to include VFNS contributions. In Table~\ref{table:Dimuon1} we show the impact on the fit quality to these data sets, comparing against the approximate NNLO theory used in previous fits, as described in the corresponding CC structure function discussion in Section 4.3 of~\cite{MSTW}. We can see that the NNLO theory leads to some  improvement in the fit quality to the CCFR data, while for the NuTeV data there is a very mild deterioration, but such that the fit quality to the combined data still improves. However, in all cases the global fit quality (not shown) is essentially unchanged, and hence this mild improvement in the description of the dimuon data has been matched by a mild deterioration in the fit quality to other data sets. This deterioration is spread relatively evenly, and in particular there is no obvious deterioration (or improvement) in the fit quality to the ATLAS $W$, $Z$ production data at 7, 8~TeV.  However, interestingly we can see that with the addition of the NNLO theory the preferred value of the dimuon branching ratio is higher by $\sim 1\sigma$ in comparison to the previous result, and in better agreement with the input value of $0.092 \pm 10\%$. With a more precise determination of this quantity, it is therefore certainly possible that the impact of the NNLO corrections could be larger. In all cases, there is relatively little difference between the pure FFNS and VFNS cases, as one might expect since the data is at low $Q^2$ and relatively high $x$.

\begin{table}
\begin{center}
\begin{tabular}{|c|c|c|c|}
\hline
& Approx. NNLO & Full NNLO (FFNS) & Full NNLO (VFNS) \\ \hline
  CCFR &0.85&0.76&0.78\\ \hline
 NuTeV  &0.67&0.71&0.69 \\ \hline
 CCFR + NuTeV &0.76&0.74&0.74 \\ \hline
 Dimuon BR ($D \to \mu$).  & 0.079&0.088&0.089\\ \hline
\end{tabular}
\caption{\sf $\chi^2/N_{\rm pts}$ for neutrino--induced dimuon production data in MSHT NNLO fits ($\alpha_S$ free), as well as the corresponding $D\to \mu$ branching ratio (with input value $0.092 \pm 0.1$). Results for different theoretical treatments are given; full NNLO (VFNS) corresponds to the default MSHT fit.}
\label{table:Dimuon1}
\end{center}
\end{table}

\begin{figure}[t]
\begin{center}
\includegraphics[scale=0.24]{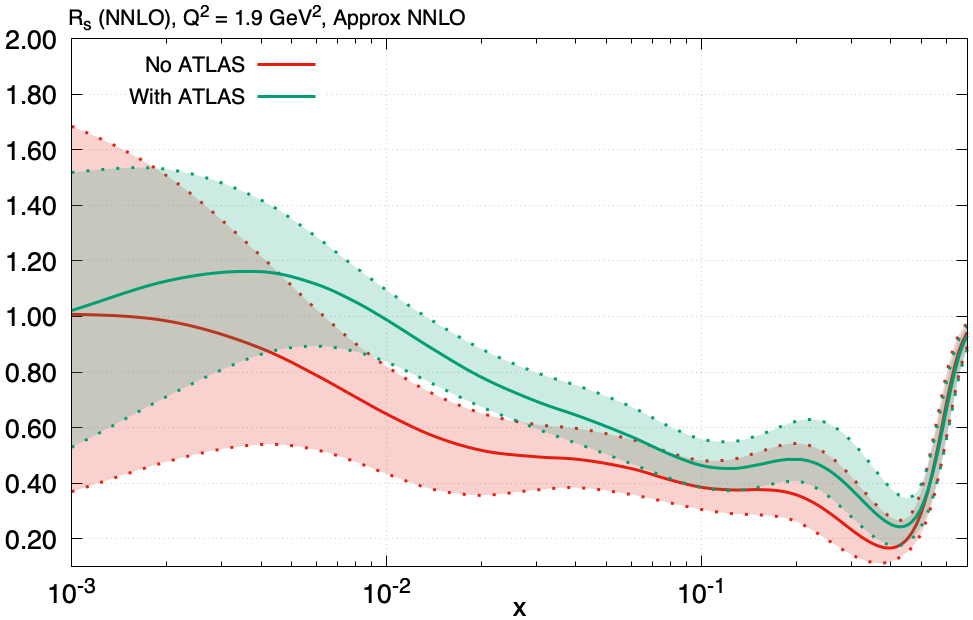}
\includegraphics[scale=0.24]{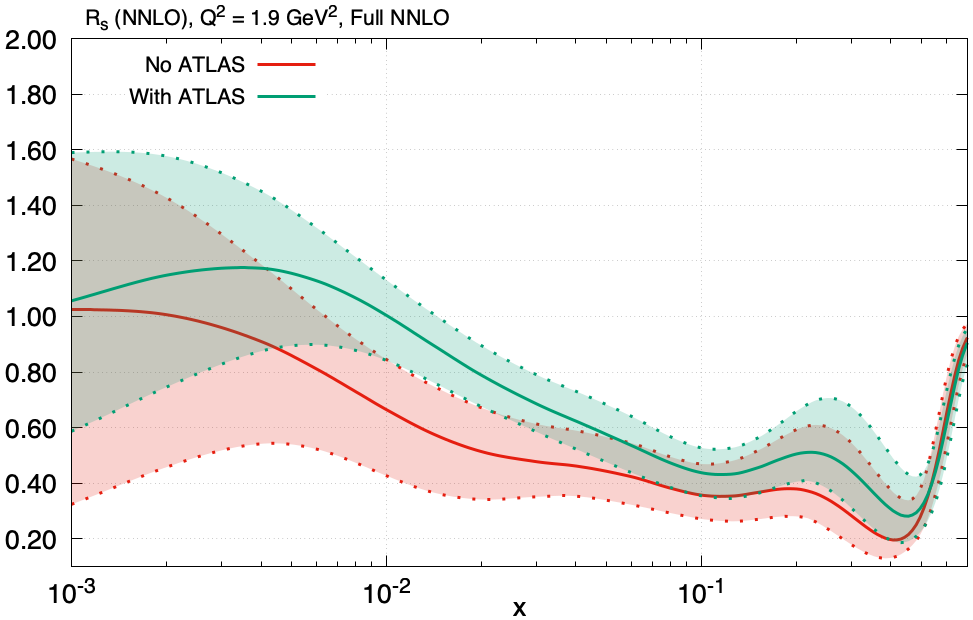}
\caption{\sf The strangeness ratio $R_s=(s+\overline{s})/(\overline{u} +\overline{d})$, at $Q^2=1.9$ ${\rm GeV}^2$, for variants of the MSHT NNLO fits ($\alpha_S$ free). The result of the previous approximate NNLO theory, and the full NNLO theory for the dimuon data are shown in the left and right figures, respectively. In both cases the result of the global fit including and excluding  the ATLAS 7 and 8~TeV precision $W$, $Z$ data are shown.}
\label{fig:dimuon2}
\end{center}
\end{figure} 

\begin{figure}[t]
\begin{center}
\includegraphics[scale=0.24, trim = 50 0 0 0 , clip]{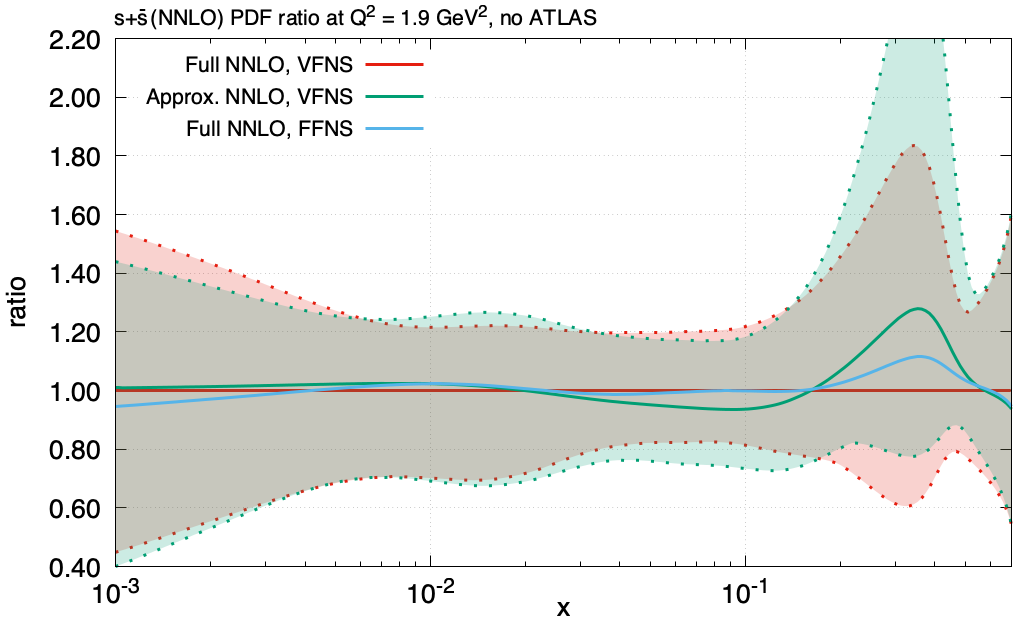}
\includegraphics[scale=0.24, trim = 50 0 0 0 , clip]{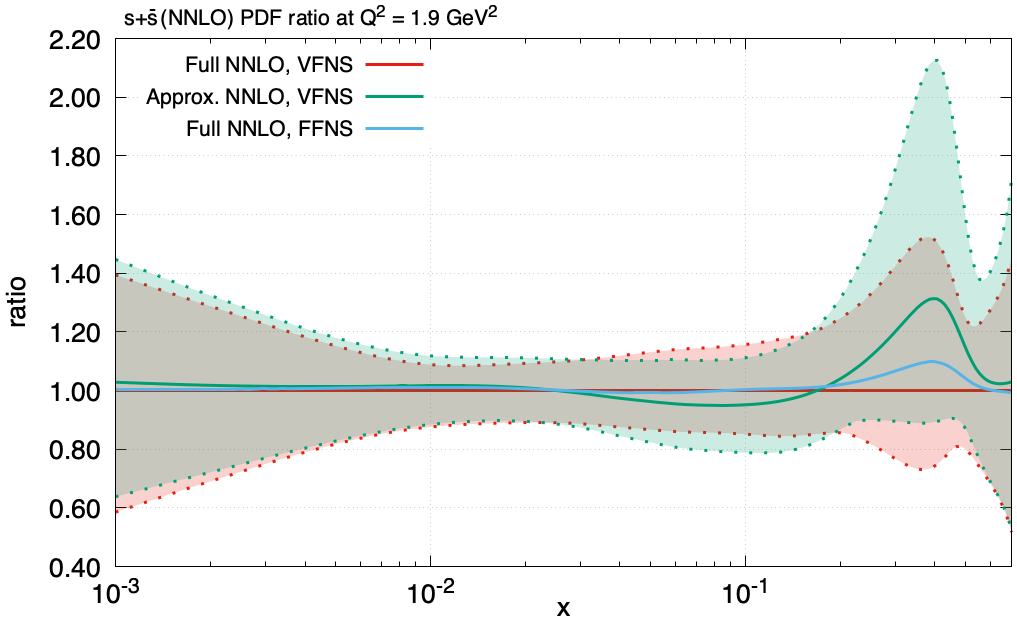}
\caption{\sf The strangeness at $Q^2=1.9$ ${\rm GeV}^2$, for variants of the MSHT NNLO fits ($\alpha_S$ free). The ratio of the result with the previous approximate NNLO theory and the full NNLO theory (FFNS only) to the updated full NNLO theory (VFNS) is shown. The two plots show the result of the global fit excluding (left) and including (right) the ATLAS 7 and 8~TeV precision $W$, $Z$ data.}
\label{fig:dimuon3}
\end{center}
\end{figure} 

In Fig.~\ref{fig:dimuon2} we show the impact of the NNLO corrections on the strangeness ratio $R_s=(s+\overline{s})/(\overline{u} +\overline{d})$, at $Q^2=1.9$ ${\rm GeV}^2$. We recall that the expectation is that these corrections, which tend to reduce the cross section prediction, will prefer a slightly larger value of $R_s$, and hence that this might alleviate some of the tension between the more recent ATLAS 7 and 8~TeV precision $W$, $Z$ data and the dimuon data. We can see that indeed the difference between fits including and excluding the ATLAS data is generally somewhat less after including the NNLO corrections to the dimuon theory, though clearly this is a very mild effect. In more detail, in Fig.~\ref{fig:dimuon3} we show the impact of the NNLO corrections on the strangeness, at $Q^2=1.9$ ${\rm GeV}^2$, for the VFNS and pure FFNS cases as well as the previous approximate NNLO treatment. In Fig.~\ref{fig:dimuon3} left (right) we show the results of fits excluding (including) the ATLAS 7 and 8~TeV precision $W$, $Z$ data. We can see that in both cases indeed the NNLO corrections lead to a somewhat larger strangeness in the $x \sim 0.02 - 0.2$ region, where the ATLAS data also place constraints. Interestingly, at high $x$ this leads to a decrease in the strangeness, albeit within quite large errors. This may be due to an interplay between the NNLO corrections and the dimuon branching ratio discussed above. In particular, for a higher branching ratio a smaller strangeness is preferred, and while at lower $x$ this is compensated by the negative NNLO corrections, these are milder at high $x$ and hence may not compensate this increased branching ratio. Exactly the same increase in the strangeness in the intermediate $x$ region is seen if only the pure FFNS theory is used, while the decrease at high $x$ becomes slightly milder.

\subsection{Impact of D{\O} $W$ asymmetry data.} \label{D0Wasymeffects}

The D{\O} $W$ asymmetry is one of the key changes in MSHT20. We now interpret this as a $W$ asymmetry, rather than a lepton asymmetry, in contrast to other global fitting groups \cite{NNPDF3.1,ABMP16,CT18}. We find that the impact of this data set is significant: it is found to constrain the $d_V$ distribution strongly, but also the $\bar d$ and to some extent $u_V$ and $\bar u$, and even has some indirect effect on the gluon and strange quarks. Therefore in this section we examine in more detail both the effects of regarding it as a $W$ asymmetry rather than an electron asymmetry and its effects in constraining the MSHT20 PDF uncertainty bands. 

\begin{figure}[t]
\begin{center}
\includegraphics[scale=0.24, trim = 50 0 0 0 , clip]{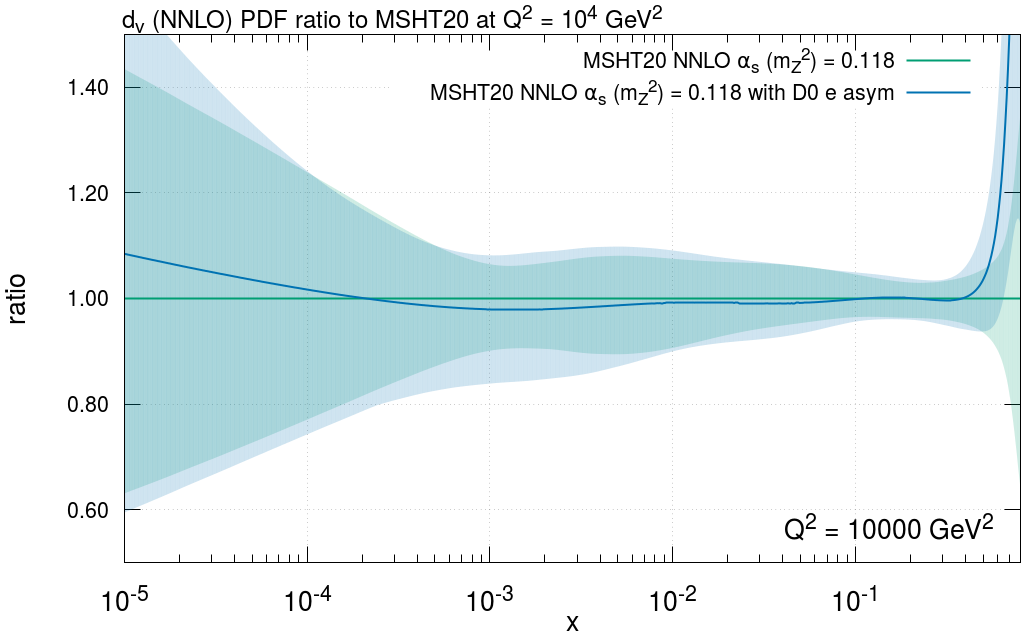}
\includegraphics[scale=0.24, trim = 50 0 0 0 , clip]{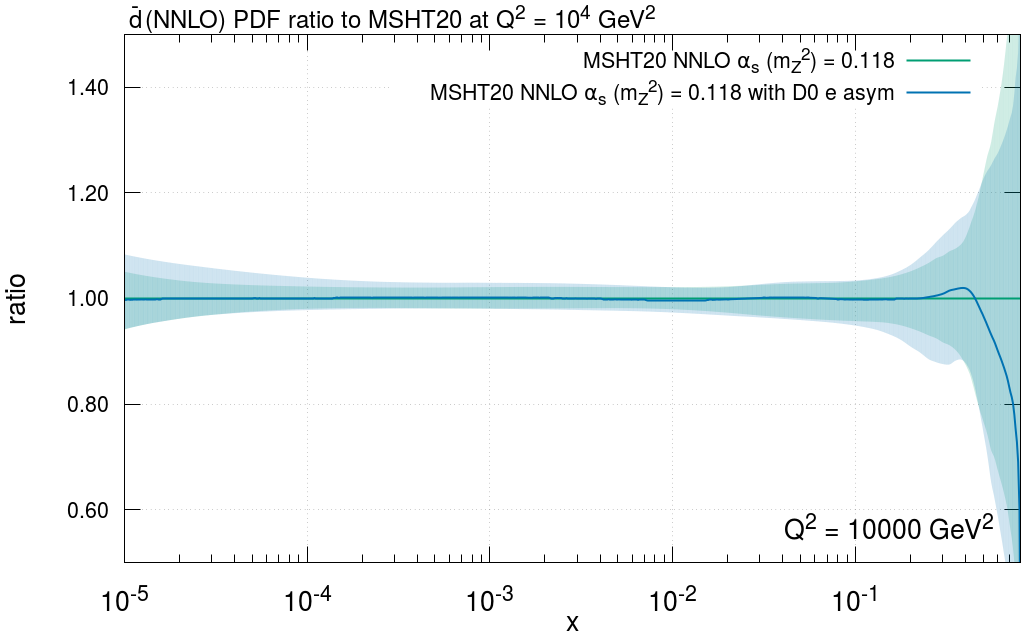}
\caption{\sf  (Left) $d_V$ PDF and (right) $\bar{d}$ PDF ratios to the MSHT20 default at $Q^2=10^4~\GeV^2$ at NNLO showing the effect of including the D{\O} data as an electron asymmetry rather than the  $W$ asymmetry.}
\label{MSHT20_NNLO_D0elasym_downratios}
\end{center}
\end{figure}

\begin{figure}[t]
\begin{center}
\includegraphics[scale=0.24, trim = 50 0 0 0 , clip]{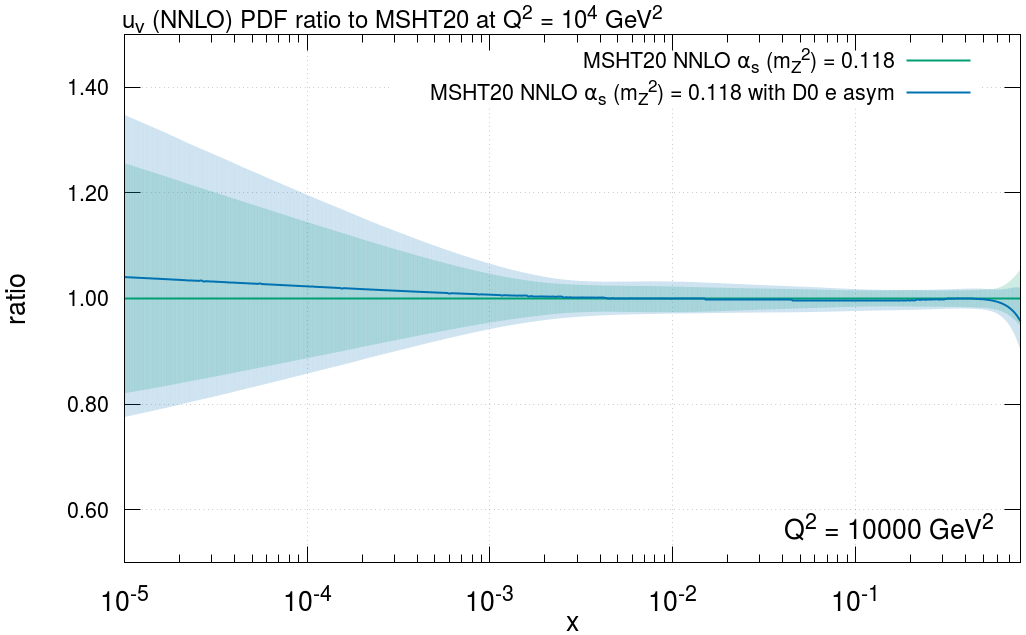}
\includegraphics[scale=0.24, trim = 50 0 0 0 , clip]{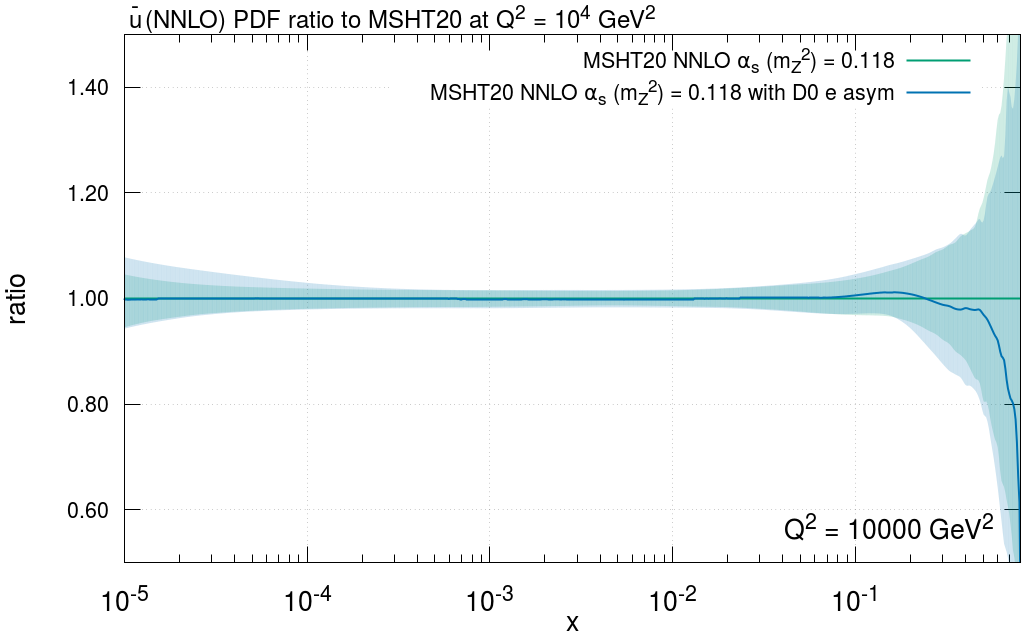}
\caption{\sf  (Left) $u_V$ PDF and (right) $\bar{u}$ PDF ratios to the MSHT20 default at $Q^2=10^4~\GeV^2$ at NNLO showing the effect of including the D{\O} data as an electron asymmetry rather than the  $W$ asymmetry.}
\label{MSHT20_NNLO_D0elasym_upratios}
\end{center}
\end{figure}

First, we explore the issue of fitting the $W$ asymmetry rather than the lepton asymmetry. The rationale for this has been explained in Section~\ref{Tevatronasymdata}, nonetheless here we provide comparisons with the global fit extracted with it treated instead as an electron asymmetry. As this is a small data set, whilst it is very constraining on several of the error bands, replacing the D{\O} $W$ asymmetry with the previous D{\O} electron asymmetry has only a limited effect on the fit quality of the other data sets in the MSHT20 central fit. The use of the electron asymmetry instead improves the fit to the other MSHT20 data sets by $\Delta \chi^2 = 8.8$, with the older D{\O} electron asymmetry data~\cite{D0Wnue} in the fit showing a notable improvement in $\chi^2$ of 2.6. Nonetheless, fitting the $W$ asymmetry does result in an increased impact of the data set.
Fig.~\ref{MSHT20_NNLO_D0elasym_downratios} presents the ratio of the full global fit with the data as an electron asymmetry to the standard default MSHT20 fit, at NNLO. As is clear, the default fit has both a lower down valence PDF and slightly higher down antiquark PDF at high $x$, where the difference between a $W$ asymmetry and electron asymmetry measurement is most pronounced. There is no major constraint on the $\bar{d}$ (or $\bar{u}$) in this region, and hence the increase in the down valence will be largely driven by the constraint on the $d$. In addition, the error bands across the whole $x$ range, but predominantly at high $x$, are also noticeably reduced by fitting the $W$ asymmetry. Fig.~\ref{MSHT20_NNLO_D0elasym_upratios} presents the analogous plots for the $u_V$ and $\overline{u}$, with the former better constrained PDF clearly being less affected than the $d_V$. The change in $u_V$ at high $x$ is in the opposite direction to $d_V$, as it compensates for the large increase in the  latter in this region to maintain the charge-weighted sum of up and down quarks. The $\bar{u}$ on the other hand shows the same behaviour at high $x$ as the $\bar{d}$.

\begin{figure} 
\begin{center}
\includegraphics[scale=0.24, trim = 50 0 0 0 , clip]{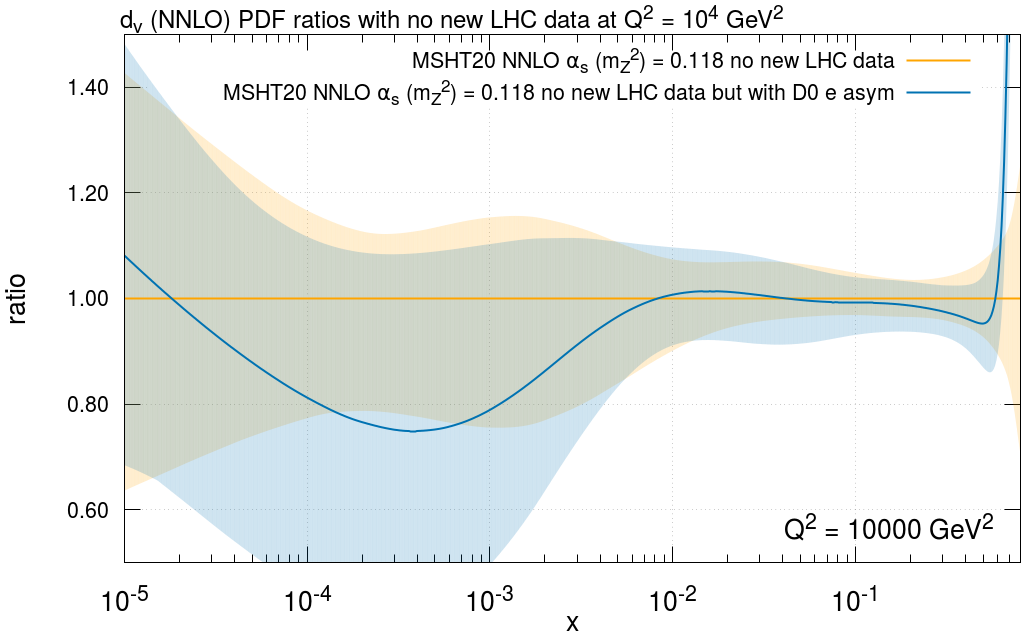}
\includegraphics[scale=0.24, trim = 50 0 0 0 , clip]{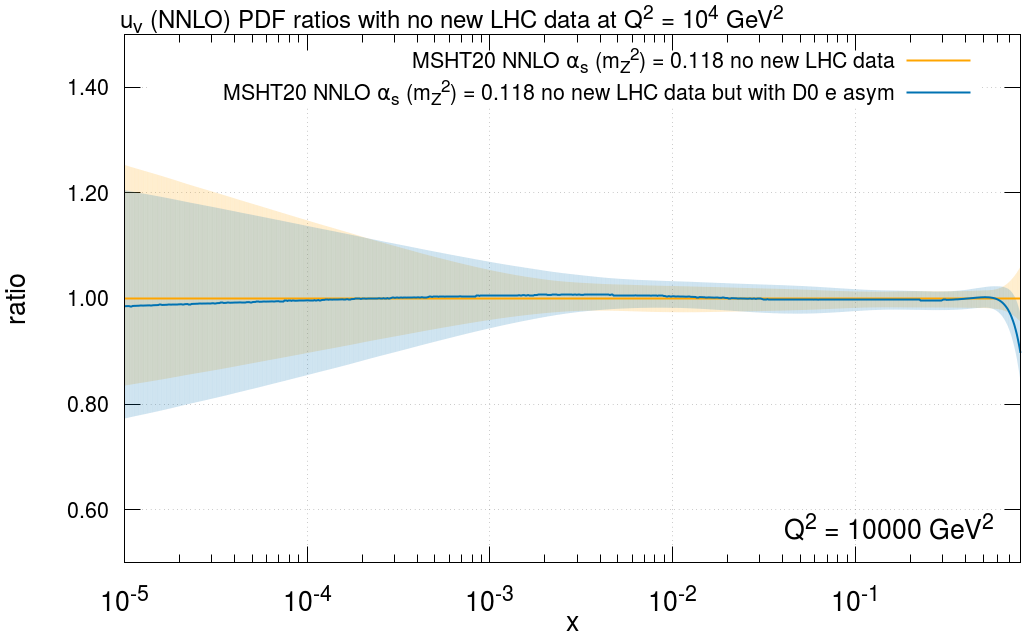}
\caption{\sf  (Left) $d_V$ PDF and (right) $u_V$ PDF ratios to the MSHT20 default at $Q^2=10^4~\GeV^2$ at NNLO without any new LHC data included, showing the effect of including the D{\O} data as an electron asymmetry rather than the $W$ asymmetry.}
\label{MSHT20_NNLO_nonewLHCD0elasym_valenceratios}
\end{center}
\end{figure} 

In order to highlight the above differences further, we can also compare results when the new LHC data are excluded, as in Section~\ref{nonewLHCdatacomp}.  As one might expect, in Fig.~\ref{MSHT20_NNLO_nonewLHCD0elasym_valenceratios} the differences at high $x$ are greater, with the down valence showing a particularly large reduction in this region. It is also evident in Fig.~\ref{MSHT20_NNLO_nonewLHCD0elasym_valenceratios} (left) that the down valence shape in the low $x$ region changes. This mirrors larger changes seen in the down valence shape resulting from the parameterisation in section~\ref{newparameterisationeffects} in Fig.~\ref{dnvratio_q210000_NNLOoldparamtonewparam}, the inclusion of new LHC data in Fig.~\ref{MSHT20_NNLO_nonewLHC_downratios} (left) and in the overall MSHT20 default fit in Fig.~\ref{uvdvratios} (right). Given these changes in shape are not present in the comparison of the lepton and $W$ asymmetry in the full fit in Fig.~\ref{MSHT20_NNLO_D0elasym_downratios} (left) it is clear that the shape changes resulting from the D{\O} $W$ asymmetry, the new LHC data and the parameterisation are complementary and result in the overall large change in the down valence in MSHT20 relative to MMHT14.

\begin{figure} 
\begin{center}
\includegraphics[scale=0.24, trim = 50 0 0 0 , clip]{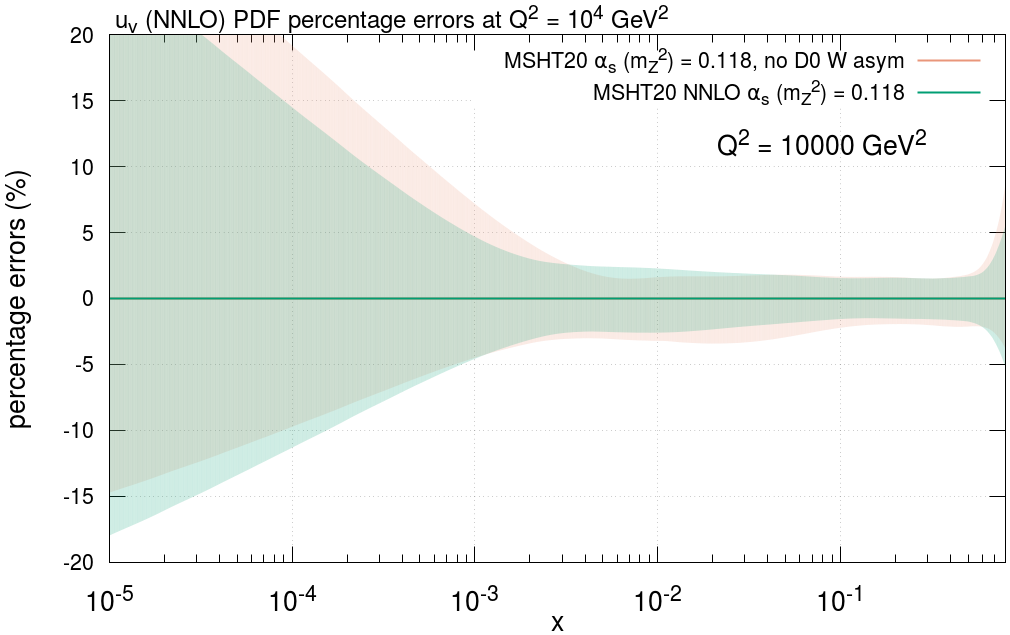}
\includegraphics[scale=0.24, trim = 50 0 0 0 , clip]{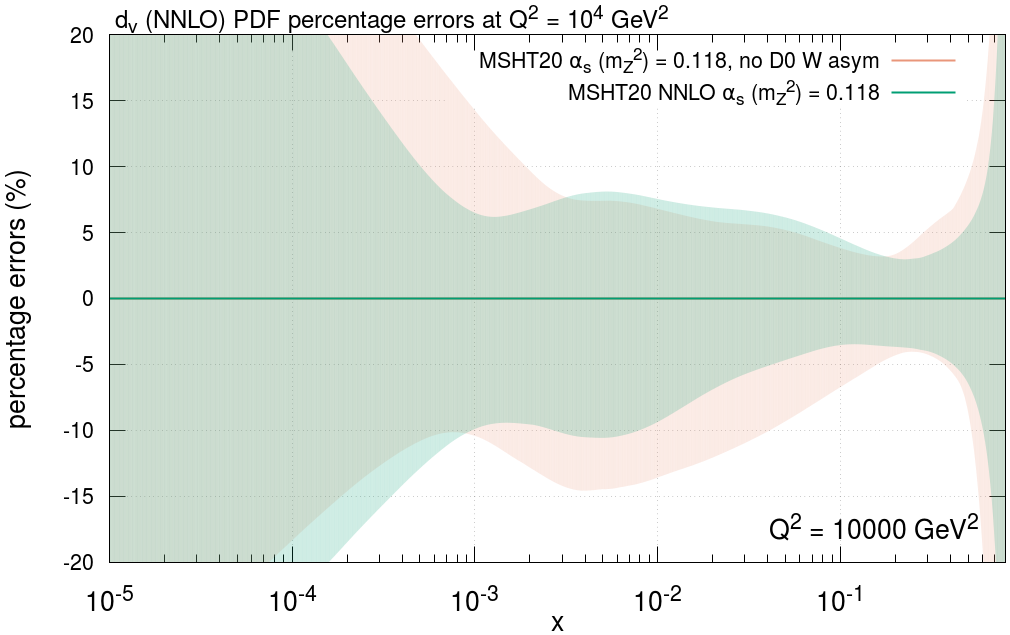}
\caption{\sf  (Left) $u_V$ and (right) $d_V$ PDF percentage errors at $Q^2=10^4~\GeV^2$ at NNLO in both the MSHT20 default fit and the same with the removal of the D{\O} $W$ asymmetry data, and without fitting the electron asymmetry.}
\label{MSHT20_NNLO_noD0Wasym_upvanddnvpercentageerrors}
\end{center}
\end{figure} 

\begin{figure} 
\begin{center}
\includegraphics[scale=0.24, trim = 50 0 0 0 , clip]{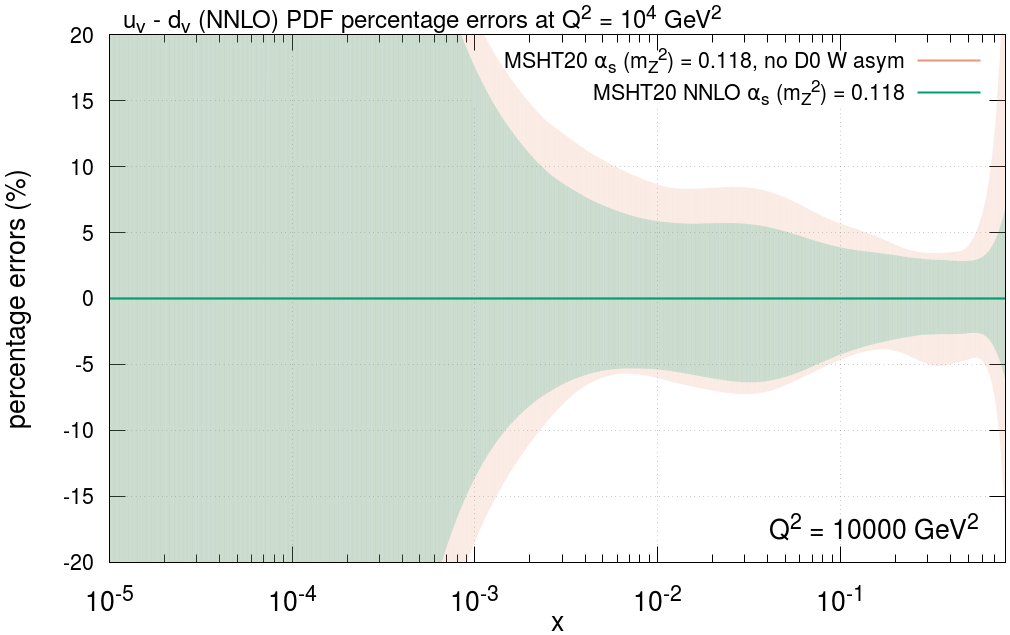}
\caption{\sf  $u_V - d_V$ PDF percentage errors at $Q^2=10^4~\GeV^2$ at NNLO in both the MSHT20 default fit and the same with the removal of the D{\O} $W$ asymmetry data, and without fitting the electron asymmetry.}
\label{uvminusdv_q210000_NNLOnoD0Wasym_percentageerrors}
\end{center}
\end{figure} 

\begin{figure} 
\begin{center}
\includegraphics[scale=0.23, trim = 50 0 0 0 , clip]{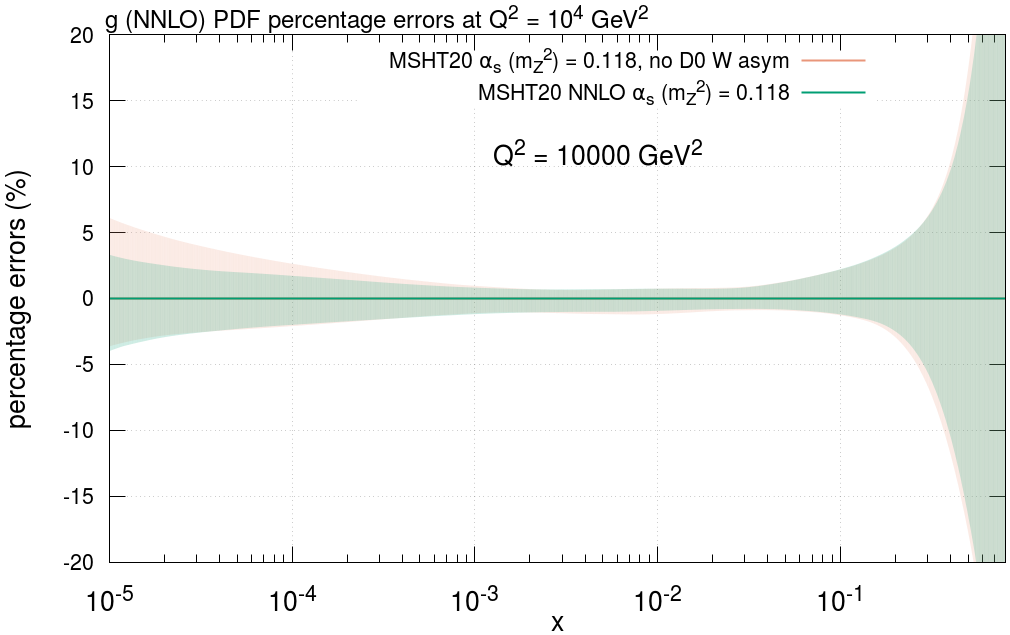}
\includegraphics[scale=0.23, trim = 50 0 0 0 , clip]{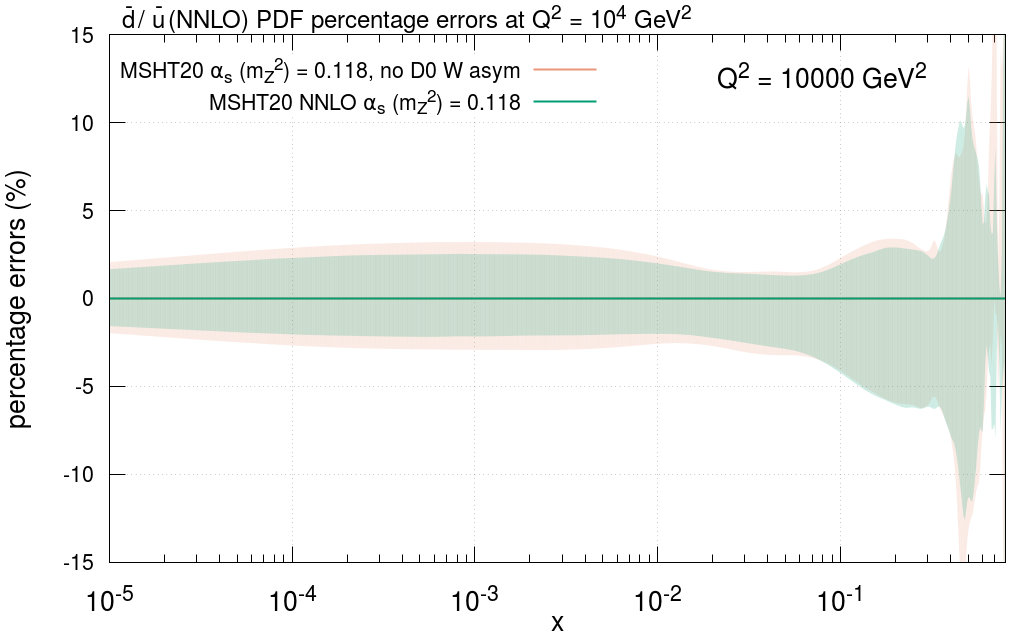}
\caption{\sf  (Left) $g$ and (right) $\bar{d}/\bar{u}$ PDF percentage errors at $Q^2=10^4~\GeV^2$ at NNLO in both the MSHT20 default fit and the same with the removal of the D{\O} $W$ asymmetry data, and without fitting the electron asymmetry.}
\label{MSHT20_NNLO_noD0Wasym_gluonanddbaroverubarpercentageerror}
\end{center}
\end{figure} 

\begin{figure} [t]
\begin{center}
\includegraphics[scale=0.23, trim = 50 0 0 0 , clip]{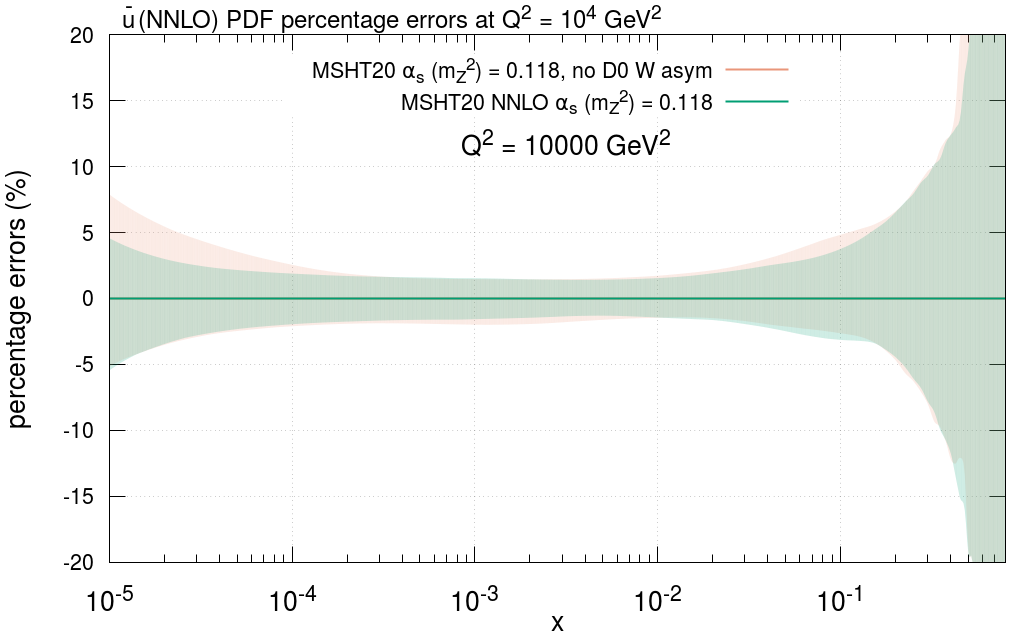}
\includegraphics[scale=0.23, trim = 50 0 0 0 , clip]{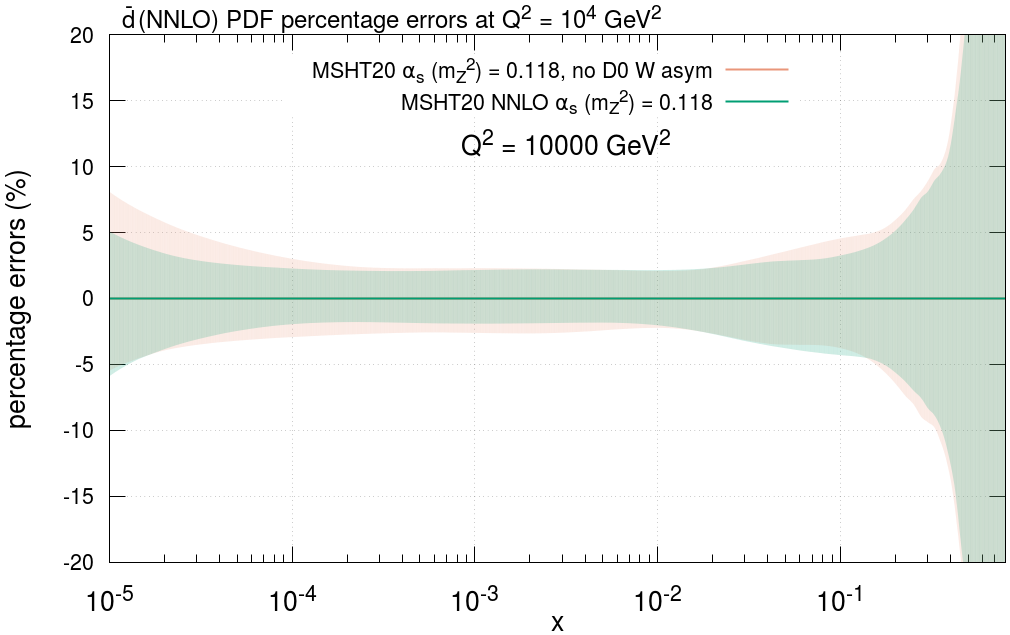}
\caption{\sf  (Left) $\bar{u}$ and (right) $\bar{d}$ PDF percentage errors at $Q^2=10^4~\GeV^2$ at NNLO in both the MSHT20 default fit and the same with the removal of the D{\O} $W$ asymmetry data, and without fitting the electron asymmetry.}
\label{MSHT20_NNLO_noD0Wasym_ubaranddbarpercentageerrors}
\end{center}
\end{figure} 

\begin{figure}[t]
\begin{center}
\includegraphics[scale=0.23, trim = 50 0 0 0 , clip]{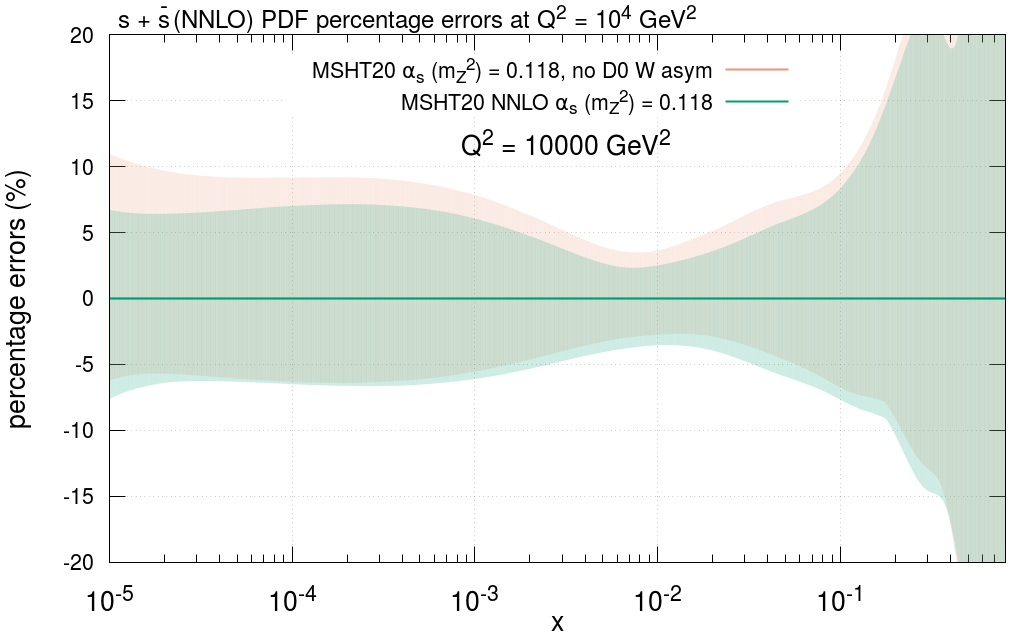}
\includegraphics[scale=0.23, trim = 50 0 0 0 , clip]{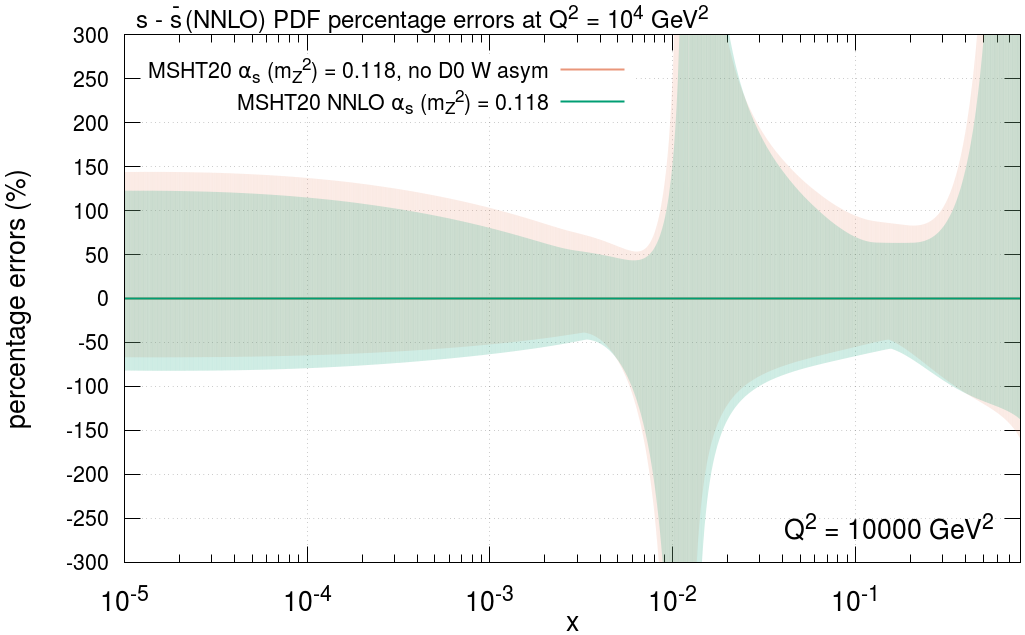}
\caption{\sf  (Left) $s+\bar{s}$ and (right) $s-\bar{s}$ PDF percentage errors at $Q^2=10^4~\GeV^2$ at NNLO in both the MSHT20 default fit and the same with the removal of the D{\O} $W$ asymmetry data, and without fitting the electron asymmetry.}
\label{MSHT20_NNLO_noD0Wasym_strangenesspercentageerrors}
\end{center}
\end{figure} 

We now consider the effect of this data set on the error bands in the global fit. We find the D{\O} $W$ asymmetry to be the most constraining data set in 13 of the 64 eigenvector directions  (see Section~\ref{PDFuncertainties}). Hence by this measure it is one of the most constraining data sets in the fit, though as discussed in detail in Section~\ref{PDFuncertainties} this provides a rather over--simplified picture. Indeed, in the case of 4 eigenvector directions, this data set is only just the most constraining one.
Nonetheless, if we fit the electron asymmetry then this only constrains 2 of the eigenvector directions, and hence the $W$ asymmetry clearly has a more significant impact on the PDF uncertainties. It constrains eigenvectors related primarily to the down valence (particularly at high $x$) and $\bar{d}/\bar{u}$ (4 eigenvectors each) with also some impact on the strangeness (3 across the strangeness total and asymmetry) and 2 eigenvectors relating partly to the gluon. The effect of this data set on the error bands can be gauged by comparing the percentage errors with and without the D{\O} $W$ asymmetry data set, as shown in Figs.~\ref{MSHT20_NNLO_noD0Wasym_upvanddnvpercentageerrors}-\ref{MSHT20_NNLO_noD0Wasym_strangenesspercentageerrors}.

We begin by analysing its effects on the valence quark error bands in Fig.~\ref{MSHT20_NNLO_noD0Wasym_upvanddnvpercentageerrors}, as these show both the largest differences. Removing the D{\O} $W$ asymmetry data results in enlarged error bands across the $x$ range for both valence quarks, with greater constraints on the less well-constrained down valence. This is not surprising given the sensitivity of the data set to the central values at high $x$ of the valence quarks. Given the constraints on the uncertainties around $x \sim 10^{-2}$  (where it provides a lower bound), due to the $u_V$ and $d_V$ sum rules there are also consequent upper bounds on the low $x$ valence quarks. In the case of the down valence there are also constraints on both upper and lower bounds at very high $x$. These constraints on the valence quarks translate into the error bands on the valence quark difference, $u_V - d_V$. Fig.~\ref{uvminusdv_q210000_NNLOnoD0Wasym_percentageerrors} displays the reduction on the error bands due to fitting the D{\O} $W$ asymmetry data across the entire $x$ range and on both upper and lower bounds.

As a result of these direct constraints on the valence quarks, there are also indirect consequences for the uncertainties of other PDFs. Specifically, as a result of constraining the $d_V$ lower error bound where it peaks, the momentum sum rule causes it to provide an upper bound on the gluon total momentum. The principle place the gluon can still vary (i.e. is less well constrained) is at low $x$ and so the D{\O} $W$ asymmetry provides an indirect upper bound in that region, via the momentum sum rule. This can be seen in Fig.~\ref{MSHT20_NNLO_noD0Wasym_gluonanddbaroverubarpercentageerror} (left). This upper bound at low $x$ on the gluon transfers into an analogous upper bound at low $x$ on the $\bar{u}$ and $\bar{d}$ at high $Q^2$ where they are dominantly driven by the gluon, as evident in Fig.~\ref{MSHT20_NNLO_noD0Wasym_ubaranddbarpercentageerrors}. Furthermore, the $W$ asymmetry has direct constraints on these too, mainly near $x=0.1$ and more so for $\bar d$. It also constrains their ratio $\bar{d}/\bar{u}$ over much of the $x$ range, although again more at low $x$, see Fig.~\ref{MSHT20_NNLO_noD0Wasym_gluonanddbaroverubarpercentageerror} (right).

The constraint on the upper bound of the gluon at low $x$ also translates into the total strangeness in Fig.~\ref{MSHT20_NNLO_noD0Wasym_strangenesspercentageerrors} (left). The D{\O} $W$ asymmetry nonetheless does have some constraint on the total strangeness at high $x$ through its contribution to the cross section. The $s-\bar{s}$ is also constrained, albeit weakly. Here the sum rule for zero strangeness asymmetry ensures that where there is an upper bound at low $x$ there is a lower bound at high $x$ (and vice versa). In the percentage errors in Fig.~\ref{MSHT20_NNLO_noD0Wasym_strangenesspercentageerrors} (right) this appears as a consistent upper bound across the whole $x$ range, as there is also a sign change on the strangeness asymmetry that further flips these bands with respect to one another.

\subsection{Impact of ATLAS $W$, $Z$ production data} \label{ATLASDYeffects}

\begin{figure} [t]
\begin{center}
\includegraphics[scale=0.24, trim = 60 0 0 0 , clip]{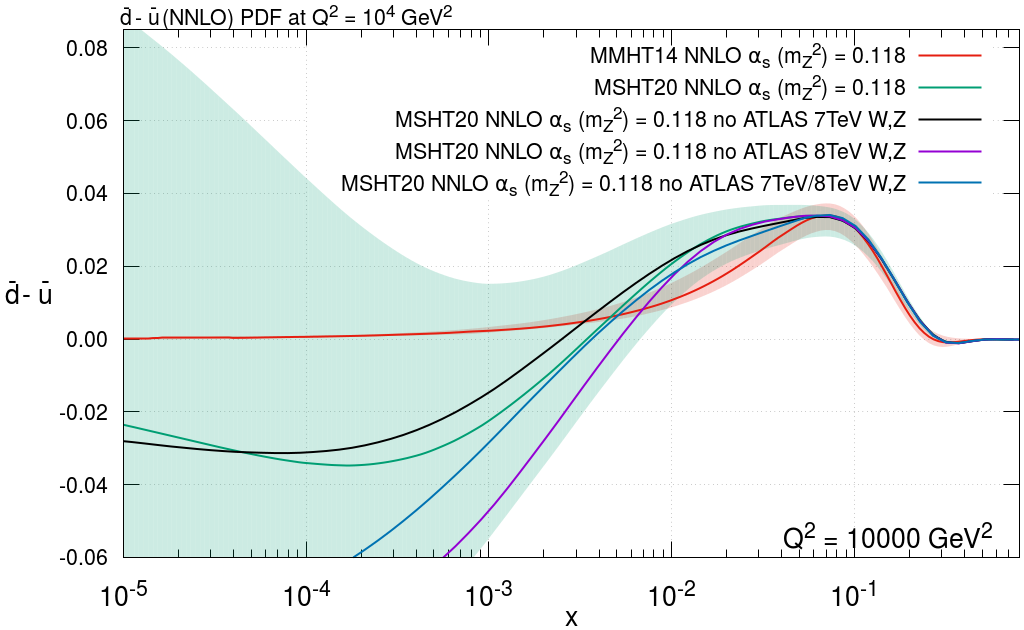}
\includegraphics[scale=0.24, trim = 60 0 0 0 , clip]{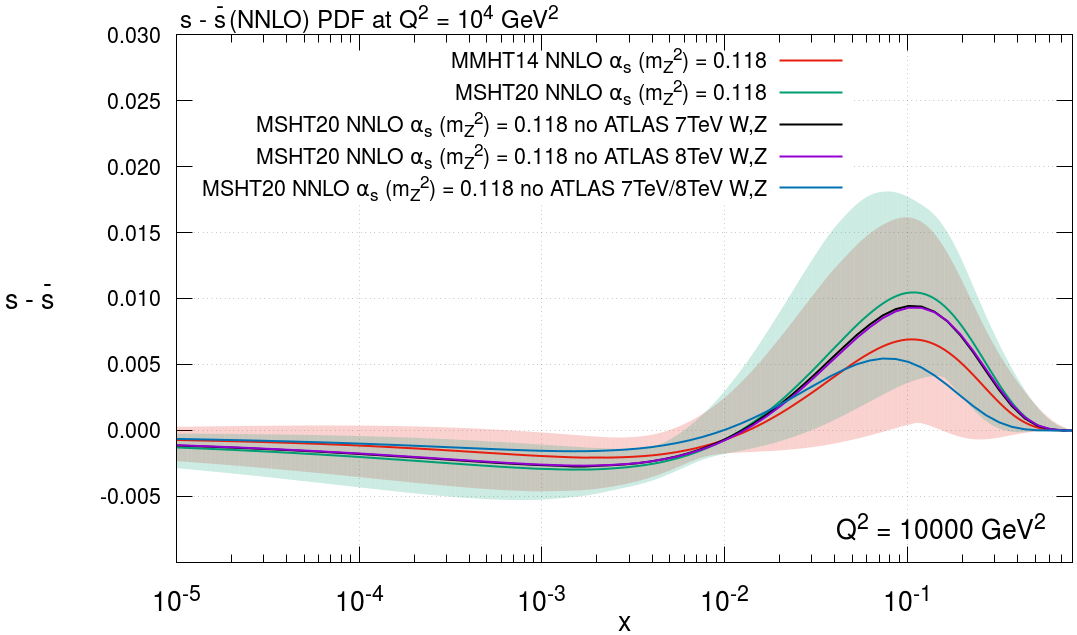}
\includegraphics[scale=0.24, trim = 50 0 0 0 , clip]{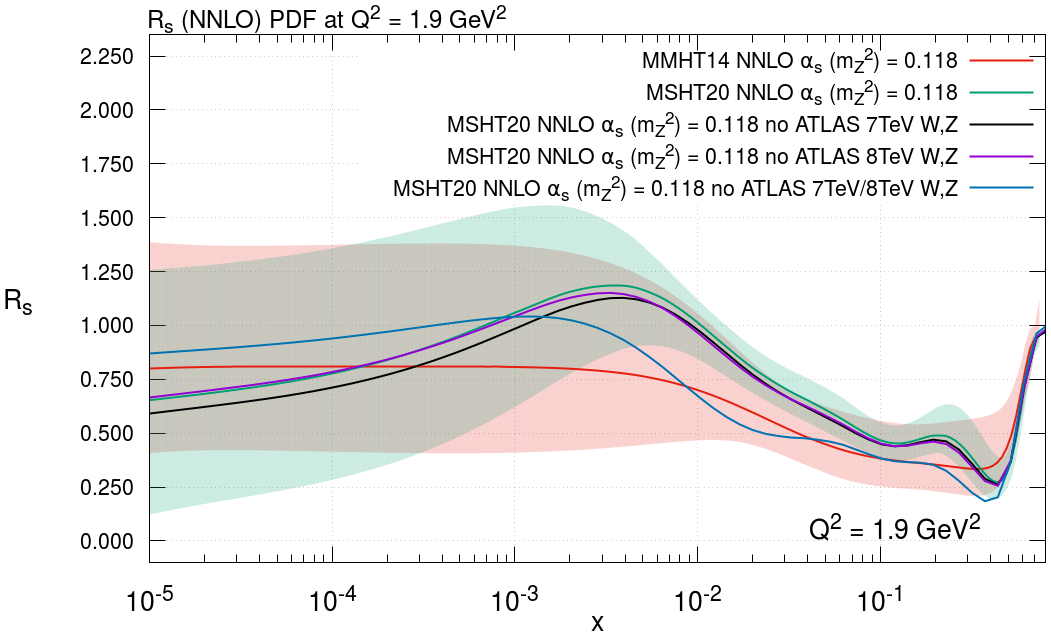}
\caption{\sf (Left) $\bar{d}-\bar{u}$ PDF, (right) $s-\bar{s}$ PDF and (bottom) $R_s$ at NNLO, the former two at $Q^2=10^4~\GeV^2$ and the latter at $Q^2=1.9~\GeV^2$, all showing the effects of removing just the 7~TeV ATLAS $W$, $Z$ data or the combination of the ATLAS 8~TeV $W^{\pm}$ data and ATLAS 8~TeV $Z$ double differential data and also of removing both of these together. Central values and uncertainty bands are included for the default MSHT20 and MMHT14 plots, whilst only central values are given for the cases with ATLAS Drell-Yan data removed from MSHT20.}
\label{MSHT20_NNLO_noATDY}
\end{center}
\end{figure} 

The ATLAS Drell-Yan data at 7 and 8~TeV are amongst the highest precision LHC data included in the fit, and the 7~TeV data are well-known to be in some tension with previous dimuon data, see Sections~\ref{AT7WZdata} and \ref{strangeness}. It is therefore interesting to investigate the effects of these data sets on the MSHT20 PDFs further.  Drell-Yan data at ATLAS are sensitive to the light quark and antiquark flavour separation over a range of intermediate $x$ values through the rapidity spectrum, whilst correlations between $Z$ and $W$ data are particularly able to constrain the strangeness. It should be noted that, in contrast to the 7~TeV data, the 8~TeV data are reported as $W$ and $Z$ measurements separately so correlations between the two cannot be incorporated. In this section we analyse the effects of these data sets on the MSHT20 PDFs by removing them from the overall fit. In particular we focus upon the asymmetry of the down and up antiquarks, the total strangeness in the form of the ratio $R_s$, and the strangeness asymmetry, as these are the quantities where the ATLAS data sets have the most noticeable impact.

We begin with the effects of removing either the 7~TeV or 8~TeV data sets, or both, from the overall MSHT20 default fit, as shown in Fig.~\ref{MSHT20_NNLO_noATDY}. Considering first the $\bar{d}-\bar{u}$ in Fig.~\ref{MSHT20_NNLO_noATDY} (left), the rapidity spectrum covers the region $ 10^{-3} \lesssim x \lesssim 10^{-1}$ with the central rapidity at $x \approx 10^{-2}$. There are clear effects on the $\bar{d} - \bar{u}$ PDF in both the intermediate and low $x$ regions. In the former case, removing both the 7 and 8~TeV data causes the PDF to be lowered, suggesting a pull of these data upwards on the $\bar{d}-\bar{u}$ in this region. This is part of the reason for the broadened peak at $10^{-2} \lesssim x \lesssim 10^{-1}$ in $\bar{d}-\bar{u}$ seen in MSHT20. As we move to lower $x$ there are competing effects on the asymmetry from the 7 and 8~TeV data. Removing the 7~TeV data alone pulls the $\bar{d}-\bar{u}$ up in the $ 10^{-3} \lesssim x \lesssim 10^{-2}$ interval, whilst removing the 8~TeV data alone pulls the PDF down within this region (to near the edge of the error bands). Consequently, removing both the 7 and 8~TeV data results in a compromise between these two pulls with the $\bar{d}-\bar{u}$ similar to the default MSHT20 around $x \approx 3 \times 10^{-3}$ before dropping lower at low $x$. The result of these effects in the intermediate to low $x$ region is then extrapolated into the very low $x$ region. Here, whilst the default MSHT20 $\bar{d}-\bar{u}$ PDF clearly tends upwards towards 0 at very low $x$, this effect is not present as visibly when the 8~TeV data are removed. On the other hand, the removal of the 7~TeV data has only a limited effect in this region. Therefore it seems that the effects of the 8~TeV ATLAS Drell-Yan data sets are to pull the $\bar{d}-\bar{u}$ asymmetry back up at low $x$, which then causes the asymmetry to tend to 0 at very low $x$. This is, at least partly, responsible for the ratio $\bar{d}/\bar{u}$ then tending to 1 in this region, as noted in Section~\ref{dbarminusubarMSHT20}.

\begin{figure} 
\begin{center}
\includegraphics[scale=0.24, trim = 60 0 0 0 , clip]{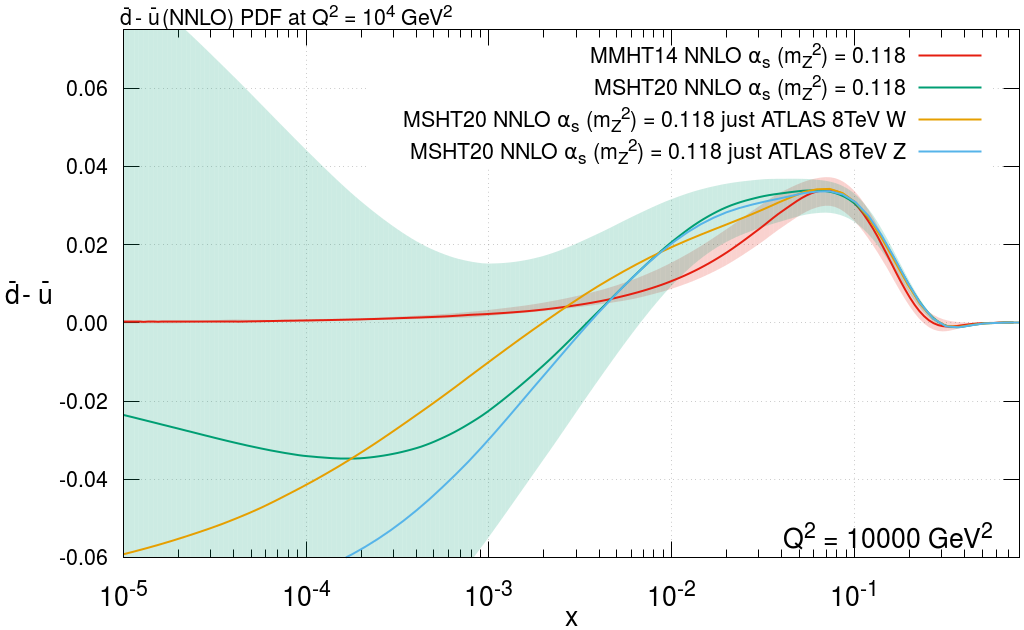}
\includegraphics[scale=0.24, trim = 60 0 0 0 , clip]{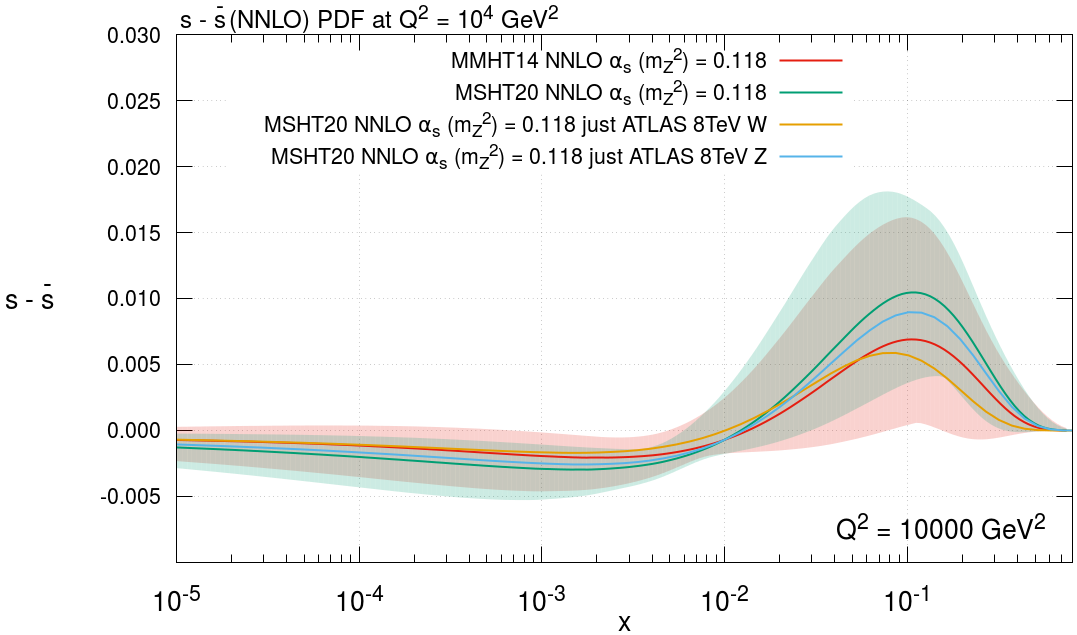}
\includegraphics[scale=0.24, trim = 50 0 0 0 , clip]{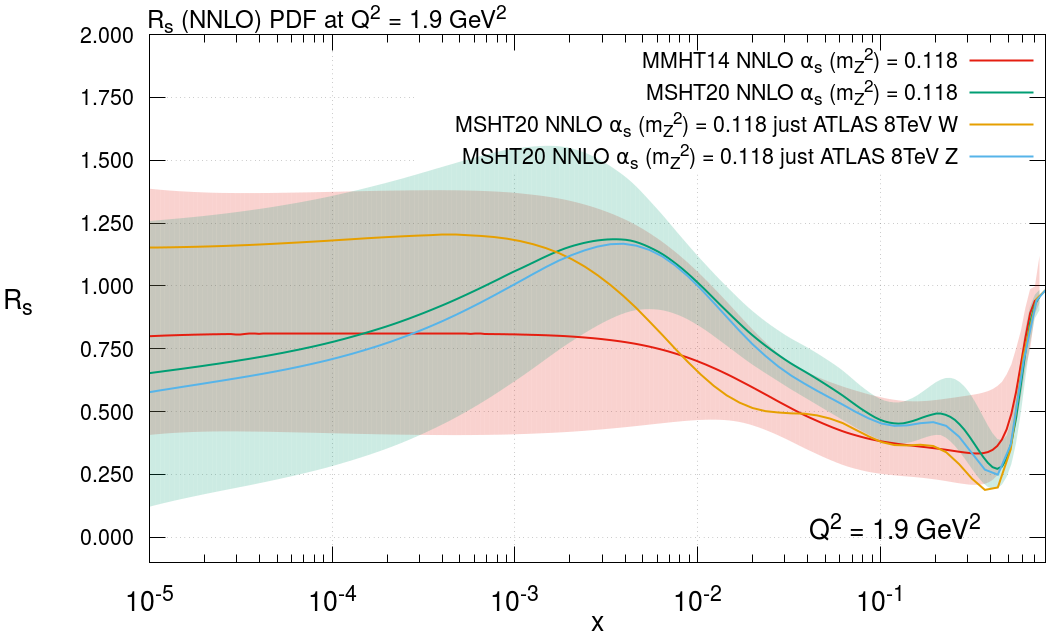}\label{Rs_q21p9_NNLOnoATDY_extra}
\caption{\sf As in Fig.~\ref{MSHT20_NNLO_noATDY}, but showing the effects of removing separately just the ATLAS 8 TeV $W$ or the $Z$ data set.}
\label{MSHT20_NNLO_noATDY_extra}
\end{center}
\end{figure} 

\begin{figure} 
\begin{center}
\includegraphics[scale=0.24, trim = 60 0 0 0 , clip]{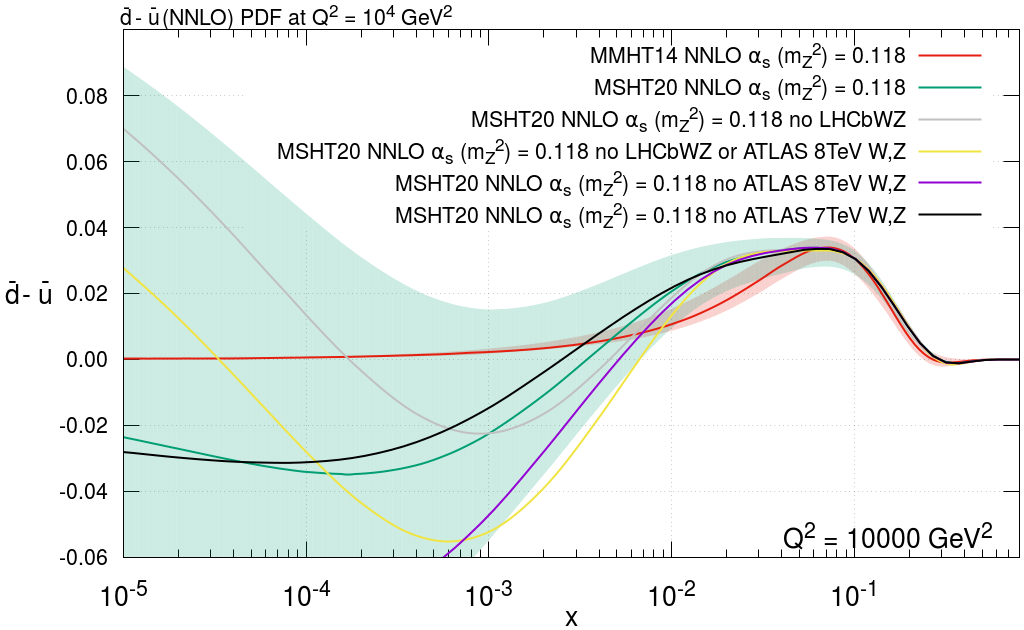}
\caption{\sf $\bar{d}-\bar{u}$ PDF at $Q^2=10^4~\GeV^2$, showing the effects of removing separately various of the ATLAS and LHCb $W$, $Z$ data sets. Central values and uncertainty bands are included in for the default MSHT20 and MMHT14 plots, whilst only central values are given for the cases with data sets removed from MSHT20.}
\label{dbarminusubar_q210000_NNLOnoATDY_extranoLHCb}
\end{center}
\end{figure} 

As well as the more obvious sensitivities of the precision ATLAS Drell-Yan data to the up and down antiquarks, it is also sensitive to the strange quark and antiquark. Given the contributions from the strange quark/antiquarks differ between the $W$ and $Z$ bosons, the combination of these measurements and correlations between them (at 7~TeV) enable constraints to be placed on the strangeness and its asymmetry. Fig.~\ref{MSHT20_NNLO_noATDY} (right) shows the impact of the 7 and 8~TeV data on the poorly constrained strangeness asymmetry, with the removal of both the 7 and 8~TeV data lowering its amplitude notably. This shows the impact of these data is to favour an enhanced $s - \bar{s}$, with both the 7 and 8~TeV data sets seemingly preferring this. Nonetheless, even with all these data included, the strangeness asymmetry remains poorly constrained, although in MSHT20 it is now clearly non-zero outside of the error bands, as a result of the sensitivity of the high precision ATLAS Drell-Yan data. The effects of these data sets on the total strangeness and on $R_s$ are shown in Fig.~\ref{MSHT20_NNLO_noATDY} (bottom). The impact of the 7~TeV data to enhance the strangeness in the region of $x \sim 10^{-2}$ is now well-known, however the impact of the 8~TeV data has so far not been studied; indeed we are the first PDF fitters to include these data sets. Interestingly, we observe that removing either the 7~TeV or the 8~TeV data has a similar effect on the values of $R_s$, lowering the PDF slightly relative to the overall MSHT20 global fit over much of the $x$ range below 0.1. This is particularly noteworthy as it shows that, like the 7~TeV data, the 8~TeV $W$ and $Z$ data sets observe the same strangeness enhancement. Moreover, this occurs even though the data sets are added separately, with no correlated errors included between the two. It seems that, though not explicitly correlated in this way, the global fit itself forces them to be so in order to fit them. This correlation in the fit can be observed through the fact that similar shifts in the normalisation are seen in the separate 8~TeV $W$ and $Z$ data sets, with each requiring a near identical shift of 2.15\% between data and theory in the luminosity. This is also similar to the comparable 1.58\% shift seen for 7~TeV $W$, $Z$ data. Finally, as might be expected, removing both the 7 and 8~TeV data removes the ``strangeness unsuppression'' and lowers the spectrum over the $x$ range to which the data sets are sensitive, doing so beyond the MSHT20 error bands. 

Given the impact of the 8~TeV $W$ and $Z$ data on the PDFs, it is interesting to determine their individual effects. This is shown in Fig.~\ref{MSHT20_NNLO_noATDY_extra}, where the 8~TeV $W$ and $Z$ data sets are removed separately from the global fit. Returning again to the $\bar{d}-\bar{u}$ in Fig.~\ref{MSHT20_NNLO_noATDY_extra} (left), the $W$ data favour a reduced peak around $x \sim 10^{-2}$. At lower $x$ it raises the PDF relative to both the default MSHT20 global fit and to the fit to the $Z$ data alone, with the $W$ and $Z$ data pulling the PDF up and down respectively around $x\sim 10^{-3}$, resulting in the balance seen in MSHT20. Finally, at very low $x$ both the $Z$ and in particular the $W$ data begin to see a slight upturn in the difference of $\bar{d}-\bar{u}$. Nonetheless separately their effects are much less than when combined, as observed in Fig.~\ref{MSHT20_NNLO_noATDY} (left) in the ``no ATLAS 7~TeV $W$, $Z$'' line.

The effect of the separate 8~TeV $W$ and $Z$ data sets on the strangeness is given in Figs.~\ref{MSHT20_NNLO_noATDY_extra} (right) and \ref{MSHT20_NNLO_noATDY_extra} (bottom). In both cases it is clear that the $Z$ data have a much more significant impact on the strangeness than the $W$ data, which is to be expected given the $Z$ has larger down and strange couplings than the $W$ boson. As a result, the $Z$ data are responsible for the rises in both the strangeness asymmetry and $R_s$ originating from the 8~TeV data. Typically in the $Z$ data earlier fits undershot at small rapidity (corresponding to $x \lesssim 0.01$) which then requires the normalisation to be raised in the fit, which may then explain the rise in $R_s$ as a consequence.

As an aside, we also show in Fig.~\ref{dbarminusubar_q210000_NNLOnoATDY_extranoLHCb} the impact of the ATLAS 7 or 8 TeV individually, as well as the  LHCb $W$, $Z$ data \cite{LHCbZ7,LHCbWZ8}, on the $\bar{d}-\bar{u}$. In contrast to the 8~TeV ATLAS $W$, $Z$ data, the effect of the removal of this LHCb $W$, $Z$ data is to raise substantially this PDF at very low $x$ with it observed to turnover and tend back towards positive values of $\bar{d}-\bar{u}$ around $x \sim 10^{-3}$. The effect of removing both the LHCb $W$, $Z$ and ATLAS 8~TeV $W$, $Z$ data is also shown, and whilst the combination is now not as high at low $x$, its behaviour is still very different from the MSHT20 default fit. However, the effect of the removal 7~TeV ATLAS $W$, $Z$ data was largely opposite, with it raising $\bar{d}-\bar{u}$ at intermediate to low $x$ before lowering slightly at very low $x$. Therefore it is the combination of all four of these data sets that results in the tendency for the $\bar{d}-\bar{u}$ to tend to 1. It should be noted however, there are very large error bands at very low $x$ and the effects of the removal of any of these data sets all fall well within the error bands, whilst the precise behaviour is an extrapolation of their effects at low $x$.

\begin{figure}[t]
\begin{center}
\includegraphics[scale=0.24]{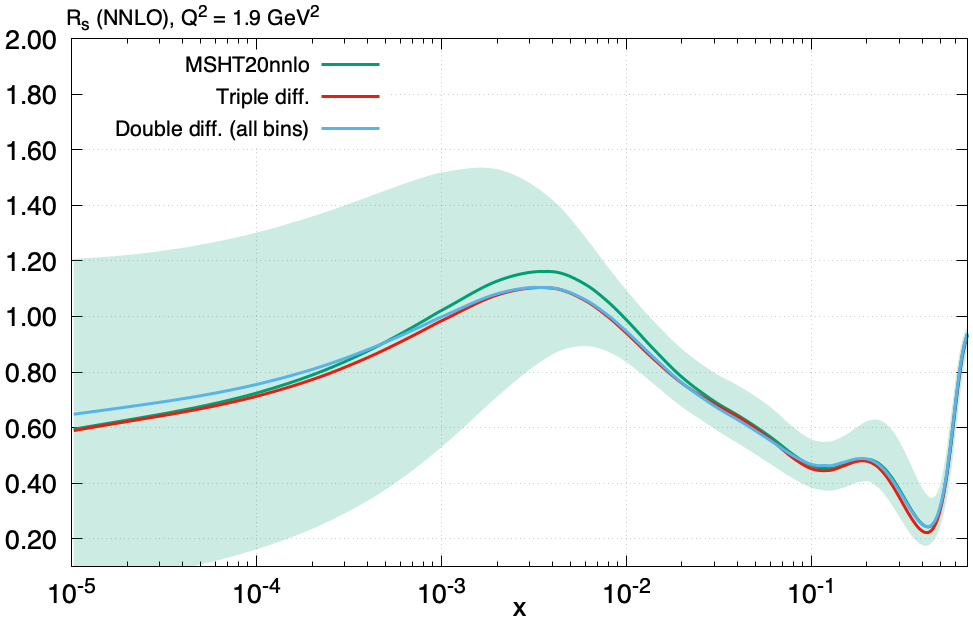}
\includegraphics[scale=0.24]{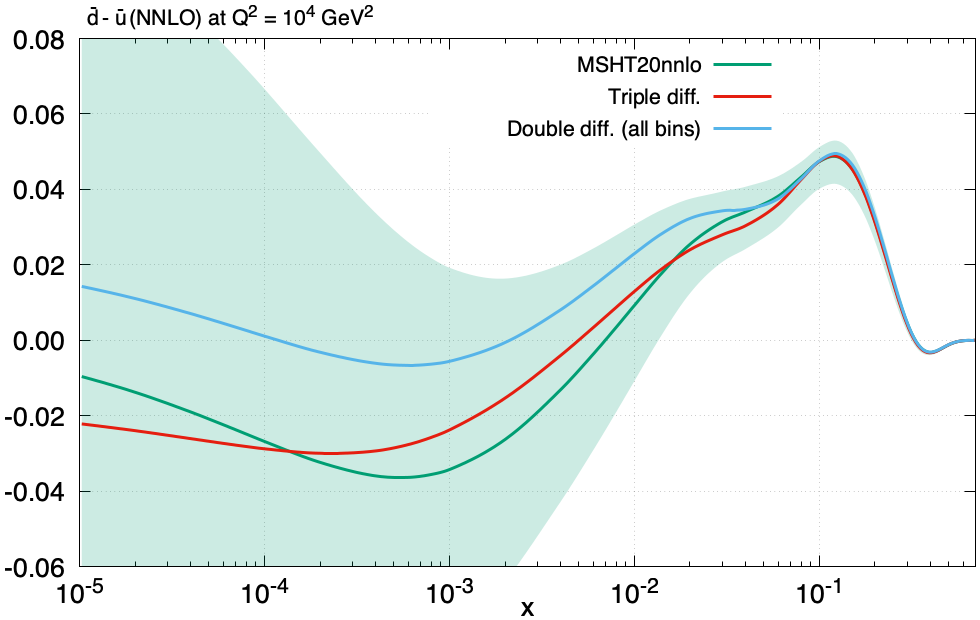}
\caption{\sf The strangeness ratio, $R_s$ at $Q^2=1.9$ ${\rm GeV}^2$, and $\overline{d}-\overline{u}$, at  $Q^2=10^4$ ${\rm GeV}^2$ for variants of the MSHT NNLO fits ($\alpha_S$ free). Results with the ATLAS 8~TeV double differential data replaced by the triple differential data are given, as well as that of removing the $ > 95\%$ acceptance requirement in the former case, are shown.}
\label{fig:Z3Dcomp}
\end{center}
\end{figure}

Finally, we briefly consider the impact of including the fully triple differential (3D) ATLAS 8~TeV data on DY lepton pair production; we recall that by default we fit the data integrated over $\cos\theta^*$, in order to limit any sensitivity to $\sin^2 \theta_W$. To do this, we repeat the default NNLO fit (with $\alpha_S$ free), but replace the 2D data with the full 3D data set, with the same constraint that only bins with greater than $95\%$ acceptance are included. In addition, we consider the impact of removing the requirement that only $\cos\theta^*$ bins with $> 95\%$ acceptance are integrated over to produce the 2D data set. We use the $G_{\mu}$ scheme, with the EW parameters set to the most recent PDG values. The fit quality to the triple differential data is good, and indeed slightly better at $\sim 1.4$ per point in comparison to the default, whereas the effect of removing the $> 95\%$ acceptance requirement in the double differential case is to give a worse fit quality of $\sim 2.1$ per point. Results are shown in Fig.~\ref{fig:Z3Dcomp} for two representative PDF choices, and we can see the effect is rather mild, though not completely negligible.

\begin{figure} [t]
\begin{center}
\includegraphics[scale=0.24, trim = 50 0 0 0 , clip]{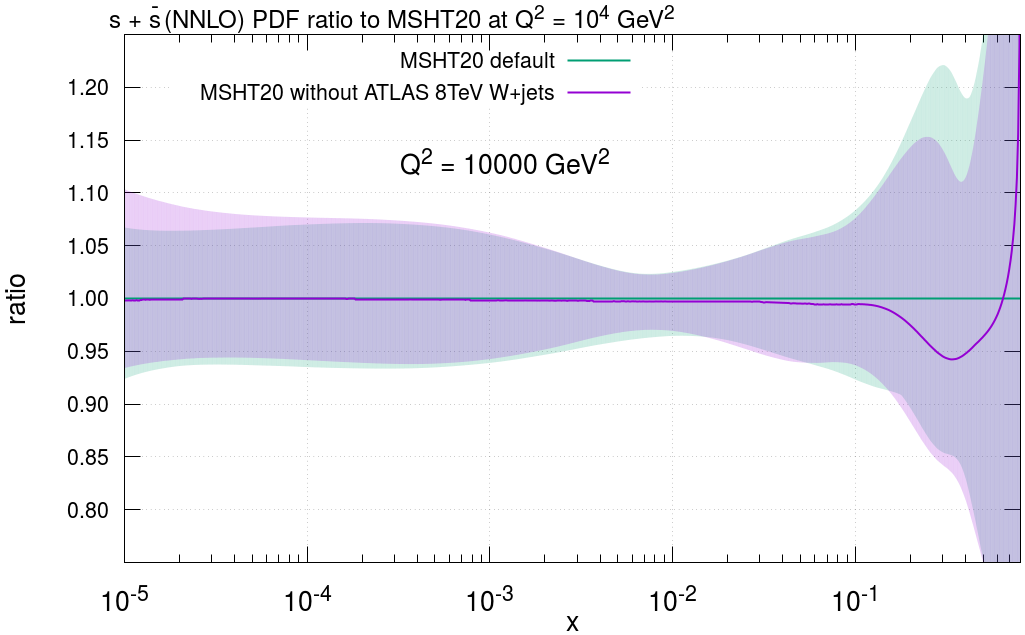}
\caption{\sf $s + \bar{s}$ PDF ratio to MSHT20 at $Q^2=10^4~\GeV^2$ at NNLO in the absence of the ATLAS 8~TeV $W$+jets data set.}
\label{splussbar_q210000_NNLO_ratio_AT8Wjets}
\end{center}
\end{figure} 

We also note that, as well as the LHC high precision Drell-Yan and $W+c$ data sets, the ATLAS 8~TeV $W+\text{jets}$ data \cite{ATLASWjet} may constrain the strangeness at higher $x$, see \cite{Faura:2020oom} and \cite{ATLAS:2019ext}. Therefore we have investigated the effect of removing this data set from the fit. Upon removing the ATLAS 8~TeV $W+\text{jets}$ data set we find a negligible change in the $\chi^2$ of the other data sets in the fit, suggesting no tension with the global fit and little sensitivity in the global fit to this data set. The same is true if, rather than removing the first $p_T^{W}$ bin for the $W^+$ and $W^-$ as is default in MSHT20 (due to their poor fit qualities and the large corrections present in these bins) we include all the $W+\text{jets}$ bins. In that case, whilst the overall $\chi^2/N$ changes from 18.1/30 to 41.0/32 (the latter compared to 41.2/32 without refitting - showing the negligible changes), the $\Delta\chi^2$ of the other data sets in the global fit changes upon refitting by just $+0.9$ and the PDFs show negligible differences. Given the limited sensitivity to the presence of the ATLAS 8~TeV $W+\text{jets}$ data set, it is unsurprising that we find only very limited changes in the strangeness PDFs when it is not included, with slight reductions in the strangeness at high $x$, well within the large uncertainty bands, as shown in Fig.~\ref{splussbar_q210000_NNLO_ratio_AT8Wjets}. Similarly, the strangeness asymmetry and the overall percentage error bands on the total strangeness also show only limited changes in the absence of this data set. Its impact is much reduced relative to that reported by ATLAS~\cite{ATLAS:2019ext,1830590}, however given the inclusion of many more data sets in our global fit this is not surprising. In particular, the MSHT20 global fit already includes the CCFR and NuTeV dimuon data sets, which both favour reduced strangeness at high $x$, and therefore offer a significant constraint in the region where these ATLAS 8~TeV $W+\text{jets}$ data are sensitive. Nonetheless, our conclusions are consistent in that we also find this ATLAS 8~TeV $W+\text{jets}$ data set is well fit with reduced strangeness in the high $x$ region, and so is not in tension with other data in our fit. It is therefore consistent with the overall strangeness reported in MSHT20. That is with strangeness unsuppressed at low $x$, strangeness intermediate to that of the dimuon, $W+c$ data sets on the one hand and the ATLAS high precision Drell-Yan data sets at 7 and 8~TeV on the other at intermediate $x$, and reduced strangeness at high $x$, with the ATLAS 8~TeV $W+\text{jets}$ data largely sensitive to this high $x$ region.

\subsection{Strangeness asymmetry} \label{strangeness_asymmetry}

\begin{table}
\begin{center}
\def\arraystretch{1.0}
\begin{tabular}{|>{\arraybackslash}m{7.6cm}|>{\centering\arraybackslash}m{1.5cm}|>{\centering\arraybackslash}m{4.5cm}|} \hline
  Data set & $N_{pts}$ & $\Delta \chi^2$ with $s=\bar{s}$ \newline relative to MSHT20 \\ \hline
  CCFR $\nu N \rightarrow \mu \mu X$ \cite{Dimuon} & 86 & +1.0\\
  NuTeV $\nu N \rightarrow \mu \mu X$ \cite{Dimuon} & 84 & +24.0\\
  D{\O} II $W\rightarrow \nu e$ asym. \cite{D0Wnue} & 12 & -1.8\\
  D{\O} II $W \rightarrow \nu \mu$ asym. \cite{D0Wnumu} & 10 & +1.4 \\
  D{\O} $W$ asym. \cite{D0Wasym} & 14 & -0.8 \\
  CMS 7~TeV $W+c$ \cite{CMS7Wpc} & 10 & -0.5\\
  ATLAS 7~TeV high precision $W$, $Z$ \cite{ATLASWZ7f} & 61 & +2.7\\
  ATLAS 8~TeV High-mass Drell-Yan \cite{ATLASHMDY8} & 48 & +0.7\\
  ATLAS 8~TeV $W$\cite{ATLASW8} & 22 & +1.3\\
  ATLAS 8~TeV double differential $Z$ \cite{ATLAS8Z3D} & 59 & +2.3\\ \hline
  Total & 4363 & +26.9\\ \hline
\end{tabular}
\end{center}
\caption{\sf The change in $\chi^2$ (with negative indicating an improvement in the fit quality) relative to the MSHT20 default fit for a selection of data sets when $s=\bar{s}$ is enforced at the input scale. We list in this table all data sets which have clear sensitivity to the strange asymmetry.}
\label{tab:seqsbar_chisqs}
\end{table}

\begin{figure} [t]
\begin{center}
\includegraphics[scale=0.24, trim = 60 0 0 0 , clip]{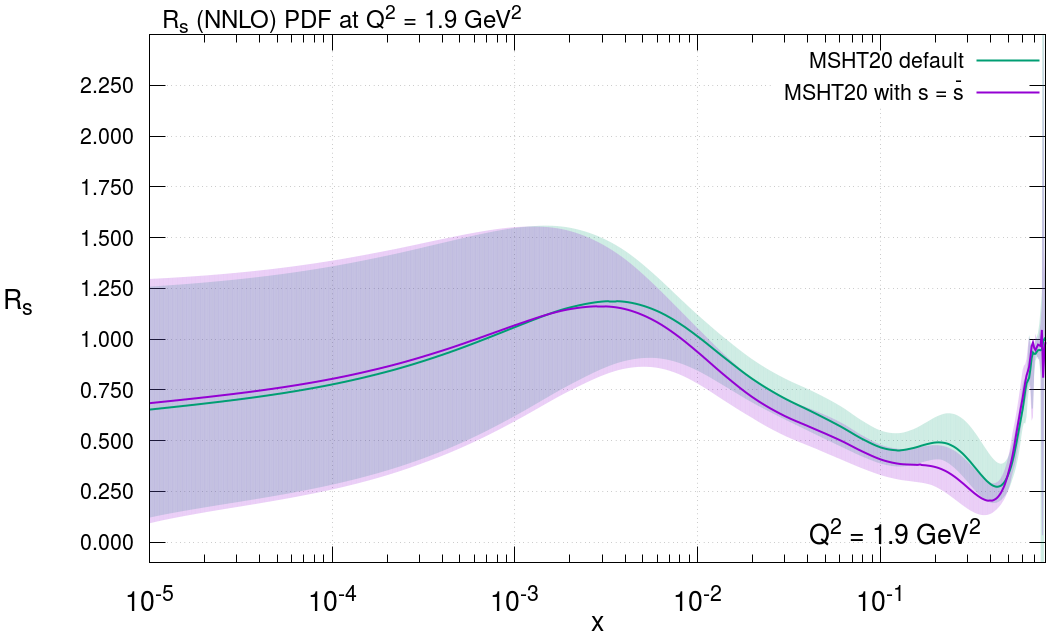}
\includegraphics[scale=0.24, trim = 60 0 0 0 , clip]{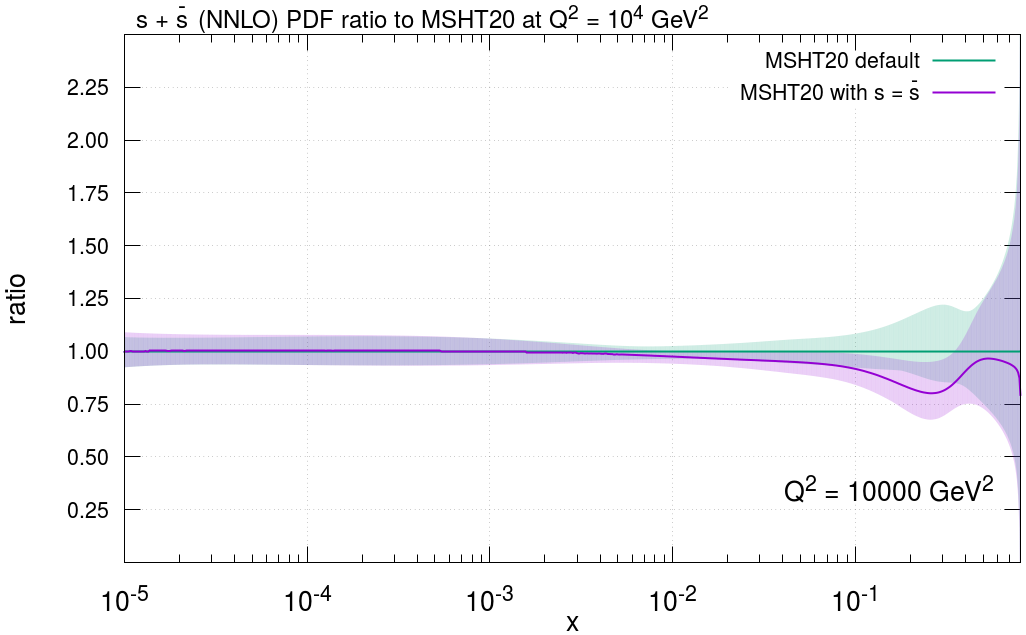}
\caption{\sf The total strangeness at NNLO when $s=\bar{s}$ is enforced at the input scale in the global fit. (Left) the values of $R_s$ at $Q^2=1.9$ ${\rm GeV}^2$ compared to the MSHT20 default fit. (Right) the ratio of the $s+\bar{s}$ at $Q^2=10^4$ ${\rm GeV}^2$.}
\label{fig:totalstrangeness_ratios_seqsbar}
\end{center}
\end{figure} 

\begin{figure} [t]
\begin{center}
\includegraphics[scale=0.24, trim = 60 0 0 0 , clip]{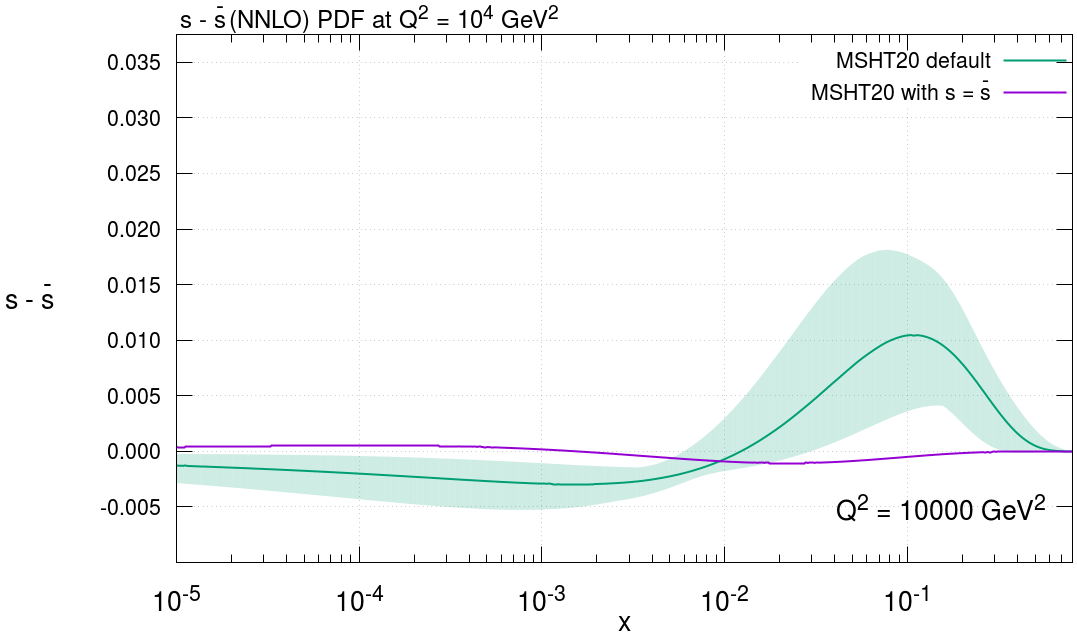}
\caption{\sf $s - \bar{s}$ PDF at NNLO at $Q^2=10^4~\GeV^2$,  comparing the MSHT20 global fit with the same fit when $s=\bar{s}$ is enforced at the input scale. Note a small asymmetry still results from the NNLO DGLAP evolution.}
\label{sminussbar_q210000_NNLO_seqsbar}
\end{center}
\end{figure} 

As discussed in the previous section, both this high precision ATLAS Drell-Yan data and the original dimuon data have sensitivity to the asymmetry in the strangeness content as well as to the total strangeness. The integral of the distribution for the strange content of the proton, $s_- = s -\bar{s}$, is constrained by a number sum rule to be zero. Nonetheless, the distribution itself may be non--zero, as may the integral of the momentum distribution. Therefore, with the additional data now in the global MSHT20 fit, the asymmetry in the distribution of the strange and the antistrange quark content of the proton may also be examined in more detail. This is an important question, as whilst DGLAP evolution at NNLO will generate a small asymmetry even from a symmetric distribution at the input scale (see for example \cite{Catani:2004nc}), such effects are small compared to the level of strangeness asymmetry favoured by the global fit, for example in Fig.~\ref{sminussbarfigs} (left), with the fit now favouring a non-zero value of the strangeness asymmetry clearly beyond the uncertainty bands. This can therefore be interpreted as clear evidence of a non-perturbative effect in the underlying PDFs. We note that both the original ATLAS PDF analysis \cite{ATLASWZ7f} and the CT18 PDF fits \cite{CT18} assume zero strangeness asymmetry.

In order to investigate the effects of the strangeness asymmetry, we therefore perform a further global fit, identical to MSHT20, but now with $s=\bar{s}$ enforced at the input scale. As a result the strangeness asymmetry observed results purely from perturbative evolution and can be seen to be much smaller than that in the default MSHT20 fit in Fig.~\ref{sminussbar_q210000_NNLO_seqsbar}. In Table~\ref{tab:seqsbar_chisqs} the clear reduction in the fit quality when the strange asymmetry is set to 0 is given, with the global fit worsening by $\Delta\chi^2 = 26.9$. The deterioration in fit quality is, unsurprisingly, focused overwhelming on the NuTeV dimuon data set, which worsens by 24 points in $\chi^2$, with the ATLAS 7~TeV $W$, $Z$ data sets and 8~TeV data sets also worsening but only by a total of 7 points. These effects result partly from the knock-on effect of the strangeness asymmetry on the total strangeness. As can be seen in Fig.~\ref{fig:totalstrangeness_ratios_seqsbar}, the effect of enforcing the strange and antistrange contents being equal is that the overall total strangeness in the $10^{-2} < x < 10^{-1}$ region, where enhancement is favoured by the ATLAS data, is reduced as the balance between the NuTeV and ATLAS data is altered. One notable feature related to this reduction is the value of the dimuon branching ratio, which is increased in the $s=\bar{s}$ fit from the default MSHT20 value of 0.089 to 0.100, i.e.  close to the edge of the 10\% uncertainty band on the branching ratio. This increase is  a result of the fit to the dimuon data and then allows the overall total strangeness to be reduced, as seen in Fig.~\ref{fig:totalstrangeness_ratios_seqsbar}.

\subsection{Fit excluding HERA data} \label{noHERA}

\begin{table}
\begin{center}
\def\arraystretch{1.0}
\begin{tabular}{|>{\arraybackslash}m{5.1cm}|>{\centering\arraybackslash}m{0.6cm}|>{\centering\arraybackslash}m{0.9cm}|>{\arraybackslash}m{6.7cm}|>{\centering\arraybackslash}m{0.75cm}|>{\centering\arraybackslash}m{0.9cm}|} \hline
  Data set & $N_{pts}$ & $\Delta \chi^2$ & Data set (continued) & $N_{pts}$ & $\Delta \chi^2$ \\ \hline
  BCDMS $\mu p$ $F_2$ \cite{BCDMS} & 163 & -5.2 & CMS $W$ asym. $p_T > 35$~GeV \cite{CMS-easym} & 11 & -1.2\\ 
  BCDMS $\mu d$ $F_2$ \cite{BCDMS} & 151 & -2.6 & CMS asym. $p_T > 25, 30$~GeV \cite{CMS-Wasymm} & 24 & +0.0\\ 
  NMC $\mu p$ $F_2$ \cite{NMC} & 123 & -4.5 & LHCb $Z\rightarrow e^+e^-$ \cite{LHCb-Zee} & 9 & +1.3\\
  NMC $\mu d$ $F_2$ \cite{NMC} & 123 & -16.2 & LHCb $W$ asym. $p_T > 20$~GeV \cite{LHCb-WZ} & 10 & -0.3\\
  NMC $\mu n/\mu p$ \cite{NMCn/p} & 148 & +1.7 & CMS $Z\rightarrow e^+e^-$ \cite{CMS-Zee} & 35 & -0.6\\ 
  E665 $\mu p$ $F_2$ \cite{E665} & 53 & +4.4 & ATLAS High-mass Drell-Yan \cite{ATLAShighmass} & 13 & -2.0\\
  E665 $\mu d$ $F_2$ \cite{E665} & 53 & +4.0 & CMS double diff. Drell-Yan \cite{CMS-ddDY} & 132 & -10.0\\
  SLAC $e p$ $F_2$ \cite{SLAC,SLAC1990} & 37 & +0.9 & Tevatron, ATLAS, CMS $\sigma_{t\bar{t}}$ \cite{Tevatron-top}-\cite{CMS-top8} & 17 & -0.3\\
  SLAC $e d$ $F_2$ \cite{SLAC,SLAC1990} & 38 & +1.2 & LHCb 8~TeV $Z\rightarrow ee$ \cite{LHCbZ8} & 17 & -1.7\\
  E866/NuSea $pp$ DY \cite{E866DY} & 184 & +3.4 & LHCb 2015 $W$, $Z$ \cite{LHCbZ7,LHCbWZ8} & 67 & -1.9\\
  E866/NuSea $pd/pp$ DY \cite{E866DYrat} & 15 & -1.1 & CMS 8~TeV $W$ \cite{CMSW8} & 22 & -0.2\\
  NuTeV $\nu N$ $F_2$ \cite{NuTeV} & 53 & -0.7 & ATLAS 7~TeV jets \cite{ATLAS7jets} & 140 & +6.5\\
  CHORUS $\nu N$ $F_2$ \cite{CHORUS} & 42 & -0.3 & CMS 7~TeV $W+c$ \cite{CMS7Wpc} & 10 & +0.7\\
  NuTeV $\nu N$ $x F_3$ \cite{NuTeV} & 42 & -3.0 & ATLAS 7~TeV high prec. $W$, $Z$ \cite{ATLASWZ7f} & 61 & -0.0\\  
  CHORUS $\nu N$ $x F_3$ \cite{CHORUS} & 28 & -0.4 & CMS 7~TeV jets \cite{CMS7jetsfinal} & 158 & +4.1\\
  CCFR $\nu N \rightarrow \mu \mu X$ \cite{Dimuon} & 86 & -1.5 & CMS 8~TeV jets \cite{CMS8jets} & 174 & -1.5\\
  NuTeV $\nu N \rightarrow \mu \mu X$ \cite{Dimuon} & 84 & -9.5 & CMS 2.76~TeV jet \cite{CMS276jets} & 81 & -0.2\\
  D{\O} II $p\bar{p}$ incl. jets \cite{D0jet} & 110 & -0.8 & ATLAS 8~TeV $Z$ $p_T$ \cite{ATLASZpT} & 104 & -40.3\\
  CDF II $p\bar{p}$ incl. jets \cite{CDFjet} & 76 & +0.6 & ATLAS 8~TeV single diff. $t\bar{t}$ \cite{ATLASsdtop} & 25 & -1.2\\
  CDF II $W$ asym. \cite{CDF-Wasym} & 13 & +0.2 & ATLAS 8~TeV single diff. $t\bar{t}$ dilep. \cite{ATLASttbarDilep08_ytt} & 5 & -1.1\\
  D{\O} II $W\rightarrow \nu e$ asym. \cite{D0Wnue} & 12 & -3.9 & CMS 8~TeV double diff. $t\bar{t}$ \cite{CMS8ttDD} & 15 & +0.9\\
  D{\O} II $W \rightarrow \nu \mu$ asym. \cite{D0Wnumu} & 10 & +0.3 & CMS 8~TeV single diff. $t\bar{t}$ \cite{CMSttbar08_ytt} & 9 & -2.5\\
  D{\O} II $Z$ rap. \cite{D0Zrap} & 28 & +0.3 & ATLAS 8~TeV High-mass DY \cite{ATLASHMDY8} & 48 & +3.8\\
  CDF II $Z$ rap. \cite{CDFZrap} & 28 & +1.5 & ATLAS 8~TeV $W$\cite{ATLASW8} & 22 & -3.2\\
  D{\O} $W$ asym. \cite{D0Wasym} & 14 & -1.1 & ATLAS 8~TeV $W+\text{jets}$ \cite{ATLASWjet} & 30 & -1.7\\
  ATLAS $W^+$, $W^-$, $Z$ \cite{ATLASWZ} & 30 & -0.3 & ATLAS 8~TeV double diff. $Z$ \cite{ATLAS8Z3D} & 59 & +22.2\\ \hline
  \multicolumn{4}{|r|}{Total} & 2919 & -63.0 \\ \hline
  \end{tabular}
\end{center}
\vspace{-0.5cm}
\caption{\sf The change in $\chi^2$ (with negative indicating an improvement in the fit quality) when the combined HERA data sets including $F_L$ and heavy flavour data are removed, illustrating the tensions of these data sets with several of the other data sets in the global fit.}
\label{tab:noHERA_delchisqtable}
\end{table}

\begin{figure} [t]
\begin{center}
\includegraphics[scale=0.24, trim = 50 0 0 0 , clip]{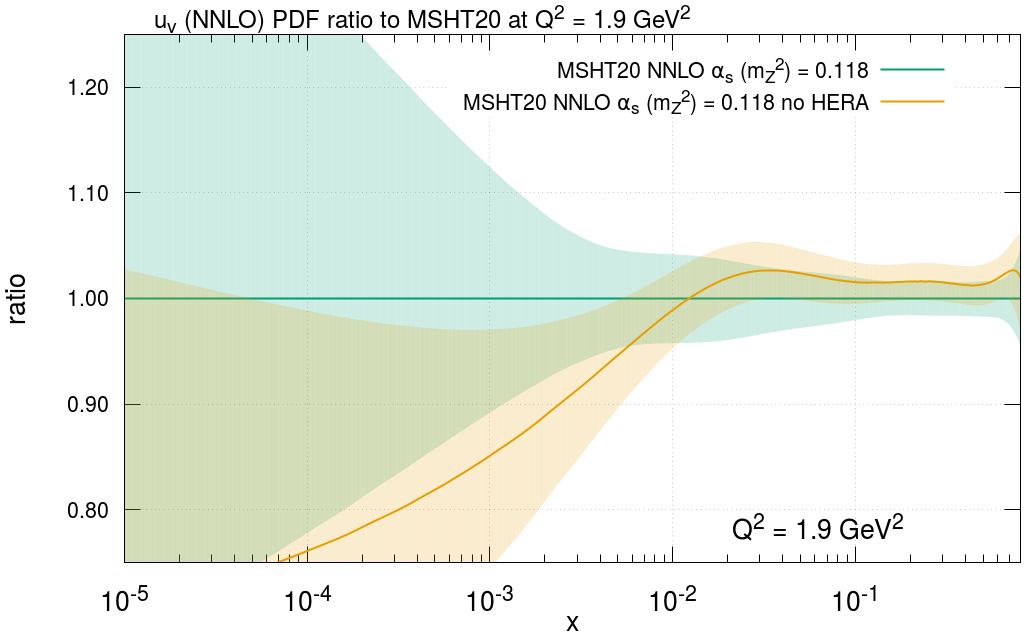}
\includegraphics[scale=0.24, trim = 50 0 0 0 , clip]{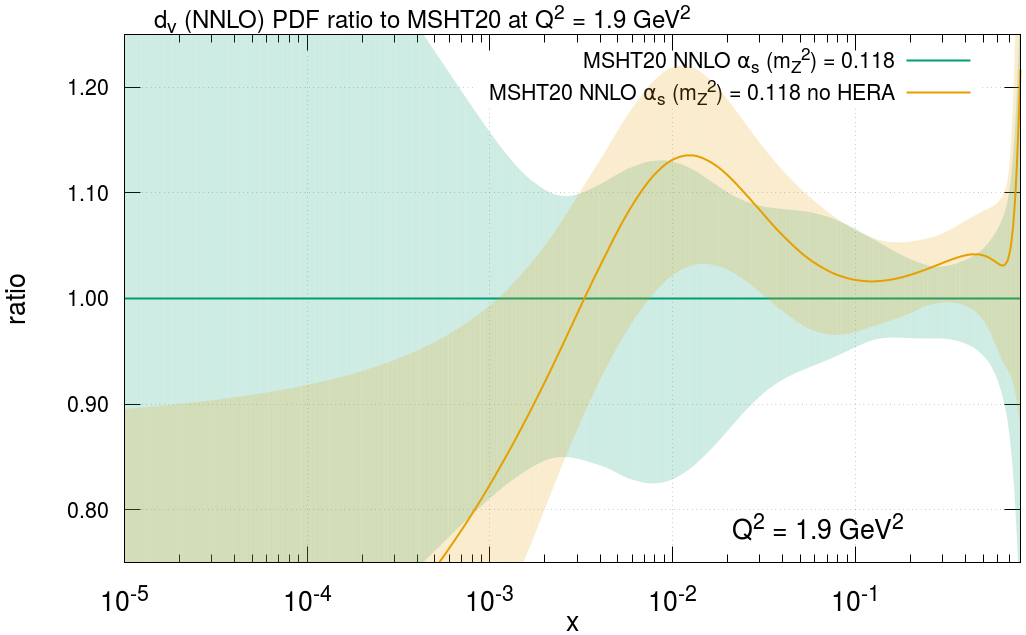}
\caption{\sf (Left) $u_V$ PDF and (right) $d_V$ PDF ratios to the MSHT20 default at $Q^2=1.9~\GeV^2$ at NNLO showing the effect of removing the HERA data from the MSHT20 default global fit.}
\label{MSHT20_NNLO_noHERA_updownvalenceratios}
\end{center}
\end{figure}

\begin{figure} [t]
\begin{center}
\includegraphics[scale=0.24, trim = 50 0 0 0 , clip]{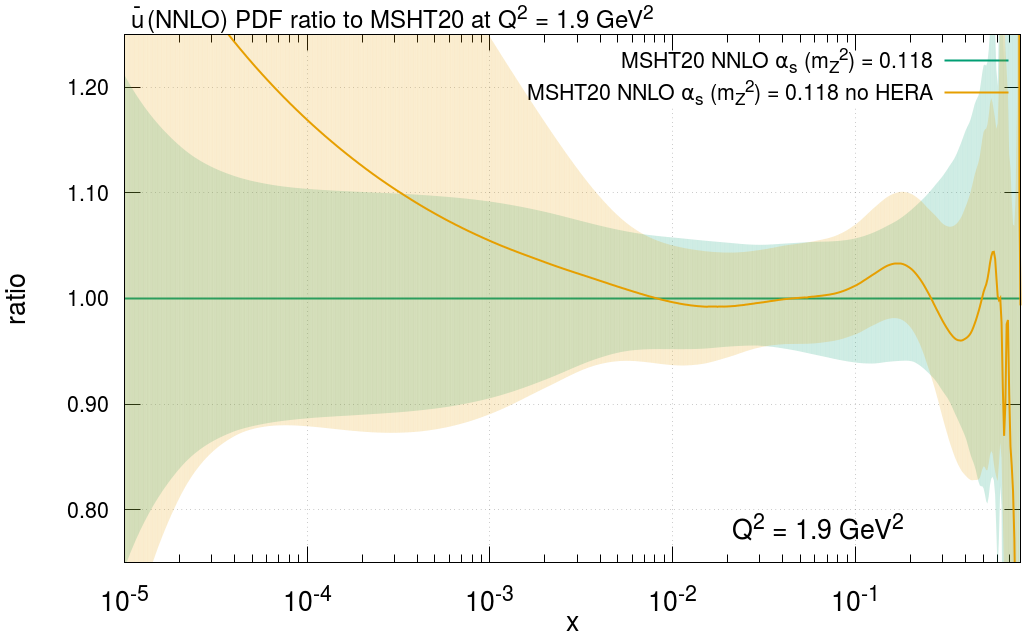}
\includegraphics[scale=0.24, trim = 50 0 0 0 , clip]{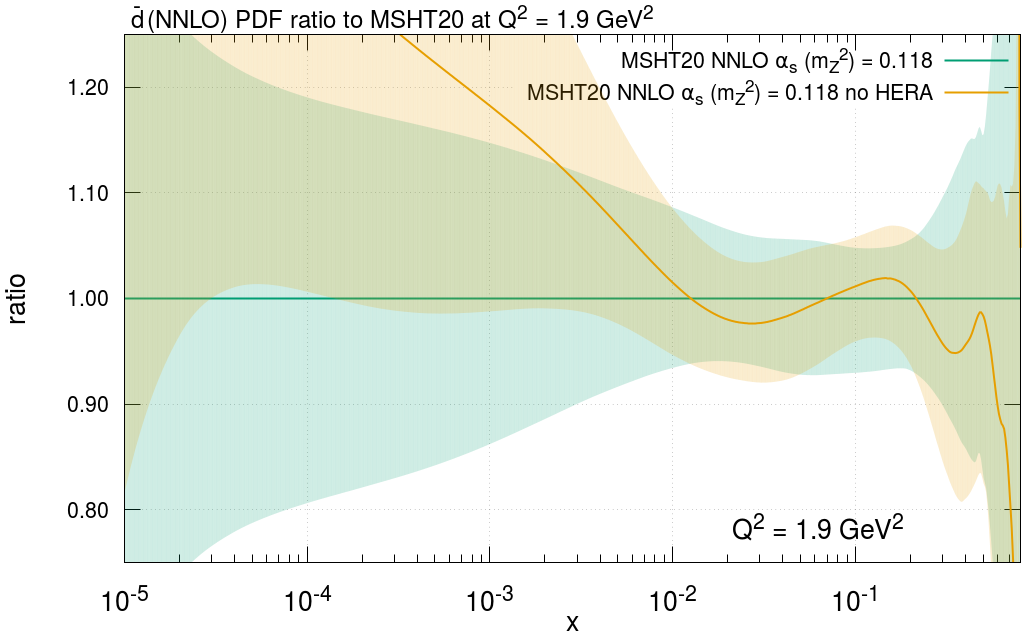}
\caption{\sf (Left) $\bar{u}$ PDF and (right) $\bar{d}$ PDF ratios to the MSHT20 default at $Q^2=1.9~\GeV^2$ at NNLO showing the effect of removing the HERA data from the MSHT20 default global fit.}
\label{MSHT20_NNLO_noHERA_updownantiquarkratios}
\end{center}
\end{figure}

\begin{figure} 
\begin{center}
\includegraphics[scale=0.24, trim = 50 0 0 0 , clip]{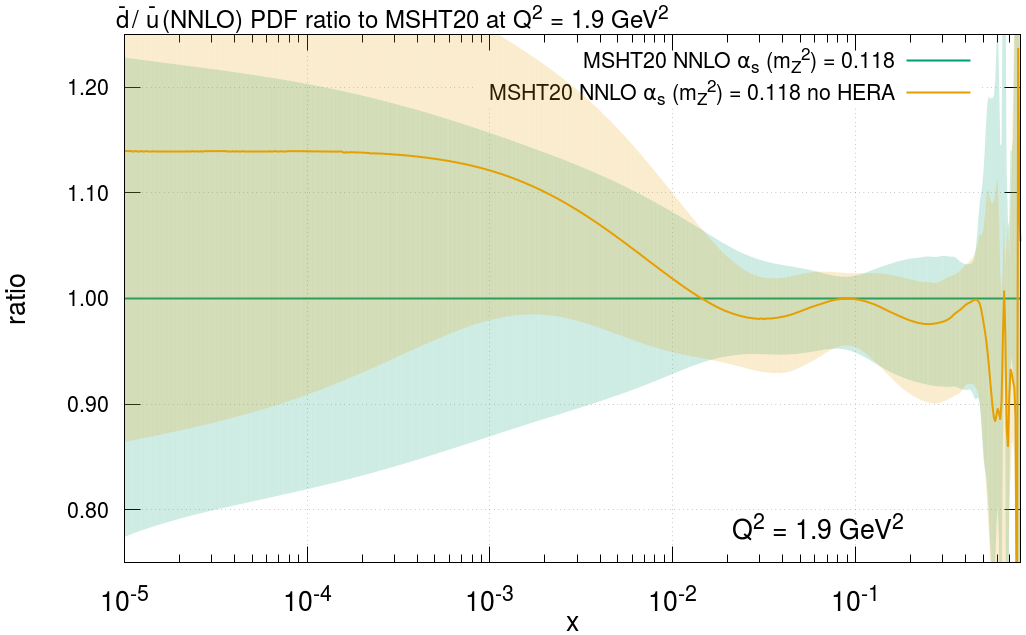}
\includegraphics[scale=0.24, trim = 80 0 0 0 , clip]{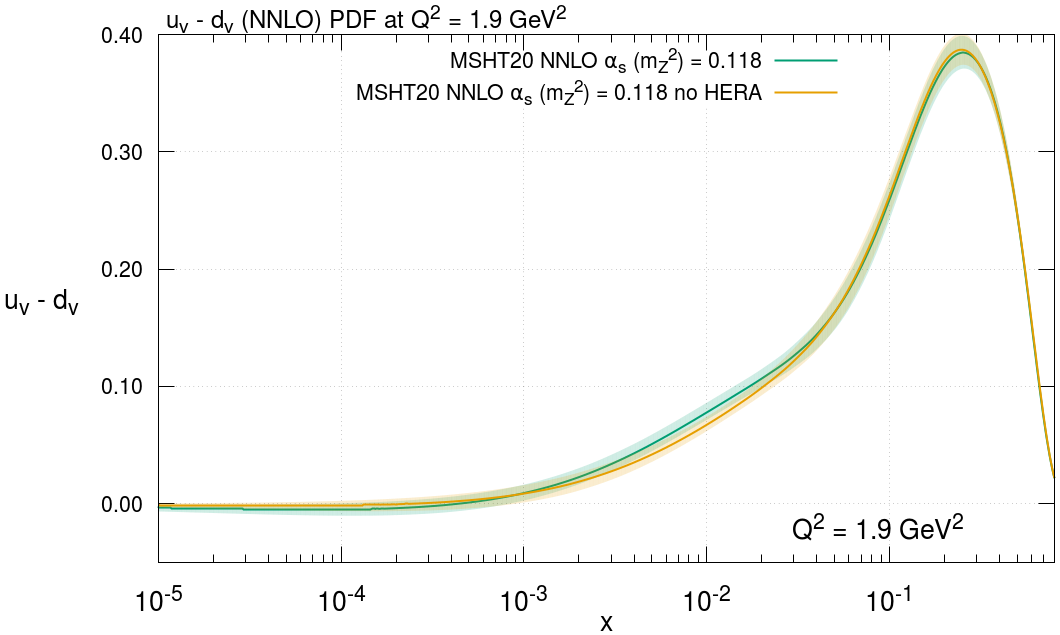}
\caption{\sf (Left) $\bar{d} / \bar{u}$ PDF ratio and (right) $u_V - d_V$ PDF absolute value compared to the MSHT20 default at $Q^2=1.9~\GeV^2$ at NNLO showing the effect of removing the HERA data from the MSHT20 default global fit.}
\label{MSHT20_NNLO_noHERA_dbaroverubaranduvminusdvratios}
\end{center}
\end{figure}

\begin{figure} [t]
\begin{center}
\includegraphics[scale=0.24, trim = 50 0 0 0 , clip]{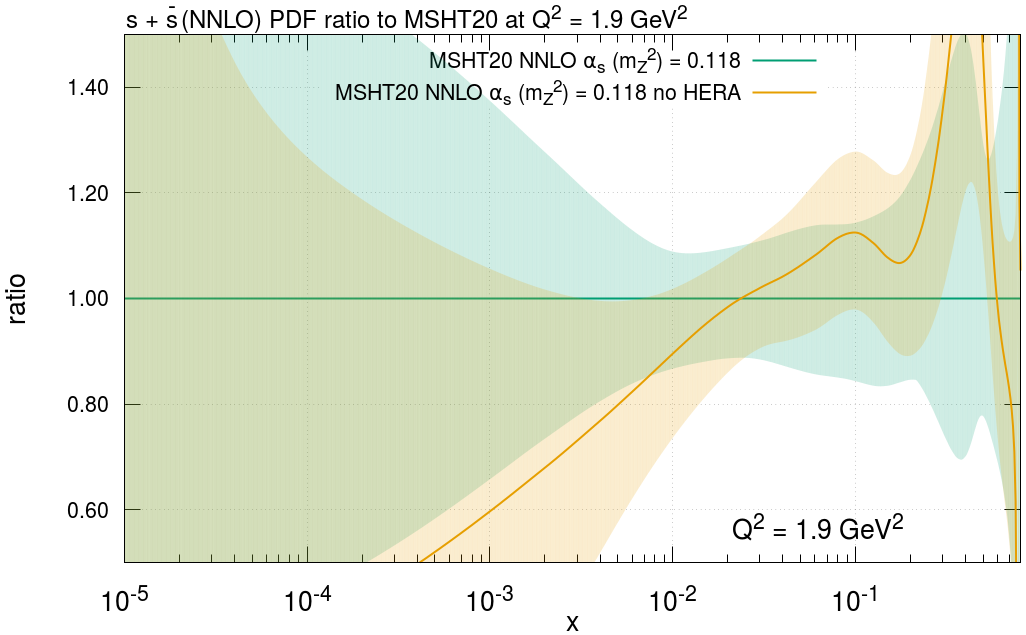}
\includegraphics[scale=0.24, trim = 50 0 0 0 , clip]{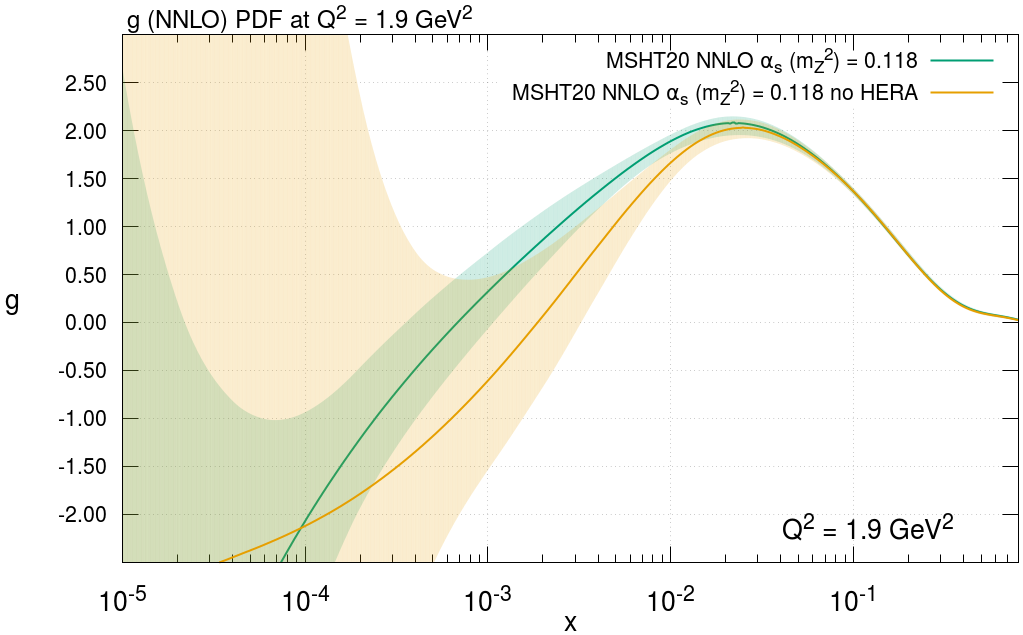}
\caption{\sf (Left) $s + \bar{s}$ PDF ratio and (right) $g$ PDF absolute value compared to the MSHT20 default at $Q^2=1.9~\GeV^2$ at NNLO showing the effect of removing the HERA data from the MSHT20 default global fit.}
\label{MSHT20_NNLO_noHERA_strangenessandgluonratios}
\end{center}
\end{figure}

\begin{figure} [t]
\begin{center}
\includegraphics[scale=0.24, trim = 50 0 0 0 , clip]{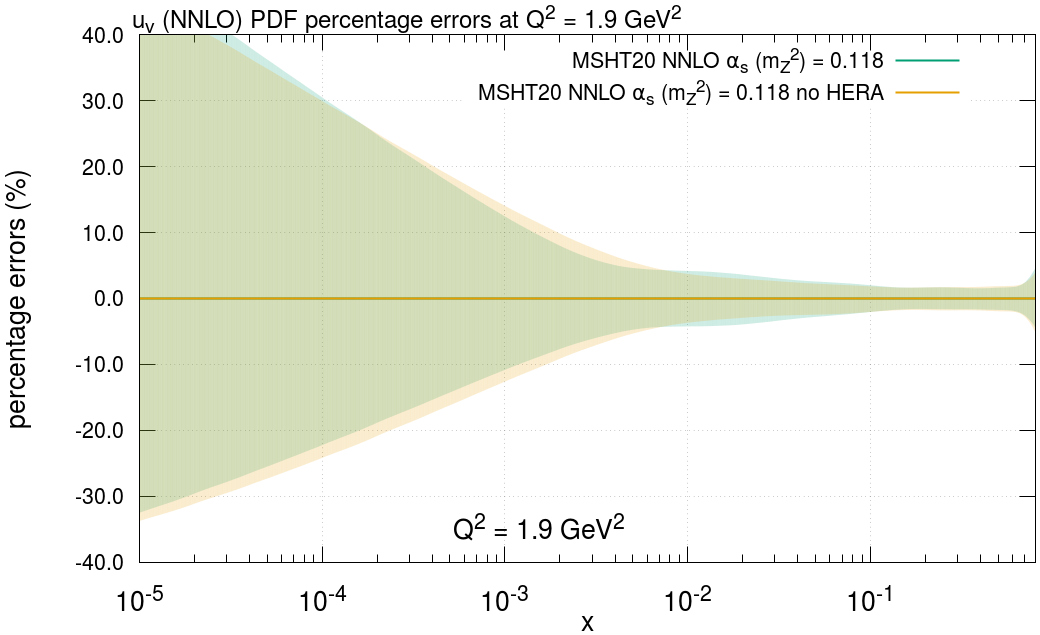}
\includegraphics[scale=0.24, trim = 50 0 0 0 , clip]{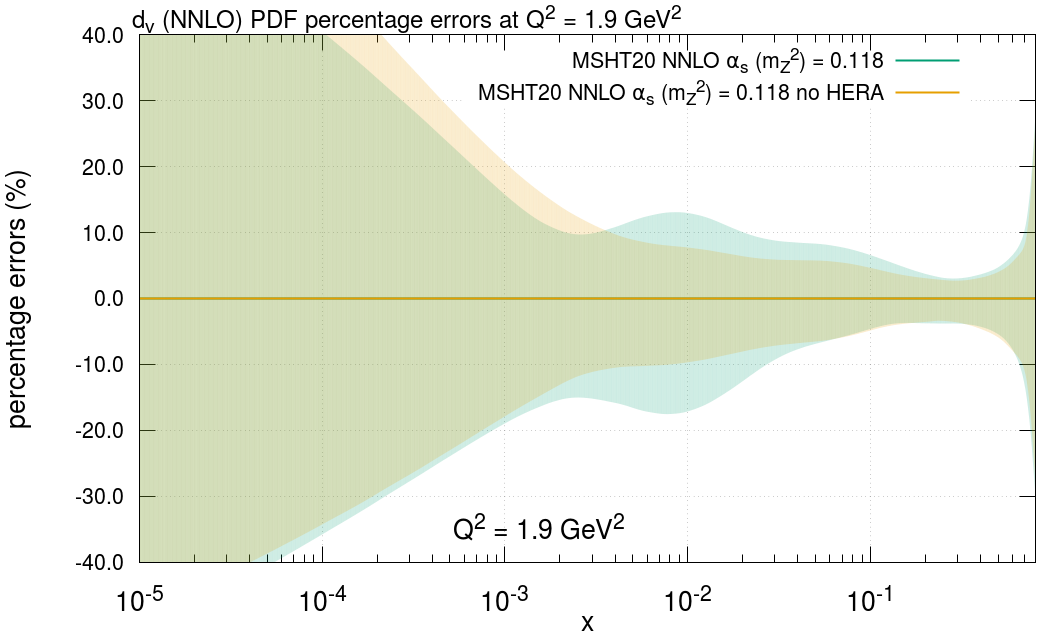}
\caption{\sf (Left) $u_V$ PDF and (right) $d_V$ PDF percentage errors at $Q^2=1.9~\GeV^2$ at NNLO showing the effect of removing the HERA data from the MSHT20 default global fit.}
\label{MSHT20_NNLO_noHERA_updownvalencepercentageerrors}
\end{center}
\end{figure}

\begin{figure} 
\begin{center}
\includegraphics[scale=0.24, trim = 50 0 0 0 , clip]{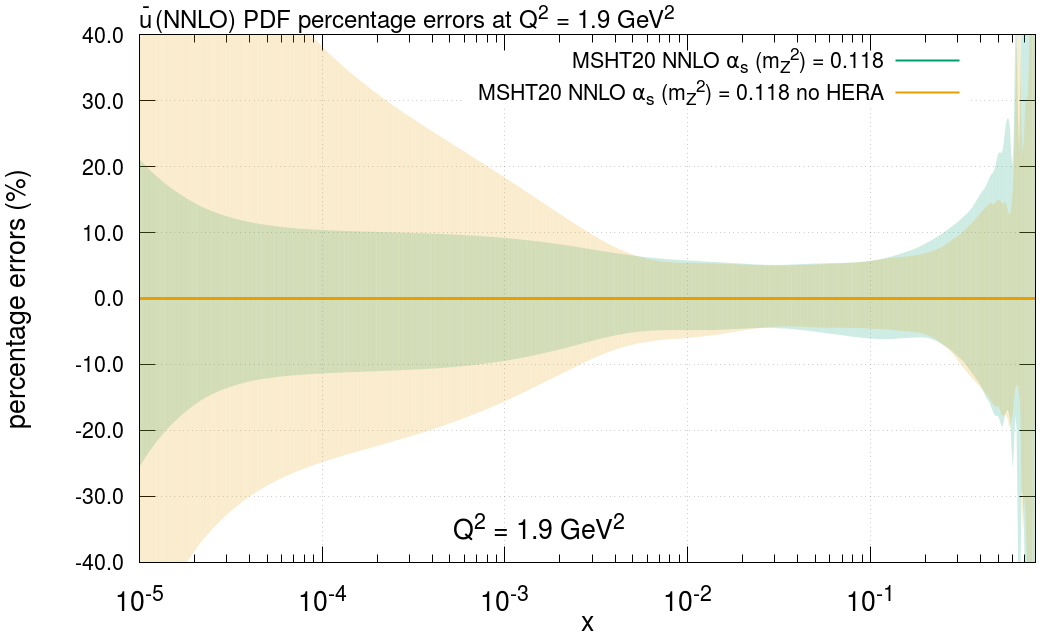}
\includegraphics[scale=0.24, trim = 50 0 0 0 , clip]{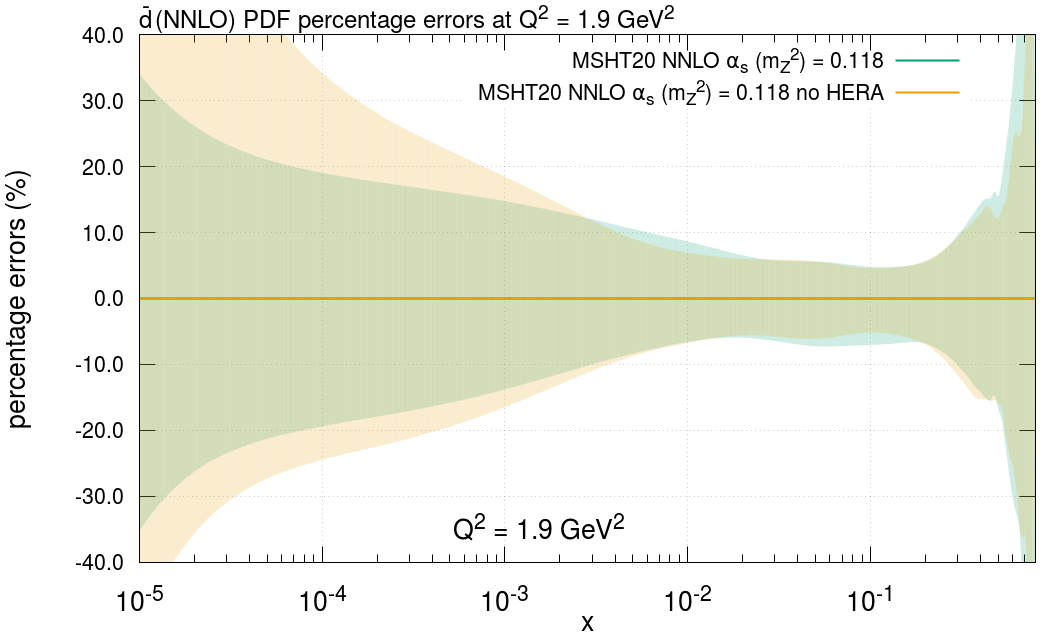}
\caption{\sf (Left) $\bar{u}$ PDF and (right) $\bar{d}$ PDF percentage errors at $Q^2=1.9~\GeV^2$ at NNLO showing the effect of removing the HERA data from the MSHT20 default global fit.}
\label{MSHT20_NNLO_noHERA_updownantiquarkpercentageerrors}
\end{center}
\end{figure}

\begin{figure} [t]
\begin{center}
\includegraphics[scale=0.238, trim = 50 0 0 0 , clip]{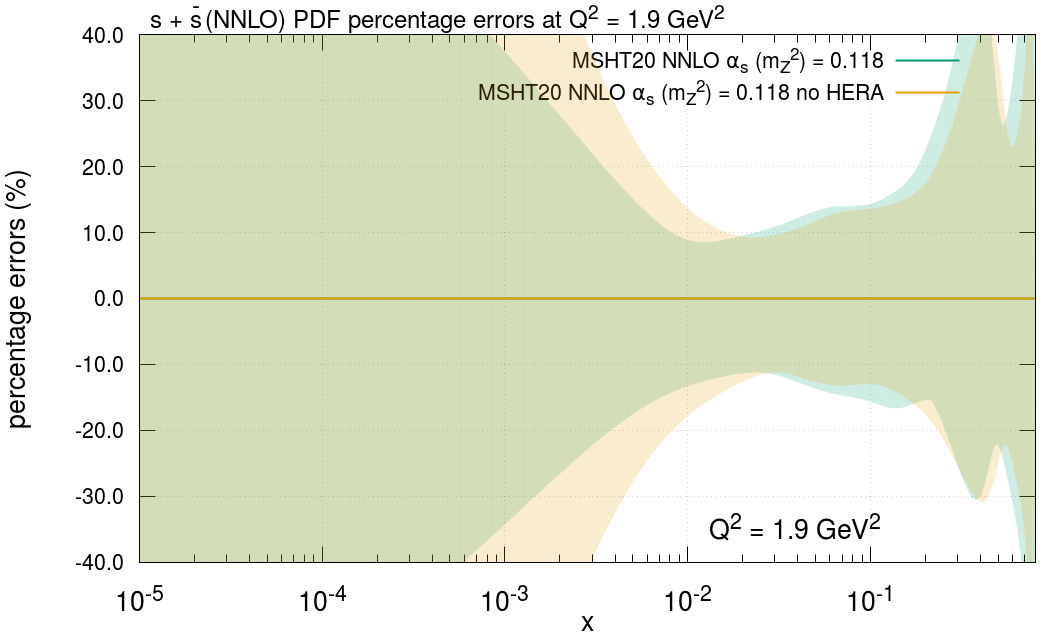}
\includegraphics[scale=0.238, trim = 50 0 0 0 , clip]{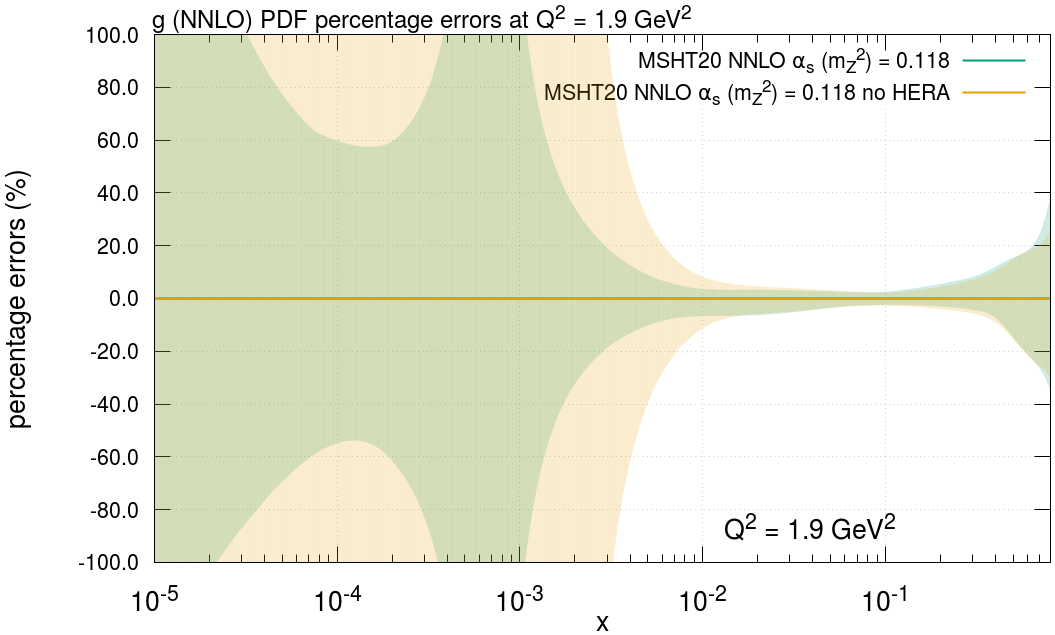}
\caption{\sf (Left) $s + \bar{s}$ PDF and (right) $g$ PDF percentage errors at $Q^2=1.9~\GeV^2$ at NNLO showing the effect of removing the HERA data from the MSHT20 default global fit.}
\label{MSHT20_NNLO_noHERA_strangenessandgluonpercentagerrors}
\end{center}
\end{figure}

The combined HERA data, whilst not included in the MMHT14 fit, are not a completely new addition to the MSHT20 fit, having been added soon after MMHT14 \cite{MMHT2015}. Nonetheless, given that the HERA data, whilst still important, are now playing a reduced role in the PDF global fit as both higher precision and a greater variety of LHC data sets are added, it is worth investigating their impact in more detail. Therefore, we have performed a fit identical to the MSHT20 default fit but with the HERA data (including that on heavy flavour and $F_L$) removed, in order to determine the effects of the HERA data sets within the MSHT20 global fit. This allows us to comment on the effect of the HERA data on the overall central fit and the uncertainties in a more detailed, less simplified manner than simply noting that none of the eigenvector directions were constrained primarily by the HERA data in Section~\ref{PDFuncertainties}. The result is that we observe, as expected, notable changes in the PDF central values and also in the uncertainties, particularly in the low $x$ region. Nonetheless, even with these significant changes, the error bands of both fits overlap in all cases. This illustrates the consistency between the fits even when such a large number of data points are dropped, and hence is a confirmation of the applicability of the dynamical tolerance procedure. In particular, if we were to use a simple $\Delta \chi^2=1$ procedure we would not find such compatibility between the PDFs. In Figs.~\ref{MSHT20_NNLO_noHERA_updownvalenceratios} to \ref{MSHT20_NNLO_noHERA_strangenessandgluonpercentagerrors} we present the PDFs both as ratios to the default MSHT20 fit and their percentage errors, shown at low $Q^2$ in order to remove evolution effects. In Table~\ref{tab:noHERA_delchisqtable} the changes in the fit $\chi^2$ to the data sets in the global fit when HERA data are removed are given.

We begin with the central values, and given that the combined HERA data set is expected to constrain the $u_V$, $d_V$, $\bar{u}$, $\bar{d}$ and $g$ most strongly, we start with these. The ratios of the up and down valence, their antiquarks, the gluon and other relevant PDFs are given in Figs.~\ref{MSHT20_NNLO_noHERA_updownvalenceratios} to \ref{MSHT20_NNLO_noHERA_strangenessandgluonratios}. 
First, examining the ratios of the valence quarks we immediately observe that both the up and down valence are slightly higher than MSHT20 at high $x$ once the HERA data sets are removed. Consequently there are also changes at low $x$, where there are much weaker data constraints, with the valence quarks significantly reduced relative to MSHT20 in order to maintain the number sum rule. The down valence also shows notable shape changes, of roughly the same form, i.e. higher at high $x$ and lower at low $x$, albeit within errors across the majority of the $x$ range. The antiquarks, on the other hand, show a different trend, being largely unchanged compared to the global fit uncertainty at higher $x$, but much increased below $x \sim 10^{-2}$. The ratio $\bar{d}/\bar{u}$ is quite stable at higher $x$, but increased at low $x$ (again consistent with the default MSHT20 within error bands). 
The valence quark difference
$u_V - d_V$ is also stable at high $x$ but changes significantly below $x=0.01$, being first reduced, and then becoming larger than MSHT20 below $x \sim 10^{-3}$. This is best seen in a plot of the PDFs directly, rather than their ratio, as shown in Fig.~\ref{MSHT20_NNLO_noHERA_dbaroverubaranduvminusdvratios} (right). There are also significant changes in the total strangeness $s+\bar{s}$, with a rather different shape present once the HERA data are removed: the total strangeness is raised at high $x$ and lowered at low $x$, below $x=0.01$, although again largely within uncertainties. The gluon comparison in Fig.~\ref{MSHT20_NNLO_noHERA_strangenessandgluonratios} (right) is perhaps the most illuminating, remaining very close to the MSHT20 default fit in the $ 10^{-2} \lesssim x \lesssim 10^{-1}$ region (albeit a little below at very high $x$), but dropping significantly relative to MSHT20 below $10^{-2}$. As a result it passes through 0 at higher $x$ than MSHT20 and has a different shape, bending up relative to MSHT20 at very low $x$. Overall, there is a net reduction in the momentum in the gluon of $1.3\%$ due to this deficit at very low $x$, and compensating increases in the momentum of the both the valence and sea quarks. This indicates a tension in the momentum sum rule, with the HERA data preferring more low to intermediate $x$ PDFs whereas other data sets (particularly the fixed target data sets) favour more high $x$ PDFs.

In order to verify this broad interpretation, the changes in $\chi^2$ of the non-HERA data sets once they are removed, and a refit is performed, are presented in Table~\ref{tab:noHERA_delchisqtable}. As expected there are many changes in the individual fit qualities, but perhaps the most relevant are those in the fixed target data sets, with the BCDMS, NMC and E665 showing significant changes of $\Delta\chi^2 = -7.8, -19.0, 8.4$ respectively. The NMC is particularly noteworthy here as it is known that there is a slight tension between the NMC data below about $x=0.05$ and the HERA data in the same $x$ range, with the former being undershot if the HERA data are fit. The HERA data in this region constrain the quarks to be smaller than is favoured by the NMC (once they are evolved between their scales). Consequently, once the combined HERA data set is removed both the valence quarks and the overall light sea in the $x>0.01$ region are allowed to increase, and for the NMC data a significant improvement in $\chi^2$ is observed. These changes are also seen in the fixed-target data set normalisations, which are all increased by about 1.5\% once the HERA data are removed. As a result of these changes, the valence quarks are also forced to reduce at low $x$ by the number sum rule, as observed in Fig.~\ref{MSHT20_NNLO_noHERA_updownvalenceratios}. 

Any differences in the valence quarks momentum distribution (by having more of their PDFs at high $x$) must then be reflected in the gluon by momentum conservation. Therefore the gluon falls at intermediate to very low $x$ relative to MSHT20, whilst remaining the same at intermediate to high $x$ to continue to match the other data sets in the fit. The total strangeness is reduced significantly at low $x$, where there is no real direct constraint. However, even a little below $x=0.01$ the NMC data also constrain the up and down antiquarks through the structure function; whereas the valence quarks are significant or dominant in the $10^{-2} \lesssim x \lesssim 10^{-1}$ interval, the antiquarks dominate in the $10^{-3} \lesssim x \lesssim 10^{-2}$ region. Similarly to the valence quarks at higher $x$, the up and down antiquarks are therefore raised to fit the fixed target data better once the HERA data are removed, but now in this intermediate $x$ region. This is then extrapolated down to low $x$ where the large increases in the antiquarks, albeit with a very large uncertainty, are observed. Finally, whilst the reduction in the total strangeness at low $x$ reflects the reduced gluon and a reduction in the momentum sum rule, the increase at high $x$  results in an increase in the momentum of the sea quarks. Once the gluon is allowed to have less momentum, the sea is also able to increase, enabling the strangeness at high $x$ to be increased and thereby allows the ATLAS 7~TeV $W$, $Z$ high precision data to be fit equally well, even with the $\bar{d}$ reduced in the region of strangeness non--suppression. In order to compensate for this increased strangeness whilst still fitting the dimuon data, the dimuon branching ratio is then reduced by approximately 8\%. This allows the $\chi^2$ of the NuTeV dimuon data to improve by 9.5 points. Looking more closely at the $\Delta\chi^2$ in Table~\ref{tab:noHERA_delchisqtable} we can note that the CMS 7~TeV $W+c$ worsens slightly ($\Delta\chi^2=0.7$ for 10 points) as you would expect from increased strangeness.

Other data sets which change notably in their fit qualities upon removal of the HERA data are the CMS double differential Drell-Yan and ATLAS 8~TeV double differential $Z$ data, which are both sensitive to the changes in the gluon at lower $x$ through their low mass bins, and to changes in the sea quarks. The former of these improves ($\Delta\chi^2=-10.0$) upon removal of the HERA data whilst the latter worsens ($\Delta\chi^2=+22.2$), showing that the CMS double differential Drell-Yan is in tension with the HERA data with respect to the gluon whilst the ATLAS 8~TeV double differential data are in agreement with it. Data sets which are sensitive to the gluon at high $x$ are also altered, with the ATLAS and CMS 7~TeV jets both worsening upon removal of the HERA data. On the other hand, the ATLAS 8~TeV $Z$ $p_T$ data set improves greatly when the HERA data are removed (the same effect the other way around is seen in Table~\ref{tab:BCDMSD0WasymATLASZpt_delchisqtable} in the next section). The ATLAS $Z$ $p_T$ data set is in significant tension with many data sets in the fit, as discussed in Section~\ref{tensions}, and therefore removing a large number of points in the global fit allows it to be fit better. Finally, we note that, as one might expect, precise Drell-Yan data sets such as the ATLAS 8~TeV $W^{\pm}$ and ATLAS 8~TeV High-mass Drell-Yan also change in $\chi^2$ as a result of the changes in the up and down quarks and antiquarks.

Given these changes in the central values of the PDFs, it is interesting to examine the effects of removing the large HERA data set on the error bands. We saw in Section~\ref{PDFuncertainties} that the HERA data sets are not the dominant constraint on any one direction of the eigenvectors for the uncertainties. This however does not preclude them having significant effects on the error bands, as they may still be one of the sub-dominant constraints. In practice, by altering the central fit the variation of the $\chi^2$ of all the data sets around this new central fit is changed, and so the constraining data sets on each eigenvector (as well as the composition of the eigenvectors themselves) are modified. In Figs.~\ref{MSHT20_NNLO_noHERA_updownvalencepercentageerrors} to \ref{MSHT20_NNLO_noHERA_strangenessandgluonpercentagerrors} we compare the uncertainties on the PDFs relative to the MSHT20 default case. It is clear that on the whole the PDF uncertainties increase once the HERA data are removed, as expected, however the differences outside of the low $x$ region are indeed small, demonstrating they have less effect on the overall PDF uncertainties themselves, at least in regions most relevant for phenomenology. The up and down valence show only very small increases in their percentage uncertainties below $x \approx 10^{-3}$, despite significant changes in their central values at $x$ values larger than this. Similar behaviour at intermediate to high $x$ is seen in the up and down antiquarks, however in these cases the error bands at low $x$ are now substantially increased, being more than double in size by $x \sim 10^{-5}$. This reflects the larger uncertainty due to the the change in central values of these PDFs, and illustrates the lack of constraint 
on very small-$x$ quarks and antiquarks in the absence of HERA data. There is a similar increase of uncertainty on the gluon at small-$x$, with the evolution of the HERA structure functions and cross sections, driven by the gluon, being the overwhelming constraint on the very small-$x$ gluon distribution. The PDFs remain largely consistent with MSHT20 within these enlarged error bands over all $x$. The absence of the HERA data has very little impact on the error bands of the ratio of the down and up antiquarks, or on the difference in the valence quarks $u_V -d_V$  until below $x \approx 10^{-3}$, where we see some increase, and we do not show these plots. The total strangeness has a similar increase in uncertainties to other sea quarks at low $x$. These increased uncertainties at low $x$ in the light sea all result from the same lack of constraint previously having been supplied directly by the HERA data at low $x$ and $Q^2$.

\begin{figure} [t]
\begin{center}
\includegraphics[scale=0.24, trim = 50 0 0 0 , clip]{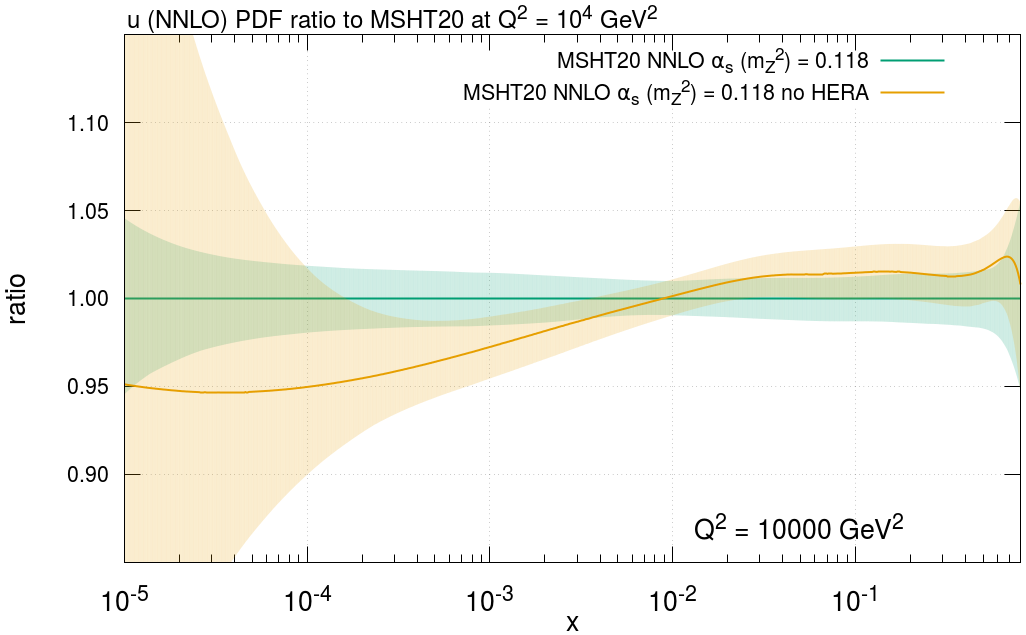}
\includegraphics[scale=0.24, trim = 50 0 0 0 , clip]{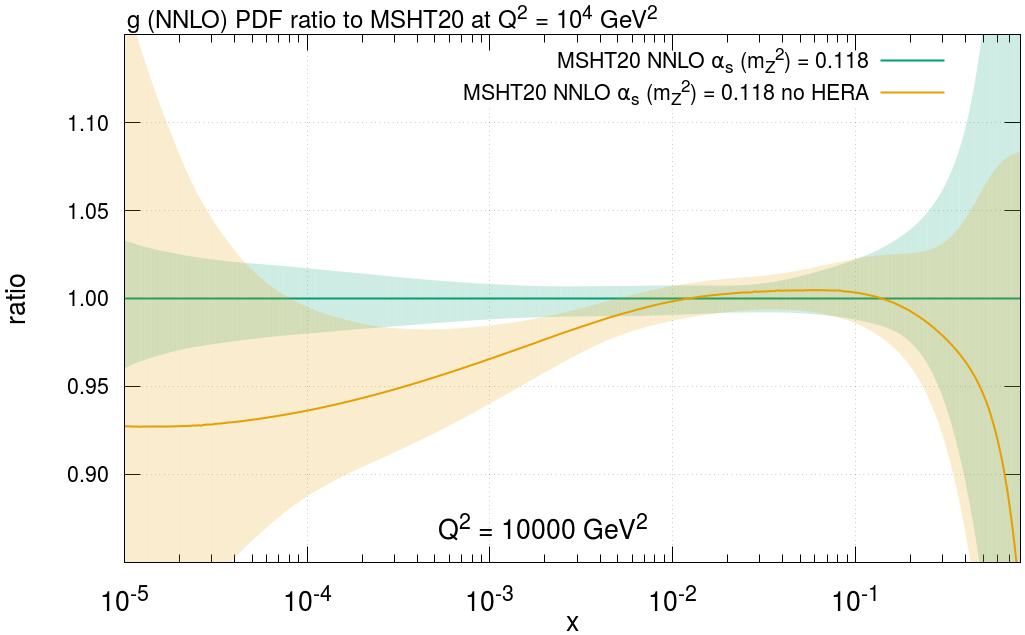}
\caption{\sf (Left) $u$ PDF ratio and (right) $g$ PDF ratio at $Q^2=10^4~\GeV^2$ at NNLO showing the effect of removing the HERA data from the MSHT20 default global fit.}
\label{MSHT20_NNLO_noHERA_highQupandgluon}
\end{center}
\end{figure}

In order to illustrate the effect of removal of the HERA data in a slightly different manner, in 
Fig.~\ref{MSHT20_NNLO_noHERA_highQupandgluon} we also display the the ratio of the up quark (left) and gluon (right) at $Q^2=10^4~\GeV^2$, a scale more relevant for LHC physics. The up quark at high $x$ displays the same 
change already evident in the $u_V$ at lower $Q^2$, i.e. it is a little higher in normalisation, with largely
unchanged uncertainty. Also, as at low $Q^2$, the gluon distributions above $x=0.01$ in MSHT20 and in the 
fit without HERA data are very similar, both in terms of central values and uncertainty. Below $x=0.01$ we see
the gluon in the fit without HERA data falling significantly below that of MSHT20, and the relative size of 
the uncertainties increases until at $x=10^{-5}$ it has become more than 5 times bigger. Clearly this 
alternative gluon distribution would give a rather different prediction for Higgs boson production, 
particularly as a function of rapidity, with much bigger uncertainties. Below $x=0.01$ the relative shape of 
the ratio of the up quark from the fit without HERA data to that in MSHT20 follows that of the gluon rather
closely, as does the increase in the uncertainty. This is simply a consequence of the fact that at very small 
$x$ the evolution of quarks, and hence their values at high $Q^2$, is driven very largely by the gluon, 
and the plots would look similar in this region for all quarks and antiquarks. 

At this stage, now we know the HERA data have a direct impact on the uncertainties of some of the PDFs at small $x$, we can examine the eigenvectors of the full fit in more detail to look for signs of their sensitivity to the HERA data. This goes beyond the simplified analysis of which data set most constrains each eigenvector in Section~\ref{PDFuncertainties} and Table~\ref{tab:NNLOevecdatasetstable}. Specifically, given we know which eigenvectors contribute to which PDFs in which regions of $x$ from Table~\ref{tab:NNLOevecparamstable}, we can identify which eigenvectors are relevant to the low $x$ uncertainties and analyse their dependence on the HERA data. However, as discussed previously, with the HERA data significantly altering the central values of some of the PDFs, the eigenvectors alter once the HERA data sets are removed, and so this analysis still has some limitations. In any case, from Table~\ref{tab:NNLOevecparamstable} we can see that the eigenvector with the largest contribution to the gluon uncertainties at low $x$ in our default MSHT20 fit is eigenvector 21 (which is sensible as this eigenvector is largely made up of the low $x$ gluon power $\delta_g$). Whilst this eigenvector is constrained by the ATLAS 8~TeV $Z$ $p_T$ and ATLAS 8~TeV single differential $t\bar{t}$ dilepton data sets in each direction, examining the changes in $\chi^2$ as we move along the eigenvector in each direction reveals that the HERA data (specifically the HERA $e^+ p $ NC $ 820~\GeV$ data set) is the next most constraining data set in the `+' direction. This illustrates the sensitivity of the HERA data set to the low $x$ gluon, and hence its pull on the central fit and effect in reducing the uncertainties. It is also interesting to note that this eigenvector demonstrates how a direction previously most constrained by the HERA data is now most constrained by new LHC data. In the `-' direction along this eigenvector however the HERA data are less sensitive, with many of the new LHC data sets providing more stringent constraints. Performing a similar analysis for the up and down antiquarks or total strangeness, which also have notably increased error bands at low $x$ in the absence of the HERA data, is more difficult, as in the case of the antiquarks they are not one of the basis PDFs. For the strangeness several different eigenvectors at low $x$ contribute.

We may also use such a method to  understand why the HERA data  have an effect on the central values of the valence quarks at low $x$, without affecting their uncertainties. For the up valence quark it is eigenvector 28 which is most relevant at low $x$. Studying its variations in $\chi^2$ as we move from the central fit along either eigenvector direction indicates that the HERA data are indeed sensitive to moving away from the central PDFs, particularly in the `+' direction. However, whilst the HERA data therefore favour moving in one direction away from the central fit here, several of the other data sets in the fit, including the ATLAS 8~TeV $Z$ $p_T$, ATLAS 8~TeV $W^{\pm}$ and CMS 8~TeV $W$ data sets, all constrain the uncertainties in both directions, preventing the uncertainties from showing any impact of the effect of the HERA data. Again, in the absence of these newer LHC data sets the HERA data would be one of the most constraining data sets on this eigenvector in the `+' direction. A similar picture also emerges for the down valence PDF and eigenvector 22, although it is less clear-cut, with more data sets constraining the uncertainties and the effects of fixed target data sets also relevant, whilst eigenvector 20 also contributes to the down valence at low $x$ too. It is at least interesting to note that the latter is constrained by the NMC $d$ data in the `-' direction, indicating the effect of the fixed target data sets on the PDFs in this region and reinforcing the suggestion that it is these which drive some of the PDF changes in the fit with the HERA data sets removed.

Finally, as we suggested that much of the improvement in the global fit quality when the HERA data are removed is due to extra momentum that can be given to PDFs when less momentum is required at small $x$ (particularly in the gluon), we also investigate a NNLO fit where the total momentum at input is not required to be exactly 1. On releasing this constraint the global fit quality improves by about 15 units and the input momentum is 1.011. Hence, given our typical tolerance value we can conclude that the fit is consistent with the momentum sum rule being preserved at the one sigma level, with some small preference for a slightly larger value of the total momentum.
The data sets which improve most in $\chi^2$ are similar to those
when the HERA data are omitted, i.e. BCDMS and NMC fixed-target DIS data and ATLAS 8~TeV $Z~p_T$ data improve most, followed by ATLAS 7~TeV $W$, $Z$ data. There is an increase in momentum mainly in the gluon 
distribution, and rather less so in the quarks. Hence, a general increase in the gluon, i.e. some increase at high $x$ without a compensating drop at low $x$ is able to resolve some of the tension between 
fixed target DIS data and ATLAS 8~TeV $Z$ $p_T$ data without a very significant change in quark and antiquark PDFs. On this basis we conclude that there is generally consistency with the conclusions drawn above, and that some of the tension between HERA data and other sets can be alleviated by artificially increased input momentum, which can translate into a modification of quark and antiquark evolution resulting from a change in the gluon or to an increase in gluon initiated cross sections. However, clearly much of the tension remains. This may be due to further theoretical corrections beyond NNLO which 
are not well mimicked by an increase in momentum, due to genuine data tensions, or most likely some combination of both.   

\subsection{Further tensions between data sets} \label{tensions}

In any global fit with the breadth and detail of the MSHT20 PDFs there will be tensions between data sets. In addition to those already discussed, such as that between the E866 Drell-Yan ratio data and the high precision ATLAS 7~TeV $W$, $Z$ data in Section~\ref{newparameterisationeffects}, or the small tension between some of the jet data sets in Section~\ref{LHCjets}, there are a number of further, sometimes larger tensions. We comment on the most significant of these in this section.

\begin{figure} 
\begin{center}
\includegraphics[scale=0.24, trim = 50 0 0 0 , clip]{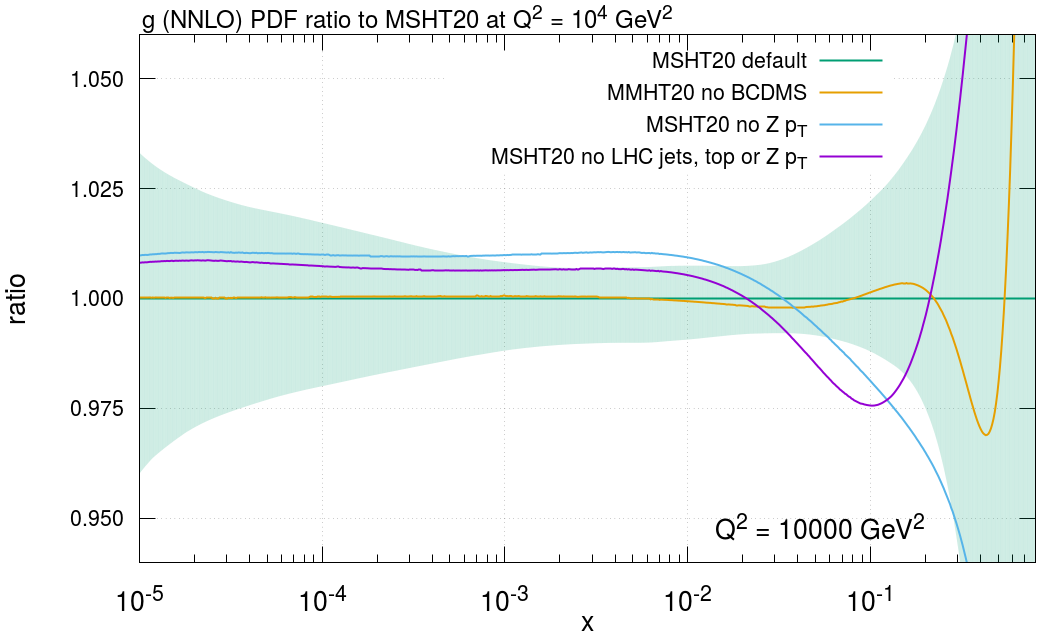}
\includegraphics[scale=0.24, trim = 50 0 0 0 , clip]{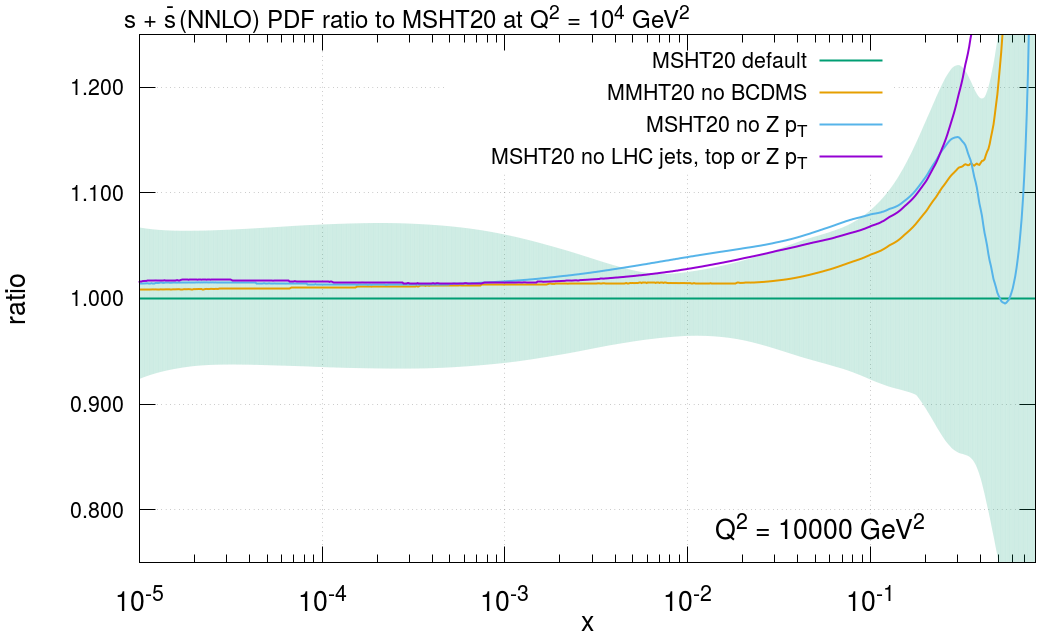}
\caption{\sf The MSHT20 (left) gluon and (right) $s + \bar{s}$  PDFs at NNLO and $Q^2=10^4~\GeV^2$, compared to the same fit upon removal of the BCDMS data, removal of the ATLAS 8~TeV $Z$ $p_T$  data and removal of the LHC jets, top and $Z$ $p_T$ data.}
\label{gluon_ssbar_q210000_NNLO_noBCDMSZptLHCjetstopZpt}
\end{center}
\end{figure}

One of the most prevalent tensions within the global fit is that between BCDMS, D{\O} $W$ asymmetry and ATLAS 8~TeV $Z$ $p_T$ data and other data sets in the fit, in particular with the ATLAS Drell-Yan data sets. In order to demonstrate and investigate this, the changes in the fit quality of many of the individual data sets in the fit, and the overall fit quality, when the former three data sets are removed, are given in Table~\ref{tab:BCDMSD0WasymATLASZpt_delchisqtable}. In Fig~\ref{gluon_ssbar_q210000_NNLO_noBCDMSZptLHCjetstopZpt} we then show the impact their removal has on the gluon and strangeness, respectively.

We first consider the BCDMS data, which comprise both the proton and deuteron data \cite{BCDMS}. The first column of Table~\ref{tab:BCDMSD0WasymATLASZpt_delchisqtable} presents the effect of removing just this pair of data sets. Upon its removal the overall $\chi^2$ of the other 60 data sets improves significantly by 53 points, demonstrating a very clear tension between the BCDMS data and other data sets in the global fit. In particular, analysing the changes in fit quality of the individual data sets, the three NMC data sets (proton, deuteron and $d/p$ ratio) all improve by several points in $\chi^2$ with the ratio improving by 7 points. Similarly, the SLAC $p$ and $d$ structure function data improve by a total of nearly 10 points. Therefore there is a notable tension between the BCDMS data and other structure function data sets. Beyond this, the E866 Drell-Yan data improve by 11 points and the combined HERA data by approximately $\Delta \chi^2 = 20$, though in the latter case this absolute improvement is enlarged by the high number of points the HERA data includes. Other data sets also show significant sensitivity to the presence or absence of the BCDMS data, such as the D{\O} II $W \rightarrow \nu \mu$ asymmetry data and the LHCb $Z\rightarrow e^+e^-$ \cite{LHCb-Zee}, which both improve by more than 2 points upon removal of the BCDMS data despite only having 10 and 9 data points respectively. Whilst this analysis is done at fixed $\alpha_S(M_Z^2) = 0.118$, some of the tensions of the BCDMS data with other data sets may be related in essence to the differences in pull on $\alpha_S$, which can be compensated for elsewhere, such as in the gluon due to effects on the DGLAP evolution. In particular, the BCDMS data prefer a low value of $\alpha_S(M_Z^2)$, but when this is fixed at $\alpha_S(M_Z^2)=0.118$, it instead favours a generally larger high $x$ gluon than the global fit prefers, in order to slow the fall of the structure function with $Q^2$ by increasing the positive contribution from gluon to quark-antiquark splitting. This is seen clearly in in Fig.~\ref{gluon_ssbar_q210000_NNLO_noBCDMSZptLHCjetstopZpt} (left). Hence, the BCDMS data may be in tension with other data sets, which favour a smaller high $x$ gluon and/or, in some cases, larger values of the strong coupling. The pulls on the strong couplings of different data sets may therefore indicate potential tensions with the BCDMS data if they prefer a larger value of the strong coupling. The pulls on $\alpha_S(M_Z^2)$ will indeed be analysed in a future publication \cite{Cridge:2020}, but we already see some consequences of it even in the
fixed $\alpha_S(M_Z^2)$ used in this article. 

The ATLAS 7~TeV high precision Drell-Yan $W$, $Z$ data set and the 8~TeV $W^{\pm}$ data set also both display tensions with the BCDMS data, with each improving by 4.7 and 1.5 points in $\chi^2$ respectively, again they both favour larger values of $\alpha_S(M_Z^2)$ in contrast to the lower values favoured by the BCDMS data. However, the ATLAS 8~TeV double differential $Z$ data set also favours larger values of $\alpha_S$ but shows very little change in $\chi^2$ upon the removal of the BCDMS data. This demonstrates that the tension can clearly be more complicated than this and may be due to, for example, the shape of the valence quark distributions preferred by BCDMS data, and the separation into up and down quarks. Despite these numerous tensions, other data sets nonetheless favour the inclusion of the BCDMS data in their fit quality, with some of the jet data sets (although not all) such as the D{\O} II jets, ATLAS 7~TeV jets and the CMS 7~TeV jet data sets worsening a little by $\Delta \chi^2 = 1.4, 1.2, 4.1$ respectively when the BCDMS data are removed. On the other hand the CMS 8~TeV jets shows very little impact of the BCDMS data removal on its $\chi^2$. Given the 7~TeV jet data (from both ATLAS and CMS) favour lower values of the strong coupling (like the BCDMS data) whilst the 8~TeV data do not, again perhaps this has an impact. The exact reasons for this agreement with the LHC jet data is unclear as the latter favours a lower high $x$ gluon (see for example Fig.~\ref{gluon_q210000_NNLO_novariousjets} where the effect of the removal of the LHC jet data is to raise the gluon here) in contrast to the BCDMS data. A further data set which worsens noticeably upon the removal of the BCDMS data sets is the LHCb 2015 $W$, $Z$ data set which deteriorates by 4.9 points for 67 points, implying compatibility with the valence quarks preferred by BCDMS data. 

The D{\O} $W$ asymmetry data \cite{D0Wasym} have already been analysed in section~\ref{D0Wasymeffects} and shown to have significant effects on the global fit, particularly on its uncertainties, despite being a small data set. Therefore perhaps it is unsurprising that it also has both tensions and consonances with several of the other data sets in the fit. Nonetheless its global effect across all the remaining data sets is small with only a marginal improvement of less than 5 points across the other global fit data sets, perhaps as a result of its small number of points. Its effects on individual data sets are however larger, and in particular it affects several of the other electroweak data sets. As pointed out in section~\ref{D0Wasymeffects}, replacing this data set with the same data regarded as an electron asymmetry improved the $\chi^2$ of the remaining data sets by $\Delta \chi^2 = 8.8$, with the older D{\O} electron asymmetry data in the fit showing a notable improvement in $\chi^2$ of 2.6. Therefore, as one might anticipate, once the D{\O} $W$ asymmetry data are removed completely the older D{\O} II electron asymmetry data \cite{D0Wnue} improve substantially by $\Delta \chi^2 = 5.8$ for just 12 points. On the other hand, the similar D{\O} II muon asymmetry data set \cite{D0Wnumu} worsens in $\chi^2$ by 3.1 so some of this improvement is lost in the overall global fit. At the same time, the LHCb 2015 $W$, $Z$ data, ATLAS 7~TeV high precision $W$, $Z$ data set and ATLAS 8~TeV $W^{\pm}$ data also improve by 2.8, 1.0 and 1.6, respectively, when this data set is taken out, but the ATLAS 8~TeV double differential $Z$ data worsens by 1.4; this reflects the effects the D{\O} $W$ asymmetry data has on the up and down quarks and antiquarks. The effect of the D{\O} $W$ asymmetry data on the BCDMS data are also given in Table~\ref{tab:BCDMSD0WasymATLASZpt_delchisqtable}. The D{\O} $W$ asymmetry worsens notably when the BCDMS data is left out, implying that they have similar pulls, most likely on the quarks.

\begin{table}
\begin{center}
\def\arraystretch{0.9}
\begin{tabular}{|>{\arraybackslash}m{7.4cm}|>{\centering\arraybackslash}m{1.5cm}|>{\centering\arraybackslash}m{1.4cm}|>{\centering\arraybackslash}m{1.5cm}|>{\centering\arraybackslash}m{1.1cm}|>{\centering\arraybackslash}m{1.65cm}|} \hline
  Data set & $N_{pts}$ & \multicolumn{4}{c|}{$\Delta \chi^2$ relative to MSHT20} \\ \cline{3-6}
  & & no BCDMS & no D{\O} $W$ asym & no $Z$ $p_T$  & All 3 removed \\ \hline
  BCDMS $\mu p$ $F_2$ \cite{BCDMS} & 163 & - & +2.3 & +1.1 & - \\ 
  BCDMS $\mu d$ $F_2$ \cite{BCDMS} & 151 & - & -1.9 & -0.3 & - \\ 
  NMC $\mu p$ $F_2$ \cite{NMC} & 123 & -3.2 & -0.4 & -1.2 & -3.6\\
  NMC $\mu d$ $F_2$ \cite{NMC} & 123 & -3.2 & +0.2 & -4.6 & -7.0\\
  NMC $\mu n/\mu p$ \cite{NMCn/p} & 148 & -7.4 & -0.2 & -1.2 & -7.6\\ 
  E665 $\mu p$ $F_2$ \cite{E665} & 53 & -0.1 & -0.2 & +1.6 & +1.2\\
  E665 $\mu d$ $F_2$ \cite{E665} & 53 & -0.3 & -0.2 & +2.5 & +2.2\\
  SLAC $e p$ $F_2$ \cite{SLAC,SLAC1990} & 37 & -3.6 & -0.2 & -0.2 & -4.7 \\
  SLAC $e d$ $F_2$ \cite{SLAC,SLAC1990} & 38 & -5.9 & +0.4 & +0.6 & -5.8 \\
  E866/NuSea $pp$ DY \cite{E866DY} & 184 & -11.1 & -0.7 & +0.6 & -13.7\\
  E866/NuSea $pd/pp$ DY \cite{E866DYrat} & 15 & -0.5 & -0.8 & -1.6 & -2.1\\
  CCFR $\nu N \rightarrow \mu \mu X$ \cite{Dimuon} & 86 & +0.7 & +0.3 & -1.7 & -1.0\\
  NuTeV $\nu N \rightarrow \mu \mu X$ \cite{Dimuon} & 84 & -1.9 & -0.5 & -1.4 & -7.2\\
  HERA $e^+ p$ NC $ 920$~GeV \cite{H1+ZEUS} & 402 & -8.5 & +0.4 & -2.6 & -4.3\\
  HERA $e^- p$ NC $ 460$~GeV \cite{H1+ZEUS} & 209 & -2.0 & -0.5 & -5.4 & -12.2\\
  HERA $e^- p$ NC $ 575$~GeV \cite{H1+ZEUS} & 259 & -1.4 & -0.3 & -2.9 & -4.5\\
  HERA $e^- p$ NC $ 920$~GeV \cite{H1+ZEUS} & 159 & -6.7& -0.4 & -2.2 & -10.3\\
  D{\O} II $p\bar{p}$ incl. jets \cite{D0jet} & 110 & +1.4 & +0.2 & +1.9 & +3.4\\
  CDF II $p\bar{p}$ incl. jets \cite{CDFjet} & 76 & +0.1 & -0.7 & +3.1 & +2.7\\
  CDF II $W$ asym. \cite{CDF-Wasym} & 13 & -1.4 & -0.3 & -0.3 & +2.0\\
  D{\O} II $W\rightarrow \nu e$ asym. \cite{D0Wnue} & 12 & -0.1 & -5.8 & +0.3 & -5.2\\
  D{\O} II $W \rightarrow \nu \mu$ asym. \cite{D0Wnumu} & 10 & -2.7 & +3.1 & -1.6 & -1.4\\
  D{\O} $W$ asym. \cite{D0Wasym} & 14 & +4.5 & - & -1.6 & - \\
  LHCb $Z\rightarrow e^+e^-$ \cite{LHCb-Zee} & 9 & -2.1 & -0.7 & +1.0 & -2.1\\
  LHCb $W$ asym. $p_T > 20$~GeV \cite{LHCb-WZ} & 10 & +0.3 & +0.2 & +1.2 & +1.5\\
  CMS double diff. Drell-Yan \cite{CMS-ddDY} & 132 & -0.5 & -0.5 & +2.6 & +2.3\\
  LHCb 8~TeV $Z\rightarrow ee$ \cite{LHCbZ8} & 17 & +1.1 & -0.8 & +1.2 & +1.6\\
  LHCb 2015 $W$, $Z$ \cite{LHCbZ7,LHCbWZ8} & 67 & +4.9 & -2.8 & +1.6 & +0.1\\
  ATLAS 7~TeV jets \cite{ATLAS7jets} & 140 & +1.2 & -0.6 & -2.3 & -1.9\\
  CMS 7~TeV $W+c$ \cite{CMS7Wpc} & 10 & -0.4 & 0.0 & +1.6 & +2.1\\
  ATLAS 7~TeV high precision $W$, $Z$ \cite{ATLASWZ7f} & 61 & -4.7 & -1.0 & -7.5 & -11.8\\
  CMS 7~TeV jets \cite{CMS7jetsfinal} & 158 & +4.1 & -0.9 & +0.1 & +1.8 \\
  CMS 8~TeV jets \cite{CMS8jets} & 174 & -0.3 & -1.6 & -7.0 & -7.2\\
  CMS 2.76~TeV jet \cite{CMS276jets} & 81 & -0.7 & +0.3 & -1.6 & -1.8\\
  ATLAS 8~TeV $Z$ $p_T$ \cite{ATLASZpT} & 104 & -0.7 & +8.1 & - & - \\
  ATLAS 8~TeV single diff $t\bar{t}$ \cite{ATLASsdtop} & 25 & -0.6 & 0.0 & +3.2 & +2.4\\
  ATLAS 8~TeV single diff $t\bar{t}$ dilepton \cite{ATLASttbarDilep08_ytt} & 5 & -0.1 & -0.2 & -0.6 & -0.9\\
  CMS 8~TeV double differential $t\bar{t}$ \cite{CMS8ttDD} & 15 & +0.3 & 0.0 & -2.0 & -1.6\\
  CMS 8~TeV single differential $t\bar{t}$ \cite{CMSttbar08_ytt} & 9 & -0.3 & -0.4 & -2.7 & -3.6\\
  ATLAS 8~TeV High-mass Drell-Yan \cite{ATLASHMDY8} & 48 & +0.6 & +0.7 & -3.4 & -2.2\\
  ATLAS 8~TeV $W$ \cite{ATLASW8} & 22 & -1.5 & -1.6 & -4.6 & -6.5\\
  ATLAS 8~TeV double differential $Z$ \cite{ATLAS8Z3D} & 59 & -0.1 & +1.4 & -5.4 & -6.6\\ \hline
  Total common data sets & 4363 & -53.0 & -4.7 & -41.3 & -116.2 \\ \hline
\end{tabular}
\end{center}
\vspace{-0.5cm}
\caption{\sf The change in $\chi^2$ (with negative indicating an improvement in the fit quality) for a selection of data sets once the BCDMS \cite{BCDMS}, D{\O} $W$ asymmetry \cite{D0Wasym} and ATLAS 8~TeV $Z$ $p_T$  \cite{ATLASZpT} data sets are removed, illustrating the tensions of these data sets with several of the other data sets in the global fit.}
\label{tab:BCDMSD0WasymATLASZpt_delchisqtable}
\end{table}

Finally, the ATLAS 8~TeV $Z$ $p_T$ data show significant impacts on the fit qualities of other data sets with the overall fit improving by $\Delta\chi^2 = 41.3$ when this data set is not included. This is also a reflection of its overall poorer fit quality and relatively large number of points. This overall improvement in the $\chi^2$ masks significant variation between the data sets with several improving significantly and several worsening notably, demonstrating both tensions and consonances with the ATLAS $Z$ $p_T$ data and other data sets in the fit. This is true within the structure function data sets, which vary noticeably in $\chi^2$ when it is removed, with the NMC data sets all improving whilst the E665 data fit quality worsens. This again may partly reflect the fact that the NMC $p$ and $d$ data both favour higher values of $\alpha_S$, in contrast to the ATLAS 8~TeV $Z$ $p_T$ data which favour lower values of the strong coupling, as does the E665 data. The ATLAS $Z$ $p_T$ data are also in tension with the dimuon data sets, which improve by 3.1 across the NuTeV and CCFR data sets, whereas the CMS 7~TeV $W+c$ data set favours its inclusion as it worsens by 1.6. 
The sensitivity of the $Z$ to the strangeness results in the $Z$ $p_T$ data altering its shape and size, as seen in Fig.~\ref{gluon_ssbar_q210000_NNLO_noBCDMSZptLHCjetstopZpt} (right). Here, the absence of the $Z$ $p_T$ data increases the strangeness, with the largest effect being 
near $x=0.3$. In general a lower strangeness is favoured by the older dimuon data sets but disfavoured by the strangeness enhancements preferred in newer LHC Drell-Yan data. However, the details in 
the change in shape on omission of the  $Z$ $p_T$ data seem to improve the Dimuon data fit, despite the increase in magnitude. The HERA combined data in the table also improve by 13.1, albeit for a very large number of data points.

The impact of the ATLAS $Z$ $p_T$ data is also clear on the electroweak boson, top and jet data sets. In regard to the first of these, there is an appreciable difference in how the slightly older LHC electroweak data sets tend to worsen a little once it is removed (including all 4 of the LHCb electroweak data sets included as well as the CMS double differential Drell-Yan data), whilst the new ATLAS Drell-Yan data sets all improve significantly when the ATLAS $Z$ $p_T$ data are removed. The 7~TeV $W$, $Z$ and 8~TeV high mass Drell-Yan, $W^{\pm}$ and double differential $Z$ data sets improve by $\Delta\chi^2 = -7.5, -3.4, -4.6$ and $-5.4$ respectively, showing a noteworthy tension with the 8~TeV $Z$ $p_T$ data. At least part of this tension or improvement is due to the change in total strangeness, with the ATLAS 7 and 8~TeV precision Drell-Yan data indeed favouring strangeness enhancement in the $10^{-2} < x < 10^{-1}$ region. The effect on the jets data is also somewhat split between the older and newer data with the D{\O} jets and CDF jets data each worsening once it is removed whilst the newer ATLAS and CMS jet data sets across 2.76, 7 and 8~TeV improve by a total of $\Delta\chi^2=-10.8$. The tension with the new LHC jet data is a reflection of the different pulls of the $Z$ $p_T$ and new LHC jet data on the high $x$ gluon, as seen in Fig.~\ref{MSHT20_NNLO_gluonpulls_q210000_NNLO}, with the LHC jet data preferring a lower gluon in this region in contrast to the $Z$ $p_T$ data. The same effects on the high $x$ gluon also apply for the top data with the new LHC top data also favouring a lower gluon and improving by 2.1 once the ATLAS $Z$ $p_T$ data are removed. However the detailed picture is a little more varied, with the ATLAS 8~TeV single differential $t\bar{t}$ data worsening by 3.2 points in $\chi^2$ for 25 data points, whilst the other top data, including the ATLAS 8~TeV single differential $t\bar{t}$ dilepton data and the CMS 8~TeV single and double differential top data  improve by 5.3 across 29 data points. One interesting detail to notice on the topic of the ATLAS $Z$ $p_T$ data and its tensions with the newer LHC data sets is that in Table~\ref{tab:TabnewLHC} the ATLAS 8~TeV $Z$ $p_T$ data set is one of the few data sets fit worse in the MSHT20 global fit than it would be using the MMHT14 global fit, again in contrast to the ATLAS Drell-Yan data sets which were all poorly fit in predictions using MMHT14. Finally, the effects of removing the $Z$ $p_T$ data on the $\chi^2$ of the BCDMS and D{\O} $W$ asymmetry data are notable. The BCDMS data show little change, perhaps indicating the data sets have either compatible or orthogonal effects, whilst the D{\O} $W$ asymmetry actually improves slightly. In contrast the $Z$ $p_T$ fit quality worsens by 8.1 points in $\chi^2$ when the D{\O} $W$ asymmetry is removed, clearly favouring its inclusion.

If all 3 data sets are removed together, then the effects seen when the data sets are individually removed are also apparent, as given in the final column of Table~\ref{tab:BCDMSD0WasymATLASZpt_delchisqtable}. In some places, such as for the E866 Drell-Yan $pp$ data, some of the HERA data, NuTeV dimuon data or the ATLAS 8~TeV double differential $Z$ data, the improvement upon removing all 3 data sets is notably larger than the sum of removing the three individually. This is also true of the overall $\Delta\chi^2~=~-116.2$ improvement and indicates the magnification of some of the tensions inherent by the inclusion of subsequent data sets. This reflects the 3 data sets having similar pulls to one another in a number of respects. In other places, in contrast, the tensions are more similar to, or even less than the sum of the individual $\chi^2$ changes upon removal of the three data sets, perhaps indicating these represent separate tensions. This is for example true for the NMC $n/p$, CDF $W$ asymmetry, LHCb 2015 $W$, $Z$ and others. In the case of the improvements in the fit quality of the LHC jets and LHC top data sets, this is perhaps expected given that both the BCDMS and ATLAS $Z$ $p_T$ data favour a larger high $x$ gluon and lower $\alpha_S$ value, which is in tension with these LHC data sets. The removal of all three data sets also makes clear the tensions of these with the new LHC Drell-Yan data sets, with the fit quality of the ATLAS 7~TeV $W$, $Z$ improving significantly by 11.8 (i.e. 0.19 per data point); similarly the ATLAS 8~TeV $W^{\pm}$ and double differential $Z$ also improve noticeably by $\Delta \chi^2 = -6.5, -6.6$ (or equivalently 0.30 and 0.12 per data point).

\section{Predictions for benchmark processes \label{sec:9}}

\begin{figure} 
\begin{center}
\includegraphics[scale=0.7]{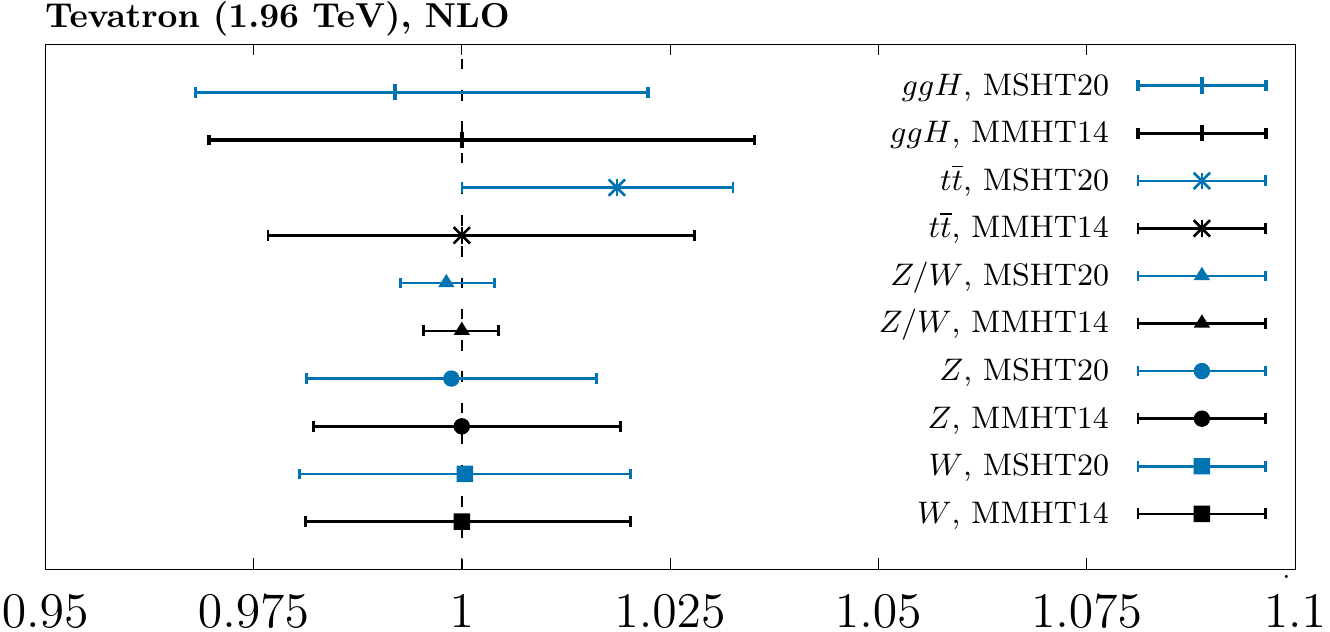}
\includegraphics[scale=0.7]{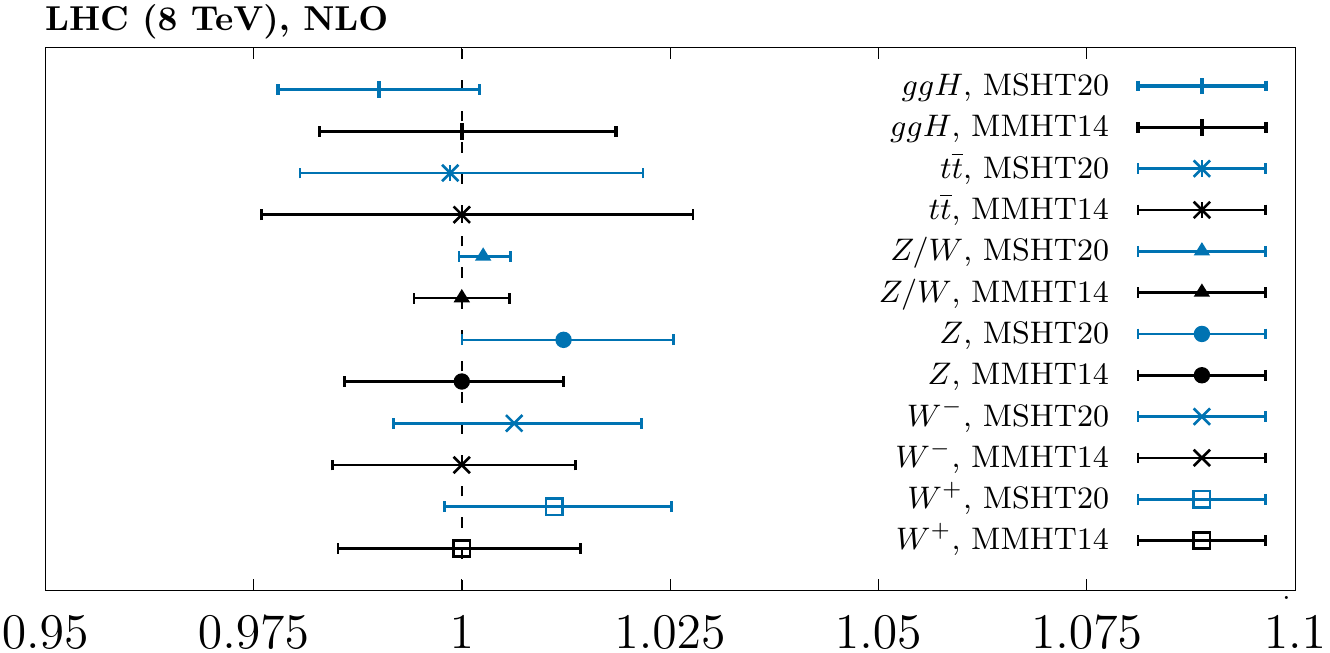}
\includegraphics[scale=0.7]{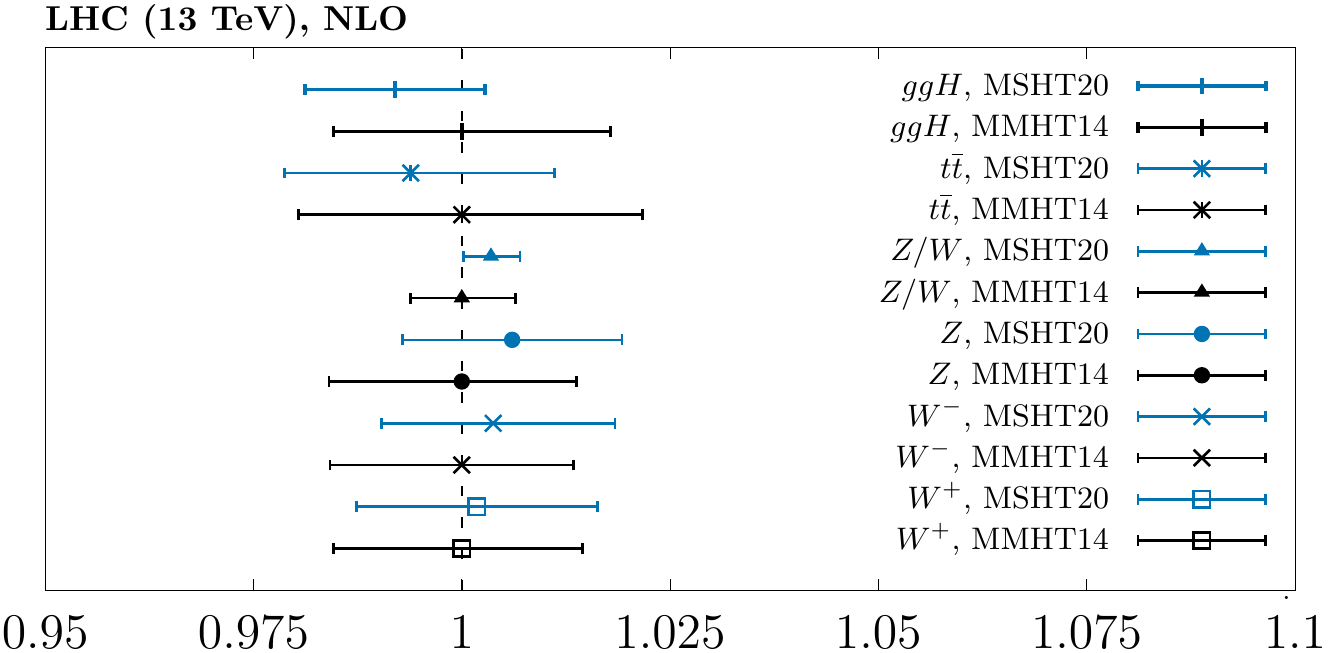}
\includegraphics[scale=0.7]{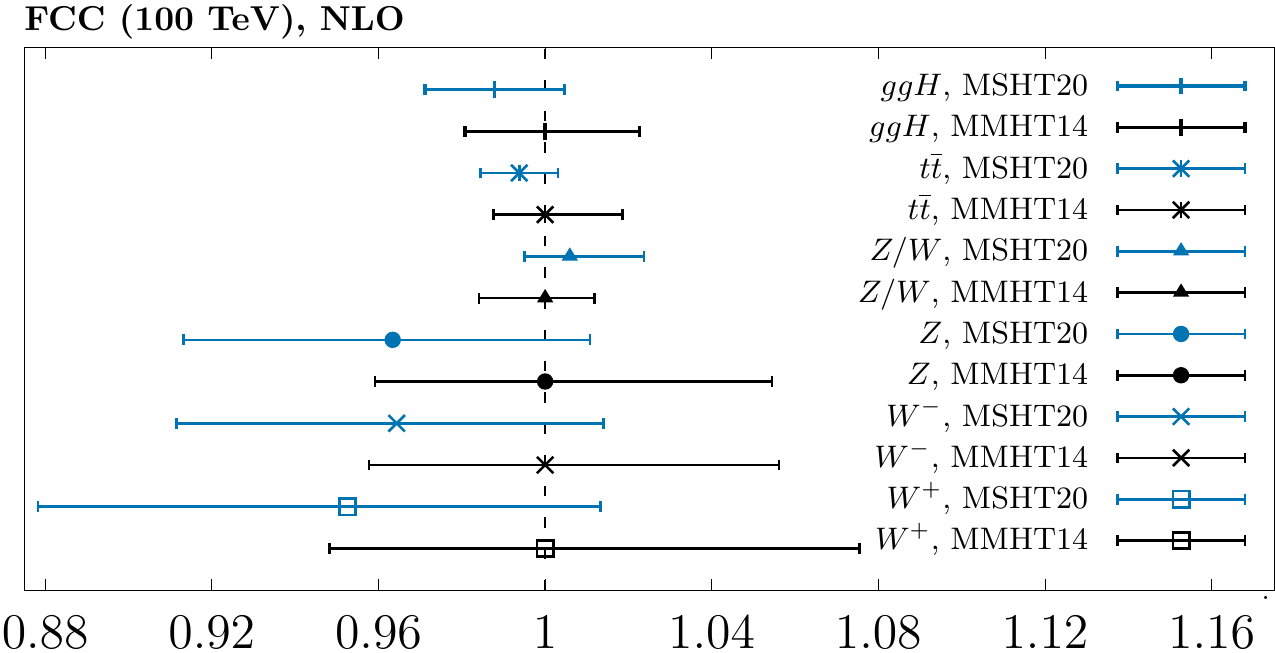}
\caption{\sf Benchmark cross sections obtained with the NLO MMHT14~\cite{MMHT14} and the NLO MSHT20 PDFs. Results are normalized to the central value of the MMHT14 prediction, and PDF uncertainties only are shown.}
\label{fig:BenchmarksNLO}
\end{center}
\end{figure}

\begin{figure} 
\begin{center}
\includegraphics[scale=0.7]{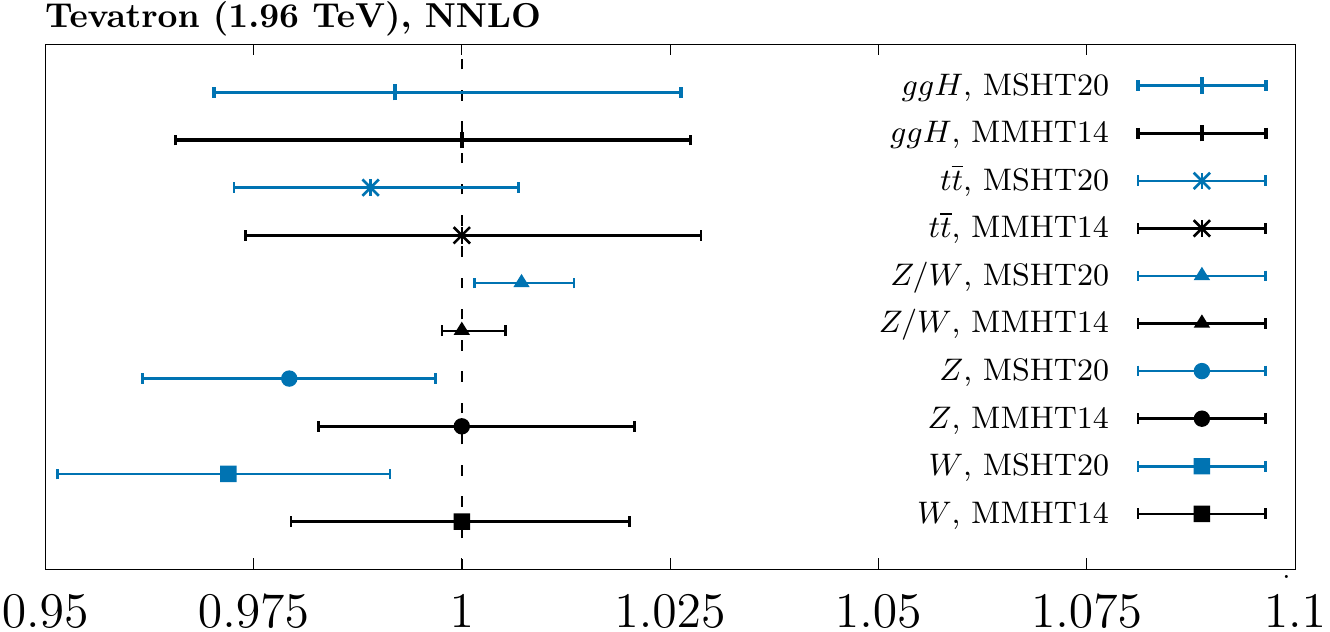}
\includegraphics[scale=0.7]{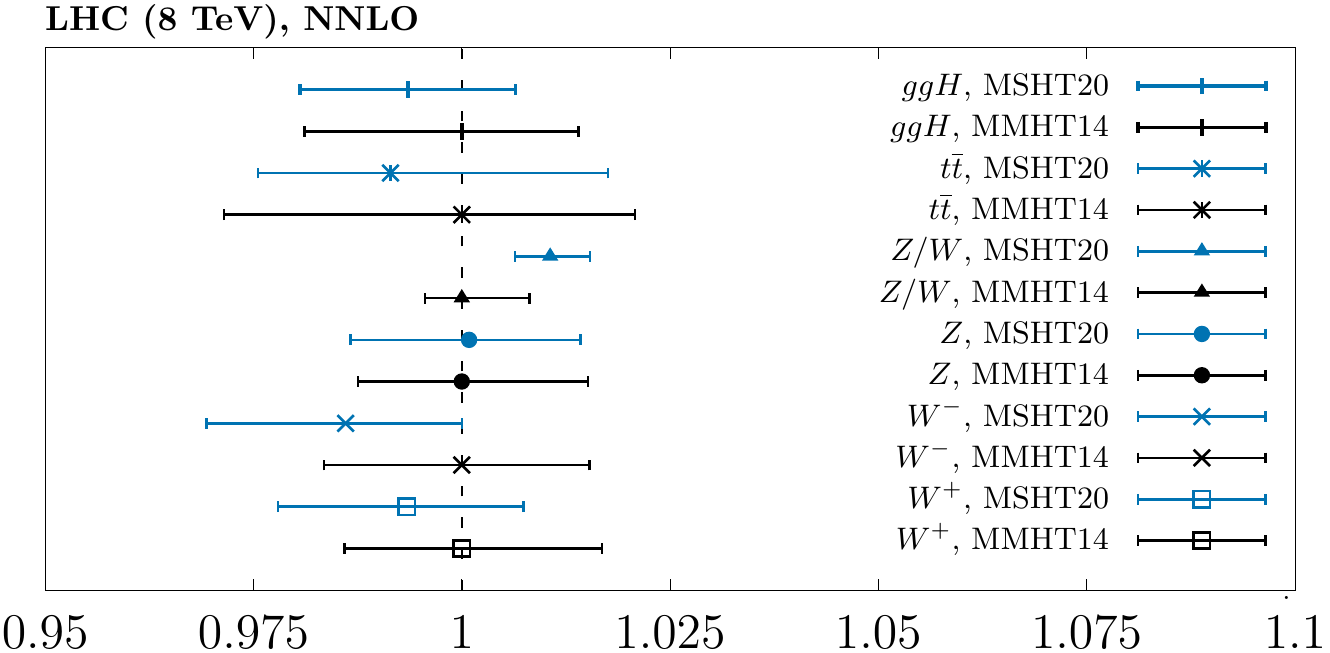}
\includegraphics[scale=0.7]{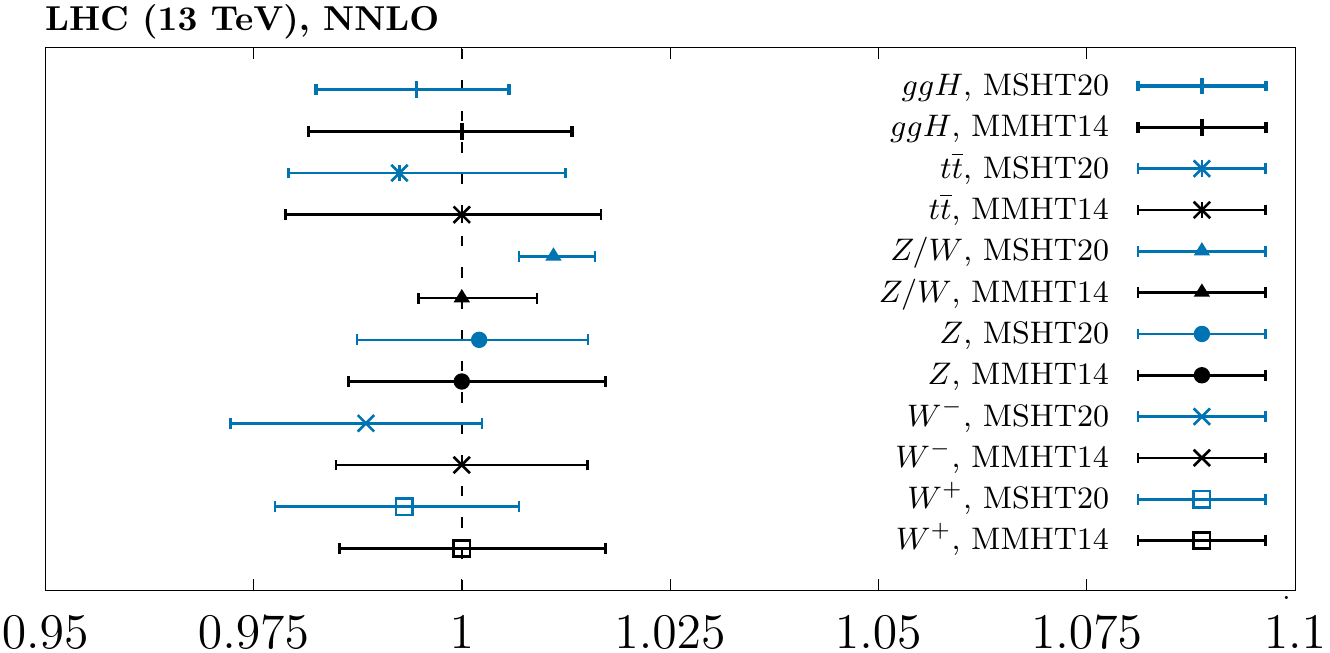}
\includegraphics[scale=0.7]{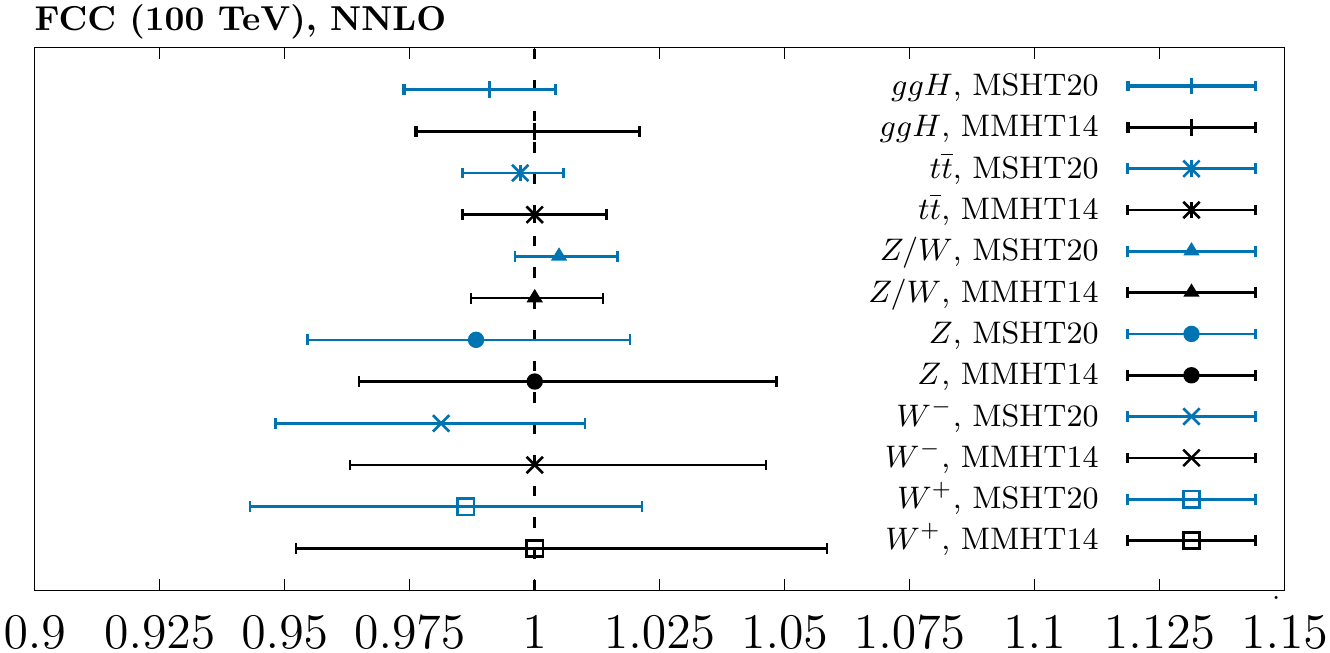}
\caption{\sf Benchmark cross sections obtained with the NNLO MMHT14~\cite{MMHT14} and the NNLO MSHT20 PDFs. Results are normalized to the central value of the MMHT14 prediction, and PDF uncertainties only are shown.}
\label{fig:BenchmarksNNLO}
\end{center}
\end{figure}

In Figures~\ref{fig:BenchmarksNLO}  and~\ref{fig:BenchmarksNNLO} we show predictions 
for various benchmark processes at the Tevatron (1.96~TeV), LHC (8 and 13~TeV) and a FCC--pp (100~TeV) for the 
MMHT14 and MSHT20 PDF sets, at NLO and NNLO. In all cases we show the ratio to the MMHT14 central prediction, for ease of comparison.
We use LO electroweak perturbation theory, with the $qqW$ and $qqZ$ 
couplings defined by 
\begin{equation}
  g_W^2 =  G_F M_W^2 / \sqrt{2}, \qquad g_Z^2 = G_F M_Z^2 \sqrt{2}, 
\end{equation}
and other electroweak parameters as in \cite{MSTW}. We take the Higgs mass to be 
$m_H=125~\GeV$,  and the top pole mass $m_t=172.5~\GeV$. The Higgs cross section corresponds to the gluon fusion channel only.  For the $t\overline{t}$ cross section we use \texttt{top++}~\cite{NNLOtop}.

The main purpose of the presentation is to investigate how both the central values and the uncertainties of the predictions have changed in going from MMHT14 to the MSHT20 PDFs. We therefore provide results for the Tevatron, and the LHC at 8 and 13 TeV, as well as for a 100~TeV $pp$ FCC, to give a representative spread in energies. We do not intend to present definite predictions or compare in detail to other PDF sets, as both these results are frequently provided in the literature with very specific choices of codes, scales and parameters which may differ from those used here.

Considering the NNLO case first, we can see that the central value of the $W^\pm$ cross sections are generally lower in the MSHT20 case in comparison to the MMHT14, by $\sim 1 \sigma$ or less with the exception of the Tevatron, where the change is $\sim 1.5 \sigma$. The $Z$ boson cross section is generally rather unchanged, and hence the ratio of $Z$ to $W$ is higher by $\sim 1\%$. This  is principally due to the large range of precision LHC data on $W$, $Z$ production and the resultant changes these induce in the quark flavour decomposition. The uncertainties are somewhat lower in the MSHT case, in particular at larger energies. For the $t\overline{t}$ and Higgs cross sections, these are slightly lower in the MSHT case due to the smaller gluon, with a non--negligible reduction in uncertainty, due principally to the impact of new LHC data and their constraint on the gluon at intermediate to high $x$. It is interesting to note though that the central value of the 13~TeV Higgs cross section remains relatively stable.

At NLO, the $W$ and $Z$ cross sections generally change by less, being even slightly larger in the MSHT case, with the exception of the 100~TeV case, where the trend matches the NNLO comparison. The resulting ratio of $Z$ to $W$ cross sections is roughly unchanged. We may expect a qualitatively different behaviour in comparison to the NNLO fit, due e.g. to the difficulty the NLO fit has in accounting for the range of precision LHC DY data. We can see for example in Section~\ref{sec:7} that the strangeness in the NLO fit is rather different from the NNLO case, potentially as a result of this. For the $t\overline{t}$ and Higgs cross sections, the trend of lower cross sections and smaller uncertainties for the MSHT PDF set is rather similar, with the exception of the Tevatron $t\overline{t}$ cross section, which is higher in the MSHT case.

\section{Predictions for a selection of LHC processes}\label{sec:10}

\begin{figure}[t]
\centerline{\includegraphics[scale=0.25, clip,]{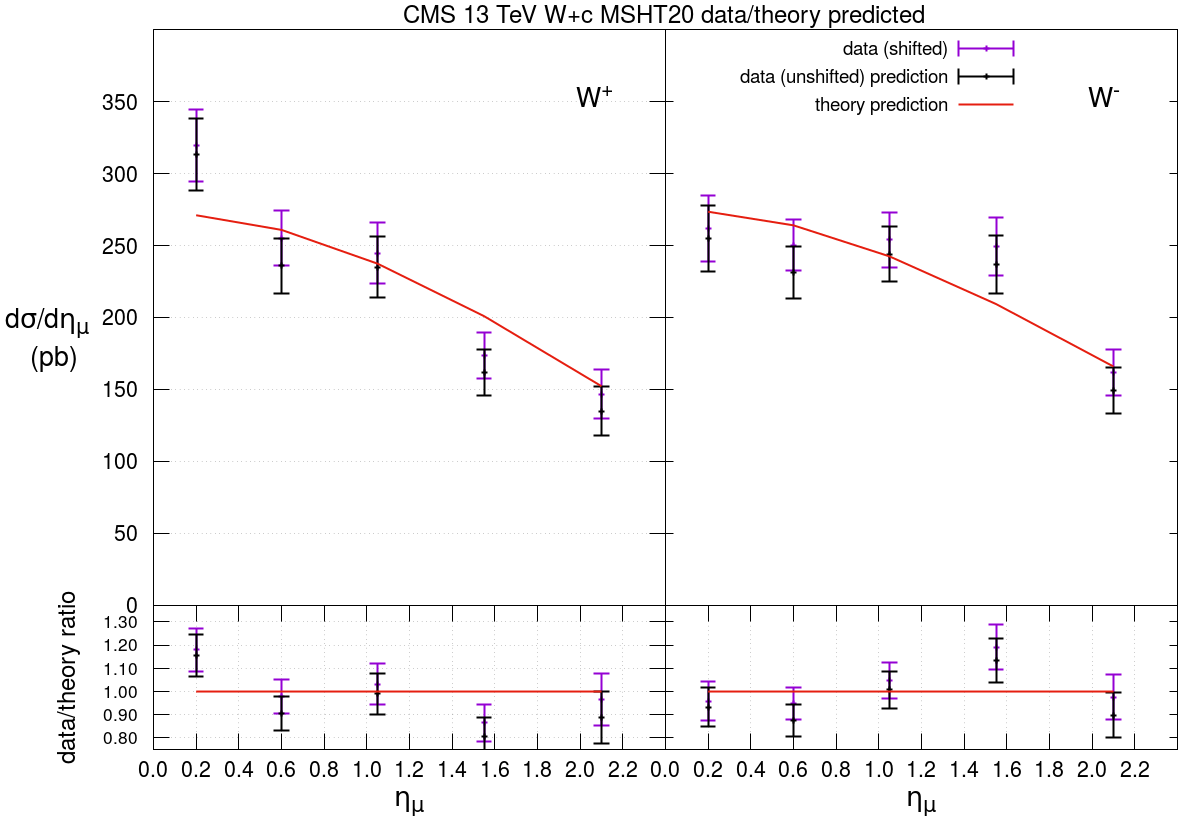}}
\caption{\sf Data and theory comparison for the CMS 13~TeV $W+c$ data set, including both shifted and unshifted data, without refitting.\label{CMS13Wpc_datavstheory_increfit} }
\end{figure}  

\begin{figure} [t]
\begin{center}
\includegraphics[scale=0.6]{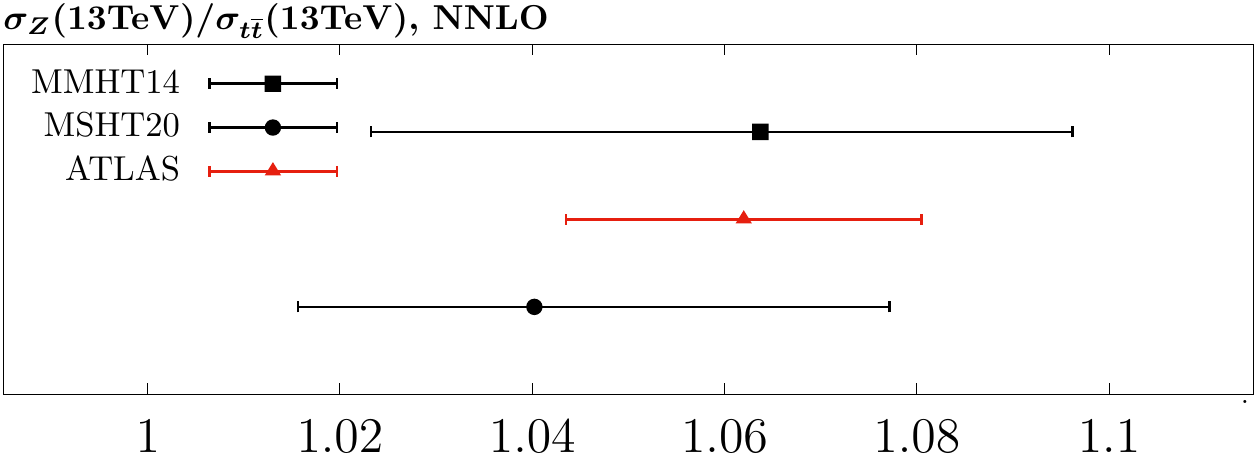}
\includegraphics[scale=0.6]{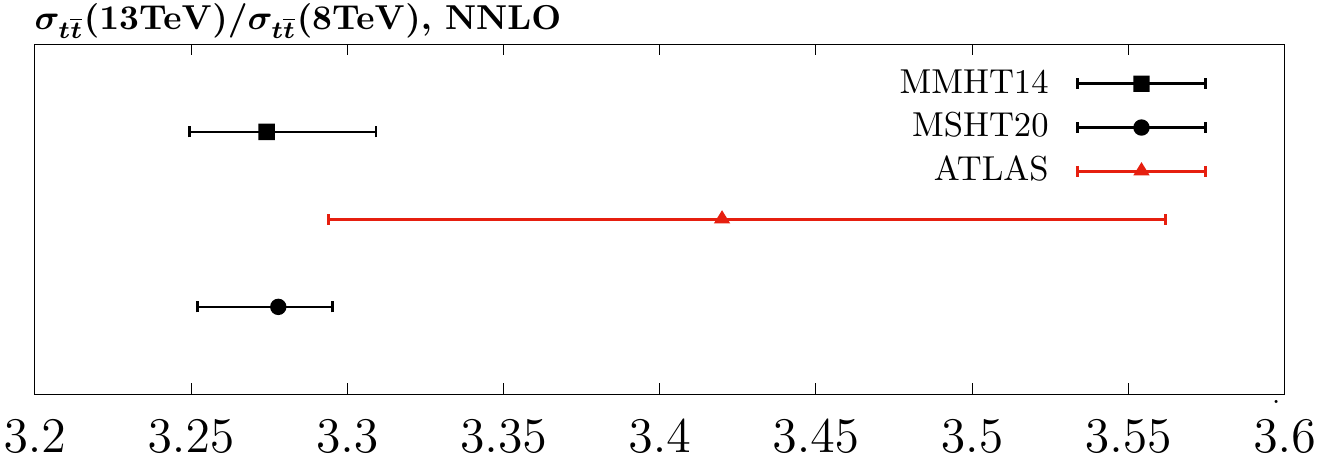}
\caption{\sf MMHT14 and MSHT20 NNLO predictions for the cross section ratios of $Z\to ll$ production (with $p_\perp^l > 25$ GeV, $|\eta_l|<2.5$) and the $t\overline{t}$ total cross section, compared to data from ATLAS~\cite{Aad:2019hzw}. PDF uncertainties alone shown in the theory, while the experimental uncertainties are added in quadrature.}
\label{fig:zt_ttbar}
\end{center}
\end{figure}

In this section we present a brief selection of predictions for LHC measurements not currently included in the fit. Namely we consider: CMS data on $W+c$ jet at 13~TeV; ratios of $Z$ and $t\overline{t}$ cross sections at 8 and 13~TeV measured by ATLAS; CMS measurements of differential and total $t$--channel single top production; and the gluon at low $x$ and $Q^2$ compared to extractions from fits to exclusive $J/\psi$  and open charm production

The CMS $W$ + charm jet data are presented differentially in the muon pseudorapidity at 13~TeV \cite{Sirunyan:2018hde}. We achieve a good description of the data with $\chi^2/N_{pts} = 13.0/10$ without refitting. This shows that these data are already well predicted by using the MSHT20 PDFs, albeit only using NLO theory (for recent results at NNLO see \cite{Czakon:2020coa}). After refitting, this $\chi^2$ is unchanged, and indeed changes little even if the data set is given a larger weight in the fit. The $\chi^2$ predicted for this data set using the MMHT14 PDFs is 14.3. Therefore it perhaps shows a slight preference for the moderately increased strangeness included in MSHT20, although it is clearly consistent with both the MMHT14 and MSHT20 global fits. In particular, regarding the strangeness content of the proton, given this data set is already well predicted by the MSHT20 default PDFs it suggests the amount of strangeness in MSHT20, which is intermediate between the lower strangeness favoured by the dimuon data and the unsuppressed strangeness favoured by the ATLAS 7~TeV $W$, $Z$ data \cite{ATLASWZ7f}, is  favoured by this 13~TeV data set. Indeed, as described in \cite{Sirunyan:2018hde}, the CMS 13~TeV $W+c$ data do not favour  unsuppressed high $x$ strangeness.  The comparison of the data and theory before refitting is shown in Fig.~\ref{CMS13Wpc_datavstheory_increfit}. There is both a difference in normalisation and perhaps shape in the theory compared to the data, with the total $\chi^2$ coming dominantly (10.7 out of 13.0) from the lowest rapidity point for the $W^+$ and the fourth rapidity point for both the $W^+$ and $W^-$, though these largely appear to be due to fluctuations. This does not change upon refitting.

 \begin{figure} [t]
\begin{center}
\includegraphics[scale=0.17]{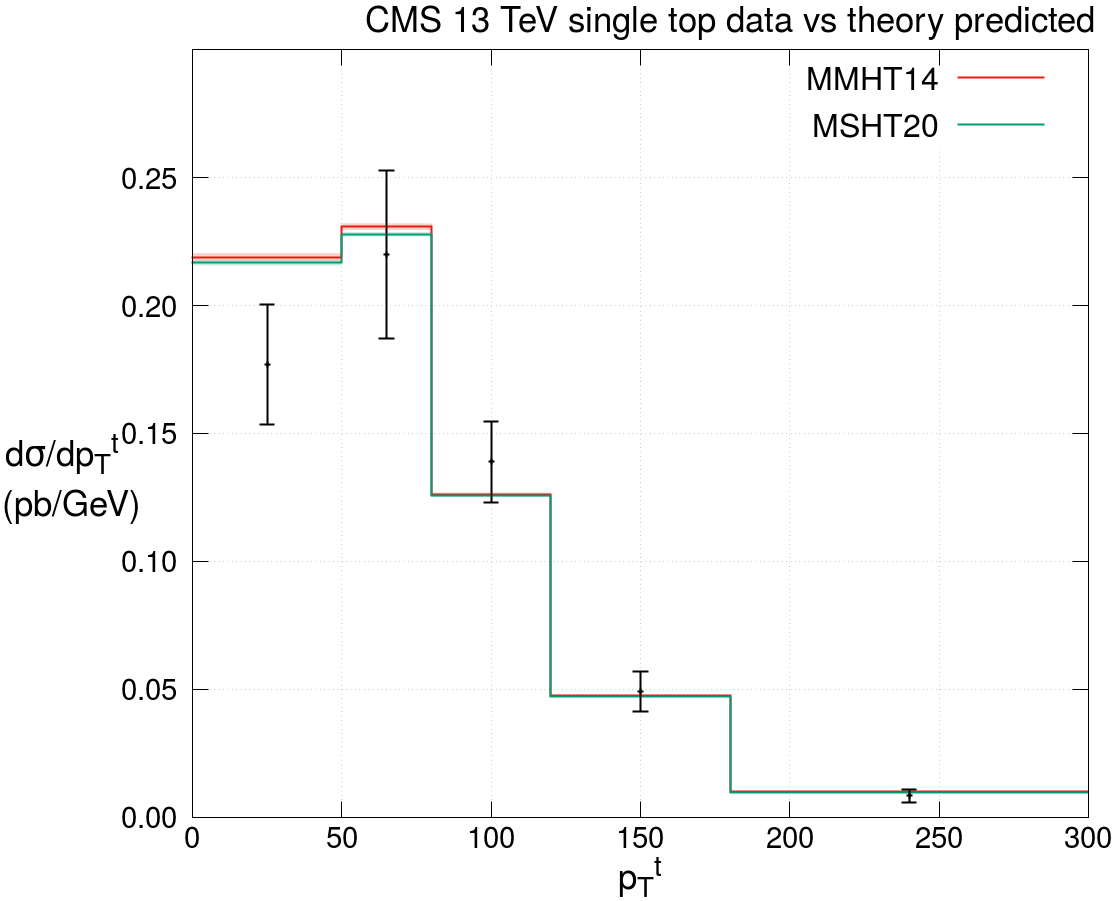}
\includegraphics[scale=0.17]{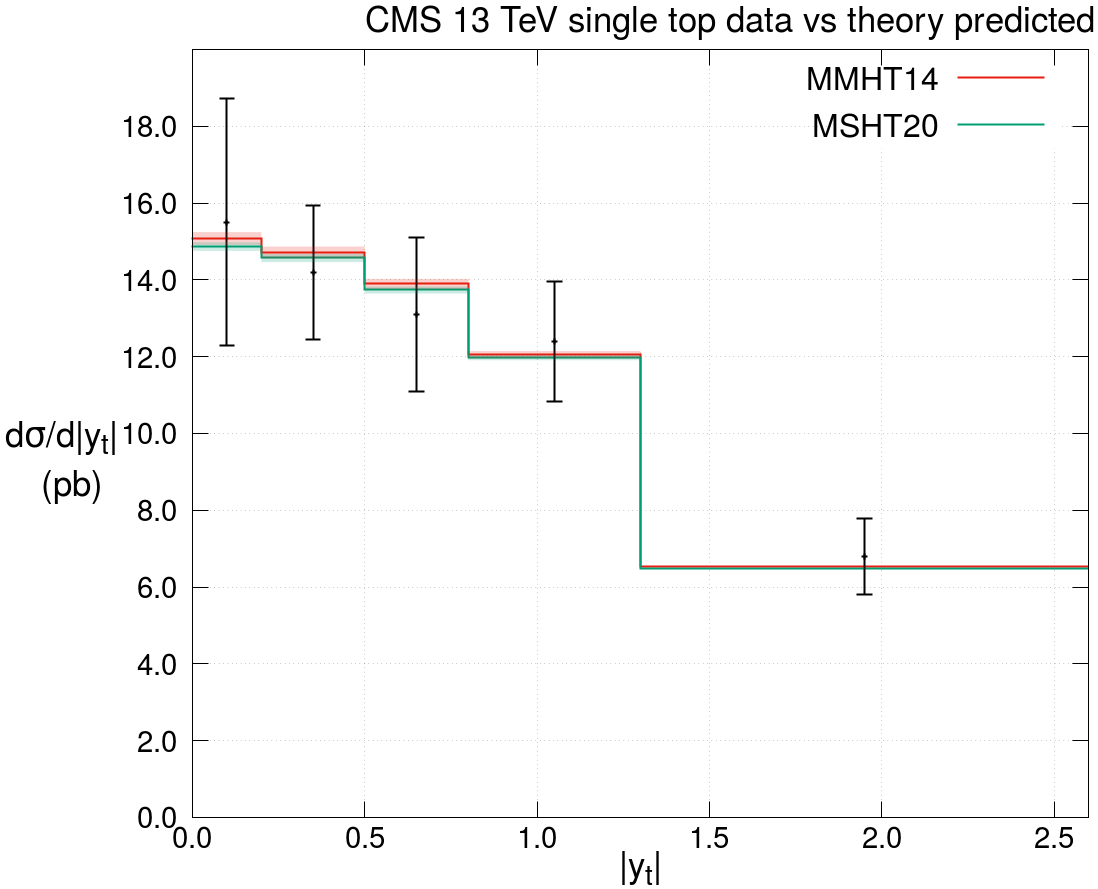}
\includegraphics[scale=0.17]{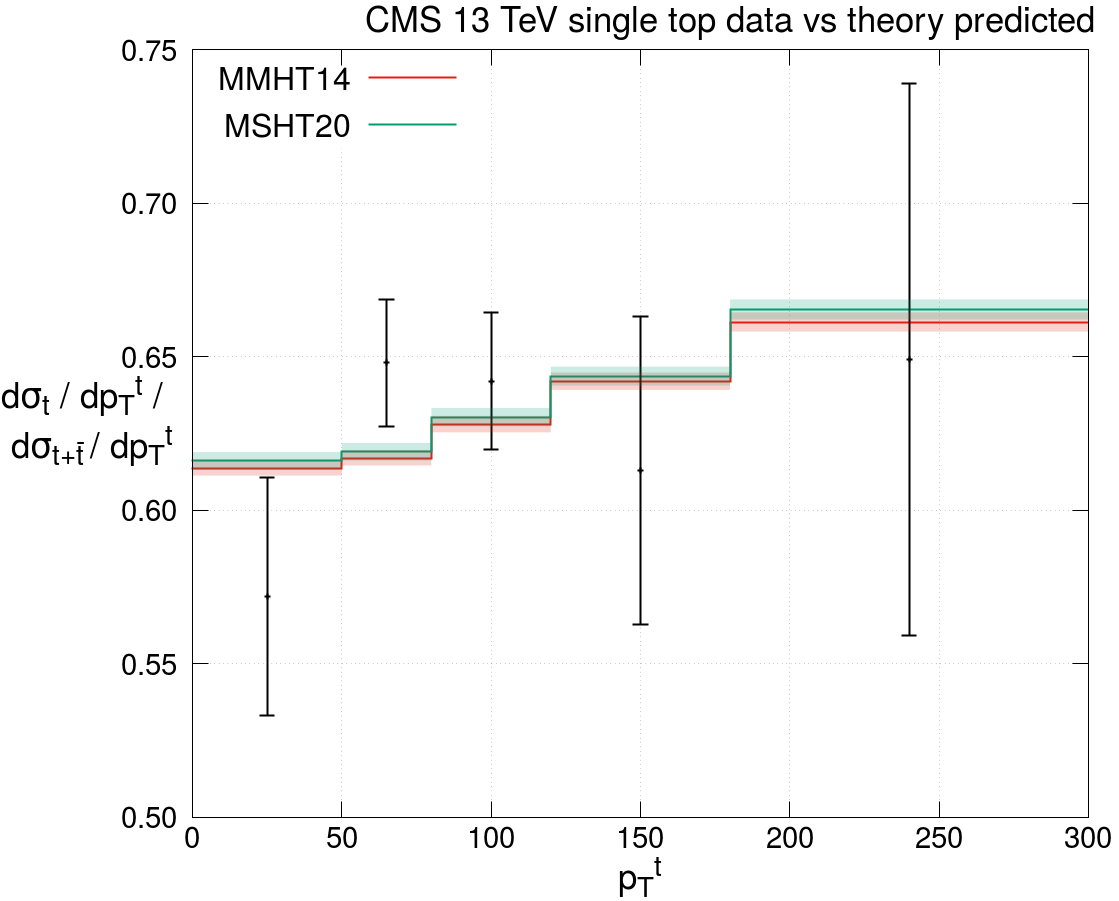}
\includegraphics[scale=0.17]{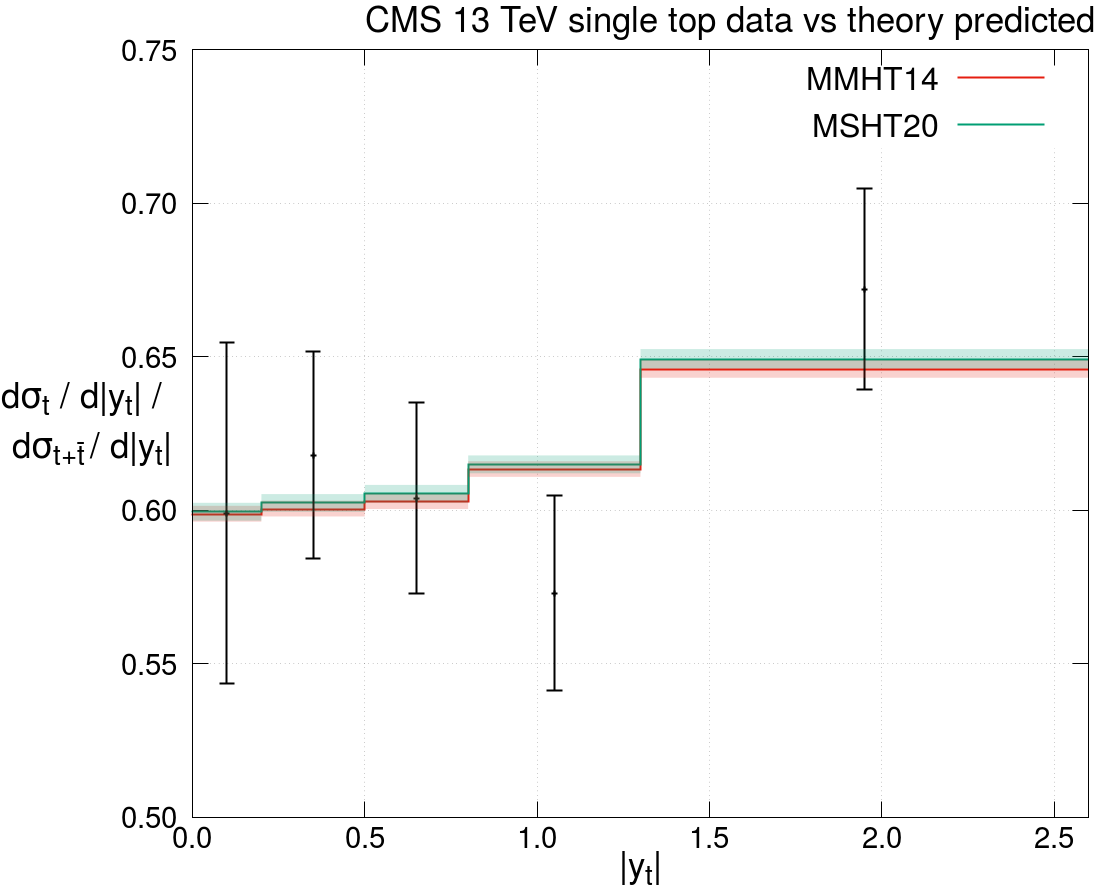}
\caption{\sf MMHT14 and MSHT20 NNLO predictions for the $t$--channel single top and anti--top production. We compare to the sum $t+\overline{t}$ (upper plots) and the ratio of $t$ to the total $t+\overline{t}$ (lower plots) at 13~TeV, differential in the top/antitop $p_\perp$ (left-hand plots) and rapidity (right-hand plots), compared to the data from CMS~\cite{Sirunyan:2019hqb}. PDF uncertainties alone are shown in the theory, while the experimental uncertainties are added in quadrature.}
\label{fig:st}
\end{center}
\end{figure}

Next, in Fig.~\ref{fig:zt_ttbar} we compare the MMHT14 and MSHT20 NNLO predictions for the cross section ratios involving $Z\to ll$ production (with $p_\perp^l > 25$ GeV, $|\eta_l|<2.5$) and the $t\overline{t}$ total cross section, to data from ATLAS~\cite{Aad:2019hzw}. Here, when ratios of different processes at the same beam energy or ratios of the same process at different energies are taken, certain experimental systematics cancel, and hence a cleaner result can be obtained. For $Z$ and $t\overline{t}$ predictions we use \texttt{MCFM 8.3}~\cite{Boughezal:2016wmq} and \texttt{top++}~\cite{NNLOtop}, respectively. The ratio of $Z$ to $t\overline{t}$ at 13~TeV is shown in the left figure, and we can see that the agreement is good for both PDF sets, with some reduction in PDF error seen in the MSHT20 case. Clearly the PDF uncertainties are somewhat larger than the experimental error, indicating that this may be a useful measurement to include in future fits. The ratio of the $t\overline{t}$ cross sections at 13 to 8~TeV are shown in the right figure, and we can observe that both MMHT14 and MSHT20 lie somewhat below the data, though within uncertainties. The MSHT20 prediction has smaller PDF uncertainties and lies a little closer to the data, due to the impact of other LHC data on the gluon PDF presumably. On the other hand, the PDF uncertainties are significantly smaller than the data uncertainty, and hence we may expect it to play less of a role in any future fit.

\begin{figure} [t]
\begin{center}
\includegraphics[scale=0.6]{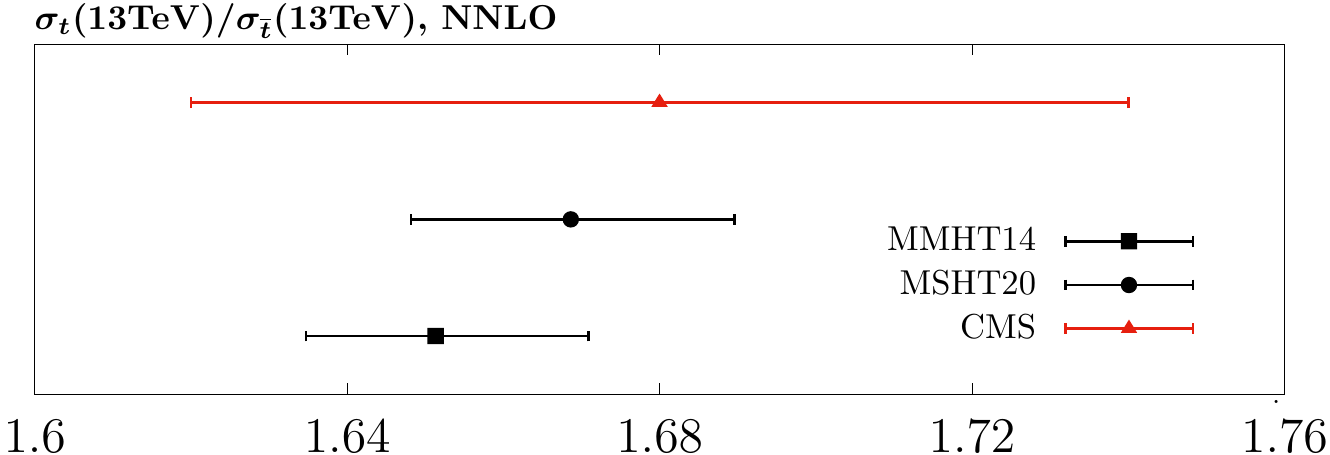}
\caption{\sf MMHT14 and MSHT20 NNLO predictions for the ratio of $t$ to $\overline{t}$ total cross sections at 13~TeV, compared to the CMS measurement~\cite{Sirunyan:2018rlu}.}
\label{fig:strt}
\end{center}
\end{figure}

\begin{figure} [t]
\begin{center}
\includegraphics[scale=0.17]{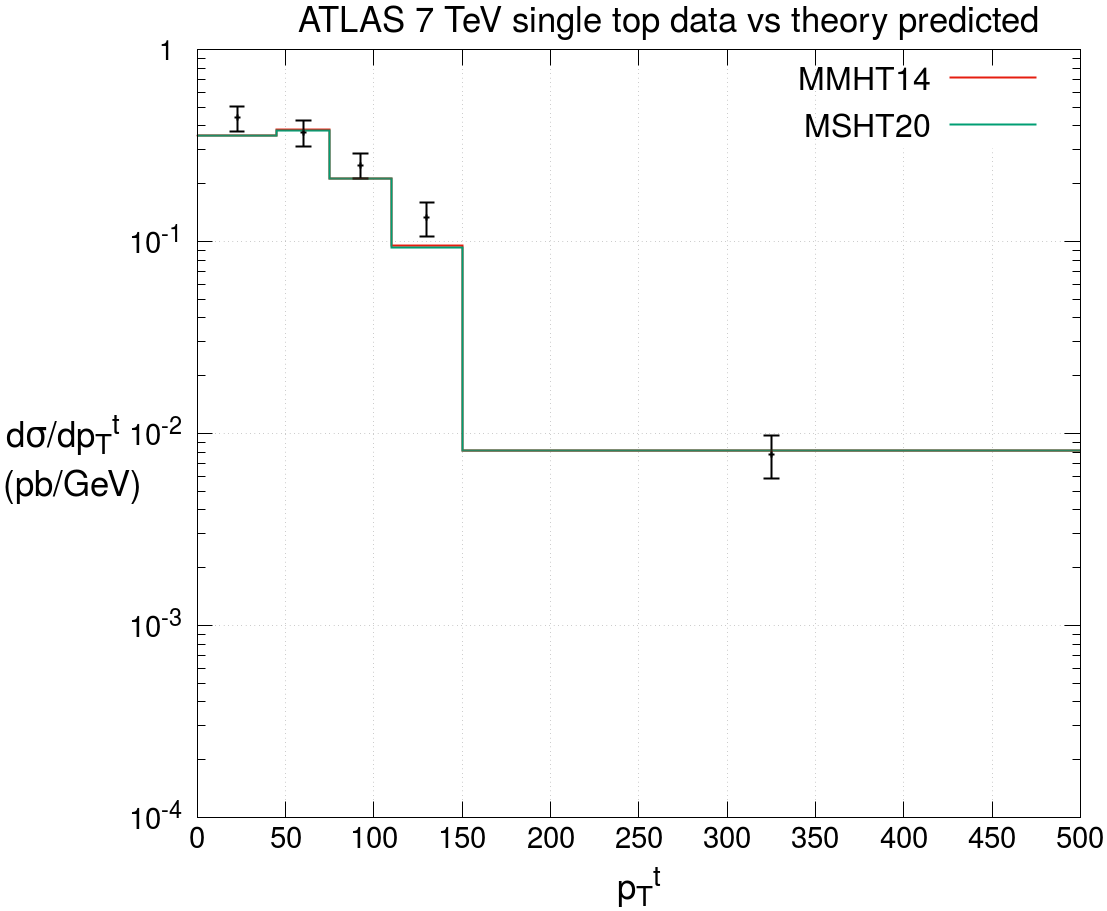}
\includegraphics[scale=0.17]{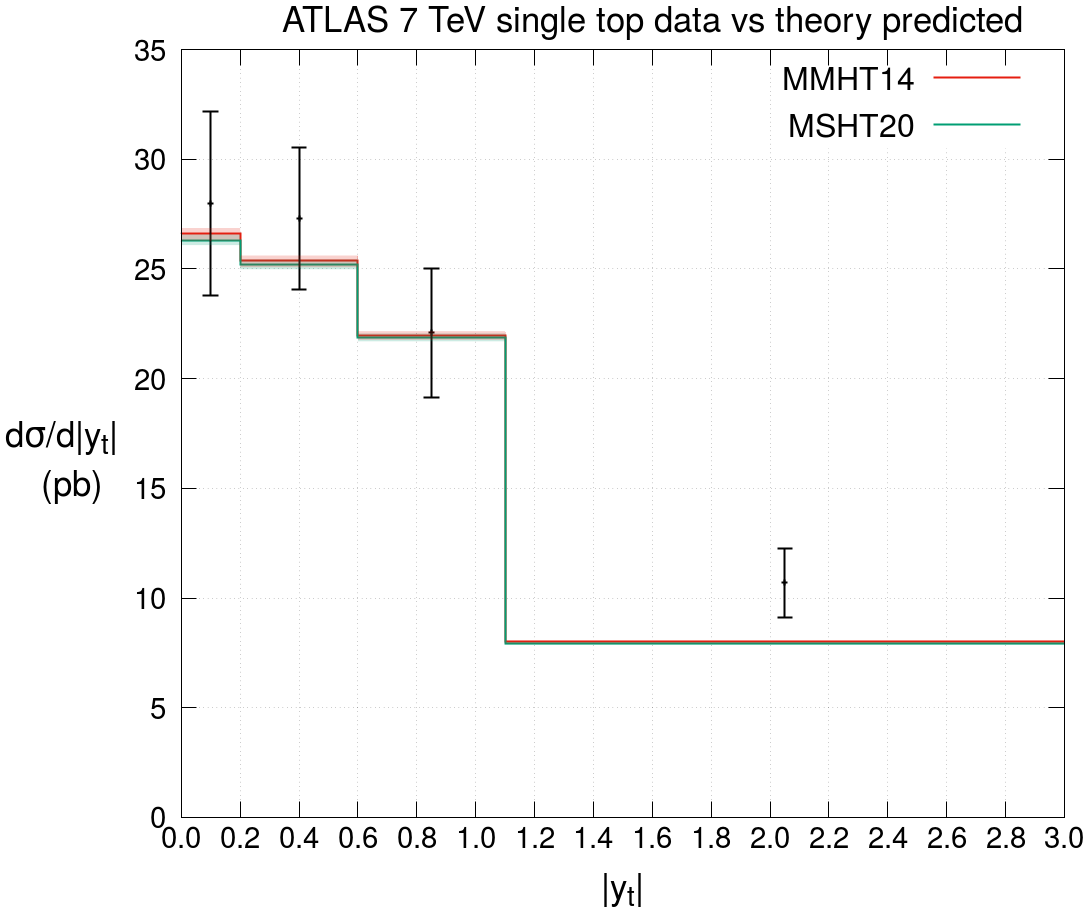}
\includegraphics[scale=0.17]{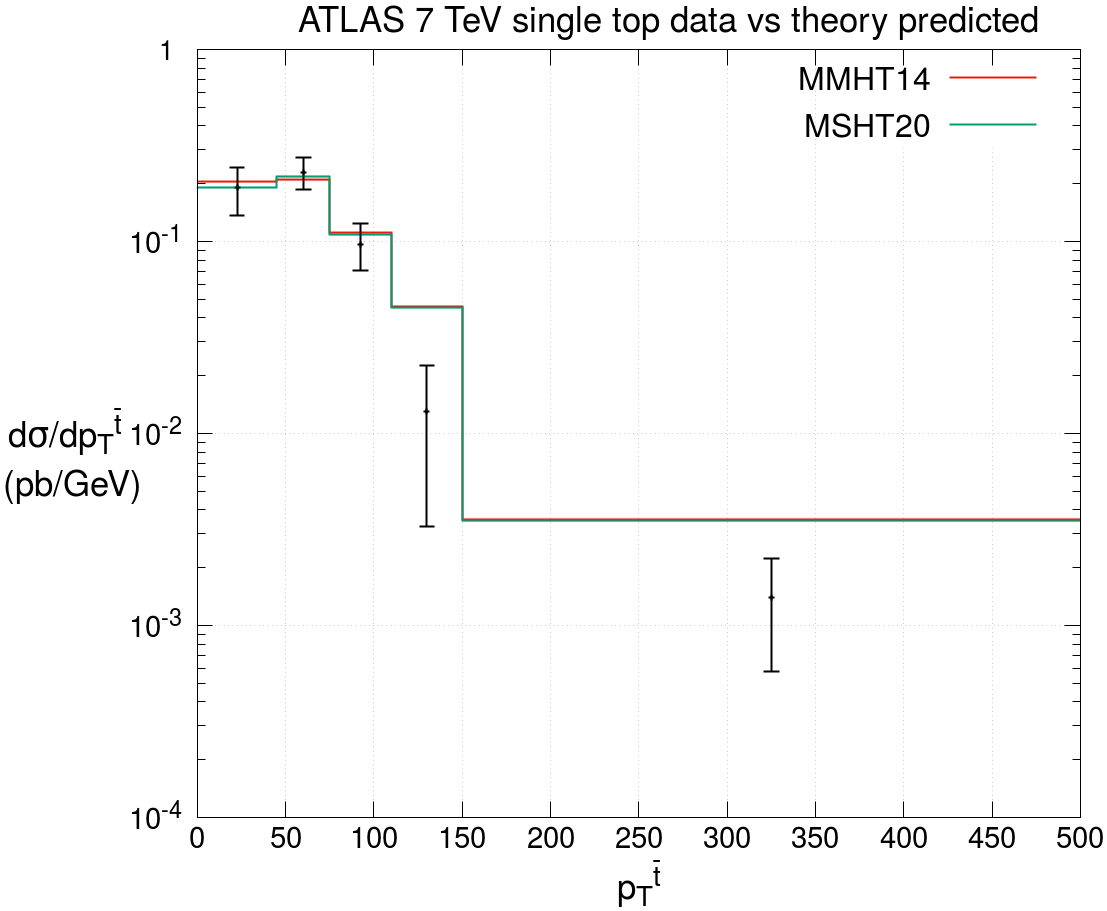}
\includegraphics[scale=0.17]{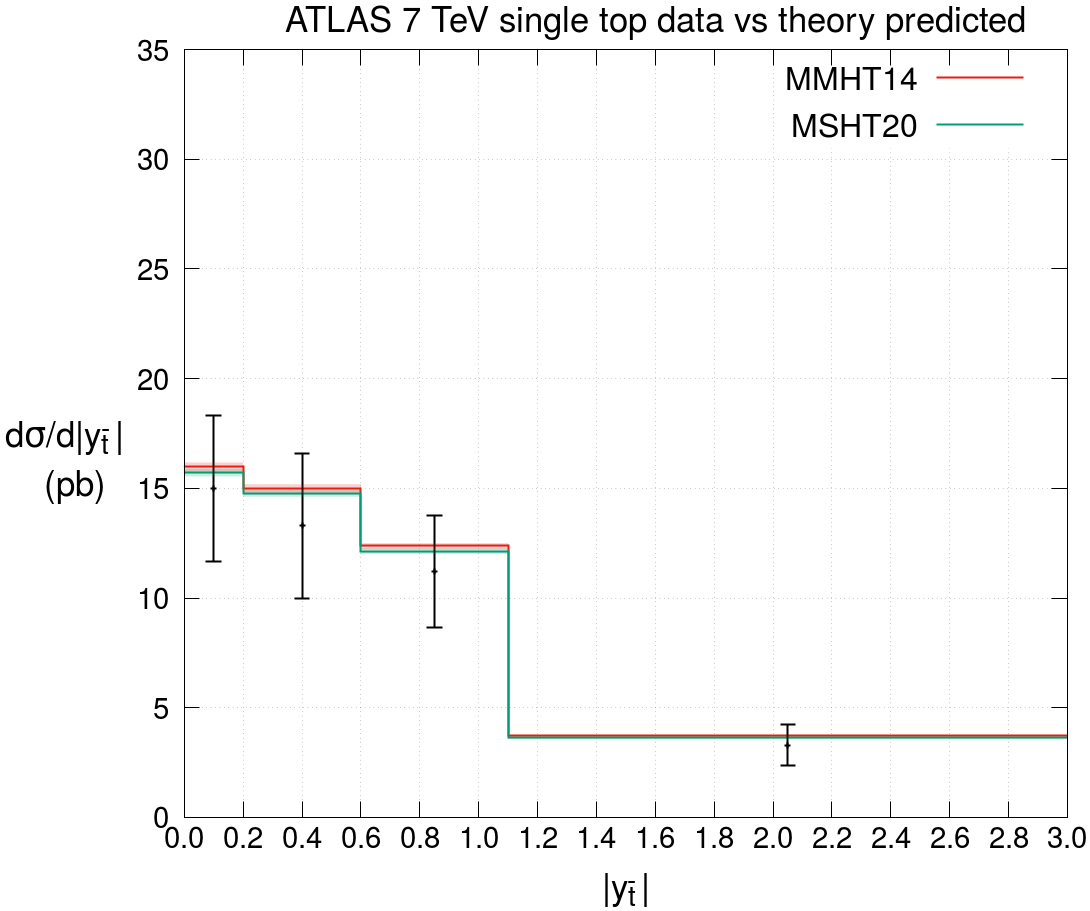}
\caption{\sf MMHT14 and MSHT20 NNLO predictions for $t$--channel single top (upper plots) and anti--top (bottom plots) production, differential in the top/antitop $p_\perp$ (left-hand plots) and rapidity (right-hand plots), compared to the data from ATLAS~\cite{Aad:2014fwa}. PDF uncertainties alone are shown in the theory, while the experimental uncertainties are added in quadrature.}
\label{fig:st7}
\end{center}
\end{figure}

We also compare to CMS 13~TeV data on $t$--channel single top production, considering in particular the sum $t+\overline{t}$ and the ratio of $t$ to $\overline{t}$, differential in the top/antitop $p_\perp$ and rapidity~\cite{Sirunyan:2019hqb}, as well as the ratio of the total $t$ to $\overline{t}$ production, $R_t$, presented in~\cite{Sirunyan:2018rlu}. We use the NNLO calculations in~\cite{Berger:2016oht,Berger:2017zof}. The differential spectra are shown in Fig.~\ref{fig:st} with the upper two figures containing the spectra for the sum of the top and antitop and the lower two figures containing the ratio. The ratio of the total cross sections for $t$ and $\bar{t}$ is given in Fig.~\ref{fig:strt}. We can see that the agreement is good in all cases, albeit within rather large experimental uncertainties. There is in particular no sign of any similar significant disagreement to that observed ~\cite{Nocera:2019wyk} in certain distributions in the ATLAS 7~TeV measurement (though not the 8~TeV). We show the corresponding comparison to the 7 TeV top and anti--top rapidity and $p_\perp$ distributions~\cite{Aad:2014fwa} in Fig.~\ref{fig:st7} and observe a similar undershooting of the $\overline{t}$ $p_\perp$ distribution in particular, at higher $p_\perp$. It is worth commenting here that the differential distributions are generally dominated by systematics, and therefore a full evaluation of the fit quality would require a careful treatment of their correlations, though this is beyond the scope of the current comparison. Nonetheless, the PDF uncertainties on these predictions are small, generally at the sub percent level, and hence given the rather large experimental uncertainties these particular data sets would not be expected to have a significant impact on the fit.
 
 Finally, we compare the NLO gluon PDF at low $x$ and $Q^2$ to specific extractions that are dedicated to constraining this region. There is an absence of precise data in global fits which are directly sensitive to the gluon in this regime, and hence the resulting uncertainties are rather large. In particular, we compare to the extraction of~\cite{Flett:2020duk}, based on a fit to LHCb exclusive $J/\psi$ production data at 7~\cite{Aaij:2014iea} and 13 TeV~\cite{Aaij:2018arx}, and the extraction of~\cite{Bertone:2018dse} based on a fit to LHCb data on $D$ meson production at 5~\cite{Aaij:2016jht}, 7~\cite{Aaij:2014iea} and 13 TeV~\cite{Aaij:2015bpa}. In more detail, the former study applies a two parameter power--law fit in the low $x$ region, while the intermediate to high $x$ region is constrained to follow the MMHT14NLO set, while the latter applies Bayesian reweighting to the NNPDF3.1 set.
 
 We note that the interpretation of both of these data sets is not completely straightforward.
 For the LHCb exclusive data, the theoretical predictions are given in terms of generalized PDFs, rather than the collinear ones directly, and the scale variation uncertainty is very large. However, a recent study~\cite{Flett:2020duk} uses a known method for relating the generalized PDFs to the collinear ones, and argues that the scale variation uncertainty is due to a double counting which can be controlled by a $Q_0 \simeq m_c$ subtraction. For the $D$ meson data, the scale variation uncertainties are again very large, but it is argued in~\cite{Bertone:2018dse} that this can be effectively overcome by considering normalized distributions.
 
\begin{figure} 
\begin{center}
\includegraphics[scale=0.24]{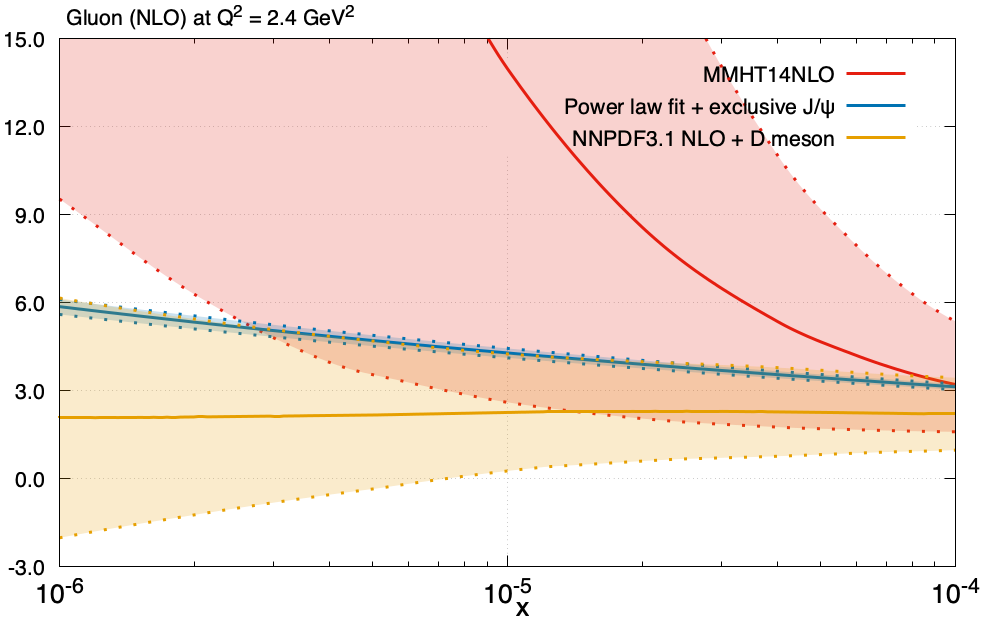}
\includegraphics[scale=0.24]{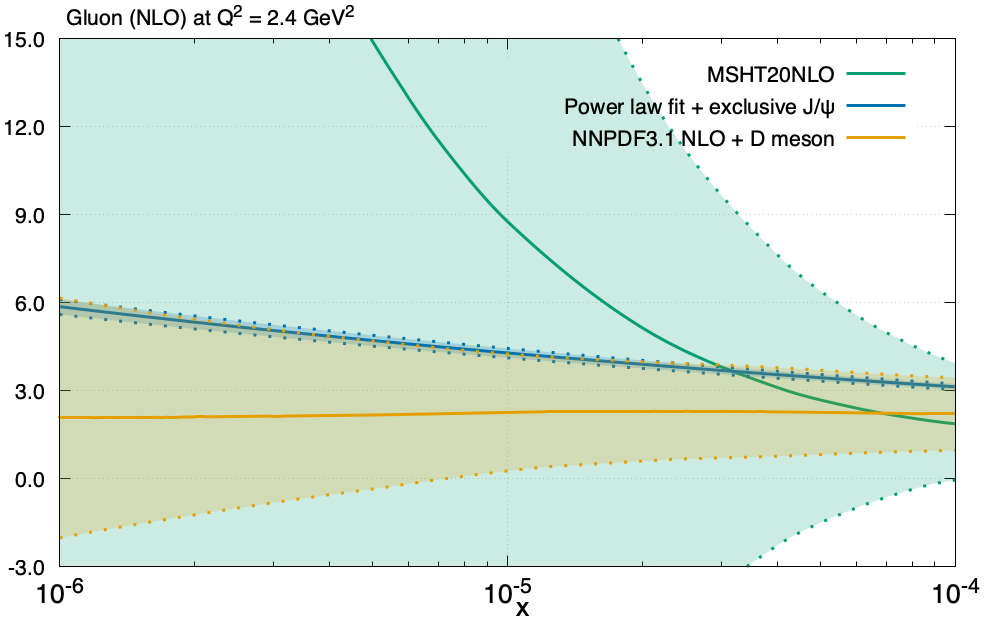}
\caption{\sf Comparison of the MMHT14 (left) and MSHT20 (right) NLO gluon PDFs at $Q^2=2.4~\GeV^2$ with a power--law fit to exclusive $J/\psi$ data~\cite{Flett:2020duk} and a NNPDF3.0 fit with the addition of LHCb $D$ meson data~\cite{Bertone:2018dse}.}
\label{fig:lowxg}
\end{center}
\end{figure}

Comparisons of the MMHT14 and MSHT20 NLO gluon PDFs at $Q^2=2.4~\GeV^2$ to the above extractions are shown in the left and right plots of Fig.~\ref{fig:lowxg}, respectively. We can first see that indeed the uncertainty on the global PDFs is very large in this region. The uncertainty resulting from the $D$ meson fit and in particular the exclusive $J/\psi$ fit is on the other hand much smaller. This reflects the significant potential constraining power of these data sets, though in the latter case the precise size of the uncertainty band will be driven by the more restrictive parameterisation of the power--law fit, and the manner of determining theoretical uncertainties is a potential source of uncertainty in both cases. The MSHT20 NLO gluon is somewhat lower at low $x$ than MMHT14, which in the latter case does not overlap entirely within uncertainties with the dedicated fits. However, given the lack of directly constraining data in this region in the fits, there is no strong reason to expect this improved agreement in the MSHT20 case, and clearly for either sets the PDF uncertainties are very large. A fit which additionally constrains by hand the gluon to be similar to that required by the $J/\psi$ analyses, by introducing high precision gluon pseudo-data at low $x$ and $Q^2$, finds that only a small deterioration in the global fit quality is required to obtain a gluon of the suggested form at NLO, as the agreement within the uncertainty bands in Fig.~\ref{fig:lowxg} (right) would suggest. However, imposing the same constraint on the gluon at NNLO leads to a deterioration of about $\Delta\chi^2 \sim 200$ in the fit quality to HERA data, implying that the NNLO corrections to the cross sections for these heavy flavour processes would have to be significant in order for them to be successfully incorporated in a global fit at NNLO.  
 
\section{Comparison of MSHT20 with other PDF sets\label{sec:11}}   

In this section we compare our PDFs with other representative available PDF sets, focussing on the most phenomenologically relevant NNLO case for brevity. The most direct comparison is with the CT18A~\cite{CT18} and NNPDF3.1~\cite{NNPDF3.1} sets, which both result from global fits that include a significant amount of LHC Run--I data, much of which we also fit, though we do fit more recent data sets not included in either of these. We will for brevity refer to these as the `global' sets in what follows.  

Both CT and NNPDF also use a GM--VFNS, which have been shown to converge with that used in our  analysis as the order increases~\cite{HERAbench}. There are nevertheless some  significant differences in the theoretical approaches. For example, NNPDF3.1 does not apply deuteron and heavy nuclear target corrections by default, though the effect has been discussed in subsequent publications~\cite{NNPDFnuc,Ball:2020xqw}, and NNPDF3.1 uses as default a fitted charm at the input scale rather than generating all heavy flavour perturbatively, as is done by ourselves and CT18. Moreover, the MSHT and NNPDF collaborations use quite a different procedure for the analysis.  The NNPDF collaboration combine a Monte Carlo representation of the probability measure in the space of PDFs, with the use of neural networks to give a set of unbiased input distributions. We instead use parameterisations of the input distributions based on Chebyshev polynomials, where the optimum order of the polynomials for the various PDFs has been explored in detail in the fit. 
As with MSHT20, CT18 use a tolerance criterion to evaluate the experimental PDF uncertainties. The CT18 fit is also based on a polynomial parameterisation (in their case Bernstein rather than Chebyshev), albeit with a rather more restrictive parameterisation in general, for example with respect to the $s - \bar s$  distribution (which is taken to be zero) and the light quarks in certain regimes. On the other hand, the CT18 tolerance criterion is now shown to be consistent with  taking account of the differences that can potentially arise from fitting with a range of modified input parameterisations -  i.e. the total uncertainty is, in some sense, designed in order to incorporate the contribution from parameterisation uncertainty. 

\begin{figure} [t]
\begin{center}
\includegraphics[scale=0.24]{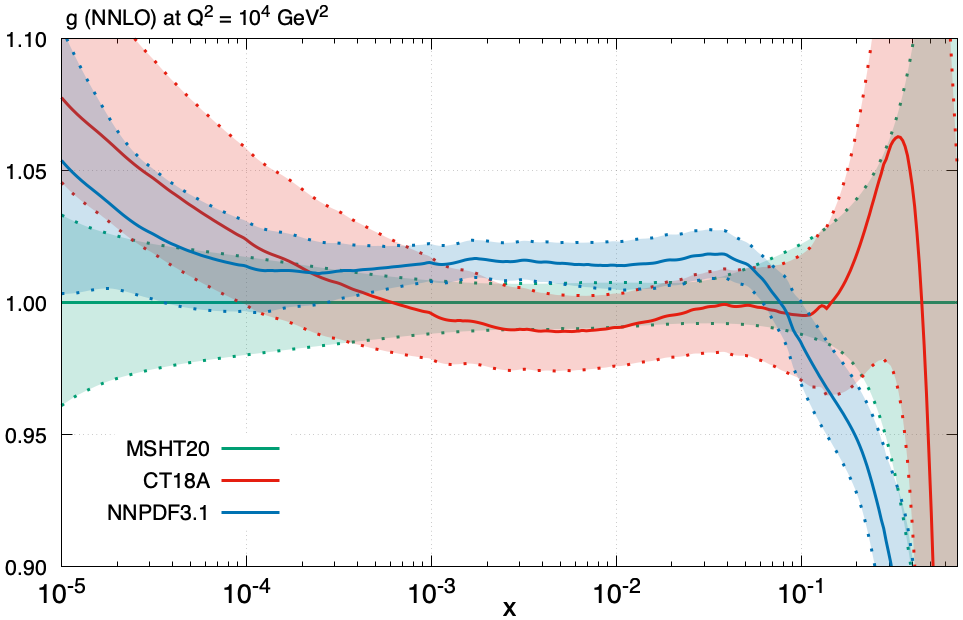}
\includegraphics[scale=0.24]{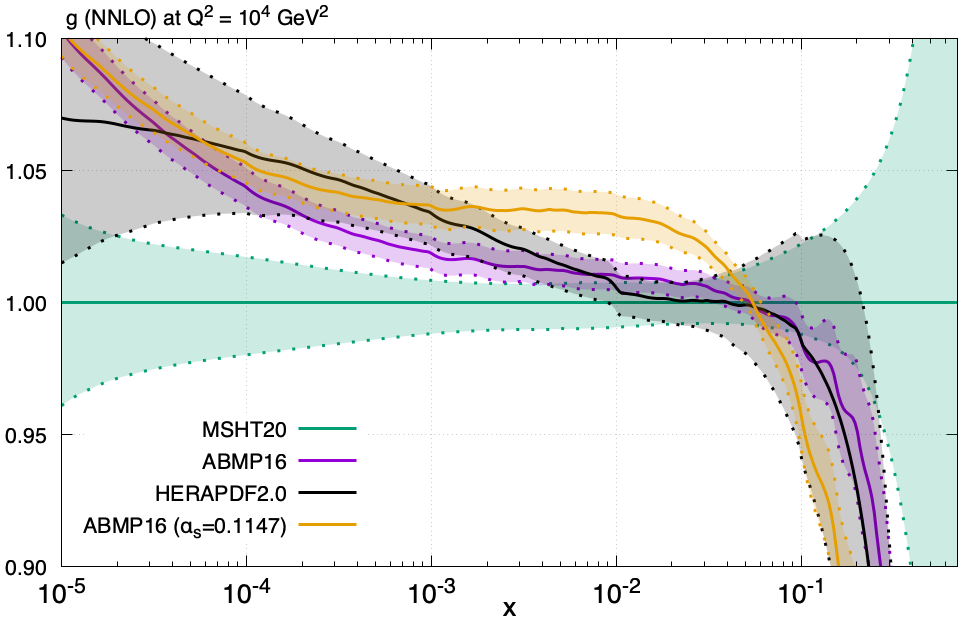}
\caption{\sf Comparison of the MSHT20 NNLO gluon PDF at $Q^2=10^4~\GeV^2$ with other representative PDF sets, presented as a ratio to the MSHT set. These correspond to $\alpha_S(M_Z^2)=0.118$ in all cases, though the set corresponding to the preferred value of  $\alpha_S(M_Z^2)=0.1147$ is also shown in the ABMP case.}
\label{fig:gluoncomp}
\end{center}
\end{figure}

\begin{figure} [t]
\begin{center}
\includegraphics[scale=0.24]{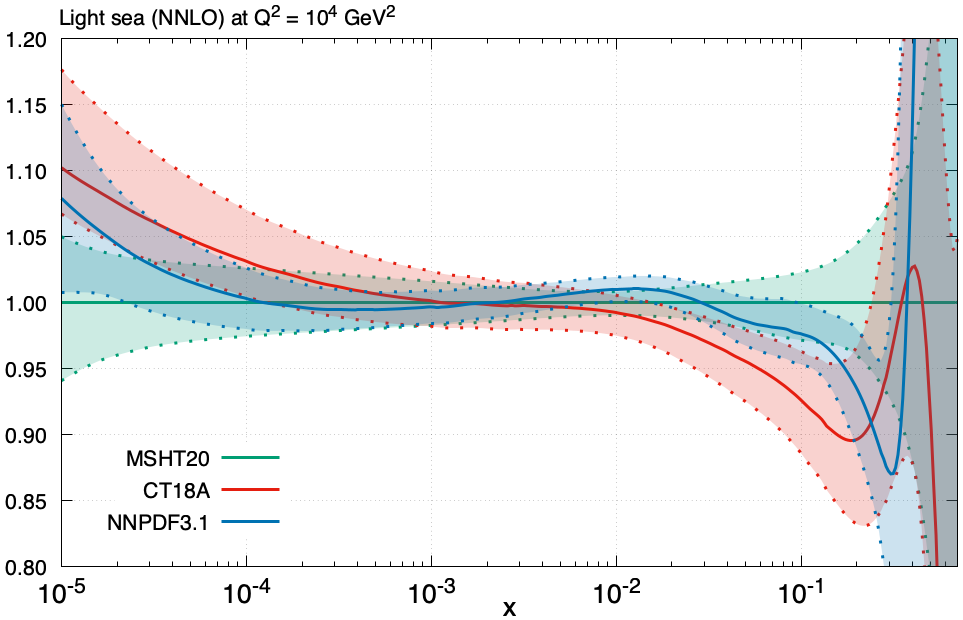}
\includegraphics[scale=0.24]{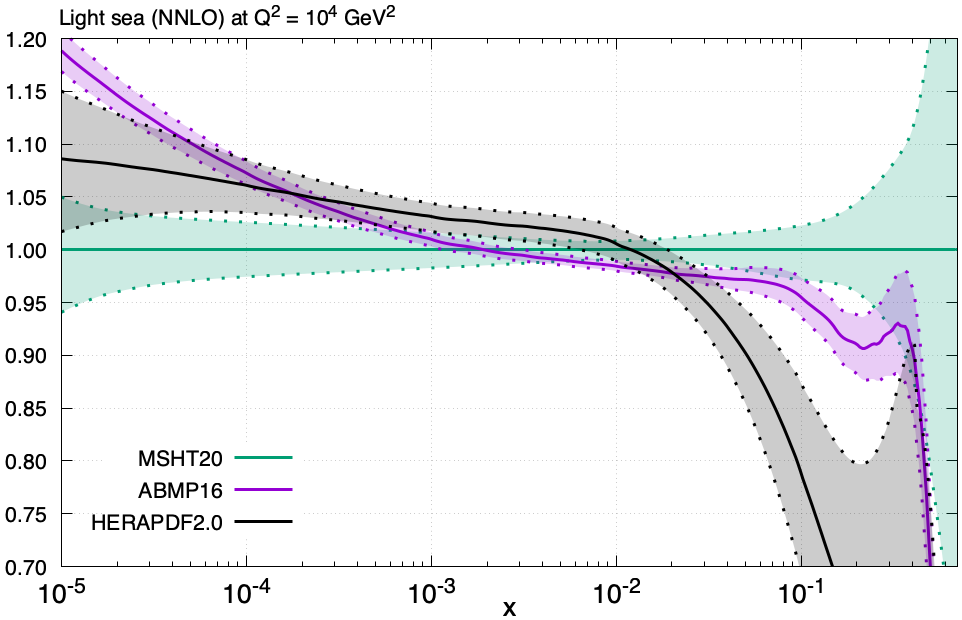}
\caption{\sf Comparison of the MSHT20 NNLO light sea PDF at $Q^2=10^4~\GeV^2$ with other representative PDF sets, presented as a ratio to the MSHT set. These correspond to $\alpha_S(M_Z^2)=0.118$ in all cases.}
\label{fig:lseacomp}
\end{center}
\end{figure}

We also compare against the ABMP16~\cite{ABMP16} and HERAPDF2.0~\cite{HERAcomb} sets. In the former case a global fit is performed, including some LHC data, though rather less than is included in the global sets above, and crucially a FFNS is used in the calculation of DIS cross sections. Also, rather lower $Q^2$ and $W^2$ cuts than other analyses are used in the fits to DIS data, with higher--twist corrections included, which have a significant impact in this additional region of parameter space. It is also the case that no deuteron DIS data are fit. The upshot of these differences in procedure is that a rather low value of $\alpha_S(M_Z^2)=0.1147$ is preferred by the fit, and the corresponding PDFs are often rather different, as we will see. A further point to note is the use of the $\Delta \chi^2=1$ criterion to evaluate the errors. This is also applied by the HERAPDF2.0, albeit with additional model and parameterisation uncertainties, which we include in our comparisons, following the prescription of~\cite{HERAcomb}. This set results from a fit to HERA DIS data alone, with a parametric form with many fewer free parameters than ours.

In all cases we consider the sets defined at $\alpha_S(M_Z^2)=0.118$. While this is not the value preferred by ABMP, we argue it will give the most direct comparison, given any observable quantities will often depend on the value of the coupling. However for the gluon (which depends on the strong coupling the most) we also show the ABMP result for their best fit $\alpha_S(M_Z^2)=0.1147$. For the CT set we take the `A' PDF, for which the ATLAS high precision $W$, $Z$ data at 7~TeV~\cite{ATLASWZ7f} are included, as these are fit in both the MSHT20 and NNPDF3.1 cases. 

\begin{figure}[t]
\begin{center}
\includegraphics[scale=0.24]{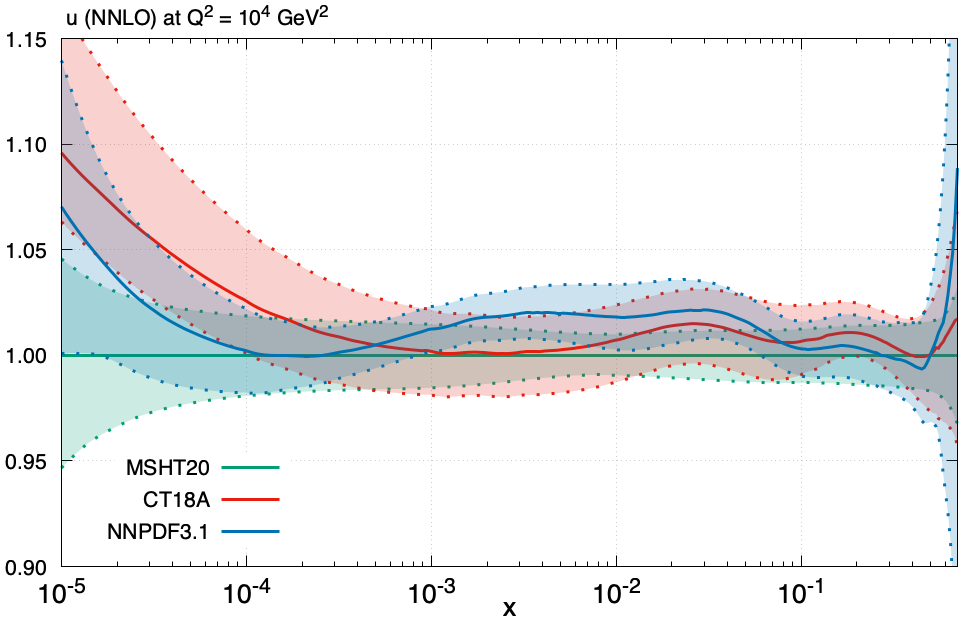}
\includegraphics[scale=0.24]{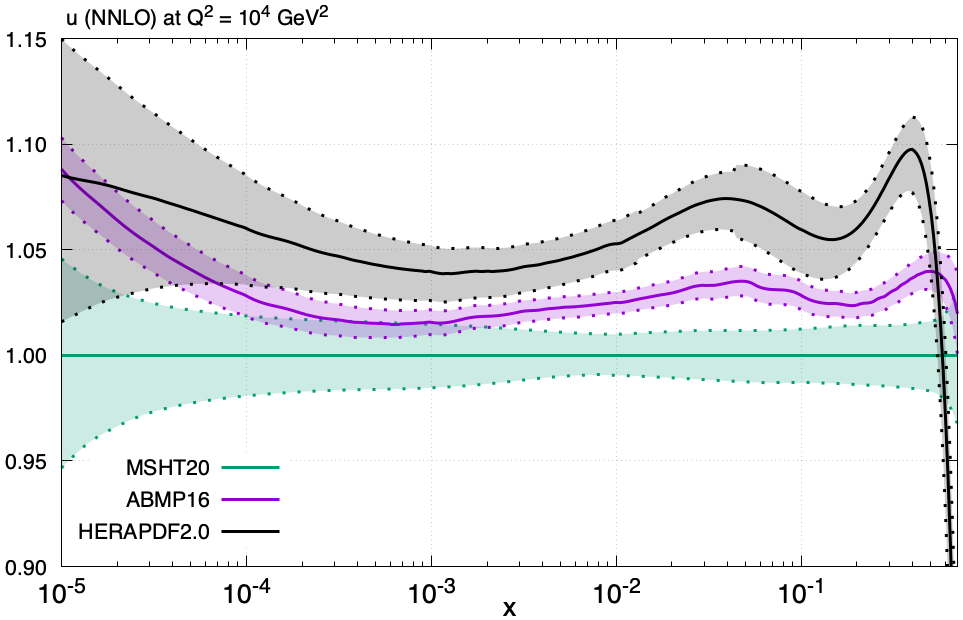}
\includegraphics[scale=0.24]{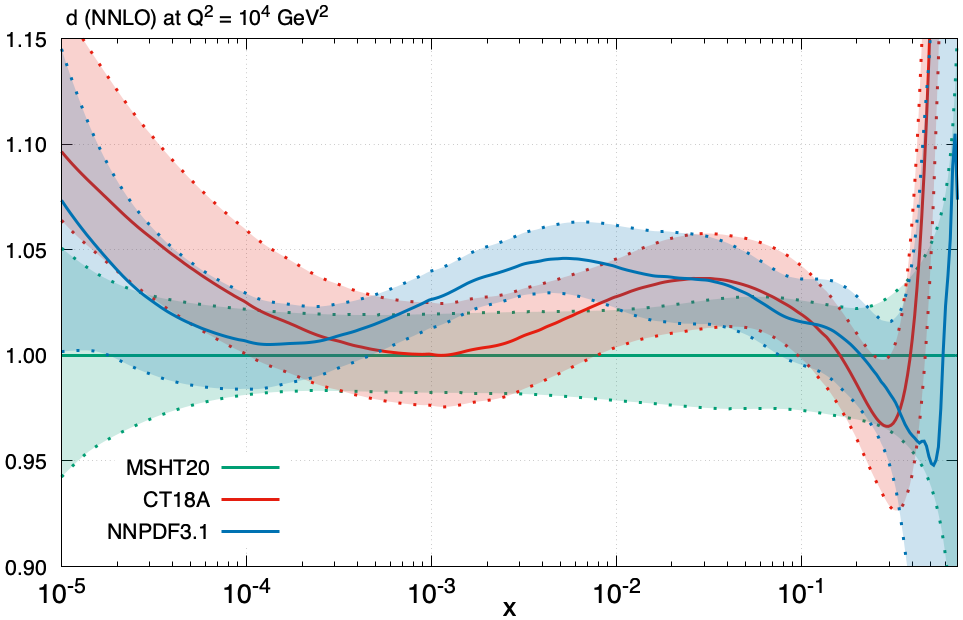}
\includegraphics[scale=0.24]{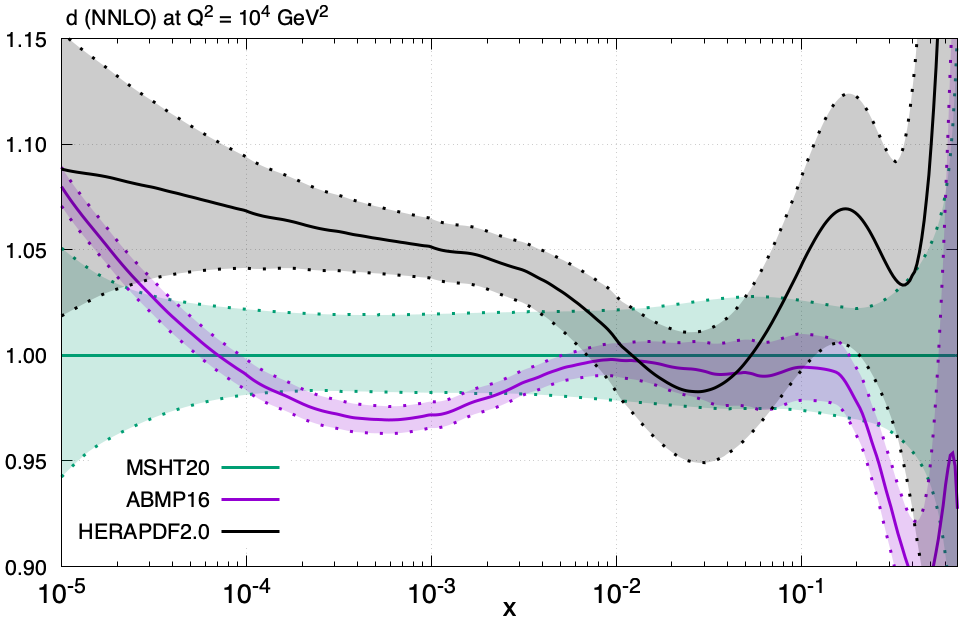}
\caption{\sf Comparison of the MSHT20 NNLO up and down quark PDFs at $Q^2=10^4~\GeV^2$ with other representative PDF sets, presented as a ratio to the MSHT set. These correspond to $\alpha_S(M_Z^2)=0.118$ in all cases.}
\label{fig:udcomp}
\end{center}
\end{figure}

We first show results for the gluon PDF in Fig.~\ref{fig:gluoncomp}. The PDF uncertainty for NNPDF3.1 is comparable in size to MSHT20, albeit a little lower at higher $x \gtrsim 0.1$, while the CT18 PDF uncertainty is larger than both of these. This difference  is also seen in the earlier CT14 set in comparison to NNPDF3.0 and MMHT14, which entered the PDF4LHC15 combination~\cite{PDF4LHC15}. We believe this larger CT18 uncertainty is due to a slightly more conservative tolerance criterion than used in our dynamical tolerance procedure. In all cases the corresponding updated PDF sets show a marked improvement in the error bands with respect to these older sets. However, in terms of the absolute values of the PDFs, we can see that the NNPDF3.1 gluon is systemically higher (lower) than  MSHT20 and CT18 at low to intermediate (high) $x$, even lying outside the corresponding uncertainty bands in some regions. This is in contrast to the earlier sets, where the agreement was rather better; thus we are in a slightly problematic situation that while the PDF uncertainties of the newer sets are improved with respect to the PDF4LHC15 sets, their spread has not necessarily decreased and has even increased in some regions. The cause of this enhancement at intermediate $x \sim 10^{-3}-10^{-2}$ is in part due to the fact that NNPDF3.1 now fit the charm quark PDF, see~\cite{NNPDF3.1} for a detailed discussion, while at larger $x \gtrsim 0.1$ the impact of new LHC data on the central value of the high $x$ gluon region appears to be somewhat larger in the NNPDF case, where it reduces from versions NNPDF3.0 to NNPDF3.1, in comparison to that seen in MSHT20 (see Fig.~\ref{gluon_q210000_ratio__percentageerrors_NNLO} (left)).  
At low $x$ the CT18 gluon is larger than MSHT20, outside their respective uncertainty bands. This effect was not seen as clearly in the PDF4LHC15 sets, but historically the CT/CTEQ gluons have been larger at very small $x$ due to the positive definite nature of their input gluon at the starting scale for evolution.  We should note, however, that the agreement between the MMHT14, NNPDF3.0 and CT14 gluons in the PDF4LHC15 combination was rather better than one might expect, and indeed better than any quark or antiquark distribution, and may be regarded as at least partially coincidental. 

For the ABMP16 set we can see that the agreement is  poor for the default value of $\alpha_S(M_Z^2)=0.1147$, with the gluon lying well outside the quoted uncertainty bands across a wide range of $x$. This is in part due to the differing value of $\alpha_S$, however we can see that while taking a consistent value of $\alpha_S(M_Z^2)=0.118$ does improve the agreement, it is still not good. This is a known result from previous comparisons, and in~\cite{Thorne:2014toa} it was argued that this is primarily due to their use of a FFNS as opposed to our use of a GM-VFNS for DIS data. The uncertainty for the gluon and all other PDF sets shown is smaller than ours, due to the use of a simple $\Delta \chi^2 =1$ criterion for the calculation of PDF uncertainties. Interestingly, the ABMP16 gluon with $\alpha_S(M_Z^2)=0.118$ is rather more similar to the NNPDF3.1 case, though given the significant differences in their methodologies and data included in the fit, this is possibly largely coincidental. In the HERAPDF2.0 case the gluon is lower (higher) at high (low) $x$ than MSHT20, lying outside the quoted uncertainty bands.

In Fig.~\ref{fig:lseacomp} the result for light quark sea $S=2(\overline{u} + \overline{d}) + s+ \overline{s}$ is shown, and the agreement between the global sets is reasonable, though not entirely within errors, and with some reduction (enhancement) at high (low) $x$ for CT18 relative to MSHT20. This combination however masks some of the true differences seen in the quark flavour decomposition, as demonstrated in Fig.~\ref{fig:udcomp}, where the up and down quark PDFs are shown. As for the gluon, there is some enhancement in the NNPDF3.1 up and in particular down quarks PDFs at intermediate $x \sim 10^{-3}-10^{-2}$, with the latter lying outside uncertainty bands. This is in contrast to the PDF4LHC sets, where the agreement was better, albeit with the NNPDF3.0 up quark PDF actually lying below the MMHT14 PDF across a wide range of $x$. A major factor in this difference can again be traced to the updated methodology in the NNPDF3.1 case, which is seen in~\cite{NNPDF3.1} to enhance the up and down in this region, due to some extent but not entirely to their fitting of charm. The trend for the $\overline{u}$ and $\overline{d}$ (not shown) is rather similar, while as we will see below the strangeness in the NNPDF case is rather lower, compensating the contribution from this enhancement to the light sea. As for the gluon, at low $x$ all three quark distributions are larger in the CT18 case in comparison to MSHT, lying outside the uncertainty bands. In general, again as for the gluon the MSHT20 and NNPDF3.1 uncertainty bands are rather comparable in size, with CT18 being somewhat larger, though this depends on the precise $x$ region being considered.

In the ABMP16 and HERAPDF2.0 cases, we can see that level of agreement with MSHT is rather poor. The HERAPDF2.0 light sea, up and down all lie above MSHT20 at low to intermediate $x$, outside the error bands, while the up quark enhancement persists to high $x$. A rather similar, though generally less pronounced, trend is seen in the ABMP16 case for the light sea and the up, while the down is in somewhat better agreement, though still not within errors across the whole $x$ region. An enhancement of the ABM-type up quark was also 
noticed and commented on in ~\cite{Thorne:2014toa}. The effect was deduced to be mainly due to the use of the FFNS scheme, with somewhat slower evolution of the charm over much of the $x$ range being compensated for by an increase in the up-quark in fitting inclusive DIS data, but at highest $x$ the use of the lower
$Q^2$ and $W^2$ cuts was also seen to play some role. 

\begin{figure}[t]
\begin{center}
\includegraphics[scale=0.24]{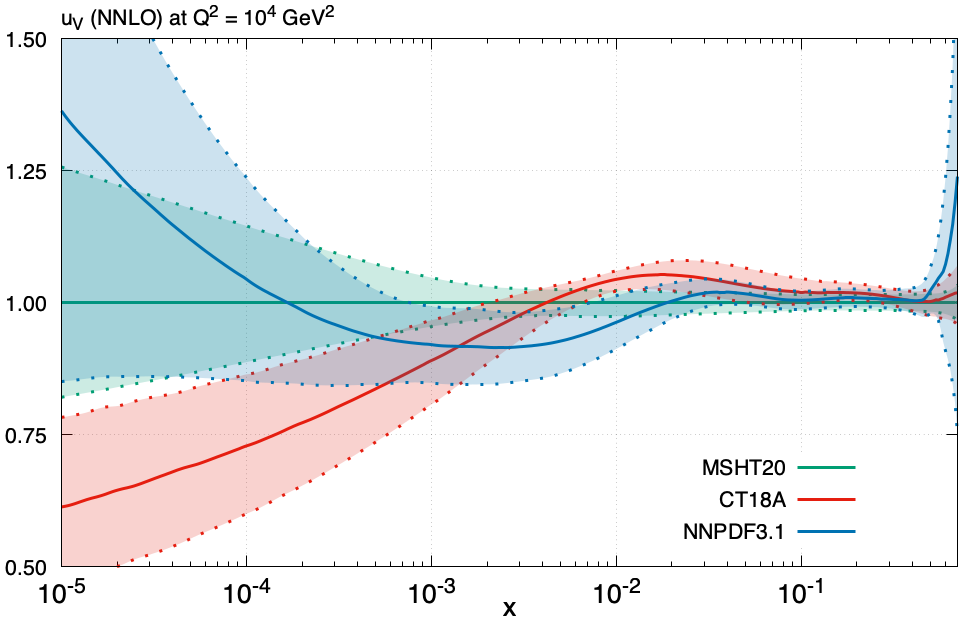}
\includegraphics[scale=0.24]{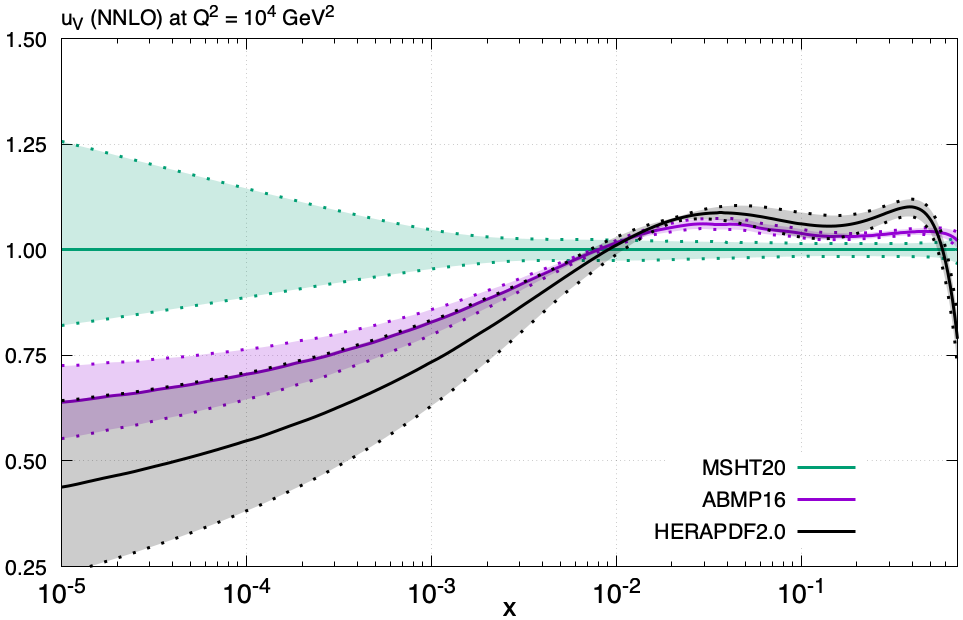}
\includegraphics[scale=0.24]{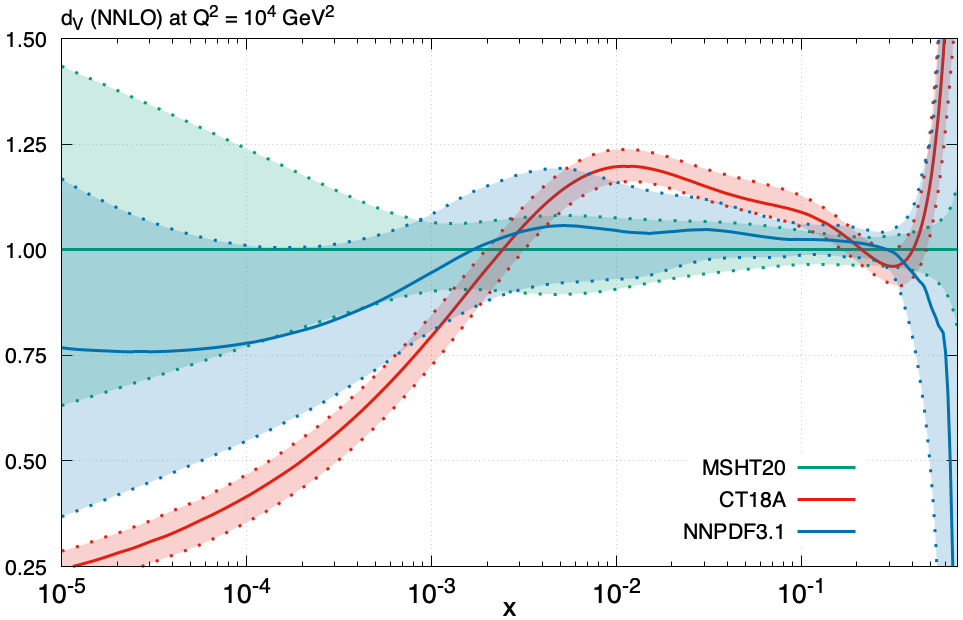}
\includegraphics[scale=0.24]{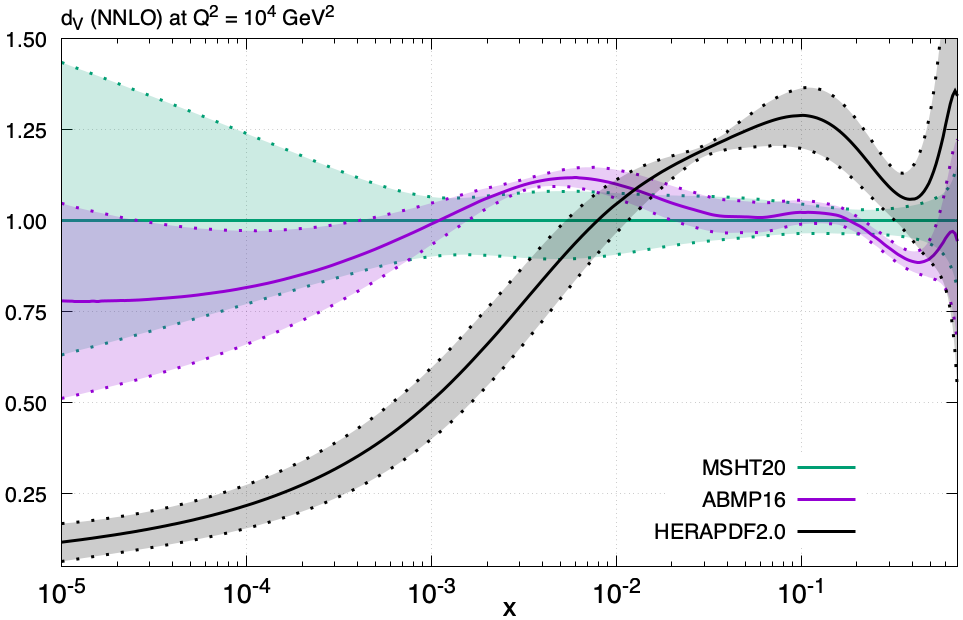}
\caption{\sf Comparison of the MSHT20 NNLO up and down quark valence PDFs at $Q^2=10^4~\GeV^2$ with other representative PDF sets, presented as a ratio to the MSHT set. These correspond to $\alpha_S(M_Z^2)=0.118$ in all cases.}
\label{fig:udvcomp}
\end{center}
\end{figure}

The up and down valence distributions are shown in Fig.~\ref{fig:udvcomp}. The agreement between NNPDF3.1 and MSHT20 is in both cases rather good across all $x$. This is a rather interesting development, as previously~\cite{MMHT14} both the up and down valence were found to be rather larger in the NNPDF3.0 case in comparison to MMHT14. This is probably due to a combination of both new LHC data placing some constraint in this region on both sets and hence tending to bring them together, as well as the more flexible parameterisation now used in MSHT. Indeed, the difference between the CT18 and MSHT20 down valence is rather striking in the intermediate $x$ region, and closely follows the trend seen in Fig.~\ref{dnvratio_q210000_NNLOoldparamtonewparam} when comparing our fit to the MSHT20 data set, but using the previous less flexible parameterisation. This may suggest some issues with parameterisation flexibility in the CT18 case, though in~\cite{CT18} it is claimed that the PDF uncertainty accounts for this source of uncertainty. It is however noticeable that the uncertainty on the CT18 down valence distribution is smaller in comparison to MSHT20 and NNPDF3.1, in contrast to most other cases. The NNPDF3.1 uncertainty is generally rather larger at low to intermediate $x$ in both cases, perhaps due to the greater parameterisation flexibility there. For the ABMP16 and HERAPDF2.0 sets, we can see that the agreement is poor for the up valence, with both distributions undershooting with respect to MSHT20 at low $x$ and overshooting at intermediate to high $x$, often well outside uncertainty bands. For the down valence, the agreement between HERAPDF2.0 and MSHT20 is equally poor, while for ABMP16 it is somewhat better. In the former case the PDF in the intermediate $x$ region is rather similar to CT18, indicating the role of parameterisation flexibility here and also the lack of LHC data in the fit.

\begin{figure} 
\begin{center}
\includegraphics[scale=0.24]{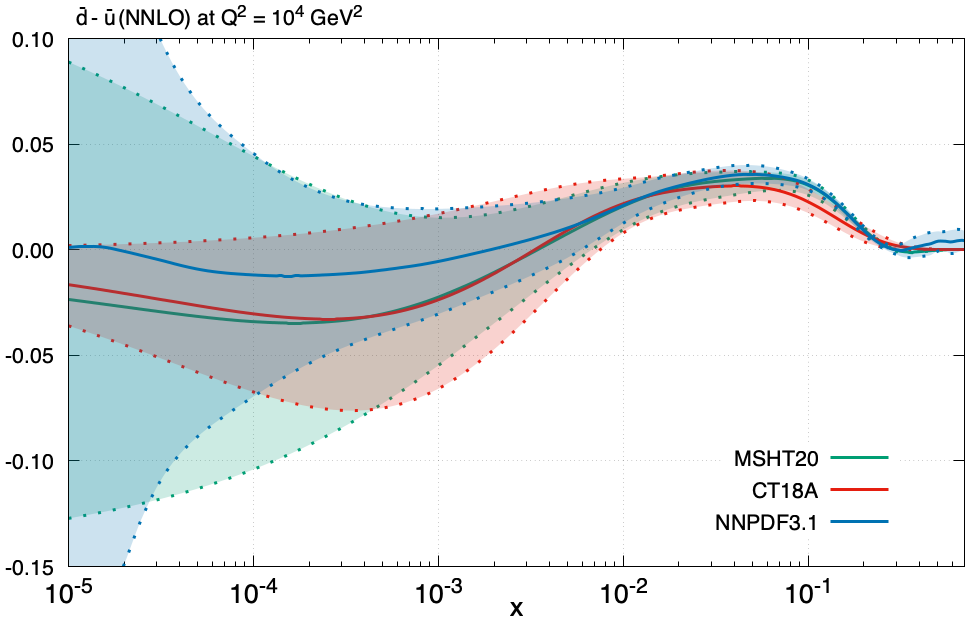}
\includegraphics[scale=0.24]{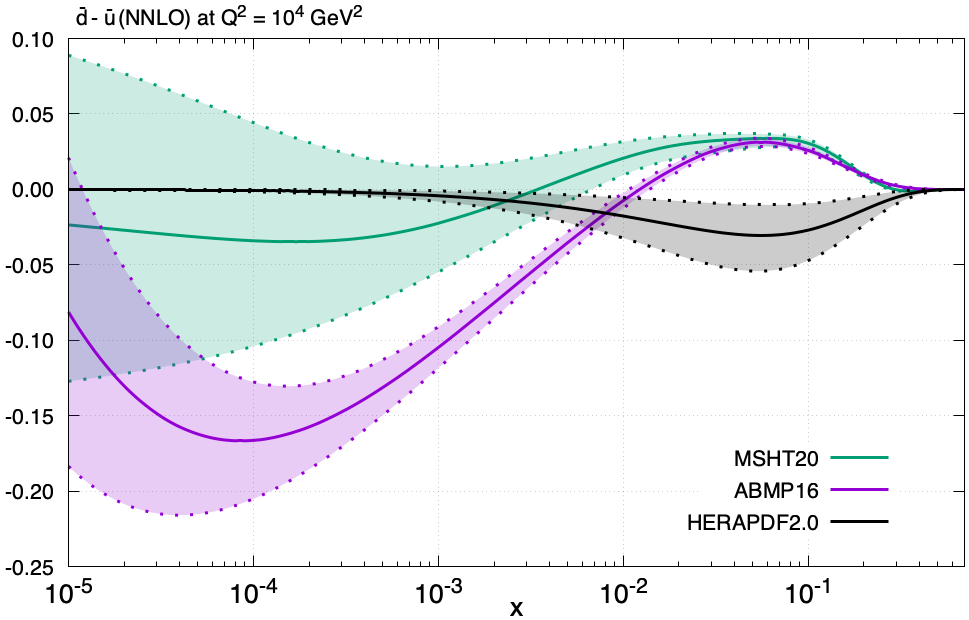}
\caption{\sf Comparison of the MSHT20 NNLO $\overline{d}-\overline{u}$ PDF at $Q^2=10^4~\GeV^2$ with other representative PDF sets. These correspond to $\alpha_S(M_Z^2)=0.118$ in all cases.}
\label{fig:dbmubcomp}
\end{center}
\end{figure}

The $\overline{d}-\overline{u}$ distribution is shown in Fig.~\ref{fig:dbmubcomp}. The global sets are broadly in agreement across the entire $x$ region, while the CT18 uncertainty is somewhat larger at intermediate $x \sim 0.01-0.1$ and lower at low $x$, than MSHT20 and NNPDF3.1, perhaps due to parameterisation. The PDF uncertainties in the NNPDF3.1 case at high $x$ are clearly larger, due to the greater parameterisation flexibility in this less constrained region, and interestingly the central value somewhat higher than the other sets. The ABMP16 distribution displays a strong preference to be negative in the $x \sim 10^{-4}-10^{-2}$ region, and lies below the MSHT20 distribution, outside uncertainty bands. We emphasise here that the parameterisation is sufficiently flexible to allow the MSHT20 distribution to also be negative in this region, and indeed it is consistent with this within error bands, but that the fit does not prefer such a negative distribution. Indeed, we note that some of the fits where we remove some data sets can move further in the direction of the ABMP16 distribution, see e.g. 
Fig.~\ref{dbarminusubar_q210000_NNLOnoATDY_extranoLHCb}. This is consistent with both NNPDF3.1 and CT18, which are also in disagreement with ABMP16 here. For HERAPDF2.0 the parameterisation at low $x$ is fixed to give zero, while at intermediate $x$ a completely opposite behaviour to all other PDFs is preferred, with the distribution being negative.
This highlights the fact that HERA data have no effective constraint on this distribution, and an improvement is noted when also including LHC data, see \cite{Sutton:2019pug}.  

\begin{figure} 
\begin{center}
\includegraphics[scale=0.24]{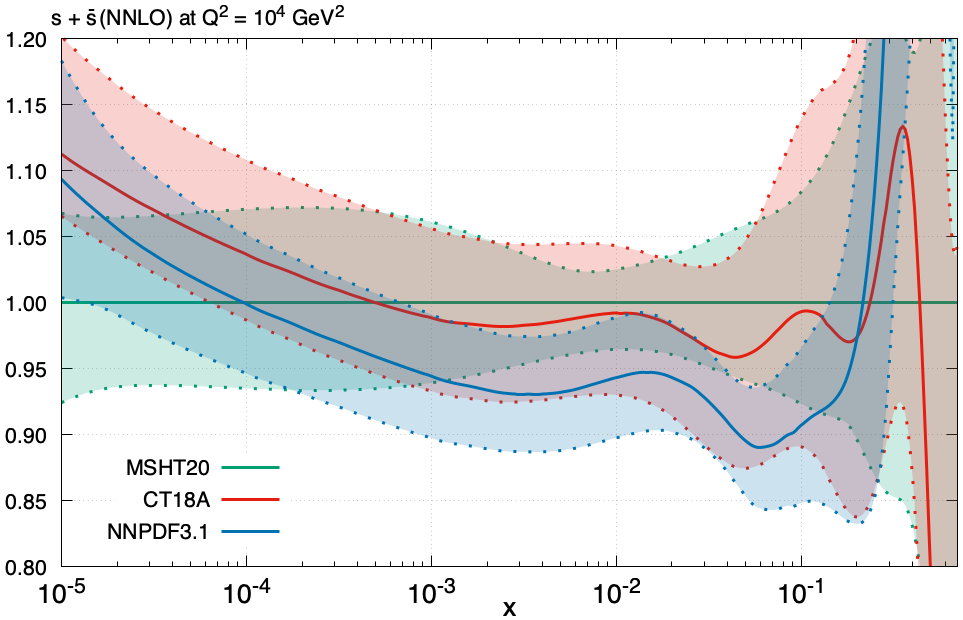}
\includegraphics[scale=0.24]{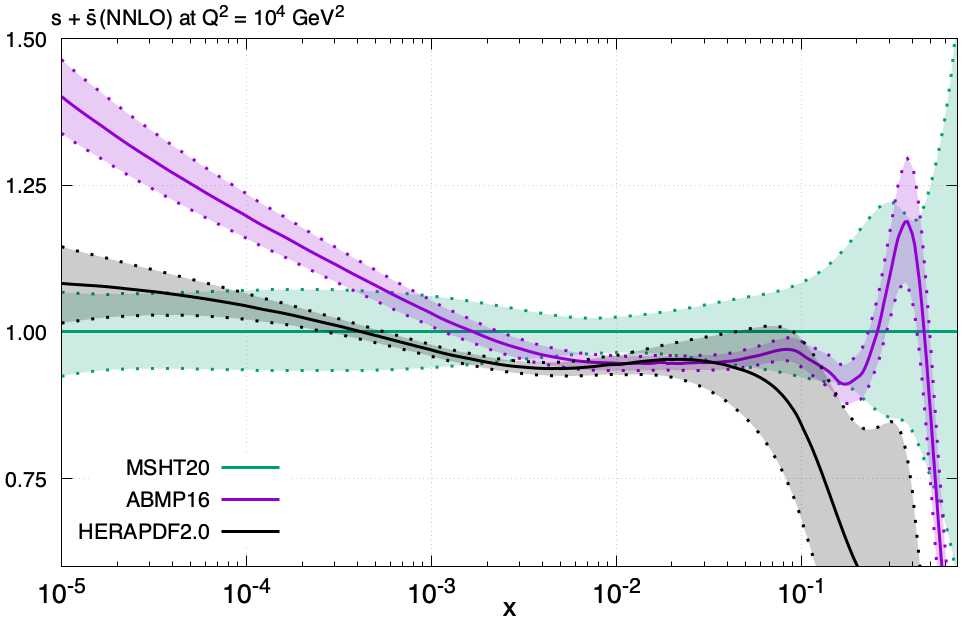}
\caption{\sf Comparison of the MSHT20 NNLO $s + \overline{s}$ PDF at $Q^2=10^4~\GeV^2$ with other representative PDF sets, presented as a ratio to the MSHT set. These correspond to $\alpha_S(M_Z^2)=0.118$ in all cases.}
\label{fig:spsbarcomp}
\end{center}
\end{figure}

We now consider the strangeness, shown in Fig.~\ref{fig:spsbarcomp}. The global sets are broadly in agreement and indeed in all cases have higher strangeness, with smaller PDF uncertainties, in comparison to the PDF4LHC sets, due to the impact of new LHC data and in particular the ATLAS 7~TeV high precision $W$, $Z$ data. However, in more detail we can see that NNPDF3.1 and to a lesser extent CT18 lie below the MSHT20 distribution in the $x\sim 10^{-3} - 10^{-1}$ region, in the former case even slightly outside the PDF uncertainty band. From Fig.~\ref{MSHT20_NNLO_noATDY} (bottom) we have seen that the ATLAS 8~TeV data on $W$ and $Z$ production also prefer a larger strangeness, consistent with the 7~TeV, and that when both are fit together the impact is more significant than when just the 7~TeV is fit. Given NNPDF3.1 and CT18 do not include these data sets, this may well explain the difference. In addition, these analyses use NLO theory for the dimuon cross sections, which may have some impact, in particular given the $D\to \mu$ branching ratio is fixed in these cases. The ABMP16 and HERAPDF2.0 strangeness distribution lie below MSHT20 in the $x\sim 0.01$ region, though are still roughly consistent within uncertainties, while at low $x$ ABMP16 lies above MSHT20, and at high $x$ HERAPDF2.0 lies below MSHT20. We note that ABMP16 are reasonably consistent with MSHT20 in the region where there are direct data constraints on the strange distribution, while at very low $x$ the difference is driven by that in the gluon distribution, which in turn drives the strange quark evolution in this region.  

\begin{figure} 
\begin{center}
\includegraphics[scale=0.24]{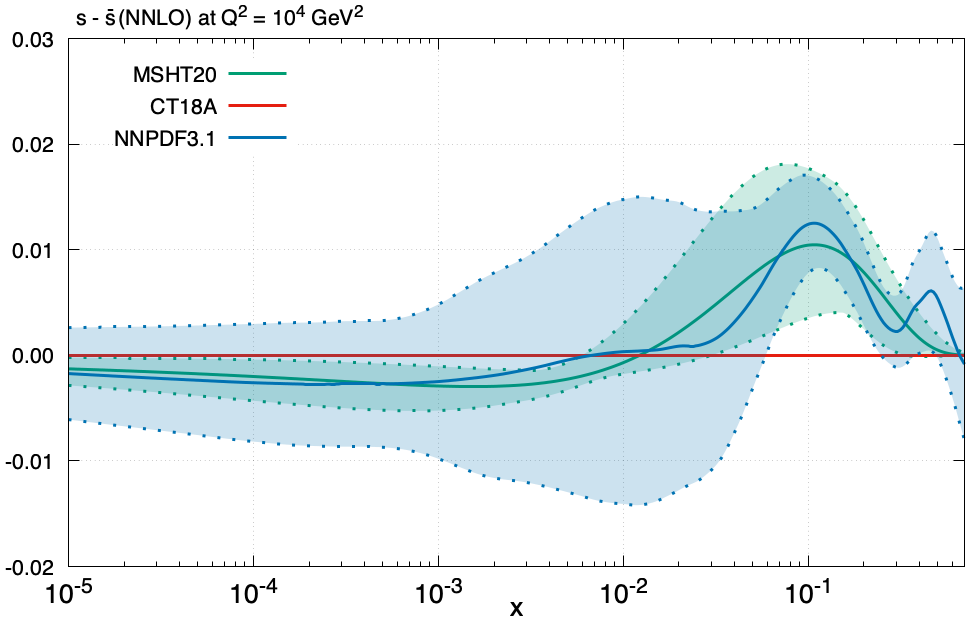}
\includegraphics[scale=0.24]{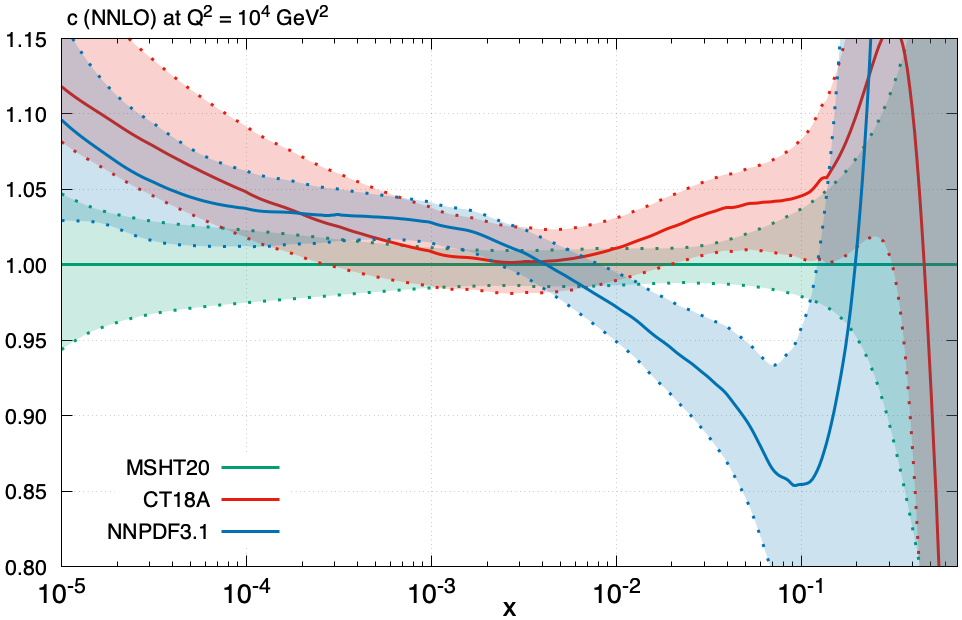}
\caption{\sf Comparison of the MSHT20 NNLO $s - \overline{s}$ and charm PDFs at $Q^2=10^4~\GeV^2$ with other representative PDF sets. These correspond to $\alpha_S(M_Z^2)=0.118$ in all cases. The charm case is presented as a ratio to the MSHT set while for the $s - \overline{s}$ the PDF values are shown.}
\label{fig:smsbar_charm_comp}
\end{center}
\end{figure}

Next, we consider the strangeness asymmetry for the global PDF sets, shown in Fig.~\ref{fig:smsbar_charm_comp} (left). We can see that NNPDF3.1 and MSHT20 are consistent across the entire $x$ region, and in particular show a clear preference for a non--zero and positive asymmetry in the $x\sim 0.1$ region. Interestingly, while the NNPDF3.1 uncertainties are larger in the unconstrained low $x$ region, in this $x\sim 0.1$ region they are actually somewhat smaller than MSHT20. Generally the NNPDF3.1 central value undergoes somewhat more variation at higher $x$. The CT18 asymmetry is fixed to be zero by construction, as is evident in the plot, though we note that at NNLO DGLAP evolution itself will generate an asymmetry \cite{Catani:2004nc}. We do not show the corresponding ABMP16 and HERAPDF2.0 distributions, as these are similarly constrained to be zero at input.

We show the charm quark PDF in Fig.~\ref{fig:smsbar_charm_comp} (right) for the global sets. These broadly follow the trend of the gluon, but the NNPDF3.1 PDF at $x \gtrsim 0.01$ lies significantly below MSHT20 and even more so relative to CT18, and with rather larger uncertainties. To some extent this is due to the lower gluon PDF at high $x$ observed above, and its contribution to the charm through evolution, but in addition the effect of fitting charm is seen in~\cite{NNPDF3.1} to reduce the charm PDF in this region.

\begin{figure} 
\begin{center}
\includegraphics[scale=0.24]{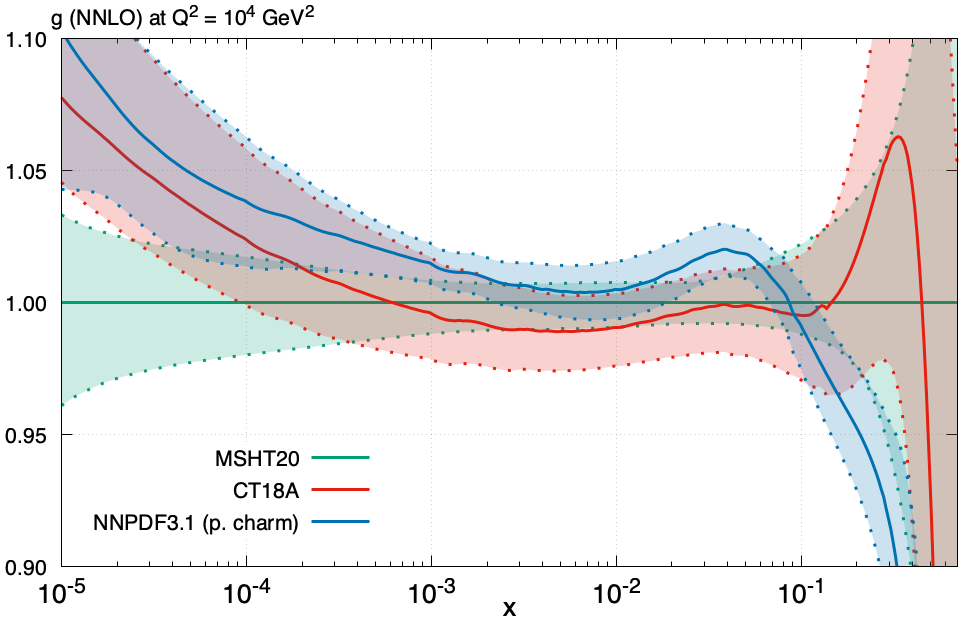}
\includegraphics[scale=0.24]{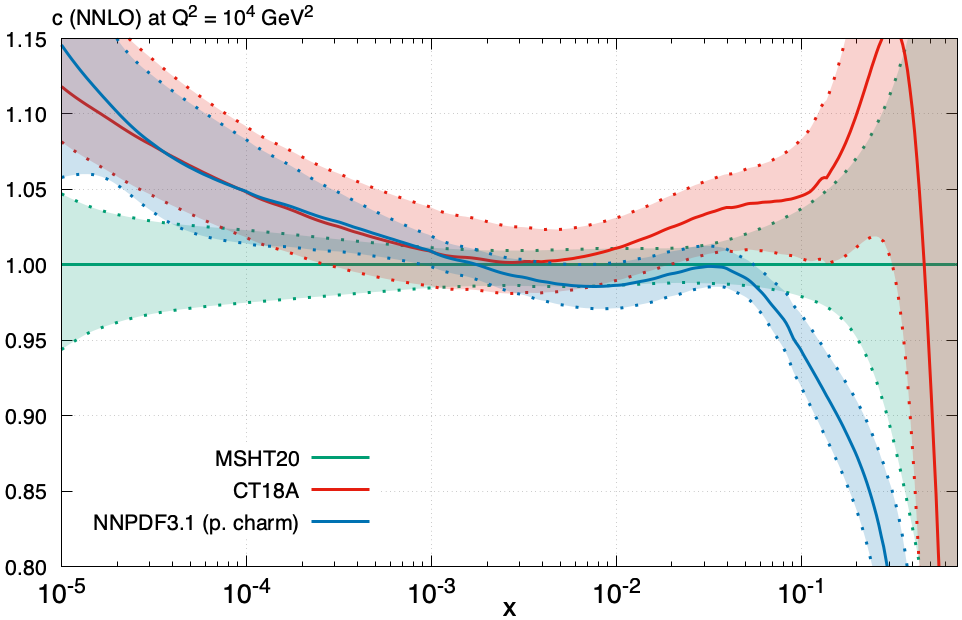}
\caption{\sf As in Figs.~\ref{fig:gluoncomp} (left) and ~\ref{fig:smsbar_charm_comp} (right), but with the NNPDF3.1 set corresponding to perturbative charm shown.}
\label{fig:pcharm_comp}
\end{center}
\end{figure}

Finally, as we have discussed the role that the fitting of charm appears to play in some regions in increasing the differences seen between NNPDF3.1 and MSHT20/CT18, we show in Fig.~\ref{fig:pcharm_comp} the same comparisons as in Figs.~\ref{fig:gluoncomp} (left) and ~\ref{fig:smsbar_charm_comp} (right), for the gluon and charm quark PDFs, but now replacing the NNPDF3.1 default set with that corresponding to perturbative charm. For the gluon, we can see that the agreement at intermediate $x \sim 10^{-3}-10^{-2}$ is indeed improved, as expected from the discussion above. The agreement at high $x$ is also somewhat better, though the NNPDF gluon remains lower than the other sets. On the other hand, at low $x$ we can see that the NNPDF gluon now lies above MSHT20, outside error bands, consistent with CT18 and reflecting the default NNPDF gluon in this region. For the charm PDF we can see that clearly the PDF uncertainty on the charm is greatly reduced, in particular at high $x$, as we expect. As with the gluon the low $x$ distribution lies somewhat above MSHT20, consistent with CT18. More significantly, at high $x$ the NNPDF 
 charm PDF lies below CT18 and MSHT20, outside of error bands, as a result of feed down from the gluon at high $x$, which as mentioned above follows a similar, though milder, trend.
The charm distributions do not mirror the gluon quite as closely as they might since the 
values of charm mass are different, with pole masses $m_c=1.4~\GeV$ for MSHT20 but 
$m_c=1.3~\GeV$ for CT18 and $m_c=1.51~\GeV$ for NNPDF3.1. This automatically raises the
charm distribution for CT18 a little compared to MSHT20 and lowers it for NNPDF3.1.

\section{Availability of MSHT20 PDFs   \label{sec:access}}

We provide the MSHT20 PDFs in the \texttt{LHAPDF} format~\cite{Buckley:2014ana}:
\\
\\
\href{http://lhapdf.hepforge.org/}{\texttt{http://lhapdf.hepforge.org/}}
\\
\\
as well as on the repository:
\\
\\ 
\href{http://www.hep.ucl.ac.uk/msht/}{\texttt{http://www.hep.ucl.ac.uk/msht/}}
\\
\\
We present NLO and NNLO eigenvector sets of PDFs at the default value of $\alpha_S(M_Z^2)=0.118$:
\\
\\
\href{http://www.hep.ucl.ac.uk/msht/Grids/MSHT20nnlo_as118.tar.gz}{\texttt{MSHT20nnlo\_as118}}\\
\href{http://www.hep.ucl.ac.uk/msht/Grids/MSHT20nlo_as118.tar.gz}{\texttt{MSHT20nlo\_as118}}
\\
\\
as well as the NLO set at the best fit value of $\alpha_S(M_Z^2)=0.120$ and a LO set at $\alpha_S(M_Z^2)=0.130$:
\\
\\
\href{http://www.hep.ucl.ac.uk/msht/Grids/MSHT20nlo_as120.tar.gz}{\texttt{MSHT20nlo\_as120}}\\
\href{http://www.hep.ucl.ac.uk/msht/Grids/MSHT20lo_as130.tar.gz}{\texttt{MSHT20lo\_as130}}
\\
\\
In addition to the above, we provide best fit NLO and NNLO sets for a short range of $\alpha_S(M_Z^2)$ values:
\\
\\
\href{http://www.hep.ucl.ac.uk/msht/Grids/MSHT20nlo_as_smallrange.tar.gz}{\texttt{MSHT20nlo\_as\_smallrange}}\\
\href{http://www.hep.ucl.ac.uk/msht/Grids/MSHT20nnlo_as_smallrange.tar.gz}{\texttt{MSHT20nnlo\_as\_smallrange}}
\\
\\
in order to allow the $\alpha_s$ + PDF uncertainty to be evaluated, according to the procedure outlined in~\cite{asuncertainty}. In particular, PDFs from $\alpha_S(M_Z^2)=0.115$ to 0.121 (0.122) at NNLO (NLO) are provided, in increments of 0.001.

\section{Conclusions \label{sec:12}}

We have performed fits to the available global hard scattering data in order 
to determine the PDFs of the 
proton at NLO and NNLO, as well as at LO. These PDF sets, denoted MSHT20, supersede the MMHT14 
sets \cite{MMHT14}, that were obtained using a similar framework. They include a very significant 
update in terms of data included, and also a number of important improvements in the 
framework of the PDF fit and in terms of more precise cross section calculations for a number of
processes. The MSHT20 PDF sets may be accessed, as functions of $x,Q^2$ in computer retrievable 
form, as described in Section \ref{sec:access}. 

First we summarize the improvements to the theoretical framework. As in MMHT14 we base the 
parameterisation of the input distributions on Chebyshev polynomials. Previously we typically used
4 polynomials for each PDF. However, it was shown in \cite{MMSTWW} that in order to 
be able to obtain comfortably sub-percent precision 6 polynomials are needed. Hence, with the
exception of the still poorly constrained strange asymmetry, our parameterisations for each set 
include 6 Chebyshev polynomials (or equivalent for the gluon). In particular, we have a
much improved parameterisation for the $\bar d, \bar u$ difference and now parameterise the ratio
directly rather than $\bar d -\bar u$. Hence, we 
use many more free parameters than in MMHT, 52 in total, and this results in an improvement 
of around 75 units in $\chi^2$ at NNLO. As there is still some redundancy in parameters, these are not all allowed to be free in the PDF error evaluation, and hence we have 32 eigenvector pairs. This is to be compared with 25 in MMHT14. Our treatment of deuteron and heavy nucleus corrections are the same 
as in MMHT14, and the deuteron correction obtained from the fit is similar to before. Our treatment of 
correlated systematic uncertainties is the same as in MMHT14, but we now additionally include 
information on statistical correlations. Our 
treatment of heavy flavour via a general-mass variable flavour scheme is identical to MMHT14, 
i.e. we use the ``optimal'' version of the Thorne-Roberts GM-VFNS \cite{Thorne}. We do now, however, 
use full NNLO cross sections for the NNLO fit to dimuon data in neutrino DIS \cite{NNLOdimuon}, and for all LHC 
inclusive jet data \cite{NNLOjets}, as well as for all the new types of collider processes in our analysis 
(with the exception of the small amount of $W+c$ data \cite{CMS7Wpc}, where the NNLO calculation is only now 
becoming available). 

Second, we briefly discuss the new data sets included, and their effects on the PDFs.
We are now able to use the final combined H1 and 
ZEUS Runs I + II HERA data for the neutral and charged current \cite{HERAcomb}, and for the heavy flavour cross sections \cite{HERAhf}. 
Neither of these has a huge impact compared to previous similar data, but they do make slight changes to the 
best-fit PDFs and a little more so to the uncertainties. We include final Tevatron data, in particular the
$W$-asymmetry data from D{\O} \cite{D0Wasym}. This has a large impact on both the central value and uncertainty of the high-$x$ down
quark, and also now have a surprisingly large effect on related PDFs. We also have very much more precise 
LHC data: for $W,Z$ rapidity distributions from ATLAS, CMS and LHCb; on high and low mass Drell-Yan production;  
for inclusive top-quark-pair production; and on inclusive jet production. The $W,Z$ data has a very significant impact on
quark flavour separation, in particular the strange quark and antiquark, and on the valence quarks at small $x$. The 
jet data adds constraints to the high-$x$ gluon. We also include a large amount of data of types that were not included in 
the MMHT14 analysis. This includes single and double differential data on top pair production, and also on the $p_T$ 
distribution of $Z$ bosons. Both of these constrain the high-$x$ gluon, but there is some tension between them, as well 
as tensions with jet data and older DIS data. We include new data on $W +$ jets and $W +$ charm jets, the latter 
providing essentially a consistency check on the strange quark, obtained as a compromise between that obtained from
inclusive $W,Z$ data and dimuon data. Overall, it is still the case that the constraints on PDFs come from  a very wide 
variety of data sets, both old and new. For the first time we make a detailed investigation into which data sets are 
most responsible for determining the central PDFs, i.e. the best fit, and also from this central point, which PDFs 
have most constraint on the uncertainty eigenvectors. We find that the best fit is, unsurprisingly, largely driven 
by those data sets with a large number of data points, particularly if the measurements are precise, e.g the 
inclusive HERA data. Then, it turns out that eigenvectors are often constrained by smaller data sets which 
are often not fit optimally within the best fit, due to their small weight, and which can therefore deteriorate quite 
quickly in certain directions away from the best fit.
Nonetheless, in most cases there are multiple data sets that place rather similar constraints on these eigenvectors, highlighting the constraining power and relative stability of a global PDF fit.
Compared to MMHT14 we find that LHC data are providing many more important constraints, 
these still being most clearly evident for quark flavour decomposition.

Some LHC data are not included in the present fits; for example single top quark production data and 
data on low scale heavy quark production at colliders. These choices are either due to the so-far limited impact 
these data would provide, or due to the lack of a calculation of the NNLO cross section and/or uncertainty in the 
existing calculations. However, these types of data  
are found to be consistent with MSHT20 predictions, and may add important constraints in future fits. 
We also avoid fitting LHC data at 13~TeV. This is partially because there is still a limited amount of data
of the form which would constrain PDFs at a similar level of precision to the 7 and 8 TeV data. However, it 
is also a conscious decision to avoid the highest energy LHC data in order for the PDFs to be free of influence from 
these data, should the PDFs be used to provide evidence for any discrepancies in the Standard Model seen in 
future analyses.    

The new MSHT20 PDFs only significantly differ from the MMHT14 PDF sets in the details of flavour
decomposition. This is seen for $u_V$, mainly at small $x$ and particularly $d_V$ for more general
$x$. It is also seen for the $\bar u, \bar d$ difference. The main new data probing valence quarks 
and light flavour asymmetry are 
the $W$ rapidity measurements at the LHC and Tevatron. The MMHT14 partons indeed give a poor 
description of some of these data, particularly the extremely precise recent LHC data. A much better fit
is obtained with MSHT20 PDFs, partially facilitated by the new parameterisation, particularly in the case of 
$\bar d/\bar u$.  The only other significant change is in the total strange quark distribution, with a significant
increase in magnitude, of the order of the size of the MMHT14 uncertainty. This is driven very largely by the
precise $W,Z$ data at the LHC, and particularly the relative difference in $Z$ and $W$ production. 
The overall light quark distributions, and most particularly the gluon distribution are not changed
significantly from the MMHT14 PDFs, or even from MSTW08 \cite{MSTW}. 
Hence, one is unlikely to obtain a dramatic change in the prediction 
for the many processes of phenomenological relevance when using MSHT20 PDFs instead of MMHT14 PDFs, though the uncertainty does decrease, sometimes quite significantly. We demonstrate this
for a variety of benchmark processes. In summary, we most definitely recommend the use of MSHT20 PDFs for the 
most reliable prediction for the central values and for both the best and, in most cases, reduced uncertainties.    

It is becoming increasingly necessary for LHC studies  for us to have PDFs determined as 
precisely as possible. This means examining not only individual PDF sets and their uncertainties, but also 
the comparison between sets obtained from similar data and using similar frameworks. Hence, we compare 
to other recent global PDF sets, in particular NNPDF3.1 \cite{NNPDF3.1} and CT18A \cite{CT18}, which are most directly 
comparable
to MSHT20, but also to the ABMP16 set \cite{ABMP16} and the final HERAPDF2.0 set \cite{HERAcomb}. We note that while MSHT20, CT18A and NNPDF3.1 in general agree
well with each other, and uncertainties have generally decreased in each compared to the previous PDF releases, 
in some respects there is more relative difference in the central values than there were for the MMHT14, CT14 and NNPDF3.0 sets. 
This is particularly the case for the gluon distribution, where NNPDF3.1 now has a rather different shape to MSHT20
and CT18A, being significantly softer at high-$x$, and also having the smallest uncertainty despite in general fitting less data than MSHT20. Some other PDF comparisons 
show some notable differences, both in central values and uncertainties. Some of these differences are due to the choice of data fitted or the treatment of these, but others are undoubtedly due to methodology, e.g parameterisation choices, or other
inherent assumptions in the fitting procedure. For example NNPDF3.1 now chooses to fit the charm by default, and
this is responsible for some systematic difference compared to MSHT20 and CT18A. It will be interesting to 
continue to investigate the comparison between the PDFs, and try to understand the reasons for differences.   

In this article we have made PDFs available at NNLO, which has become the standard, but also NLO and even LO, should these be 
useful. However we note that the most recent precision data from the LHC has finally made the inherent superiority of 
the NNLO analysis very clear, with the NLO fit struggling badly with the detailed description of some $W,Z$ data 
and the LO fit now being unable to give a truly quantitatively satisfactory description of the most precise  data.  
As well as the NLO and NNLO PDFs at the default $\alpha_S(M_Z^2)=0.118$ value we also make PDFs 
available at nearby values, in order that the PDF + $\alpha_S(M_Z^2)$ uncertainty can be calculated by treating the
PDFs defined at the upper and lower $\alpha_S(M_Z^2)$ values as additional eigenvectors \cite{asuncertainty}.     
In subsequent studies we will return to some additional issues, such as the dependence of the fit on the value of 
$\alpha_S$ and on the heavy quark ($c,b$) masses, and make PDFs available for a wide variety of values in each case. 
We will also provide a PDF set analogous to MSHT20 but including QED corrections, in particular with a photon PDF, using 
the same approach as in \cite{MMHTQED}.  We will in addition look into the theoretical uncertainties associated with the 
PDFs, noting that this is a further source of uncertainty to that presented here for PDFs defined 
at a fixed perturbative order. 

However, we conclude by noting that the MSHT20 PDFs presented in this article are obtained from 
the most up-to-date set of data currently available and using the most complete and sophisticated global analysis we have
so far performed. Hence, we present these sets as our best possible PDFs for analysing data from any experiment where 
hadrons are at least one of the colliding particles, or for making predictions for all such processes.  
  
\section*{Acknowledgements}

We would like to thank Mandy Cooper-Sarkar, Stefano Forte, Sasha Glazov, Claire Gwenlan, Jun Gao, Joey Huston, Pavel Nadolsky, Juan Rojo, Maria Ubiali and the wider PDF4LHC working group for many discussions on PDFs and related issues. We would also like to thank, in no particular order: Maria Ubiali for providing EW corrections for ATLAS $Z$ boson $p_\perp$ distribution; Claire Gwenlan, and Mandy Cooper-Sarkar and Jan Kretzschmar for helpful information related to various ATLAS data sets; Engin Eren and Katerina Lipka for help with CMS data; Chris Hays for discussions on Tevatron data and the D{\O} $W$ asymmetry in particular; Matthias Komm for clarification on the details of the CMS single top analysis; Jun Gao for providing theoretical calculations for single top and dimuon production; Sasha Glazov and Simone Amoroso for providing grids for and significant discussion relating to the ATLAS 8 TeV $Z$ production data; Alexander Huss and Joao Pires for providing \texttt{NNLOJET} predictions and helpful discussion on their implementation; Valerie Lang and Gavin Pownall for help relating to the ATLAS $W$ + jets data; Pavel Starovoitov and in particular Mark Sutton for help relating to \texttt{APPLgrid}; Bogdan Malaescu  for clarification and discussion relating to the ATLAS jet data; Francesco Giuli for  helpful information related to various ATLAS data sets and in particular the ATLAS 8 TeV $W$ data set; Emanuele Nocera for useful discussion relating to the treatment of top quark pair production data.

T. C. and R. S. T. thank the Science and Technology Facilities Council (STFC) for support via grant awards ST/P000274/1 and ST/T000856/1. S. B. acknowledges financial support from STFC. L. H. L. thanks STFC for support via grant award ST/L000377/1.

\newpage

\bibliography{references.bib}

\bibliographystyle{h-physrev}

\end{document}